\documentclass[a4paper,11pt]{article}
\pdfoutput=1

\usepackage{jheppub}
\usepackage{amsmath,amssymb}
\usepackage{graphicx}
\usepackage{subcaption}
\usepackage[percent]{overpic}
\usepackage[dvipsnames]{xcolor}
\usepackage{multirow}
\usepackage{mathtools}
\usepackage{tikz-cd}

\def\be{\begin{equation}}
\def\ee{\end{equation}}

\newcommand{\CA}{\mathcal{A}}
\newcommand{\CS}{\mathcal{S}}

\newcommand{\CE}{\mathcal{E}}

\newcommand{\CN}{\mathcal{N}}
\newcommand{\CL}{\mathcal{L}}
\newcommand{\CK}{\mathcal{K}}

\newcommand{\CW}{\mathcal{W}}
\newcommand{\tCW}{{\widetilde{\mathcal{W}}}}
\newcommand{\tSigma}{{\widetilde{\Sigma}}}
\newcommand{\tpi}{{\widetilde{\pi}}}
\newcommand{\bTheta}{{\overline{\Theta}}}

\newcommand{\IC}{\mathbb{C}}

\newcommand{\IZ}{\mathbb{Z}}
\newcommand{\IN}{\mathbb{N}}

\newcommand{\IR}{\mathbb{R}}
\newcommand{\fQ}{\mathfrak{Q}}

\newcommand{\fm}{\mathfrak{m}}
\newcommand{\fH}{\mathfrak{H}}
\newcommand{\fq}{\mathfrak{q}}

\newcommand{\eff}{\mathrm{eff}}

\newcommand{\Li}{\mathrm{Li}}
\newcommand{\rel}{\mathrm{rel}}

\renewcommand{\(}{\left(}
\renewcommand{\)}{\right)}

\title{Exploring 5d BPS Spectra with Exponential Networks}

\author[a]{Sibasish Banerjee}
\author[b]{Pietro Longhi}
\author[c]{Mauricio Romo}

\affiliation[a] {Vivatsgasse 7, Max Planck Institute for Mathematics, Bonn - 53111, Germany,  \\
  Appellstrasse 2, Riemann center of Geometry and Physics, Leibniz University, Hannover - 30167, Germany}

\affiliation[b]{
Institute for Theoretical Physics, ETH Zurich, 8093, Zurich, Switzerland\\
Department of Physics and Astronomy, Uppsala University, Box 516, 75120, Uppsala, Sweden
 }

\affiliation[c]{School of Natural Sciences, Institute for Advanced Study, Princeton, NJ 08540, USA \\
Yau Mathematical Sciences Center, Tsinghua University, Beijing, 100084, China}

\emailAdd{sbanerje@mpim-bonn.mpg.de, longhip@phys.ethz.ch, mromoj@ias.edu}

\abstract{
We develop geometric techniques for counting BPS states in five-dimensional gauge theories engineered by M theory on a toric Calabi-Yau threefold.
The problem is approached by studying framed 3d-5d wall-crossing in presence of a single M5 brane wrapping a special Lagrangian submanifold $L$.
The spectrum of 3d-5d BPS states is encoded by the geometry of the manifold of vacua of the 3d-5d system, which further coincides with the mirror curve describing moduli of the Lagrangian brane.
The information about the BPS spectrum is extracted from the geometry of the mirror curve by construction of a nonabelianization map for the exponential networks.
For the simplest Calabi-Yau, $\IC^3$ we reproduce the count of 5d BPS states encoded by the Mac Mahon function in the context of topological strings, and match predictions of 3d $tt^*$ geometry for the count of 3d-5d BPS states.
We comment on applications of our construction to the study of enumerative invariants of toric Calabi-Yau threefolds.

}

\begin{document}

\maketitle
\flushbottom

\section{Introduction}

In this paper we develop a geometric approach to count BPS states in five-dimensional gauge theories with eight supercharges compactified on a circle or finite radius.
This problem is directly related to BPS counting in Calabi-Yau compactifications of M theory, and therefore is of relevance to questions in enumerative geometry \cite{Gopakumar:1998ii,Gopakumar:1998ki, Dijkgraaf:2006um, 2003math.....12059M, 2004math......6092M}.
Progress on Wall-Crossing over the past decade has yielded new insights into these questions, and led to developments in the study of BPS spectra both in the context of supersymmetric gauge theories and in string theory  \cite{Kontsevich:2008fj, Joyce:2008pc, Gaiotto:2008cd, Gaiotto:2009hg, Denef:2007vg, Alim:2011ae, Alim:2011kw, Manschot:2010qz, Alexandrov:2008gh, Alexandrov:2011ac, Manschot:2012rx, Pioline:2011gf, Alexandrov:2018iao}.

The present work focuses on the BPS spectra of five-dimensional gauge theories engineered by M theory on a toric Calabi-Yau threefold $X$. The goal is to develop a systematic framework for studying both BPS instanton-dyons and magnetic monopole strings.
A fruitful approach to probe BPS spectra in quantum field theory and in string theory is to introduce supersymmetric defects of various types.
For this purpose we consider codimension-two defects engineered by an M5 brane on $L\times S^1\times \IR^2$, where $L$ is a special lagrangian submanifold of $X$.
The M5 brane defect can be described at low energies by a 3d $\CN=2$ theory on $S^1\times \IR^2$, whose field content and couplings are determined by the geometry of $L$. This theory, which will be denoted  $T[L]$, further couples to the bulk five-dimensional gauge theory in a way that is encoded by the embedding of $L$ in $X$.
Defects of this type provide a familiar way to embed topological strings into M theory, and establish a relation between vortex partition functions and open topological string amplitudes \cite{Dimofte:2010tz}.
In this paper we study another kind of BPS states of $T[L]$, corresponding to field configurations that reduce to BPS kinks (instead of vortices) when the circle shrinks.

The coupling of $T[L]$ to the ambient five-dimensional theory includes interactions between 3d and 5d BPS spectra, and their boundstates give rise to a hybrid  {3d-5d BPS sector}.
This new sector is rich in information, being sensitive both to the stability of 3d BPS states and of 5d BPS states, as well as to additional stability conditions specific to 3d-5d boundstates. In other words, 3d-5d BPS states contain both information about the 3d BPS spectrum and about the 5d BPS spectrum.
The first main result of this work is to develop a systematic framework for studying 3d-5d BPS spectra.
With this information is under control, we can further investigate how to extract BPS states of the pure five-dimensional theory, this is our second main result.
The construction behind both results is inspired by seminal ideas of \cite{Gaiotto:2011tf},
which involve studying entire families of configurations for the defect engineered by $L$.
From the viewpoint of the 3d-5d system this corresponds to varying certain couplings of the 3d theory.
From a geometric viewpoint it means instead studying the geometry on the moduli space of $L$. The deformation moduli of $L$ is 1 real dimensional if $b_{1}(L)=1$ \cite{Mclean96deformationsof}. 
In this work we will limit ourselves to single M5 branes wrapping a lagrangian $L$ with topology $\mathbb{R}^{2}\times S^{1}$ hence, the space of deformations plus the flat $U(1)$-bundle moduli can be described in terms of a curve.

For simplicity we take $L$ to be a {toric brane} of the type studied in \cite{Aganagic:2000gs, Aganagic:2001nx}, although our construction applies directly to a much larger class of Lagrangian branes. In this case, the moduli space of $L$ coincides with the mirror curve $\Sigma$ of $X$ (more precisely the mirror of $X$ is a conic bundle over $\Sigma$ \cite{Hori:2000kt}).
This choice has the advantage that $T[L]$ can be presented as a $U(1)$ gauge theory, with a finite number of charged chiral multiplets coupled to the bulk gauge and background connections.
On the Coulomb branch of the five-dimensional theory the 5d vectormultiplets' adjoint scalars acquire vacuum expectation values.
Since they couple to 3d chiral multiplets as twisted masses, the latter can be integrated out to yield an effective description
of the 3d-5d system \cite{Gaiotto:2013sma, Ashok:2017lko, Ashok:2017bld}.
Viewing the 3d theory as a 2d $\CN=(2,2)$ theory of KK modes, the low-energy dynamics is described by an effective theory for the twisted chiral field strength multiplet, controlled by a superpotential $\tCW_\eff(\sigma, t, u)$.
Here $\sigma$ is the 2d scalar fieldstrength, $t$ is a complex Fayet-Ilioupoulos coupling, and $u$ collectively denotes moduli of the 5d theory (both Coulomb moduli and masses).
For a generic choice of moduli $t$ and $u$, the twisted superpotential has a discrete set of massive vacua, and we study the spectrum of kink-like BPS states that interpolate between any two of them.
The critical points of $\tCW_\eff$ trace out an algebraic curve $\Sigma\subset \IC^*\times\IC^*$ with coordinates $(e^{2\pi R\,\sigma}, e^{2\pi R \, t})$, which turns out to be a Seiberg-Witten curve for the 5d theory, and further coincides with the mirror curve of $X$.

BPS kinks of 2d $\CN=(2,2)$ theories are counted by the CFIV index $Q$ which admits an infra-red expansion as a sum  with integer coefficients $\mu(a)$ that count BPS states of charge $a$ \cite{Cecotti:1992qh, Cecotti:1992rm}.
An effective way to compute $Q$ is to study the $tt^*$ connection on the vacuum bundle over the $t$ plane \cite{Cecotti:1991me}.
Spectral networks provide another powerful approach to computing the BPS degeneracies $\mu(a)$, essentially taking advantage of their interpretation as (signed sums of) intersection numbers of Lefschetz thimbles of the BPS equations \cite{Cecotti:1992rm, Gaiotto:2011tf, Gaiotto:2012rg}.
A bit more precisely, spectral networks provide a construction of a flat non-abelian connection $\CA$ on the vacuum bundle over the $t$-plane, that arises in the Lax representation of the $tt^*$ equations. This construction is known as the \emph{nonabelianization map}.

Returning to three dimensions, it is natural to expect the existence of an analogous construction, at least when the 3d theory is compactified on a circle, when it can be viewed as the 2d theory of the KK modes.
Indeed this viewpoint was adopted in \cite{Cecotti:2013mba} to formulate the $tt^*$ equations in three dimensions, making contact with work on periodic monopoles and hyperholomorphic connections.
We construct a nonabelianization map for 3d-5d systems, based on an uplift of spectral networks known as \emph{exponential networks} \cite{Eager:2016yxd}.
The counting of 3d-5d BPS states is a byproduct of our construction, since $\mu(a)$ are explicitly encoded by this map.

\medskip

\begin{figure}[h!]
\begin{center}
\includegraphics[width=0.3\textwidth]{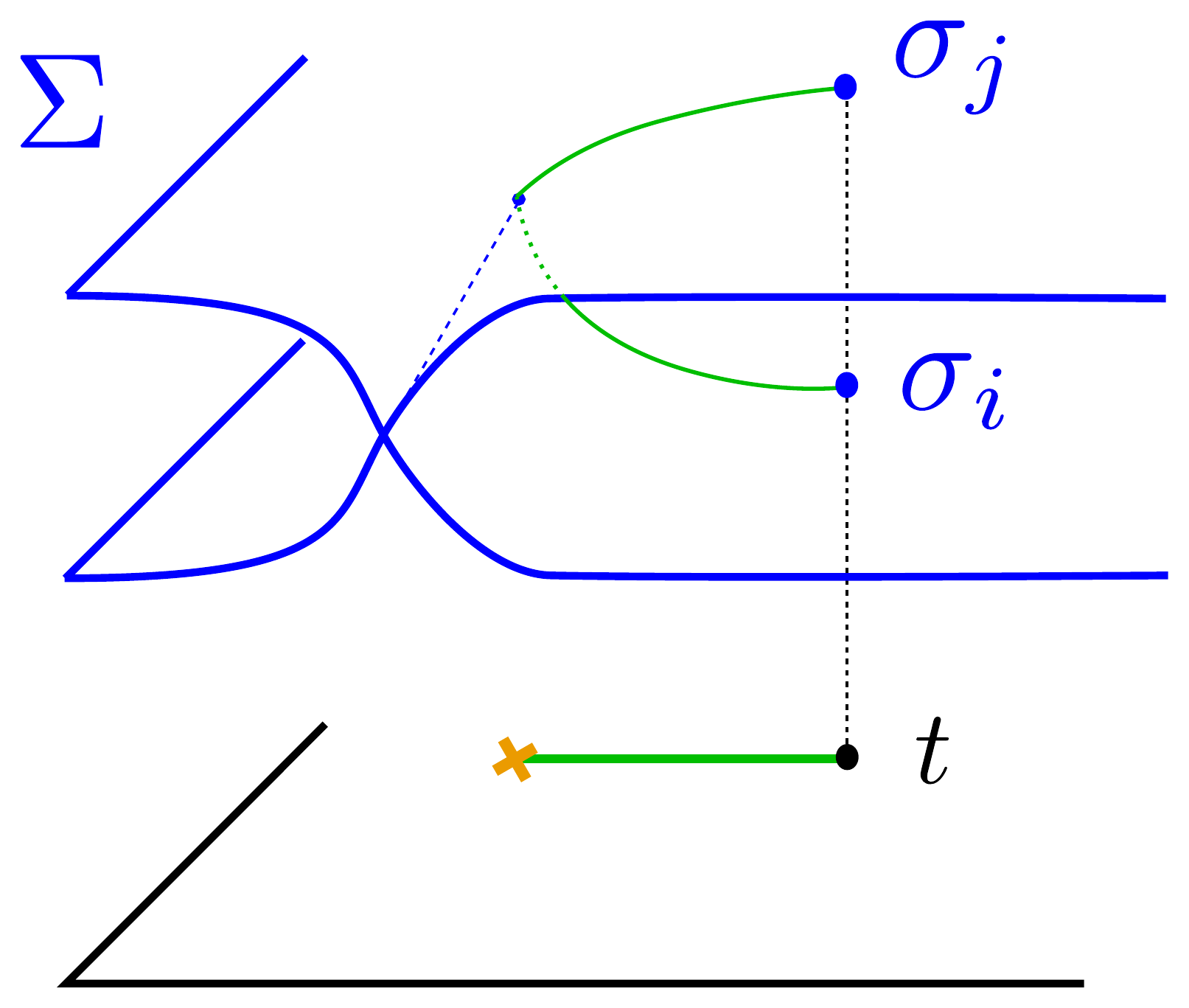}
\caption{An open path on $\Sigma$ connecting two vacua $\sigma_i(t), \sigma_j(t)$. Calibrated paths correspond to BPS states $T[L]$. Exponential networks arise from projecting BPS paths to the $t$-plane.}
\label{fig:kink}
\end{center}
\end{figure}

Exponential networks arise by projecting certain special-Lagrangian cycles of the mirror Calabi-Yau $Y$ onto the mirror curve $\Sigma$.
As already mentioned, $\Sigma$ also coincides with the manifold of vacua of $T[L]$, as well as with a Seiberg-Witten curve of the ambient 5d theory \cite{Gaiotto:2009fs, Gaiotto:2013sma, Ashok:2017bld}.
BPS solitons are static field configurations interpolating between two vacua, a standard argument shows that they correspond to calibrated open paths on $\Sigma$ connecting the corresponding points on the curve,  see Figure \ref{fig:kink}.
Projecting these BPS paths to the $t$-plane gives trajectories that correspond to the geometric data of the exponential network.

An important difference from spectral networks arising in 2d-4d systems, is that in the context of 3d $\CN=2$ theories $\sigma$ is periodic, due to invariance under large gauge transformations on the compactification circle.
In part because  of this, the twisted superpotential turns out to be multi-valued on the vacuum manifold $\Sigma$, and it is often convenient to introduce a $\IZ$-covering $\tSigma$ with logarithmic branching at points where $\sigma(t)$ goes to zero or infinity. Since $\Sigma$ was presented as a finite ramified covering of the $t$-plane, $\tSigma$ is an infinite-sheeted covering of the latter.
At each $t$ there are towers of vacua labelled by $|i,N\rangle$, where $i$ is one of a finite set of critical points of $\tCW_\eff$ (a point on $\Sigma$), and $N \in \IZ$ denotes a choice of branch for the logarithm (fixing a point on $\tSigma$).
One can trade infinite towers of vacua for an extra continuous modulus by passing to a ``Bloch basis'', this would lead to periodic monopole equations of 3d $tt^*$ geometry  \cite{Cecotti:2010fi, Cecotti:2013mba}.
However it is more convenient for us to  work with infinite towers of vacua. Our nonbalienization map then constructs a formal flat connection for this infinite-dimensional vacuum bundle.
Its construction involves counting calibrated open paths on $\tSigma$ running between two vacua $|i,N\rangle$ and $|j,M\rangle$. 
Physically these paths are interpreted as charges of 3d-5d BPS states.
Therefore we collectively refer to this set of open paths, counted by appropriate BPS indices $\mu(a)$, as the \emph{soliton data} of the exponential network.
For a given pair $(i,j)$, physical properties of 3d-5d BPS states (such as their central charges) are sensitive only to the difference $N-M$.
This property is indeed reflected by a $\IZ$-symmetry of the soliton data that we compute, providing a nontrivial check of our construction.

Another salient novelty that arises in three dimensions is the presence of \emph{collinear solitons}, BPS states interpolating between vacua $|i,N\rangle$ and $|i,N+n\rangle$ (in short, $(ii,n)$-solitons).
These BPS states are collinear in the sense that their central charges  have the same phase, and their existence leads to multi-particle contributions to the nonabelianization map with \emph{fractional} coefficients.
Similar phenomena have been observed in the general context of periodic $tt^*$ geometry, in the form of multi-particle contributions to the CFIV index \cite{Cecotti:2010qn}.
We reproduce and extend these results by studying wall-crossing for solitons of types $(ij,N,N+n)$ and $(ji,M,M+m)$.
The nonabelianization map provides a powerful tool to compute the spectrum of boundstates, for which we obtain a closed-form expression.
We find BPS states with infinitely many different charges, including collinear solitons and their multi-particle boundstates.\footnote{
Details on this result can be found in subsection \ref{sec:ij-ji-joints}.
}

\medskip

The second main goal of this paper,  five-dimensional BPS counting, is accomplished by studying jumps of the nonabelianization map when the topology of the exponential network degenerates.
These jumps arise as a natural uplift of the \emph{$\CK$-wall} phenomenon of 2d-4d wall-crossing \cite{Gaiotto:2011tf, Gaiotto:2012rg}
and capture boundstates of 3d-5d BPS states that carry purely five-dimensional charges.
Our study of these jumps focuses on their influence on the soliton data of nonabelianization. We leave a field-theoretic description of this phenomenon to future work.
We argue that the nonabelianization map changes by certain morphisms of the Kontsevich-Soibelman type, in line with overall expectations from analogies with spectral networks.
This should also be expected on physical grounds, since our nonabelianization map is built to capture the KK modes of 3d-5d BPS states, whose physics is essentially that of 2d-4d states captured by spectral networks.
In fact, more precisely, our $\CK$-wall formula is expected to compute indices of 4d BPS states corresponding to  KK modes of 5d BPS states.
We hasten to stress that this \emph{does not} imply that the 5d spectrum should be a trivial Kaluza-Klein uplift of a 4d BPS spectrum.
It is in principle possible that, in certain regions of the moduli space, only part of a KK tower of 5d BPS states decays. In fact, the central charges of different KK modes need not have the same phase for finite compactification radius, and in principle may experience wall-crossing separately from each other.
We expect this phenomenon to occur generically in models where $X$ has at least one two-cycle.
Similar questions on the relation between 4d and 5d BPS states were recently raised in \cite{Kachru:2018nck}, we believe that the nonabelianization map provides a tool to address these questions in detail.

As an illustration of our construction, we consider exponential networks for the mirror curve of the toric brane in $\IC^3$.
We plot the exponential network at various phases, and direct application of our nonabelianization formulae provide explicit counts for the 3d-5d BPS states captured by the soliton data.
Already in this simple example we observe several interesting phenomena, including the occurrence of wall-crossing of Cecotti-Vafa type as well as wall-crossing of the $ij/ji$-type described above, with an infinite number of boundstates, including collinear solitons and multi-particle states.
We check that the 3d-5d BPS spectrum computed by the nonabelianization map is compatible with  predictions about the CFIV index from 3d $tt^*$ geometry.
The exponential network for $\IC^3$ exhibits a single $\CK$-wall jump (together with its conjugate) caused by a tower of bulk BPS particles with charges $n\gamma$.
This is in line with physical expectations, since the five-dimensional theory should only contain BPS states arising from boundstates of D0 branes.
We find  a particle of charge $n\gamma$ for each $n\geq 1$, with  index $\Omega(n\gamma)=-1$.
The BPS central charge is  $Z_{n\gamma} = 2\pi n/ R$ confirming the physical expectation that these particles are D0 boundstates, or equivalently the Kaluza-Klein tower of the massless modes of M theory on $\IC^3$ \cite{Gopakumar:1998ii}.
%

\subsection{A mathematically oriented summary}

The results of this paper may be of interest to mathematicians, as they provide concrete predictions and computational tools for enumerative geometry on Calabi-Yau threefolds.
However some of the language used may obscure key ideas and claims, here we attempt to bridge this gap with a self-contained discussion.

The main subject of this paper is the development of a framework to study the connection between certain problems in enumerative geometry.
On one side, we wish to count generalized Donaldson-Thomas invariants \cite{Joyce:2008pc}  of toric Calabi-Yau threefolds, which arise in physics as BPS states of five-dimensional gauge theories engineered by M theory on a Calabi-Yau threefold $X$. In fact, in appropriate regions of the complexified Kahler moduli space, these BPS states may be identified with Donaldson-Thomas or with Gromov-Witten invariants \cite{Aganagic:2009kf}. 
A standard physics description of generalized DT involves D6-D4-D2-D0 boundstates in type IIA string theory on $X$. The relation to BPS counting in M theory was explained in \cite{Dijkgraaf:2006um} (see also \cite{Gaiotto:2005gf, Jafferis:2008uf, Aganagic:2009kf, 2003math.....12059M, 2004math......6092M, 2008arXiv0809.3976M, Morrison:2011rz}). 
On the other side, we will repackage these enumerative invariants as counting special Lagrangians in the mirror Calabi-Yau $Y$, a problem that arises naturally once we consider the insertion of defects in the 5d gauge theory, engineered by an M5 brane wrapped on a Lagrangian submanifold $L\subset X$. 
Therefore we approach the original counting problem by embedding it into another one: we enrich the geometry by including certain Lagrangian defects in $X$, and count BPS states in their presence. The BPS spectrum of this system includes the previously mentioned generalized DT invariants, as well as new sectors related to the presence of the defect.
To our understanding, there is currently no mathematical interpretation of these new BPS states in the language of enumerative geometry. 
The main motivation for considering defects is that it is easier to compute the defect BPS states, which moreover encode information about the previously mentioned generalized DT invariants.

Let us proceed to use the physical intuition of brane constructions to describe these results, since we can draw in parallel a mathematical interpretation. Consider the following system of D-branes in type IIA string theory on $\mathbb{R}^{3}\times S^{1} \times X$:
\begin{center}
\begin{tabular}{c|c|c|c}
  & $\mathbb{R}^{3}$ & $S^{1}$ & $X$  \\
  \hline
  $D6$: & pt. & $S^{1}$ & $X$ \\
  $D2$: & pt. & $S^{1}$ & $\mathcal{C}_{2}$ \\
  $kD0$: & pt. & $S^{1}$ & $x$ \\
\end{tabular}
\end{center}
where $x\in X$, $\mathcal{C}_{2}$ is a 2-cycle in $X$.
Counting BPS states on the 4d theory on $\mathbb{R}^{3}\times S^{1}$ corresponds to computations of DT invariants with a single $D6$, $D2$-brane charge given by $[\mathcal{C}_{2}]$ and $k$ units of $D0$-brane charge \cite{Iqbal:2003ds}. 
Lifting this configuration to M theory, the D6-brane becomes pure geometry, that is Taub-Nut space $TN$ which is topologically a circle $S^{1}_{TN}$ (the M theory circle) fibered over $\mathbb{R}^{3}$, and the D2 and D0-branes becomes $M2$ branes wrapping $\mathcal{C}_{2}$ with $k$ units of momentum along $S^{1}$. 
The $M2$'s are localized near the origin of $TN$. If we take the radius of $S^{1}_{TN}$ to infinity, then $TN\sim \mathbb{R}^{4}$ and the M2 branes form a diluted gas \cite{Dijkgraaf:2006um}. 
The large $S^{1}_{TN}$ radius limit gives a configuration equivalent to the Gopakumar-Vafa counting of BPS states in 5d \cite{Gopakumar:1998ii, Gopakumar:1998jq}, hence the BPS states on the 5d theory on $\mathbb{R}^{4}\times S^{1}$ are M2 branes carrying $k$ units of momentum, this is equivalent to GV invariants of type $(k,[\mathcal{C}_{2}])$. 
Then, to summarize, after taking the radius of $S^{1}_{TN}$ to infinity, we are left with M theory on $X\times \IR^4\times S^1$ and we count M2 branes wrapping ${\cal C}_2 \times S^1$. 
Let us mention that our framework also captures BPS states including D4 branes, which in M theory would lift to M5 branes wrapping a four-cycle of $X$, however we omit this from the present discussion for simplicity.

As previously mentioned, a novelty of our approach consists of employing certain defects to solve this counting problem.
In the M theory setting we consider an M5 brane wrapping a Lagrangian submanifold $L$ of $X$ and extending along $S^1\times \IR^2$ introducing several new sectors for the BPS spectrum.
A well-studied class of such BPS states are M2 branes wrapping a curve ${\cal C}_2$ and wrapping the circle $S^1$, with $\partial {\cal C}_2$ ending on $L$ along a non-trivial one-cycle. The counting of these BPS states is related to open Gromov-Witten invariants \cite{Aganagic:2000gs, Aganagic:2001nx, Ooguri:1999bv}. 
From the viewpoint of the 3d $\CN=2$ theory arising on the M5 brane on $S^1\times \IR^2$ these M2 branes give rise to  BPS vortices located at a point in $\IR^2$ and extending along $S^1$ \cite{Dimofte:2010tz}.
There is another, less studied class of BPS states, on which we will focus.
When the circle $S^1$ shrinks to zero radius, M5 descends to a D4 brane wrapping $L\times \IR^2$ and these BPS states correspond to kinks on the 2d theory supported on $\IR^2$. 
Kinks are field configurations that do not depend on the time coordinate $\IR_t\subset \IR^2$, but evolve along the spatial direction $\IR_x\subset \IR^2$, and such that that fields at $x\to\pm\infty$ approach certain vacuum configurations. 
Restoring the circle at finite radius, these BPS states will be located at a point along the circle, therefore locally they may look like vortices on a small patch of the cylinder $S^1\times \IR_x$, but globally fields must still approach constant (vacuum) configurations along $S^1\times \IR_t$ when $x\to \pm\infty$. Therefore these BPS states have some features of 3d vortices at short distance as well as some features of 2d kinks at long distance, we will refer to them as ``kinky vortices''. 
From the viewpoint of the 3d theory on $S^1\times \IR^2 $, these BPS states have not been studied to  our knowledge.

To illustrate the geometric meaning of kinky vortices, note that unlike  standard vortices, they are field configurations on $S^1\times\IR^2$ that interpolate \emph{two distinct vacua} at infinity. 
Each vacuum corresponds to a BPS configuration for the M5 brane on $L\times S^1\times \IR^2$. Therefore, each vacuum is a point in the moduli space of A-branes. 
The moduli space encodes both geometric moduli for the Lagrangian submanifold $L$, as well as the moduli for a $U(1)$ line bundle $\CL$ on $L$. 
Since we are considering toric lagrangians which have topology $\mathbb{C}\times S^{1}$, then, this moduli space is one complex dimensional and parametrized by the mirror curve $\Sigma\subset \mathbb{C}_x^{*}\times \mathbb{C}_y^{*}$ of $X$. 
Let  $p_{1},p_{2}\in {\Sigma}$ be two points corresponding to branes $(L_{},\CL_1),(L_{},\CL_2)$ with the same underlying Lagriangian $L$ but with different bundles. 
A kinky vortex corresponds to a configuration of $A$-branes fibered over $S^1\times \IR^2$ such that it approaches  $(L,\CL_1)$ along  $S^1\times \IR_t$ for $x\to -\infty$ and  $(L,\CL_2)$ along  $S^1\times \IR_t$ for $x\to +\infty$.
Note that the underlying Lagrangian $L$ is fixed, as in related counts of open strings with a single boundary ending on $L$ \cite{Aganagic:2001nx, Walcher:2006rs}.

Through physical arguments, we set up a combinatorial construction of the generalized DT invariants based entirely on the geometry of the mirror curve $\Sigma$.
Motivated by the physical interpretation of the $x$ and $y$ variables on $\IC^*_x\times\IC^*_y$, we view $\Sigma$ as a ramified covering of the $x$-plane.
We then consider networks on the $x$-plane arising from projecting special Lagrangian cycles $S$ of $Y$ down on $\Sigma$, and further down on $\IC^*_x$.
By a such a cycle $S$ we mean a Lagrangian submanifold of $Y$ along which the holomorphic three-form has constant phase $\Omega|_{S}\sim e^{i \vartheta} |\Omega|$.
We focus on cycles having the form of a $S^2$ fibration  over a one-cycle $\gamma$ on $\Sigma$.
For these the special-Lagrangian condition can be reformulated as the requirement that $\log y \, d\log x$ has a constant phase along $\gamma$. 
The network $\CW$ is made of trajectories subject to this constraint, projected to the $x$-plane and with boundary conditions that trajectories must be generated either at branch points or at intersections of other trajectories.
The geometric definition of these networks was originally introduced in \cite{Klemm:1996bj}. 
This construction was further developed in \cite{Eager:2016yxd} where it was shown that D2-D0 states in the resolved conifold and D4-D2-D0 states in local $\mathbb{P}^2$ are captured by the geometry. We take this as a proof of principle for the validity of this geometric approach.

The appropriate count of special Lagrangian cycles is the main new result of this paper, and it is determined by a combinatorial problem formulated in terms of the topology of $\CW$.
In a nutshell, one takes a flat line bundle on $\Sigma$ and considers its trivial pushforward in each (simply) connected component of $\IC^*_x \setminus \CW$, then one glues back together different patches using certain combinatorial data on the trajectories of $\CW$.
The combinatorial data of each trajectory consists of relative homology classes of open paths on $\Sigma$, each path connects two points in a canonical lift of the trajectory to the mirror curve. 
This data is uniquely fixed by demanding that the parallel transport along a path $\wp$ from one patch to another of $\IC^*_x\setminus \CW$ only depends on the homotopy class of $\wp$.
This construction of higher-rank flat bundles from the data of a covering flat line bundle was originally introduced in \cite{Gaiotto:2012rg} in the context of Spectral Networks associated to Hitchin systems. The construction is also deeply connected to the subject of $tt^*$ geometry \cite{Cecotti:1991me, Dubrovin:1992yd}.
One crucial difference with our case is that $\Sigma$ is not a spectral curve of a Hitchin system.
Our construction is motivated from a different perspective, brought about by the Lagrangian defect engineered by the M5 brane, which leads naturally to a three-dimensional version of the $tt^*$ equations considered in \cite{Cecotti:2013mba}.
We derive the flatness equations that determine the combinatorial data, and therefore establish a formal \emph{nonabelianization map} that takes a flat abelian connection on $\Sigma$ and produces a flat nonabelian connection on $\IC^*_x$. 
The physical interpretation of the combinatorial data derives from identifying the open paths $\Sigma$ running from $p_1$ to $p_2$ with charges of kinky vortices interpolating between bundles $\CL_1, \CL_2$ on $L$.
This identification, together with the fact that kinky vortices reduce to BPS kinks when $S^1$ shrinks, strongly suggests that the nonabelian connection we construct should be closely related to  the $tt^{*}$ connection computed in a massive Landau-Ginzburg model.
The model should be characterized by a superpotential $W$ such that $\exp (\partial W/\partial\log y)=1$ coincides with the algebraic equation for the mirror curve $F(x,y)=0$ and $x$ parametrizes $W$. 
Massive models are anomalous but still have a sensible $tt^{*}$ connection \cite{Cecotti:1992rm}. These anomalous cases of  $tt^{*}$ geometry have also been studied in the mathematical literature \cite{2009arXiv0906.1307I, 2011arXiv1107.1290F}. 

While the network's shape is sensitive to the geometry of $\Sigma$, the construction of the nonabelian connection is topological. In other words we may slightly vary the shape of $\CW$, but the combinatorial data remains essentially constant.
The geometry of $\CW$ also depends on the phase $\vartheta$ (in addition to the geometry of $\Sigma$), in fact varying this phase by finite amounts may produce jumps in the topology of $\CW$.
The jumps occur when $\vartheta$ is the phase of (the holomorphic volume of) a special Lagrangian cycle, and are identified with the 5d BPS states of $X$. 
It is the degeneracies of these states that correspond to the enumerative invariants that we set out to study.
The topological jump of $\CW$ induces a change in the combinatorial data on trajectories, and this takes the form of a ``Kontsevich-Soibelman morphism'' associated to ${\cal C}_2$.
Comparing the combinatorial data before and after the jump therefore provides information on the appropriate enumerative invariant $\Omega({\cal C}_2)$.
We provide explicit formulae for computing the combinatorial data that defines a nonabelianization map, as well as for the enumerative invariants of $X$ that describe its jumps.
The formulae rely entirely on the combinatorial data extracted from $\Sigma$ through the network $\CW$.
Finally, we illustrate the application of this framework in the example $X=\IC^3$,
 reproducing the count of zero-cycles that is encoded by the Mac Mahon function of topological string theory.

\subsection{Organization of the paper}

This paper is organized as follows. In the section \ref{sec:phys}, we start by setting up the physical perspectives. We briefly also recapitulate several facts from the existing literature.
In particular, we review $tt^*$ geometry in some detail and discuss the basics of the spectral and exponential networks 
In section \ref{sec:expnet}, we set up the construction of the nonabelianization map for exponential networks. We give a detailed algorithmic description about how to determine
various BPS indices in our context, namely the 3d and 5d BPS indices.
Then in section \ref{sec:C3} we realize our formalism explicitly and in particular solve the flatness equations explicitly to compute closed 5d BPS
states.
Section \ref{sec:discussion}  new questions and suggestions for future work.

\smallskip

\emph{Note added: while this paper was in the final stages of preparation, a relation between spectral networks and topological strings appeared in the interesting work \cite{cpt}.}

\section*{Acknowledgements}

We thank Sergei Alexandrov, Murad Alim, Pierrick Bousseau, Richard Eager, Jan Manschot, Joe Minahan,    Rahul Pandharipande, Boris Pioline and Johannes Walcher for discussions.
We also thank Sergei Alexandrov for related joint work (with two of the authors) which sparked some of the motivation for this project.
SB acknowledges the facilities of Max Planck Institute for Mathematics in Bonn where the majority of the work was completed and  Riemann Center for Geometry and Physics in Hannover where the rest was finished, especially the write-up.
SB also thanks  St.\ Petersburg state university, Higher School of Economics in Moscow and Uppsala University for providing excellent working conditions.
PL and MR thank the Aspen Center for Physics for hospitality during part of this work.
PL also thanks Aarhus QGM, ENS Paris, Caltech, The University of California at Berkeley, and Trinity College Dublin for hospitality during completion of this work.
MR acknowledges hospitality from Harvard University and Rutgers University.
The work of PL is supported by NCCR SwissMAP, funded by the Swiss National Science Foundation.
PL also acknowledges support from grants ``Geometry and Physics''and ``Exact Results in Gauge and String Theories'' from the Knut and Alice Wallenberg Foundation during part of this work.
MR gratefully acknowledges the support of the Institute for Advanced
Study, DOE grant DE-SC0009988 and the Adler Family Fund.

\section{Lagrangian defects and $tt^*$ geometry}\label{sec:phys}

This section contains background material that supports the physical interpretation and motivation of our main construction.
It does not contain new material, except for possibly new viewpoints on connections among several topics.
Readers interested in the construction of the nonabelianization map alone may safely skip to the next Section.

In the following we motivate the use of the exponential networks for computing the BPS indices for 3d-5d systems.
This goal prompted us to review a somewhat broad spectrum of topics, in order to provide the physical underpinnings of various objects that will be discussed later.
We have not tried to be exhaustive in details, but hopefully the following review is self-contained to a certain extent, at least on the conceptual level.

The starting point is the fruitful viewpoint of geometric engineering of quantum field theories and their BPS states.
We then move on to include co-dimension two defects in the picture, and touch on their properties as probes of BPS spectra and wall-crossing.
The next step is to introduce the geometric viewpoint on soliton spectra of defect theories, via the framework of $tt^*$ geometry.
In the context of 2d-4d systems this provides the physical interpretation for the construction of the nonabelianization map of spectral networks developed in \cite{Gaiotto:2012rg}. Likewise the 3d-5d version of $tt^*$ geometry will provide the physical interpretation for the nonabelianization map that we will construct in Section  \ref{sec:expnet} for exponential networks..
We then briefly review the framework of spectral- and exponential-networks with an emphasis on their ties to 2d-4d/3d-5d systems, their BPS spectra and their $tt^*$ geometries.

\subsection{Geometric approaches to BPS states from string theory} \label{sec:geom-approach}

The advantage of geometric approaches to BPS counting in 4d $\CN=2$ quantum field theories was advocated long ago by \cite{Klemm:1996bj}.
This philosophy was turned into a truly powerful framework more recently in \cite{Gaiotto:2012rg}, and in follow-ups \cite{Longhi:2016rjt,Galakhov:2014xba,Galakhov:2013oja,Hollands:2016kgm,Longhi:2016wtv,Gabella:2017hxq}.
The application of geometric techniques to BPS counting in 3d-5d systems was advocated more recently in \cite{Eager:2016yxd}.

The broad idea of geometric approaches relies on two pieces of data: a Riemann surface $\Sigma$ and a differential $\lambda$ on it. Roughly speaking, BPS states can be obtained by studying which homology cycles on the Riemann surface support BPS geodesics defined by the differential.
There are two different settings which naturally provide this data. One is that of class $\CS$ theories in which $\Sigma$ is the spectral curve of a Hitchin system, and $\lambda$ is determined by the Higgs field, this is the setting in which spectral networks were originally developed \cite{Gaiotto:2009hg,Gaiotto:2012rg}.
The other setting is that of Type IIB string theory on a ALE-fibered Calabi-Yau threefold, in this case $\Sigma$ arises from the monodromy properties of the fibration, and the differential on $\Sigma$ descends from the holomorphic three-form.
The two settings are in fact related to each other, see \cite{Klemm:1996bj}.

The type IIB setup will turn out to be more closely related to the approach of our paper, so we now review a few basic facts about BPS states in this setting.
In the Seiberg-Witten regime the Calabi-Yau threefold $Y$ can be taken to be a fibration by  ALE spaces (of ADE-type) over the complex $z$-plane, with singularities at isolated points.
BPS states of the 4d $\CN=2$ theory arise from D3 branes wrapping three cycles, which correspond roughly to two-cycles in the ALE space fibered over a one-cycle from the $z$-plane.
For illustration let us consider a Calabi-Yau of the following type,
\be
\prod_{i = 1}^n (x-a_i(z) ) + y^2 + w^2 = 0,
\ee
where each of the $a_i(z)$ can be multi-valued function, in principle.
The fiber over the $z$-plane is an $A_{n-1}$-type ALE space in this case, the two-cycles are associated to pairs $a_i, a_j$ which correspond to weights of $A_{n-1}$.
Varying $z$, the two-cycles in the ALE space vanish somewhere on the $z$-plane, whenever two weights $a_i = a_j$ collide.
Taking a one-parameter family of two-cycles fibered over two such points gives a compact three-cycle in the Calabi-Yau, whose topology may be $S^2\times S^1, S^3$, or a more complicated fibration.

The Riemann surface $\Sigma$ in this case is defined by the algebraic equation
\be\label{eq:string-SW-curve}
\Sigma :  \prod_{i = 1}^n (x-a_i(z) ) \, = \, 0\,.
\ee
This is an $n$-fold ramified covering of the $z$-plane with ramification points associated to roots of $A_{n-1}$.
Given a fixed $z$, the image of the two cycle of the ALE space is a zero cycle represented by the class
$[a_i(z)]-[a_j(z)]$. Then the image of the three cycle on the $z$-plane is a path on the $z$-plane, either open or closed.
Its lift on $\Sigma$ is a closed cycle, whose period $\oint \lambda $ can be shown to coincide with the period of the holomorphic three-form on the original three-cycle.
More precisely, there is a map from $H_3$ of the Calabi-Yau to the $H_1(\Sigma)$, and correspondingly from the holomorphic top form $\Omega$, one recovers
the canonical one form associated to $\Sigma$, denoted by $\lambda$.
Then one can view the three branes wrapped around the three cycles as respectively electric or magnetic
depending on what type of three cycles ($A$ or $B$) they wrap.
The BPS mass is then given by
\be
M = \left\vert \oint_{\gamma} \lambda \right\vert.
\ee
It is also quite clear how these objects are BPS, from the viewpoint of  string theory.
For example, a three brane partially wrapping $S^2$ of the ALE space gives rise to a string on the $z$-plane, whose tension
\be
	T = \left| a_i(x) - a_j(z) \right|
\ee
depends on the position on the $z$-plane, as the $S^2$ varies over it.\footnote{An explanation of this fact is most natural from the type IIA perspective which involves self-dual strings, see \cite{Klemm:1996bj}.}
The insight that BPS states correspond to strings which minimize the tension leads to a geometric approach to finding stable BPS states.
This consists of studying geodesics on the $z$-plane, either closed or stretching between two (or more) branch points.

Exponential networks, as defined in \cite{Eager:2016yxd}, arise naturally in this Type IIB setting.
The geometries of interest in this case are however slight generalizations of the ALE fibration considered above, and arise as mirrors of toric threefolds.
$\Sigma$ is replaced by a mirror curve, and $\lambda$ is now replaced by $\log x \, d z / z$.
On a geometric level, this leads to novelties in the types of BPS geodesics that can occur, as we will see in greater detail in later sections.

In the rest of this section we will take a different perspective, and motivate the study of exponential networks on mirror curves from the viewpoint of ``Lagrangian defects'' in 5d quantum field theory.
The advantage of the field-theoretic viewpoint will be to provide not only the geometric picture of BPS states, but also a proper setting to define their counting, which is the main subject of this work.

\subsection{Codimension-two defects in four and five dimensional theories }

In this section, we review the roles played by codimension-two surface defects in four and five dimensional supersymmetric gauge theories.
Three dimensional surface defects are most relevant for the BPS counting framework that we are going to develop with exponential networks.
However to motivate their features we first give a brief look at the 2d surface defects in 4d theories, and to their relation to spectral networks.

\subsubsection{2d-4d systems}

Four-dimensional $\CN=2$ quantum field theories can be modified in the UV in a variety of ways, so as to give rise to distinct half-BPS surface defects
preserving 2d $\CN=(2,2)$ supersymmetry.  The prototype of such surface defects is the so-called Gukov-Witten defect \cite{Gukov:2006jk}, which are characterized by imposing local monodromy constraints for the 4d gauge fields around the defect.
Another way to engineer surface defects is to couple a 2d $\CN=(2,2)$ theory to 4d gauge fields through gauging the 2d flavors.

We focus on the second type of defects following the treatment of \cite{Gaiotto:2013sma}.
We start by discussing the IR set up. As one flows to the IR, the bulk theory becomes abelian
and a holomorphic function of the Coulomb branch, the prepotential $\mathcal{F}$ characterizes the geometry of the vacua. The surface defect generically has a discrete  set of massive vacua,
which is fibered over the Coulomb branch $\mathcal{B}$ of the theory, defining the combined space of vacua of the 2d-4d system.
For each massive vacuum, the monodromy around the defect is characterized by the ``effective twisted superpotential'' (${W}$), one of the most important geometric quantities for our considerations.
The choice of the surface defects determines the geometry of the 2d-4d space of vacua and in IR, this effective superpotential.

We will use the powerful techniques from the 2d gauged linear sigma model (GLSM) technology \cite{Hanany:1997vm,Witten:1993yc}. One can compute the effective twisted superpotential
for the surface defects by coupling the gauge fields in the bulk to the flavor symmetries of the 2d GLSM's.
Below, we will review the procedure for 2d $N=(2,2)$ GLSM first and then will review the same for surface defects following \cite{Gaiotto:2013sma} and will illustrate with examples.

For 2d GLSM \cite{Hanany:1997vm,Witten:1993yc}, one first integrates out the 2d chiral matter fields. This gives rise to the twisted superpotential which is a function of the twisted masses and the 2d gauge multiplet
scalars. Extremizing the superpotential now with the 2d gauge multiplet scalars, one can derive the twisted chiral ring equations.

As an example, take the 2d $\mathbb{C}\mathbb{P}^1$ linear sigma model, defined by a $U(1)$ gauge theory with two chirals transforming as an $SU(2)$ flavor doublet. Then one can compute the twisted superpotential as
\be
{\tCW} \, = \, t\sigma - (\sigma-m) \log (\sigma-m) - (\sigma+m) \log (\sigma + m),
\ee
where $m$ is the twisted mass parameter for the $SU(2)$ flavor symmetry, $\sigma$ is the 2d gauge multiplet scalar and $t$ is the FI parameter. Then the chiral ring equation
can be derived from the extremization of the superpotentiala with respect to $\sigma$,
\be
\sigma^2 - m^2 = e^t,
\ee
from which one finds that there are two vacua from the solutions of this equation.

For the surface defects  \cite{Gaiotto:2013sma} suggests a similar procedure. One first integrates out the chiral matter fields. Then treating the bulk vectormultiplet scalars $\Phi$ as the twisted
masses, one gets 2d chiral with 2d mass $x$, in a representation $\mathcal{R}$ of the bulk gauge fields giving rise to an twisted effective superpotential
\be
{W}_{\mathcal{R}} (x,\Phi) = - {\mathrm{Tr}}_{\mathcal{R}} (x+\Phi) \log (x+\Phi).
\ee
Then one replaces ${\mathrm{Tr}}_{\mathcal{R}} \Phi^k$ coming from the expansion by a the VEV of this operator in the 4d $\CN=2$ theory. This will be a function of the Coulomb branch parameters $u_i$, and can be computed using resolvents of the related matrix models.
This gives an effective twisted superpotential $\mathcal{W} (\sigma, m,u)$, as a function of the 2d scalar vevs $\sigma$, 4d scalar vevs $u_i$ and twisted mass parameters
$m$. Then extremization with respect to $u$ gives the twisted chiral ring equations of the 2d-4d system.

As an example, couple $SU(2)$ to the two chirals of $\mathbb{C}\mathbb{P}^{1}$ model. Then one has the superpotential
\be
{W} = t\sigma  - {\mathrm{Tr}} (\sigma + \Phi) \log (\sigma + \Phi).
\ee
Then one can compute the twisted chiral ring equations as \cite{Gaiotto:2013sma}
\be
\sigma^2-u = \Lambda^4 e^{-t}  + e^t.
\ee
A crucial property of this expression is that it coincides with the Seiberg-Witten curve of the 4d bulk theory, or in other words the curve (\ref{eq:string-SW-curve}) arising in the string theoretic setting.
This fact is a key reason why studying BPS surface defects provides a good probe of the low-energy dynamics of 4d $\CN=2$ theories.

\subsubsection{3d-5d systems} \label{sec:3d-5d-systems}

Defects of codimension two in the context of five-dimensional gauge theories will be most relevant in what follows.
As for the 2d-4d case, there are several possibilities for defining these defects.
One familiar way is to include them in the M theoretic engineering, by designating a choice of special Lagrangian submanifold $L\subset X$ and considering a stack of M5 branes on $L\times S^1\times \IR^2$.
We will restrict to the case of a single fivebrane, and will further restrict $L$ to be the ``toric'' brane of \cite{Aganagic:2000gs, Aganagic:2001nx}.
The second restriction is made for simplicity, although our constructions with exponential networks can be directly applied to any other choice of Lagrangian branes whose mirror geometry is known. This includes knot conormals of \cite{Ooguri:1999bv} and we will return on this in Section \ref{sec:discussion}.

If we restrict to a class of 5d theories with a known Lagrangian description (as well as an M-theoretic engineering) we can define the 3d-5d system in a way that is analogous to the 2d-4d case.
For illustration purposes, let us assume that the bulk 5d theory is a gauge theory with gauge group $SU(N)$.
Namely, we consider a circle uplift of the 2d $\CN=(2,2)$ U(1) GLSM to a 3d $\CN=2$ U(1) gauge theory with a charged chiral multiplet transforming in the fundamental representation of $SU(N)$. We further turn on a minimal coupling for the 3d chirals to 5d vectormultiplet fields restricted to the defect.
The 3d theory may additionally feature a Chern-Simons term.
This is the field theoretic description of the 3d theory $T[L]$ coupled to the 5d theory, see for instance \cite{Dimofte:2010tz}.
The classical moduli space of vacua of the 3d theory can be seen to coincide with the positions of $L$ on the toric diagram of $X$ \cite{Dorey:1999rb}, wheres the quantum moduli space is expected to be captured by the mirror curve \cite{Aganagic:2000gs}. The 3d Chern-Simons term is then directly related to the choice of framing.

However it is not necessary to invoke mirror symmetry to study the low energy dynamics of the 3d-5d system.
We can rely directly on quantum field theory if we follow the strategy already employed for 2d-4d systems, pioneered by \cite{Gaiotto:2013sma}.
There is in fact a matrix model description of (at least, certain) 5d gauge theories, the difference is that this involves \emph{unitary} rather than \emph{hermitean} models \cite{Periwal:1990qb}. See for example \cite{Klemm:2008yu} for a study of matrix models associated to 5d $\CN=1$ gauge theory.
The argument for computing the twisted effective superpotential of $T[L]$ runs in the same way as the 2d case: the chirals acquire a $u$-dependent mass on the Coulomb branch of the 5d theory, and can be integrated out to yield an effective description in terms of the 3d fieldstrength scalar $\sigma$.
A 3d chiral multiplet with twisted mass $m$ then contributes a dilogarithm $\Li_2(2^{-2\pi R\, (\sigma+m)})$ to the twisted superpotential.\footnote{A neat way to see this is to consider its KK modes as fields in the 2d theory, and taking the regularized sum of the contribution of all KK modes studied in the 2d case, see for instance \cite{Dimofte:2011jd}.}
Now viewing $m$ as the VEV of the 5d vectormultiplet scalars, one integrates out the 5d degrees of freedom, further deforming the 3d twisted superpotential into a function $\tCW(\sigma, t, u)$ of the FI coupling $t$ and of the 5d Coulomb and mass moduli, collectively denoted by $u$. Since both the 3d and the 5d theories are taken on a circle of finite radius, both $\sigma$ and $t$ are complexified and have a periodic imaginary part.
The 5d path integral is carried out with the help of the matrix model, as in the 4d case it suffices to know the explicit form of the resolvent.
Minimizing $\tCW$ should then produce the Seiberg-Witten curve of the 5d bulk theory, which is expected to arise as the spectral curve of some relativistic integrable system \cite{Nekrasov:1996cz}. An example can be found in Section \ref{C34dlim}.

This program for the study of 3d-5d systems was carried out in recent works, confirming expectations in some explicit cases \cite{Ashok:2017lko,Ashok:2017bld}. Our working assumption will be that it extends to generic toric branes of toric Calabi-Yau threefolds.
One indication that this should be possible is that we may view the 3d-5d system on a circle as a 2d-4d system of the KK modes. Although this is may look rather involved from the  2d-4d viewpoint, one could once again run the logic used in the study of the latter, and obtain the expected results. A further hint that this should work comes from mirror symmetry, whose prediction identifies the moduli space of $L$ with the mirror curve, and further with the Seiberg-Witten curve of the 5d theory.

\subsection{$tt^*$ geometry in two and three dimensions}{\label{ttstar}}

Above we reviewed how codimension-two defects provide powerful probes into the low-energy dynamics of four- and five-dimensional gauge theories with eight supercharges. The upshot is that their moduli spaces of vacua (twisted chiral rings) coincide with the Seiberg-Witten curves of the bulk theories.
These curves are presented as ramified coverings of the FI plane, parameterized by the variable $t$ in the previous discussion.
At each $t$ the twisted superpotential exhibits a finite number of massive vacua, corresponding to sheets of $\Sigma$.
The next step will be to study the spectrum of BPS kinks connecting these vacua in the 2d-54d context, and similar kink-like BPS states in the 3d-5d setting.
A powerful approach to this problem is via $tt^*$ geometry.
Here we give a brief review of the main ideas of $tt^*$ in 2d and in 3d, and its relation to BPS counting.

\subsubsection{Solitons of 2d $N=2$ Landau-Ginzburg models}{\label{ttstarin2d}}

$tt^*$ geometry in two dimensions was introduced in \cite{Cecotti:1991me} and further explored in subsequent papers \cite{Cecotti:1992qh, Cecotti:1992rm}.
A rigorous mathematical formulation was proposed in \cite{Dubrovin:1992yd}.

To review its salient features we consider a Landau-Ginzburg theory in two dimensions, where the geometric meaning of the solitons is most transparent.
These theories are characterized by a superpotential $W(x^i)$ which is a function of $n$ chiral superfields, up to a variation of D-terms given by the the K\"ahler potential which turns out to be positive for unitary theories.
The bosonic part of the action is given by
\be
S =  \int d^2 z G_{i\bar{j}} \partial_\mu x^i \partial^\mu x^{\bar{j}} +  G^{i\bar{j}} \partial_i W {\bar\partial}_{\bar{j}} \bar{W},
\ee
where $ G_{i\bar{j}} = \partial_i {\bar\partial}_{\bar{j}} K$, and $K$ is the K\"ahler potential.

Minima of the superpotential correspond to the vacua of the theory, they are assumed to be non-degenerate.
In particular by a genericity assumption one can assume that the superpotential is quadratic near each vacuum.

Solitons are static field configurations with a profile along the spatial direction, and with boundary conditions governed by two distinct critical points $\alpha,\beta$ such that $x^i(-\infty) = \alpha^i$ and $x^i(\infty) = \beta^i$.
Stable solitons are the ones which satisfy the boundary conditions and minimize the energy, i.e. saturate the  BPS bound. The latter condition can be shown to take the following form
\be
\label{solitoneq2}
\partial_{\xi} x^i = e^{i\vartheta} G^{i\bar{j}} {\bar\partial}_{\bar{j}} \bar{W},
\ee
where $\xi$ denotes the space variable and $e^{i\vartheta} = \frac{W(\beta)-W(\alpha)}{|W(\beta)-W(\alpha)|}$ is the phase of the jump of the superpotential.
The central charge of the soliton is
\be
	Z_{\alpha\beta} = 2(W(\beta)-W(\alpha))\,.
\ee
The BPS equations imply
\be
\partial_\xi W = e^{i\vartheta} |\partial W|^2,
\ee
which means that the soliton field configuration maps to a straight segment in the $W$-plane of slope $\vartheta$ and connecting $W(\alpha)$ and $W(\beta)$.

To count the number of solitons, one asks how many solutions exist for \eqref{solitoneq2}. To answer this, one considers
the ``thimble'' of such solutions near each of the critical points and then evolves the thimbles according to the BPS equations in the field-space of $x^i$, and finally counts the number of points the thimble of $\alpha$ intersects the thimble of $\beta$ \cite{Cecotti:1992rm}.
The wavefront of all possible solutions originating from a critical point is a $S^{n-1}$ sphere for a given $W$. In fact, this coincides with the definition of the
vanishing cycle in the context of singularity theory, as the sphere vanishes upon approaching the critical point.
Denoting the two wavefronts originating at two critical points by $\Delta_\alpha$ and $\Delta_\beta$, the intersection number (computed on a hyperplane mid-way between the critical points) is given by
\be
|\mu_{\alpha\beta}| = \bigg| \sum_{\alpha\beta \,\,{\textrm{soliton}}} (-1)^F F\bigg| = |\Delta_\alpha \cdot \Delta_\beta|,
\ee
where $F$ is the fermionic number. In fact it turns out that  absolute value signs can be dropped (with possible ordering of vacua). Thus one arrives at an expression for an \emph{index} counting the number of solitons between two vacua in terms of intersections of thimbles:
\be
\label{2dsoliton}
\mu_{\alpha\beta} = \Delta_\alpha \cdot \Delta_\beta.
\ee
As a consequence of this fact there is no soliton interpolating between a vacuum and itself  \cite{Cecotti:1992rm}, in other words $\mu_{\alpha\alpha} = 0$.

As one perturbs $W$  the critical values move in the $W$-plane.
Intersection numbers are topologically stable, and do not change as long as no critical value crosses the straight line joining two other critical values.
However when a critical value $\beta$ happens to move onto the straight line connecting the critical values $\alpha$ and $\gamma$, then one can not continuously pass the wavefront originating from $\alpha$ to $\gamma$.
Thus the intersection number
$\mu_{\alpha\gamma}$ suddenly changes, and the change is captured by the automorphism of the group of vanishing cycles.

The transformation of the basis cycles in the above situation can be analyzed by the well-established techniques of Picard-Lefschetz theory. It turns out that
\be
\Delta'_\alpha = \Delta_\alpha + (\Delta_\alpha \cdot\Delta_\beta) \Delta_\beta,
\ee
from which one can compute the jump in the intersection number as
\be
\mu'_{\alpha\gamma} = \Delta'_\alpha\cdot \Delta_\gamma = \mu_{\alpha\gamma} + \mu_{\alpha\beta}\mu_{\beta\gamma}.
\ee
This is the Cecotti-Vafa wall-crossing formula for 2d BPS states, which also plays a key role in the construction of spectral networks.

\subsubsection{$tt^*$ geometry in two dimensions}

The $tt^{*}$ equations of  \cite{Cecotti:1991me} are defined in general for ${N}=2$ theories, not necessarily for SCFTs. Indeed for defining chiral rings and eventually $tt^{*}$ equations, we only need two adjoint supercharges $Q$ and $\overline{Q}$
such that $(Q)^{\dag}=\overline{Q}$ and they define a Hamiltonian $\{Q,\overline{Q}\}=H$. We also need to have at least one unbroken R-symmetry $U(1)_{R}$. We will mainly focus in  models where this happens to
be the axial R-symmetry $U(1)_{A}$, for example the LG models discussed above. For massive models flowing to SCFTs (critical models) we have also a vector R-charge $U(1)_{V}$ because of quasi-homogeneity of the superpotential
$W$. For any of these models, the Fermion number $F$ can be defined just as the $U(1)_{A}$ charge:
\begin{eqnarray}
F=F_{A}=q_{A}
\end{eqnarray}
the topological/antitopological or chiral/antichiral ring is defined as either $\overline{Q}$ or ${Q}$ cohomology.
To define this ring for any Riemann surface, one needs to perform a topological twist which leads to the Ramond (R) sector.
Consider  $\phi_{i}\in H_{\overline{Q}}$, the  ring structure  is encapsulated by the relations
\begin{eqnarray}\label{eq:chiral-ring}
\phi_{i}\phi_{j}=C_{ij}^{k}\phi_{k}\,.
\end{eqnarray}
By spectral flow one can argue that there is a distinguished vacuum $|0\rangle$, which can be used to define vacua associated with other chiral operators:\footnote{Here $|0\rangle$ is the unique Neveu-Schwarz (NS) vacuum and then flowed to R sector by the spectral flow operator (which is well defined by the fact we have $U(1)_{A}$). In the SCFT case we have two R-charges and they provide a grading for the chiral-chiral ring in the NS sector.
A vacuum is defined up to $\overline{Q}|\psi\rangle$. But, by imposing additionally $Q|i\rangle=0$, it is possible to fix a harmonic representative.}
\begin{eqnarray}
\phi_{i}|0\rangle=|i\rangle \,.
\end{eqnarray}
Just by CPT theorem, there must exist an invertible matrix $M$ such that
\begin{eqnarray}
\langle\bar{i}|=M_{\bar{i}}^{j}\langle j| \qquad M_{\bar{i}}^{j}(M^{*})_{j}^{\bar{k}}=\delta_{\bar{i}}^{\bar{k}}\qquad (M^{*})_{j}^{\bar{i}}=(M_{\bar{j}}^{i})^{*}
\end{eqnarray}
$M$ provides a real structure to the vacuum bundle. Then, a very simple computation shows:
\begin{eqnarray}\label{eq:tt-star-reality}
g_{i\bar{j}}=M_{\bar{j}}^{k}\eta_{ki}\qquad g\eta^{-1}(g\eta^{-1})^{*}= 1.
\end{eqnarray}
The topological metric is given by the topologically twisted two point function on $S^{2}$: $\langle \phi_{i}\phi_{j}\rangle_{S^{2}}$. For B-twisted LG models (which are the ones of interest to us), in the untwisted sector this has a simple expression:
\begin{eqnarray}
\eta_{ij}=\langle \phi_{i}\phi_{j}\rangle_{S^{2}}=\frac{1}{(2\pi i)^{N}}\oint_{\gamma}\frac{ \phi_{i}\phi_{j}}{\prod_{a}\partial_{a}W}
\end{eqnarray}
here $\gamma$ is a middle dimensional contour with the topology of $T^{N}$, and circling around each pole. For the one-dimensional case the contour is therefore a disjoint union of circles.
The derivative stands for $\partial_{i}=\partial/\partial x_{i}$ where $x_{i}$ are the LG fields. If critical points of $W$ are  isolated and nondegenerate, the Hessian $\fH_W$ is non-vanishing there and the formula can be reduced to
\begin{eqnarray}
\eta_{ij}=\sum_{p\in \mathrm{Crit}(W)}\frac{\phi_{i}\phi_{j}}{\fH_W(p)}\,.
\end{eqnarray}
In order to write the $tt^{*}$ equations, is convenient to work in a basis of vacua labeled by (a choice of) the chiral ring generators $|i\rangle$. In the massive case, we can use the so-called point-like basis $|p\rangle$ labeled by $p\in \mathrm{Crit}(W)$,
and then the hermitian metric $g_{i\bar{j}}$ is given by pairing $|i\rangle$ and its CPT conjugate $\langle \bar{j}|$.
In the point-basis we may define the matrix $C_{\phi}$  by
\begin{eqnarray}
(C_{\phi})_{j}^{k}:=C_{\phi j}^{k}\in \mathrm{Mat}_{N}
\end{eqnarray}
where $\phi$ is \emph{any} operator acting on a chiral $\phi_j$ as in (\ref{eq:chiral-ring}).
and naturally, the metric $g_{i\bar{j}}$ is $N\times N$ dimensional and in the chiral basis it is hermitian.

$tt^*$ geometry describes how $g_{i\bar j}$ varies with deformations of the theory.
One can introduce a natural connection on the vacuum bundle defined by
\be
	A_{i\alpha\bar\beta}  =  \langle\bar\beta|  \partial_i |\alpha\rangle
\ee
where $\partial_i$ denotes variation with respect to the coupling associated with a chiral operator $\phi_i$.
The $tt^*$ equations assert then that  the ``improved'' connection\footnote{The covariant derivative is understood to act according to the type of representation of the object to which it is applied.}
\be\label{eq:nabla-ttstar}
	\nabla = dt^i \(  \partial_i - A_i +C_i\)
	\qquad
	\overline \nabla = dt^{\bar i} \(  \bar \partial_{\bar i} - \overline{A}_{ {\bar i}} + \overline{C}_{{\bar i}}\)
\ee
is flat
\be
	\nabla^2 = {\overline\nabla}^2 = \nabla \overline\nabla+\overline\nabla\nabla = 0\,.
\ee
Coupling-dependent changes of basis on the vacuum Hilbert space correspond to gauge transformation for the $tt^*$ connection. It is often convenient to work in a \emph{holomorphic} gauge where $A_{\bar i} = 0$.
It is precisely in this basis that the connection is related to the metric $g$ as
\be\label{eq:A-from-g}
	A_i = -g\partial_i g^{-1} \,.
\ee
The $tt^*$ equations can then be formulated in terms of $g$ alone as follows
\be\label{eq:tt-star-eqs-g}
\begin{split}
	\bar \partial_i (g\partial_j g^{-1}) - [C_j, g(C_i)^\dagger g^{-1}] & = 0 \\
	\partial_i C_j - \partial_j C_i + [  g\partial_i g^{-1}, C_j ] -  [  g\partial_j g^{-1}, C_i ]  & = 0\,.
\end{split}
\ee
On one hand, the flatness equations of the $tt^*$  connection will be most useful to make contact with our nonabelianization map for exponential networks in Section \ref{sec:expnet}. On the other hand the latter form of the equations involving $g$ will be most useful to us in actual computations of $tt^*$ geometry in Section \ref{sec:C3}.

To conclude our lightning overview of $tt^*$ equations, let us sketch how they subsume the soliton counting techniques via intersection numbers of Lefschetz thimbles reviewed above, following \cite{Cecotti:1992rm}. The basic idea is to use $tt^*$ equations to
connect leading infrared behavior to the to the one in the ultraviolet.
In the infrared it turns out that the metric is (in the point basis) diagonal, and gets corrected by solitons between vacua; on the other hand in the ultraviolet regime it is related to $U(1)$ charges of the Ramond vacua.
The analogous quantity to study in more general models is the ``CFIV'' index \cite{Cecotti:1992qh}.  Its definition is
\be
Q_{ab}=\frac{i\beta}{2L}\mathrm{Tr}_{ab}(-1)^{F}Fe^{-\beta H},
\ee
where the partition function is taken over a cylinder of length $L$ and perimeter $\beta$. The boundary conditions are specified by two vacua $a,b$.
In the UV limit $\beta \to 0$, and $Q$ is coincides with the spectrum of left (or right) charges
of the Ramond sector
\be
Q_{ij}\big\vert_{\beta \to 0} = q_{ij}\,.
\ee

In the IR limit instead $Q$ admits an expansion in terms of corrections due to solitons, which come with integer coefficients corresponding to the $\mu_{ab}$ introduced earlier \cite{Cecotti:1992rm}.
For a completely massive theory, there is a natural set of coordinates in the space of coupling constants, called canonical coordinates $w_k$ with $k=1,...n$.
Given the central charge of the supersymmetry algebra $Z = \{Q^+,{\bar{Q}}^+\}$, they
are defined as
\be
w_i - w_j = \frac{1}{2} Z \big\vert_{(i,j)}.
\ee
In the canonical coordinates the CFIV index is related to the $tt^*$ metric as follows
\be
Q = -\sum_k w_k g \partial_k g^{-1} = -\frac{1}{2} g \beta \partial_\beta g^{-1}\,.
\ee
In the IR limit $\beta \to \infty$ the CFIV index admits an expansion of the following form
\be
\label{QIR}
Q_{ij}\big\vert_{\beta\to\infty} \approx - \frac{i}{2\pi} \mu_{ij} m_{ij} \beta K_1 (m_{ij}\beta),
\ee
where $m_{ij} = |Z|\big\vert_{(i,j)} = 2 |w_i-w_j|$ are soliton masses and
 $K_1$ is the modified Bessel function of the first kind.
Similarly the $tt^*$ metric admits an expansion of the form
\be\label{eq:2d-metric-IR-expansion}
	g_{i\bar{j}} = \delta_{ij} - \frac{i}{\pi} \mu_{ij} K_0 (m_{ij}\beta)\,.
\ee
Note in particular this this implies a similar expansion for the $tt^*$ connection $A_i$ as well as for the improved flat connection $\nabla$ through equations (\ref{eq:A-from-g}) and (\ref{eq:nabla-ttstar}).
This form of the flat connection in terms of ``corrections'' by solitons underlies the \emph{nonabelianization map} construction of spectral networks \cite{Gaiotto:2009hg, Gaiotto:2012rg, Gaiotto:2011tf}.
It will also provide the physical interpretation for our main construction, outlined in Section \ref{sec:expnet}, which is strongly inspired to spectral networks.

\subsubsection{$tt^*$ geometry in three dimensions}\label{sec:3d-tt-star}

Three-dimensional $\CN=2$ theories on a circle of finite radius may be viewed as the 2d $(2,2)$ theories of the Fourier modes of the 3d fields. In this way it is possible to uplift the construction of two-dimensional $tt^*$ geometry 3d \cite{Cecotti:2013mba}.
We will be brief on this topic, since an explicit example will be studied in detail below in Section \ref{sec:C3-tt-star}.
In view of this, we will tailor the present discussion on a specific set of models which includes our main example, namely GLSMs (gauge linear sigma models).
As recalled in Section \ref{sec:3d-5d-systems}, an important novelty in three dimensional GLSM is that the vectormultiplet scalar fields becomes complexified and periodic $Y_i\sim Y_i+2\pi$ when the theory is placed on a circle.
Let us consider a $U(1)$ GLSM with $N$ chiral multiplets of equal charge. Let  $Y$ be the vectormultiplet fieldstrength scalar, and let $X$ be the corresponding FI coupling. If we turn on masses $m_i$ for the chirals, they decouple and leave one-loop contributions for the  effective twisted  superpotential, which takes the generic form
\be
	\tCW = X Y + \sum_{i} \Li_2(e^{-2\pi R \, (Y+m_i) }) + \frac{\kappa}{2}  Y^2
\ee
where $R$ is the circle radius, $m_i$ are bare masses of 3d chirals  and  $\kappa$ is an effective Chern-Simons coupling.

For generic $X$ the theory has a finite set of massive (non-degenerate) vacua, indeed the critical set of $W$ can be conveniently described by an algebraic equation in the exponentiated variables $X,Y \to x,y$
\be\label{eq:3d-vacua}
	\frac{d W}{dX} = 0 \qquad \to \qquad F(x,y)=0\,.
\ee
Solving for $y$ gives a finite set of solutions $y_i(x)$ which can be identified with the vacua of the theory for a given value of the coupling.
BPS solitons are field configurations for $Y$ that interpolate between a vacuum $i$ and another vacuum $j$, while obeying the BPS equations.
One spatial direction is compactified, by ``soliton'' we mean a field configuration in which $Y$ approaches a constant value at $\pm\infty$ in the non-compact direction.
Unlike in two dimensions, physical properties of these solitons are captured not only by the pair of vacua they connect, but also by the relative homology class of the path the soliton traces on the $Y$-cylinder (not to be confused with the space-like cylinder).
Solitons that  wind around the cylinder pick up corrections to their mass due to the multi-valuedness of $\tCW$.
The BPS equations determine the precise profile for $Y(\xi)$ over space, and the mass of a soliton can be unambiguously defined as the integral of the energy density of a given field configuration over space (see Section \ref{sec:shift-sym-C3} for more detail).

A way to keep track of this phenomenon is to work on the universal covering of the $Y$-plane, by introducing a discrete index $N\in\IZ$ to label preimages of vacua
\be
	|i\rangle \quad\to\quad|i,N\rangle \,.
\ee
This introduces an infinite tower of vacua for each critical point of the vacuum manifold (\ref{eq:3d-vacua}), and refines the notion of soliton charge. Solitons are in fact now labeled by pairs $(i,N),(j,M)$.

Coming back to the question of $tt^*$ geometry, it is natural to consider an uplift of the $tt^*$ metric to an infinite-dimensional matrix $g_{(i,N),(\bar j,\bar M)}$.
Recall that the metric depends explicitly on information of the superpotential of the theory, for instance through the IR expansion (\ref{eq:2d-metric-IR-expansion}), which contains $m_{(i,N),(\bar j,\bar M)}  = 2|\Delta \tCW|$.
In \cite{Cecotti:2013mba} it was argued that certain models have a soliton spectrum that is invariant under simlutaneous shifts of $N,M$ preserving $M-N$.
This effectively means that solitons come in infinite towers and all solitons of the tower have the same mass.
We will study the realization of this  symmetry in detail in Section \ref{sec:C3-tt-star}, and will indeed find that it is present, although somewhat nontrivially.
In particular, since $\tCW$ is still multi-valued on the universal covering of the $Y$-cylinder, shift symmetry is not automatically verified: its presence  is tied to the structure of the BPS spectrum.
Shift symmetry will also play a prominent role in our construction of Section \ref{sec:expnet}. In fact our framework provides a general proof of shift symmetry, since we will argue that it arises naturally from the perspective of exponential networks and their soliton data.\footnote{This is an affine version of the Weyl symmetry of soliton data in ADE spectral networks constructed in \cite{Longhi:2016rjt}.}

Taking the shift symmetry for granted, it implies that the metric itself must be shift-symmetric
\be
	g_{(i,N),(\bar j,\bar M)} = g_{i\bar j}(M-N)\,.
\ee
It is then natural to introduce a ``Bloch'' basis
\be
	|i,\theta \rangle = \sum_{N} e^{i \,N\, \theta}  |i,N\rangle \,.
\ee
In this basis the metric can be expressed as a periodic function $g_{i\bar j}(\theta)$, and one trades infinite-dimensional matrices for differential operators in $\theta$ in the $tt^*$ equations, by replacing
\be
	N \to -i \frac{\partial}{\partial\theta}\,.
\ee
In this way $tt^*$ geometry in three-dimensions can be viewed as augmenting the parameter space of the theory by one extra periodic direction parameterized by $\theta$. The $tt^*$ equations have been shown to take the form of periodic monopole equations on the augmented space of couplings \cite{Cherkis:2000cj}.

%

\subsection{A field theoretic perspective on spectral networks and exponential networks}

One of the main goals of this paper is to provide a B-model version of the BPS states counting for local Calabi-Yau
threefolds. A detailed algorithmic approach towards this will be described in Section \ref{sec:expnet}, based on the geometric framework of exponential networks and key ideas from spectral networks.
Here we review both frameworks, from the viewpoint of quantum field theory of 2d-4d and 3d-5d systems respectively.
This perspective is especially suited to making contact with the previous discussion on $tt^*$ geometries in 2d and in 3d, which provides the physical foundations for our constructions.

In fact this leads us back to the geometric approach discussed at the beginning of this section in \ref{sec:geom-approach}.
Such constructions constitute only a part of the spectral (resp. exponential) network, namely the \emph{geometric data}.
In order to extract meaningful information about BPS states, such as BPS indices, from the geometry one needs to introduce certain \emph{combinatorial data} on the network.
This was accomplished for spectral networks in \cite{Gaiotto:2011tf, Gaiotto:2012rg} through insightful studies on the interplay between $tt^*$ geometry, Hitchin systems and 2d-4d systems. This led to a powerful framework to compute 4d BPS spectra.
On the other hand the appropriate combinatorial data for exponential networks has been missing so far, since the definition provided in \cite{Eager:2016yxd} is purely geometric.
One of our main results, developed in section \ref{sec:expnet}, will be to fill this gap and to develop a framework to compute 5d BPS spectra.
The motivation for our proposal comes from the interplay between 3d $tt^*$ geometry and 3d-5d systems. Having reviewed both above, we now set out to explain how the two are naturally tied to exponential networks.

\subsubsection{Spectral networks} \label{sec:spectral-networks-review}

Let us first review of how spectral networks fit in the world of 2d-4d systems and $tt^*$ geometry.
The BPS spectrum of $N=2$ supersymmetric gauge theories in 4d can be most conveniently repackaged in terms a spectral curve.
Following  \cite{Klemm:1996bj, Witten:1997sc, Gaiotto:2009we, Gaiotto:2009hg}, the basic idea is to embed the gauge theory into a higher dimensional one, rendering the spectral geometry
as a part of the spacetime.
BPS particles in 4d are then realized geometrically as extended objects calibrated by the Seiberg-Witten differential,
associated with the spectral curve.
BPS spectra for a large class of such theories, known as ``class $\CS$'', can be studied geometrically thanks to the pioneering work of \cite{Gaiotto:2012rg}.

Theories of class $\CS$ are defined as partially twisted dimensional reductions of the 6d $(2,0)$ ADE theories, on a punctured Riemann surface. Each such theory is completely characterized by the choice of ADE algebra $\mathfrak{g}$, a Riemann surface $C$ and by certain defect data corresponding to punctures.
Class $\CS$ theories of $A$ type may be naturally embedded into M theory, arising as the worldvolume low-energy dynamics of  rank$(\mathfrak{g})$ M5 branes on $\mathbb{R}^{1,3} \times C
\subset \mathbb{R}^{1,3} \times T^* C \times \mathbb{R}^{3}$.
At a generic point on the Coulomb branch in the IR, the stack of fivebranes merges into a single fivebrane wrapped on $\mathbb{R}^{1,3} \times \Sigma$.
Here $\Sigma \rightarrow C$ is the spectral cover
\be
\{\lambda : {\mathrm {det}} (\phi - \lambda I) = 0\} \subset T^* C,
\ee
where $\phi$ is a one form valued in $\mathfrak{g}$ parametrizing the Coulomb branch of the theory.

For convenience, let us restrict to $\mathfrak{g} = \mathfrak{su}(k)$. In 6d the string like excitations arise as boundaries of M2 branes ending on
the stack of M5 branes. When dimensionally reduced on $C$, they give rise to particles in 4d, provided they are extended along a path in $C$. Hence with respect
to a local choice of trivialization for the spectral covering map, the paths can be labeled locally by a pair $i,j$ of integers indexed in $1...,k$.
These are BPS strings, if and only if the equality holds below
\be
M = \int |\lambda_{(ij)}| \ge \big\vert \int \lambda_{(ij)} \big\vert = |Z|
\ee
where $\lambda_{(ij)} = \lambda_i - \lambda_j$, where $\lambda_i$ is the Liouville one form on $T^*C$ restricted to the $i$-th sheet of $\Sigma$.
This condition is satisfied only if Im$(e^{-i\vartheta}\lambda_{(ij)}) = 0$ (and Re$(e^{-i\vartheta}\lambda_{(ij)})$ is the volume form), which defines BPS trajectories on $C$ labeled by $ij$.
It is not hard to see that this geometric constraint arises from pulling back the BPS equations (\ref{solitoneq2}) for BPS solitons on surface defects from the $W$ plane onto the manifold of vacua $\Sigma$, and then further projecting it to $C$ \cite{Gaiotto:2009hg}. We will sometimes refer to this as the \emph{geometric BPS equation}.

Heuristically, spectral network is the evolution of such BPS trajectories.  The boundary conditions for these trajectories are of  two types. A trajectory of type $ij$ can either end on a branch point where $\lambda_{i} - \lambda_{j} = 0$ or
 on a junction where BPS strings of types $jk$ and $ki$ meet, all having the same phase $\vartheta$.
BPS states of the 4d theory in this case arise from finite webs of BPS strings. A finite web may include both junctions and branch points, but always lifts to a closed homology class on the spectral curve, therefore defining the charge of a 4d BPS particle \cite{Seiberg:1994rs, Seiberg:1994aj}.

For the BPS indices from the counting of the finite webs, one has to take into account that finite webs might exist in
continuous families, of which the spectral networks produce some critical members. The generic members do not pass through the
branch points. However, they are still calibrated locally by the one form and they also satisfy the junction rules.
The deformations modes of the finite webs geometrically realize the zero modes of the BPS particles in four dimensions \cite{Mikhailov:1997jv}.
One of the ways to determine BPS spectrum would be to quantize the zero modes, however this proves especially hard for higher spin BPS states, which have been shown to occur commonly \cite{Galakhov:2013oja, Hollands:2016kgm}.
A different route was proposed in \cite{Gaiotto:2012rg}.
The incipit of this route was  to consider BPS states in presence of various types of BPS line and surface defects.
Then the consistency with the wall-crossing behavior of such particles bound to the defects led to constraints that solved the 4d BPS spectra.

Such consistency constraints have a beautiful geometric interpretation, which leads to a connection to $tt^*$.
The curve $C$ is identified as the parameter space of UV couplings of a ``canonical'' surface defect.
The finite webs with an open endpoint at $z$
are identified with a particle bound to the surface defect $S_z$.
Line defects on the other hand are engineered by infinitely heavy particles arising from
M2 branes whose boundary stretched along the direction, as well as along a path $\wp$ on $C$.
Viewing a line defect as an interface between a surface defect and itself (choosing a basepoint on $\wp$), its spectrum of ``framed'' BPS states can be computed by counting intersections with open BPS strings of the network.
Physical arguments lead to the conclusion that the partition function $F(\wp,\vartheta)$ of framed BPS states must depend only on the homotopy class of $\wp$, since it is a UV observable.
In other words, $F(\wp,\vartheta)$ must be the holonomy of a flat connection over $C$ and this can, loosely speaking, be identified with the Lax connection of $tt^*$ geometry. Indeed  $\wp$ is a path in the moduli space of 2d theories on the surface defect, deformed by the coordinates on $C$, and may be viewed as a BPS domain wall in the 2d theory.
As $\wp$ crosses the network, its partition function gets corrected by 2d-4d states corresponding to the open webs.
This statement is analogous to the IR expansion of the CFIV index discussed above, in fact the latter is also closely related to the $tt^*$ connection and the coefficients of the expansion denoted $\mu(a)$ are identified with the ``soliton data'' of the spectral network.
Rather elegantly, invariance of the framed BPS partition function under homotopic deformations of $\wp$ imposes relations on the 2d-4d BPS states encoded by the network, which turn out to completely fix the soliton data of the spectral network.

Four-dimensional degeneracies are computed at values of $\vartheta$ where the network becomes degenerate, since this is the condition that leads finite BPS webs to appear.
The degeneration induces a  jump of the topology of the network, and therefore of the soliton data.
This is interpreted as the fact that some 2d-4d states bounded together to form a 4d BPS particle, and left the 2d-4d spectrum.
Comparing the network soliton data before and after the jump leads to information about the phase space for such boundstates/decays, and allows to compute the BPS index of the 4d BPS state responsible for the jump.

\subsubsection{Exponential networks}{\label{expnet-prim}}

Let us now recall the geometric definition of exponential networks in the context of Lagrangian branes in toric Calabi-Yau threefolds, following \cite{Eager:2016yxd}.
Here we will mainly focus on reviewing background on the geometric setting in whch exponential networks arise. Only a very brief mention of the networks will be made, since a detailed discussion will be given in section \ref{sec:expnet}, where we will also focus on the definition of the soliton data.

Just like the standard spectral networks, the geometric data of exponential networks also consists of trajectories determined by the geometry of a curve $\Sigma$.
This curve is now identified with the mirror curve of a
pair $(L,X)$ where $L$ is a Lagrangian brane in a toric Calabi-Yau threefold $X$.
The mirror geometry is described by $Y = \{uv = H(x,y)\} \subset \mathbb{C}^2 \times (\mathbb{C}^*)^2$, with $H = 0$ defining $\Sigma \subset (\IC^*)^2$.
If $\Omega$ is the
holomorphic top form on a Calabi-Yau threefold, the stable A-branes which are special Lagrangian submanifolds $S$, are real three dimensional submanifolds satisfying
\be
\Omega\vert_S = e^{i\alpha}\, \left|\Omega\right| \,,
\ee
where the phase $\alpha$ is related to the stability condition for the branes.
The power of the geometric approach to BPS spectra is that the study of the A-branes can be reduced to the mirror curve $\Sigma$, endowed with a differential $\lambda = \log y d\log x $ following ideas of \cite{Klemm:1996bj}.
This is accomplished by integrating the holomorphic top form over fibers
of the map $(x,y,u,v) \mapsto (x,y)$ which takes  $Y \mapsto  (\mathbb{C}^*)^2 $. Over each point in the $(x,y)$-plane the equation $H(x,y) = uv$ can be viewed as an equation for $(u,v)$,
describing an affine conic bundle reducible precisely when $H(x,y) = 0$.

Next we come to the important aspect of framing.  As was mentioned before, in the A-model the toric brane $L$ has the topology  $\mathbb{R}^2\times S^1$.
The toric brane which is mirror to the $v=0$ fiber is specified by a point on the toric diagram.
The vertex which is closest in the toric diagram  is surrounded by three faces corresponding to the divisors $z_1,z_2,z_3$. The brane is specified by
\be
|z_2|^2 - |z_1|^2 = 0,
\ee
and the real position modulus is given by
\be
r \sim |z_3|^2 - |z_1|^2.
\ee
In the semiclassical regime, which is when the brane sits far away from the vertex, the framing ambiguity is suppressed. However in the quantum regime the moduli space of $L$ grows becomes the mirror curve, with quantum corrections becoming more important in the vicinity of the vertex. There is then an ambiguity
\be
r \sim  |z_3|^2 - |z_1|^2 + f(|z_2|^2 - |z_1|^2 )
\ee
due to the fact that the second piece is classically vanishing, at infinity along a toric leg.

Having discussed the curve geometry, let us now come to calibrated cycles which correspond to BPS states.
The holomorphic top form on the mirror curve can be written as
\be
\Omega \, = \, {\mathrm{Res}}_{H=uv} \(\frac{dudvdxdy}{xy(uv-H)}\),
\ee
where all the periods can be reduced to the integration over $\Sigma$ of the form \cite{Forbes:2005xt},
\be
 {\mathrm{Res}}_{H= 0} \(\log H(x,y) \frac{dxdy}{xy}\) \, = \, \log y \frac{dx}{x}
\ee
preserving the symplectic form on $(\mathbb{C^*})^2$, namely the form $\frac{dxdy}{xy}$ preserved, up to a sign.
There is a well-known reparametrization group for $\Sigma$
which is SL$(2,\mathbb{Z}) \rtimes \IZ_2$ acting as
\be
\begin{pmatrix}
a & b \\
c & d
\end{pmatrix} \cdot (x,y)
\, = \, (x^ay^b, x^cy^d).
\ee
In particular, change of framing acts as
\be
\begin{pmatrix}
1 & f \\
0 & 1
\end{pmatrix} \cdot (x,y)
\, = \, (xy^f, y).
\ee

To conclude this mini-review let us consider a couple of examples.
First of all, the main example that will be analyzed later in the paper, namely $\mathbb{C}^3$. Although the simplest of all, it is perhaps the most important, since it is the building block of all toric diagrams.
In this case there is no gauge group for the 5d theory, and hence there is no group action.
The mirror curve is given by
$\sum_{i=1}^3 y_i$ which up on quotienting by $\mathbb{C}^*$, we have the well-known result $H=x+y+1$.
Changing the framing, one has
\be
H_f (x,y) = xy^f + y + 1.
\ee
A presentation of the LG superpotential which will play crucial role later is
\be
\label{C3sup}
W = (X-i\pi)Y + \frac{f}{2} Y^2 + \Li_2 (-e^Y),
\ee
where $x=e^X$ and $y=e^Y$, respectively.
Then the mirror curve from is recovered as $\exp(\partial W/\partial Y) = 1$.
In section \ref{sec:C3}, we will particularly focus on the case $f=-1$ for simplicity, although this choice can be relaxed.

For completeness let us also consider the next-simplest example.
The resolved confold geometry gives rise to a 5d gauge theory with a $U(1)$ gauge group, and there are two phases which are birationally equivalent by the Atiyah flop.
The mirror D-term
equation is given by $y_1y_2 = e^{-t} y_3 y_4$ which gives rise to again the well-known curve
\be
H(x,y) = 1+ x + y + e^{-t} xy.
\ee
Incorporating the framing, one has
\be
H_f(x,y) = 1+ xy^f + y + e^{-t} xy^{f+1},
\ee
from which one can compute the LG superpotential
\be
\label{conisup}
W= (X-i\pi ) Y + \frac{f}{2} Y^2 + \Li_2(-e^Y) - \Li_2 (- e^{-t} e^Y),
\ee
from which the mirror curve can be recovered again as $\exp(\partial W/\partial Y) = 1$.

\section{Flat connections for mirror curves and soliton counting}{\label{sec:expnet}}

\subsection{Data of exponential networks}
Here we collect some basic definitions about exponential networks, in order to set the stage for the rest of this section.
The data of exponential networks is of two types: geometric data and combinatorial data (a.k.a. soliton data).
For the definition of the former we follow \cite{Eager:2016yxd} with minor modifications, whereas the definition of the latter is novel and will be developed in greater detail.

\subsubsection{Geometric data}

Let $\Sigma$ be an algebraic curve in $\IC_x^*\times\IC_y^*$, endowed with the natural projection $\pi:\Sigma\to C\equiv C^*_x$.
This projection is $K:1$ and presents $\Sigma$ as a ramified covering of $C$. By a genericity assumption we assume that all branch points are of square-root type.
A choice of trivialization involves choosing branch cuts for $\pi$, and a labeling for sheets of $\Sigma$ away from the cuts. A choice of trivialization will henceforth be assumed, and the different sheets will be denoted by $y_i(x)$ for $i$ valued in a set of $K$ elements.

Let $\vartheta\in \IR/2\pi \IZ$, the exponential network $\CW(\vartheta)$ is a web of trajectories on $C$ related to the covering $\pi$.
Trajectories are called $\CE$-walls\footnote{The name is a variant of the original $\CS$-walls of spectral networks.}, and are labeled by an ordered pair of sheets of $\Sigma$ and an integer $(ij,n)$.
The shape of a wall is determined by a differential equation\footnote{Note that the convention is opposite to the one used by the authors of \cite{Eager:2016yxd}. Our choice is dictated by convenience when defining soliton data later on.}
\be\label{eq:E-wall-diff-eq}
	(\log y_j - \log y_i +2\pi i \, n)\frac{d\log x}{d t} \in e^{i\vartheta} \IR^+\,.
\ee
The boundary conditions for this equations provide the starting points for $\CE$-walls, and they come in two types: a wall can either start at a branch point (a \emph{primary} wall) or at the intersection of two walls (a \emph{descendant} wall).
A primary wall of type $(ij,n)$ must start at a branch point where $y_i = y_j$ and moreover it must have $n=0$.
When walls $\CE,\CE'$ intersect, they may generate one or more descendant walls $\CE''$, whose types are determined by those of $\CE,\CE'$.

The main difference between geometric data of $\CE$-walls and $\CS$-walls (walls of standard spectral networks) stems from the extra integer label $n$, which originates in the multivaluedness of the logarithms in (\ref{eq:E-wall-diff-eq}).
In order for this label to be well-defined, it is necessary to introduce a second covering map $\tpi:\tSigma\to\Sigma$ with branching at those points $(x,y)\in \Sigma$ where $y_i(x)$ approaches either zero or infinity.
A choice of trivialization for this second covering map is fixed by specifyng ``logarithmic'' branch cuts on $\Sigma$, and by labeling with $N\in\IZ$ different sheets of $\tSigma$ away from these cuts.
For convenience, we often represent the ``logarithmic cuts'' of $\tpi$ by their $\pi$-projection on $C$, however it is important to keep in mind that they actually live on $\Sigma$.
The curve $\tSigma$ can be presented as an infinite-sheeted covering of $C$ by composition $\pi\circ\tpi$
\be
	\tSigma\mathop{\longrightarrow}^{\tpi}\Sigma\mathop{\longrightarrow}^{\pi}C\,.
\ee
This is the viewpoint that we usually adopt, since exponential networks live on $C$.
Above $x\in C$ there are infinite families of sheets labeled by $(i,N)$, corresponding to points of $\tSigma$ located at $(x, \log y_i(x) + {2\pi i N})$.

When an $\CE$-wall $p$ of type $(ij,n)$ crosses a branch cut of square-root type that permutes sheets of $\Sigma$ as $i\to \sigma(i)$, its label changes in the obvious way to $(\sigma(i)\sigma(j),n)$.
On the other hand, across a logarithmic cut (more properly, its projection to  $C$) the integer label $n$ may change, the specific behavior depends on the finite labels $ij$.
To determine the behavior, one needs to consider a \emph{lift} of the wall to $\Sigma$
\be
	\pi^{-1}(p) = p^{(j)} - p^{(i)}
\ee
consisting of the preimages of the wall $p$ lifted to sheets $i$ and $j$ with opposite orientations (on sheet $j$ the orientation agrees with that of the wall on $C$).
Suppose that $p^{(j)}$ (resp. $p^{(i)}$) crosses the logarithmic cut on $\Sigma$, such that $\log y_j \to \log y_j + 2\pi i \, \delta n_j$ (resp. $\log y_i \to \log y_i + 2\pi i \, \delta n_i$) crossing the cut in the direction specified by the orientation of $p^{(j)}$ (resp. $p^{(i)}$).
Then  $n$ jumps by
\be
	n \to n + \delta n_j - \delta n_i\,.
\ee

\subsubsection{Combinatorial data}

Each $\CE$-wall carries combinatorial data, also known as \emph{soliton data}.
In order to characterize it, let us fix any $x$ on the wall and consider the (affine) lattice
\be\label{eq:open-charge-lattice}
	\Gamma_{ij,N,N+n}(x) = H_1^{\rel}(\tSigma; (i,N) , (j,N + n), \IZ) \,.
\ee
This is the relative homology lattice of open paths $a\subset \tSigma$ starting at $(i,N)$ and ending at $(j,N+n)$.
Physically each homology class corresponds to a possible charge for a 3d-5d BPS particle, an uplift of 2d-4d BPS states studied in \cite{Gaiotto:2011tf}.
We will use both terminologies ``soliton path'' and ``soliton charge'' interchangeably. We return to the physical picture of these 3d-5d BPS particles in Section \ref{sec:discussion}.
The central charge of a soliton $a$ is the chain integral
\be\label{eq:central-charge-def}
	Z_a = \frac{1}{2\pi R} \int_a Y(x) \, \frac{dx}{x}\,,
\ee
where $Y(x) = \log y(x) + 2\pi i N$, with $N$ the appropriate branch of the logarithm corresponding to an actual path on $\tSigma$ representing the homology class $a$. The denominator features the radius $R$ of the compactification circle  (see Section \ref{sec:3d-5d-systems}).

While the shape of the $\CE$-wall only depends on the fixed integer $n$, its soliton data is classified by a \emph{pair} of integers $N, N+n$ whose  difference is fixed.%
To lighten notation we also introduce the union of charge lattices for different solitons supported by the same wall:
\be\label{eq:full-open-charge-lattice}
	\Gamma_{ij,n}(x) = \bigsqcup_{N\in\IZ} \Gamma_{ij,N,N+n}(x) \,.
\ee

The soliton data of the network is the assignment of an integer $\mu(a)$ to each $a\in \Gamma_{ij,n}(x)$ on each $\CE$-wall. The rules that fix these integers will be determined in the remainder of this section.

\subsection{Nonabelianization map}\label{sec:nonabel-map}

Let  $\nabla^{ab}$ be a flat $GL(1)$ connection on $\tSigma$, a nonabelianization map is the construction of a flat $GL(N)$ flat connection $\nabla^{na}$ on $C$ in terms of the abelian connection $\nabla^{ab}$.
In this subsection we build a nonabelianization map $\Psi_\CW : \nabla^{ab}\to \nabla^{na}$ using the data of an exponential network $\CW$.
To formulate the map explicitly it is convenient to choose coordinates on the respective moduli spaces of these flat connections, such as their respective holonomies, or more generally their parallel transports (even along open paths\footnote{When working with transport along open paths we implicitly make a choice of trivialization for the vector bundles associated to either connection.}).

Let $\wp$ be an open path on $C$, the corresponding parallel transport is
\be
	F(\wp) = P\exp\int_\wp \nabla^{na}\,.
\ee
Likewise let $a$ will be an open path on $\tSigma$, the corresponding parallel transport is
\be
	X_a = P\exp\int_a \nabla^{ab}\,.
\ee
These variables obey the usual multiplication rule
\be
	X_a X_b = \left\{\begin{array}{lr}
	X_{ab}\qquad  & \text{end}(a) = \text{beg}(b)\\
	0 & \text{otherwise}
	\end{array}\right.
\ee
where endpoints of paths $a$ and $b$ must match on $\tSigma$, which requires both matching the sheet label $i$ and the log-branch label $N$ of the common endpoint.
For both types of transport, flatness of the connection ensures that these quantities only depend on the relative homotopy classes of the paths $\wp,a$.
Actually this is not entirely correct, due to a small but important subtlety.
More precisely we need to work with \emph{twisted} flat connections as in \cite{Gaiotto:2012rg}. One way to think about this is to take any path $\wp$ (resp. $a$) and consider its \emph{tangent framing lift} to a circle bundle over $C$ (resp. $\tSigma$). Let $\wp,\wp'$ (resp. $a,a'$) be two paths that differ by $k$ units of winding, then
\be\label{eq:sign-twisting}
	F(\wp') = (-1)^k F(\wp)
	\qquad
	X_{a'}= (-1)^k X_a
\ee
For more details about this twisting we refer the reader to \cite{Gaiotto:2012rg}.
We will often omit any mention of the twisting in what follows except where it plays an important role,
but it will be understood that we always work with twisted flat connections on $C,\tSigma$.

In the remainder of this subsection we shall detail how to construct $F(\wp)$ from $X_a$ explicitly. The construction follows the same general strategy as the nonabelianization map of spectral networks \cite{Gaiotto:2012rg}.

\subsubsection{Diagonal connection}
Suppose that $\wp \subset C\setminus \CW$, i.e. the path does not intersect the network.
Then
\be\label{eq:diagonal-transport}
	F(\wp) = D(\wp)
\ee
where we define
\be
	D(\wp) = \sum_{i}\sum_{N\in\IZ} X_{\wp^{(i,N)}} = \sum_{N\in\IZ} D_N(\wp)
\ee
with $\wp^{(i,N)}$ the lift of $\wp$ to sheet $(i,N)$ on $\tSigma$.
If $\partial\wp = x'-x$ we may say that
\be
	\wp^{(i,N)} \in \Gamma_{ii,N,N}(x,x')\,,
\ee
by a self-explanatory extension of notation in (\ref{eq:open-charge-lattice}).
Note that $i=j$ and $N+n=N$ for these paths, in this sense the connection is purely ``diagonal''.

This definition enjoys some obvious properties:
\begin{itemize}
\item $D(\wp)$ only depends on the relative-homotopy class of $\wp$
\item concatenation works as expected $D(\wp)D(\wp')=D(\wp\wp')$ if $\text{end}(\wp)=\text{beg}(\wp')$ on $C$
\end{itemize}

\subsubsection{Detours along $\CE$-walls}
When $\wp$ crosses an $\CE$-wall $p$  at $x$ the formal parallel transport gets corrected by soliton data.
Split $\wp$ into the concatenation $\wp_0\wp_1$ with
\be
	\partial \wp_0 = x-x_0
	\qquad
	\partial \wp_1= x_1-x\,.
\ee
Suppose $p$ is of type $(ij,n)$, and define
\be
\begin{split}
	\Xi_{ij,n}(p) & =  \sum_{a\in \Gamma_{ij,n}(x)} \mu(a) X_a
\end{split}
\ee
then the parallel transport along $\wp$ gets corrected to
\be\label{eq:detour-def}
\begin{split}
	F(\wp)
	& = D(\wp_0) \, e^{\Xi_{ij,n}(p)}  \, D(\wp_1) \,.
\end{split}
\ee
This \emph{detour rule} defines a correction of the diagonal transport in terms of soliton data $\{\mu(a)\}$ on the $\CE$-wall $p$.
If $i\neq j$ this definition reduces to a slight generalization of the usual detour rule of spectral networks, in fact
\be
\begin{split}
	F(\wp) & = \sum_{k,N} X_{\wp^{(k,N)}} +
	\sum_{N} \sum_{a\in\Gamma_{{ij,N,N+n}}} \mu(a) \ X_{\wp_0^{(i,N)}\cdot a\cdot  \wp_1^{(j,N+n)}}
	\\
	& =  D(\wp_0) \, \(1 +  \Xi_{ij,n}(p) \)  \, D(\wp_1) \,.
\end{split}
\ee
However when $i=j$ there are important differences, in fact the correction is much more complicated
\be
	F(\wp) = \sum_{k,N} X_{\wp^{(k,N)}} +
	\sum_{k\geq 0} \frac{1}{k!} \sum_{N} \sum_{{\tiny \begin{array}{c}{a_1 \in\Gamma_{{ii,N,N+n}}} \\ \vdots\\ {a_k \in\Gamma_{{ii,N+(k-1)n,N+kn}}}\end{array}}} \mu(a_1)\cdots \mu(a_k) \ X_{\wp_0^{(i,N)}\cdot a_1\cdots  a_k\cdot  \wp_1^{(j,N+n k)}}
\ee
Note in particular the infinite sum over $k$ and the appearance of fractional coefficients like $1/k!$.
These are novel features that were not encountered in standard spectral networks.
We elaborate on the meaning of these fractional coefficients from the viewpoint of soliton counting in Appendix \ref{app:collinear-vacua}, in connection to  $tt^*$ geometry and the CFIV index of \cite{Cecotti:1992qh}.

\subsubsection{Shift maps for open paths}\label{sec:shift-map}

Although the charge lattice of soliton data (\ref{eq:full-open-charge-lattice}) has infinitely many sectors, it turns out that there is a high degree of symmetry in the soliton data of exponential networks.
In this subsection we introduce two types of maps that relate soliton paths in different charge sectors, they will be important when describing symmetries of soliton data.

The first shift map is denoted by a superscript $(\,\cdot\,)^{(+\ell)}$. It acts by ``transporting'' a soliton path of type $(ij,N,N+k)$ into a path of type $(ij,N+\ell,N+\ell+k)$.
\be\label{eq:shift-map}
\begin{split}
	(+\ell): &\   \Gamma_{ij,N,N+k}(x) \longrightarrow \Gamma_{ij,N+\ell,N+\ell+k}(x)  \\
	(+\ell): &\  a \mapsto a^{(+\ell)}
\end{split}
\ee
Note that $i$ may or may not be equal to $j$, the map applies to both cases.
The map is defined by the following relations
\be\label{eq:shift-map-def}
		a^{(+\ell)} = d^{-1} a c \qquad \& \qquad Z_a = Z_{a^{(+\ell)}} \,,
\ee
where $c,d$ are relative homology classes of type\footnote{The  inverse path $d^{-1}\in\Gamma_{ii, N+\ell, N}(x)$ is obtained by reversing the orientation of $d$, it is the unique relative homology class that concatenates with $d$ to the trivial path.}
\be
	c\in\Gamma_{jj, N+k, N+\ell+k}(x)
	\qquad
	d\in\Gamma_{ii, N, N+\ell}(x) \,.
\ee

This map is simply a way of transporting both endpoints of $a$ from sheets $(i,N), (j,N+k)$ to sheets $(i,N+\ell)$ and $(j,N+k+\ell)$.
The transports are made via $c$ and $d^{-1}$, and the definition implies that the closed cycle $(a^{(+\ell)})^{-1} d^{-1} a c = 0\in H_1(\tSigma,\IZ)$ is trivial.
Together with the condition on the central charge, this ensures that the transport is independent of the specific choice of $c,d$. In fact consider another pair $c',d'$ that satisfy both conditions and define a potentially different new path
\be
	a c' = d' {a'}{} ^{(+\ell)} \qquad \& \qquad Z_a = Z_{{a'}{}^{(+\ell)}}
\ee
Under these assumptions
\be
\begin{split}
	Z_{a^{(+\ell)}\cdot ({a'}{}^{(+\ell)})^{-1}}
	 = Z_{a^{(+\ell)}} - Z_{{a'}{}^{(+\ell)}}
	 = Z_{a} - Z_{a}
	 = 0 \,.
\end{split}
\ee
Now $a^{(+\ell)}\cdot ({a'}{}^{(+\ell)})^{-1}$ is a closed path which belongs to $H_1(\tSigma,\IZ)$. Since the \emph{physical} charge lattice involves a quotient by $H_1(\tSigma,\IZ) / {\rm ker} Z$ (see e.g. \cite{Longhi:2016rjt}) this implies that  $a^{(+\ell)}\cdot ({a'}{}^{(+\ell)})^{-1}$ is trivial, and therefore
\be
	a^{(+\ell)} \simeq {a'}{}^{(+\ell)}\,.
\ee
This proves that the map does not depend on the specific choice of $c,d$.
There is one shift map like (\ref{eq:shift-map}) for each $\ell\in\IZ$, from its explicit form it follows immediately each one is invertible. It is easy to see that they obey the simple additive algebra
\be
	(+\ell) \circ (+k) = (+(k+\ell))\,.
\ee

\bigskip

The second map we wish to define is  denoted by $(\,\cdot\,)^{(i\to j,+\ell)}$, and ``transports'' a path of type $(ii,N,N+k)$ into a path of type $(jj,N+\ell,N+\ell+k)$.
\be\label{eq:shift-map-ii}
\begin{split}
	( i\to j,+\ell): &\   \Gamma_{ii,N,N+k}(x) \longrightarrow \Gamma_{jj,N+\ell,N+\ell+k}(x)  \\
	( i\to j,+\ell): &\  a \mapsto a^{( i\to j, +\ell)}
\end{split}
\ee
The map is defined as follows:
\be\label{eq:shift-map-ii-def}
	{a^{( i\to j, +\ell)} = d^{-1} a c \qquad \& \qquad Z_a = Z_{a^{( i\to j, +\ell)}}}
\ee
where  $c,d$ are now relative homology classes of paths of the following types
\be
	c\in\Gamma_{ij, N+k, N+\ell+k}(x)
	\qquad
	d\in\Gamma_{ij, N, N+\ell}(x) \,,
\ee
By the same argument as above, the definition is independent of the specific choice of $c,d$.
When $i=j$, this map reduces to a special case of the previous one. Indeed together they satisfy the following algebra
\be\label{eq:shift-map-algebra}
\begin{split}
	& (\,\cdot\,)^{(i\to j,+\ell)} \circ (\,\cdot\,)^{(+k)} = (\,\cdot\,)^{(+k)}  \circ (\,\cdot\,)^{(i\to j,+\ell)}   = (\,\cdot\,)^{(i\to j, +(\ell+k))} \\
	& (\,\cdot\,)^{(i\to j,+\ell)} \circ (\,\cdot\,)^{(j\to i,+k)} = (\,\cdot\,)^{(j\to j, +(\ell+k))} \\
	& (\,\cdot\,)^{(i\to i,+\ell)} = (\,\cdot\,)^{(+\ell)}
\end{split}
\ee
In the following we therefore drop the distinction between the two maps and  refer to them collectively as the ``shift-map''.

\subsection{Soliton data from flatness}\label{sec:flatness}

The construction of subsection \ref{sec:nonabel-map} provides a definition of the parallel transport $F(\wp)$ in terms of the abelian transport $X_a$ and of the data of an exponential network $\CW$.
What the construction still lacks is a proof that $F(\wp)$ is the parallel transport of a \emph{flat} connection, in this subsection we study this property.
Following a strategy adopted in \cite{Gaiotto:2012rg} we will show that requiring that $F(\wp)$ only depends on the homotopy class of $\wp$ uniquely fixes all soliton data $\{\mu(a)\}$ on the $\CE$-walls of $\CW$.
When $\wp$ doesn't intersect the network the parallel transport is diagonal as in (\ref{eq:diagonal-transport}), and it follows immediately that $F(\wp)$ only depends on the (twisted) homotopy class of $\wp$.
This property becomes less trivial when $\wp$ intersects the network, and we consider three different cases: moving $\wp$ across a branch point, across the intersection of walls of type $(ij,n)$ and $(jk,m)$ and finally across the intersection of walls of types $(ij,n)$ and $(ji,m)$.\footnote{We omit the study of a fourth case, which is when $\wp$ is deformed across a single $\CE$-wall, in a sort of Reidemeister II move. The proof of homotopy invariance for this case is essentially identical to the one found in \cite{Gaiotto:2012rg}.}

\subsubsection{Branch points}
\begin{figure}[h!]
\begin{center}
\includegraphics[width=0.4\textwidth]{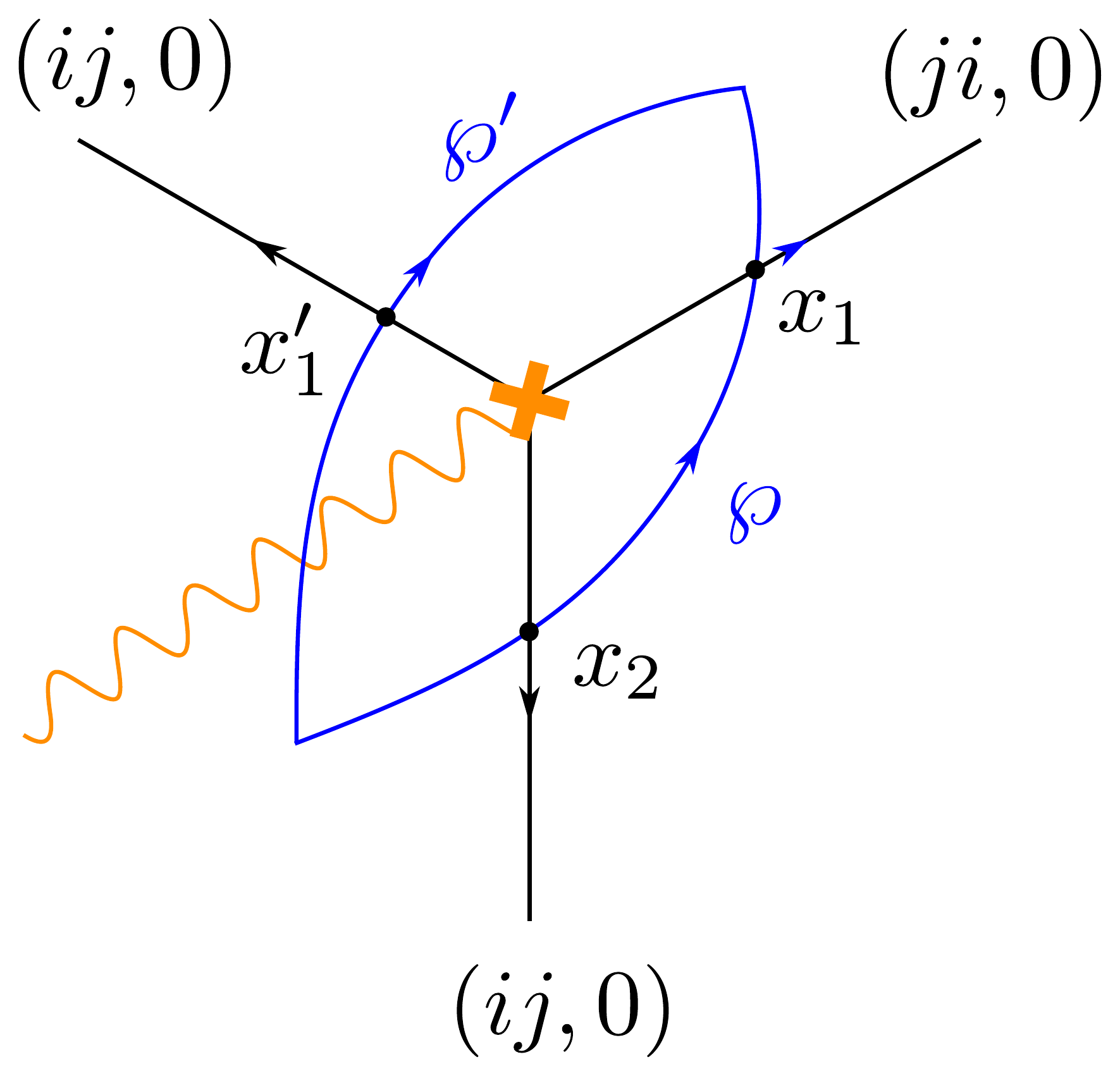}
\caption{}
\label{fig:branch-point-homotopy}
\end{center}
\end{figure}

Consider paths $\wp,\wp'$ across a branch point of type $ij$, as shown in Figure \ref{fig:branch-point-homotopy}.
Splitting each path at the intersections with walls as $\wp=\wp_0\wp_1\wp_2$ and $\wp'=\wp'_0\wp'_1$,
the two parallel transports are simply
\be
\begin{split}
	F(\wp)
	& = D(\wp_0)
	\(1 + \sum_{a\in\Gamma_{ji,0}(x_1)} \mu(a) X_a \)
	D(\wp_1)
	\(1 + \sum_{b\in\Gamma_{ij,0}(x_2)} \mu(b) X_b \)
	D(\wp_2) \\
	& = \sum_N \left[
	D_N(\wp_0)
	\(1 + \sum_{a\in\Gamma_{ji,N,N}(x_1)} \mu(a) X_a \)
	D_N(\wp_1)
	\(1 + \sum_{b\in\Gamma_{ij,N,N}(x_2)} \mu(b) X_b \)
	D_N(\wp_2)
	\right] \\
	& = \sum_N F_N(\wp)
\end{split}
\ee
and
\be
\begin{split}
	F(\wp')
	& = D(\wp'_0)
	\(1 + \sum_{c\in\Gamma_{ij,0}(x'_1)} \mu(c) X_c \)
	D(\wp'_1) \\
	& = \sum_N \left[
	D_N(\wp'_0)
	\(1 + \sum_{c\in\Gamma_{ij,N,N}(x'_1)} \mu(c) X_c \)
	D_N(\wp'_1)
	\right] \\
	& = \sum_N F_N(\wp')
\end{split}
\ee
where we slightly abuse notation for the diagonal piece, which actually contains off-diagonal $ij$ and $ji$ pieces
\be
	D_N(\wp')=X_{\wp^{(ij,N)}} +X_{\wp^{(ji,N)}} + \sum_{k\neq i,j} X_{\wp^{(k,N)}}
\ee
due to the crossing of the $ij$-branch cut.

Demanding homotopy invariance requires setting
\be
	F_N(\wp)=F_N(\wp')\,.
\ee
For each $N$ the analysis of this equation is essentially identical to the one involved in spectral networks, therefore we omit the details and refer the reader to \cite{Gaiotto:2012rg}.
The result is that the soliton data of each $\CE$-wall  consists of an \emph{infinite} tower of soliton paths: for example in $\Gamma_{ji,0}(x_1)$ there is a path $a_N \in \Gamma_{ji,N,N}(x_1)$ for each $N$, which runs from $x_1^{(j,N)}$ into the branch point and then out to $x_1^{(i,N)}$.
The degeneracy of this path is
\be\label{eq:simpleton}
	\mu(a_N) = +1 \,,
\ee
while the degeneracy vanishes for all other charges in the lattice $\Gamma_{ji,0}(x_1)$.
Moreover, central charges of these solitons are all equal because
\be
	Z_{a_N} = \int_{a_N}\log Y(x) \frac{dx}{x}  = \int_{b.p.}^{x_1} (\log y_i+2\pi i N - \log y_j -2\pi i N)\, d\log x
\ee
does not depend on $N$.

\subsubsection{$ij-jk$ joints}\label{sec:ij-jk-one-way-joint}

\begin{figure}[h!]
\begin{center}
\includegraphics[width=0.7\textwidth]{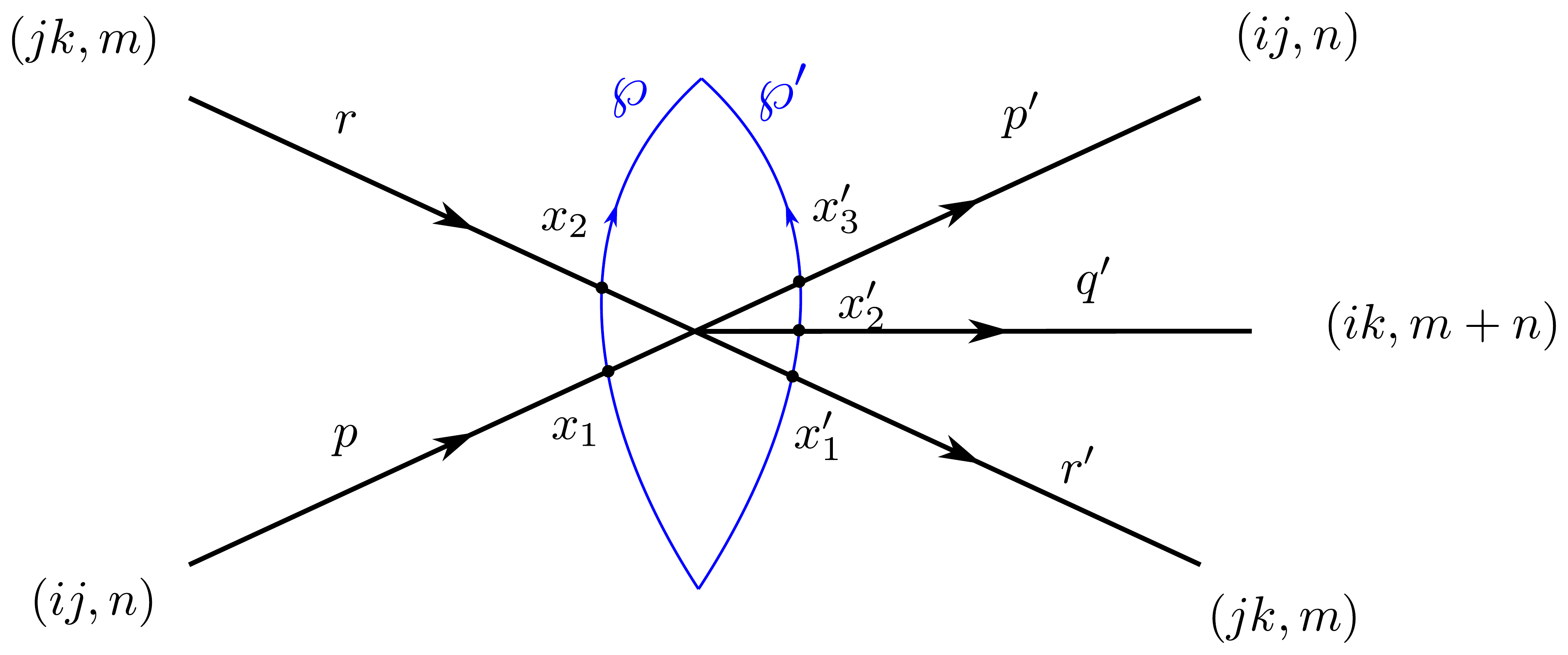}
\caption{}
\label{fig:ij-jk-joint-homotopy}
\end{center}
\end{figure}

Next let us consider homotopic paths $\wp,\wp'$ placed across an intersection of $ij$ and $jk$ $\CE$-walls, as shown in Figure \ref{fig:ij-jk-joint-homotopy}.
Splitting paths at the intersections with walls as $\wp=\wp_0\wp_1\wp_2$ and $\wp'=\wp'_0\wp'_1\wp_2'\wp_3'$, we can express the respective parallel transports read as follows.
For $\wp$ the parallel transport is
\be
\begin{split}
	F(\wp)
	& = D(\wp_0)
	\(1 + \sum_{a\in\Gamma_{ij,n}(x_1)} \mu(a,p) X_a \)
	D(\wp_1)
	\(1 + \sum_{b\in\Gamma_{jk,m}(x_2)} \mu(b,r) X_b \)
	D(\wp_2) \\
	& = \sum_N D_N(\wp) + F(\wp)_{ij, N,N+n} + F(\wp)_{jk, N,N+m} + F(\wp)_{ik, N,N+n+m}\\
	& = \sum_N F_N(\wp)
\end{split}
\ee
where
\be
\begin{split}
	F(\wp)_{ij, N,N+n} &=\sum_{a\in\Gamma_{ij,N,N+n}(x_1)} \mu(a,p) X_{\wp_0^{(i,N)} a \wp_1\wp_2^{(j,N+n)}}
	\\
	 F(\wp)_{jk, N,N+m}&=\sum_{b\in\Gamma_{jk,N,N+m}(x_2)} \mu(b,r) X_{\wp_0\wp_1^{(j,N)}  b \wp_2^{(k,N+m)}}
	\\
	F(\wp)_{ik, N,N+n+m}& = \sum_{a,b} \mu(a,p)\mu(b,r) X_{\wp_0^{(i,N)} a \wp_1^{(j,N+n)} b \wp_2^{(k,N+n+m)}}
\end{split}
\ee
Whereas for $\wp'$ the parallel transport is
\be
\begin{split}
	F(\wp')
	& = D(\wp'_0)
	\(1 + \sum_{b\in\Gamma_{jk,m}(x'_1)} \mu(b,r') X_b \)
	D(\wp'_1)
	\(1 + \sum_{c\in\Gamma_{ik,n+m}(x'_2)} \mu(c,q') X_c \) \\
	&\times
	D(\wp'_2)
	\(1 + \sum_{a\in\Gamma_{ij,n}(x'_3)} \mu(a,p') X_a \)
	D(\wp'_3)\\
	& = \sum_N D_N(\wp') + F(\wp')_{ij, N,N+n} + F(\wp')_{jk, N,N+m} + F(\wp')_{ik, N,N+n+m} \\
	& = \sum_N F_N(\wp')
\end{split}
\ee
where
\be
\begin{split}
	F(\wp')_{ij, N,N+n} & =  \sum_{a\in\Gamma_{ij,N,N+n}(x'_3)} \mu(a,p') X_{{\wp'_0\wp'_1\wp'_2}{}^{(i,N)} a {\wp'_3}{}^{(j,N+n)}}
	\\
	F(\wp')_{jk, N,N+m} & =  \sum_{b\in\Gamma_{jk,N,N+m}(x'_1)} \mu(b,r') X_{\wp'_0{}^{(j,N)}  b \wp'_1\wp'_2\wp'_3{}^{(k,N+m)}}
	\\
	 F(\wp')_{ik, N,N+n+m} & =  \sum_{c\in\Gamma_{ik,N,N+n+m}} \mu(c,q') X_{\wp'_0\wp'_1{}^{(i,N)} c \wp'_2\wp'_3{}^{(k,N+n+m)}}
\end{split}
\ee

Homotopy invariance requires $F_N(\wp)=F_N(\wp')$, and this works out as follows.
First of all it is obvious that $D_N(\wp) = D_N(\wp')$.
The respective pieces with paths connecting vacua $(i,N)$ and $(j,N+n)$ match if
\be\label{eq:ij-jk-joint-a}
	\mu(a,p') = \mu(a,p) \qquad a\in\Gamma_{ij,N,N+n}\qquad  \forall N\,.
\ee
Likewise the respective pieces with paths connecting vacua $(j,N)$ and $(k,N+m)$ match if
\be\label{eq:ij-jk-joint-b}
	\mu(b,r') = \mu(b,r) \qquad b\in\Gamma_{jk,N,N+m}\qquad  \forall N\,.
\ee
Finally the respective pieces with paths connecting vacua $(i,N)$ and $(k,N+n+m)$ match if
\be\label{eq:ij-jk-joint-c}
	\mu(c,q') =  \sum_{\left(ab\sim c \left| \mathop{}^{a\in\Gamma_{ij,N,N+n}}_{b\in\Gamma_{jk,N+n,N+n+m}}\right.\right)}  (-1)^{w(c,ab)} \mu(a,p) \mu(b,r) \qquad c\in\Gamma_{ik,N,N+n+m} \qquad  \forall N\,.
\ee
Thus the soliton data on walls $p',q',r'$ is entirely determined in terms of the soliton data of the incoming walls $p,q$ by requiring flatness of $F(\wp)$.
The equations that describe outgoing soliton data are a natural generalization of the ones found in \cite{Gaiotto:2012rg} and this is no surprise, since the underlying physics is that of the Cecotti-Vafa wall-crossing formula in both cases.
Note however that the interplay of soliton data is already a bit more complicated than we saw for the branch point, since now there are solitons connecting different logarithmic branches that mix with each other.

\subsubsection{$ij-ji$ joints}\label{sec:ij-ji-joints}

\begin{figure}[h!]
\begin{center}
\includegraphics[width=0.9\textwidth]{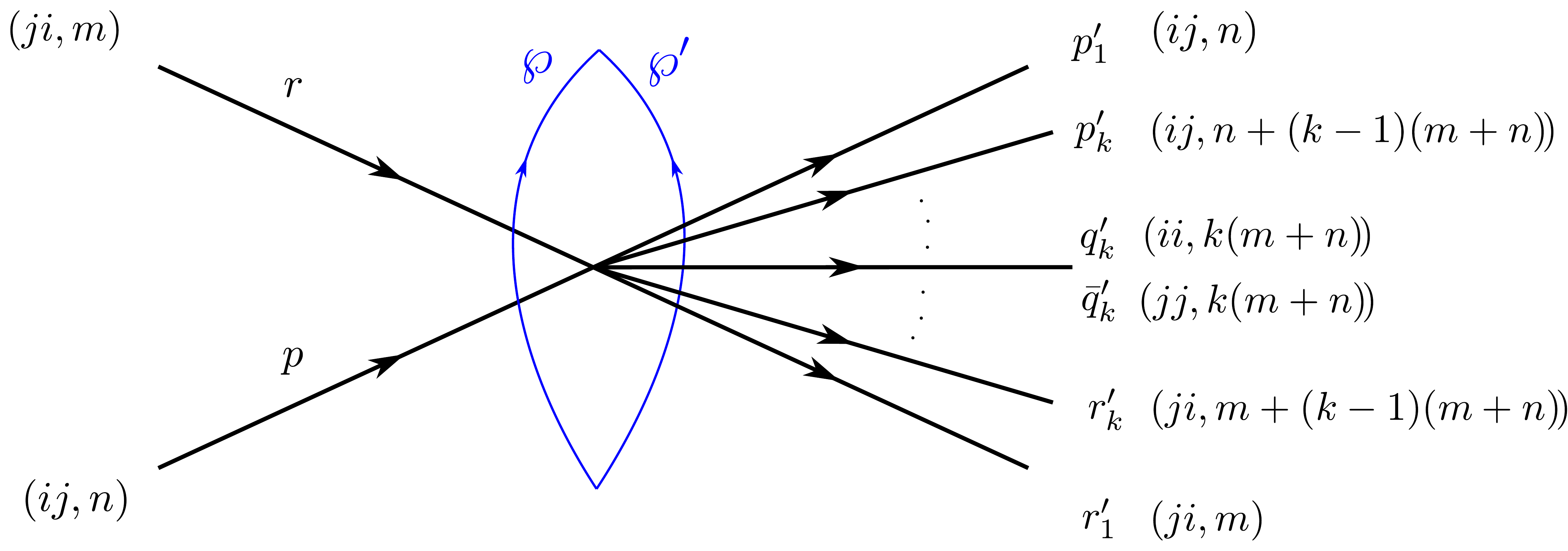}
\caption{Intersecting $\CE$-walls of types $(ij,n)$ and $(ji,m)$ give origin to an infinite family of new walls.}
\label{fig:ij-ji-joint-homotopy}
\end{center}
\end{figure}

Next we consider homotopic paths $\wp,\wp'$ placed across an intersection of $(ij,n)$ and $(ji,m)$ $\CE$-walls called $p$ and $r$, as shown in Figure \ref{fig:ij-ji-joint-homotopy}.
Note that this situation cannot arise in spectral networks, because such walls would always be anti-parallel in that setting. However such intersections can occur in exponential networks, as long as $n \neq - m$.
In the picture we draw two incoming walls $p,r$ and \emph{infinitely many} outgoing walls, organized into four main families: $p'_k, r'_k, q'_k, \bar q'_k$ of types indicated in Figure.
While $p'_k, r'_k$ are all distinct as trajectories,  walls $ q'_k, \bar q'_k$ are all overlapping with each other, but carry different soliton data.
The appearance of these four infinite families of walls will be clarified shortly by studying the requirement of flatness of the parallel transport.

Without further ado, let us study the parallel transport equations.
For the transport along $\wp$ we find
\be
\begin{split}
	F(\wp) & = (1+\Xi_{ij,n}(p)) (1+\Xi_{ji,m}(r)) \\
	& = 1+\Xi_{ij,n}(p) + \Xi_{ji,m}(r) + \Xi_{ij,n}(p) \Xi_{ji,m}(r) \,.
\end{split}
\ee
To lighten notation we suppressed trivial contributions like $D(\wp)$.
For the homotopic path $\wp'$ we obtain
\be
\begin{split}
	F(\wp')
	& = \(\prod_{k\geq 1}^{\nearrow} e^{\Xi_{ji,m+(k-1)(n+m)}(r'_k)}\)
	\(\prod_{k\geq 1}^{} e^{\Xi_{ii,k(n+m)}(q'_k)}  \)
	\\
	& \times
	\(\prod_{k\geq 1}^{} e^{\Xi_{jj,k(n+m)}(\bar q'_k)}  \)
	\(\prod_{k\geq 1}^{\searrow} e^{\Xi_{ij,n+(k-1)(n+m)}(p'_k)} \) \\
	& = \(1 + \sum_{k\geq 1} \Xi_{ji,m+(k-1)(n+m)}(r'_k)\)
	\(\prod_{k\geq 1}^{} e^{\Xi_{ii,k(n+m)}(q'_k)}  \)
	\\
	& \times
	\(\prod_{k\geq 1}^{} e^{\Xi_{jj,k(n+m)}(\bar q'_k)}  \)
	\(1 + \sum_{k\geq 1} \Xi_{ij,n+(k-1)(n+m)}(p'_k) \) \\
	& = \(1 + \Sigma_{ji}\)
	\(1 +  \Sigma_{ii} \)
	\(1 +  \Sigma_{jj} \)
	\(1 +  \Sigma_{ij} \) \\
	& = 1
	+ \Sigma_{ii}
	+ \left[\Sigma_{ji} + \Sigma_{ji}\Sigma_{ii}\right]
	+ \left[\Sigma_{ij} + \Sigma_{ii}\Sigma_{ij}\right]
	+ \left[\Sigma_{jj} + \Sigma_{ji}\Sigma_{ij} + \Sigma_{ji}\Sigma_{ii}\Sigma_{ij}\right]
\end{split}	
\ee
Note that we didn't specify the ordering of factors of types $ii/jj$, this will be justified later on in \ref{sec:factoring-Q} when we show that all such factors actually commute.\footnote{Commutativity of $ii$ factors with $jj$ factors is obvious, because the respective paths don't concatenate. However mutual commutativity of $ii$ factors among themselves is nontrivial (similarly for $jj$ factors).}

For convenience let us introduce
\be\label{eq:Theta-bTheta}
	\Theta:=\Xi_{ij,n}(p)\Xi_{ji,m}(r)\qquad \bTheta := \Xi_{ji,m}(r)\Xi_{ij,n}(p)
\ee
note that the former counts $(ii,n+m)$ solitons, while the latter counts $(jj,n+m)$ solitons.

Comparing the $ii$ components of $F(\wp),F(\wp')$ gives
\be\label{eq:ij-ji-eqs-ii}
\begin{split}
	\Sigma_{ii} & = \Xi_{ij,n}(p)\Xi_{ji,m}(r) =\Theta \,.
\end{split}
\ee
Since we defined
\be
	\Sigma_{ii}  =  \prod_{k\geq 1}^{} e^{\Xi_{k(n+m)}(q'_k)} - 1\,,
\ee
it follows that the soliton data on the $q_k$ walls must be\footnote{
This is easily seen from the identity
\be
	\Sigma_{ii} = e^{x_1+x_2+x_3+\dots} -1 = \Theta = \exp\log(1+\Theta) - 1 = \exp\(\Theta-\frac{\Theta^2}{2} + \frac{\Theta^3}{3} + \dots\) -1\,.
\ee
}
\be\label{eq:Sigma-ii-exp-detour}
\begin{split}
	\Xi_{ii,k(n+m)}(q'_k) &= -\frac{1}{k} \(-\Theta\)^k  \,.
\end{split}
\ee

Comparing the $ij$ components gives
\be
\begin{split}
	\Sigma_{ij} + \Sigma_{ii}\Sigma_{ij} & = \Xi_{ij,n}(p)  \,,
\end{split}
\ee
which implies $\Sigma_{ij} = (1+\Theta)^{-1} \cdot \Xi_{ij,n}(p) $ and therefore
\be\label{eq:ij-ji-eqs-ij}
\begin{split}
	  \Xi_{ij,n+(k-1)(n+m)}(p'_k)  & =  (-\Theta)^{k-1} \cdot \Xi_{ij,n}(p)  =   \Xi_{ij,n}(r) \cdot (-\bTheta)^{k-1} \,.
\end{split}
\ee
Similarly comparing the $ji$ components gives
\be
\begin{split}
	\Sigma_{ji} + \Sigma_{ji}\Sigma_{ii} & = \Xi_{ji,m}(r) \,,
\end{split}
\ee
which leads to
\be\label{eq:ij-ji-eqs-ji}
	\Xi_{ji,m+(k-1)(n+m)}(r'_k)  =   \Xi_{ji,m}(r)  \cdot (-\Theta)^{k-1} =   (-\bTheta)^{k-1} \Xi_{ji,m}(p) \,.
\ee

Turning to the $jj$-component, this vanishes for $F(\wp)$ therefore
\be
\begin{split}
	\Sigma_{jj} + \Sigma_{ji}\Sigma_{ij} + \Sigma_{ji}\Sigma_{ii}\Sigma_{ij} & = 0 \,,
\end{split}
\ee
substituting what we found previously for $\Sigma_{ij}, \Sigma_{ji}$ and $\Sigma_{ii}$ we obtain
\be\label{eq:Sigma-jj}
\begin{split}
	\Sigma_{jj} & = - \Xi_{ji,m}(r)\cdot  (1+\Theta)^{-1} (1+\Theta) (1+\Theta)^{-1} \cdot \Xi_{ij,n}(p)  \\
	& = - \Xi_{ji,m}(r)\cdot  \(\sum_{k\geq 0} (-\Theta)^k\) \cdot \Xi_{ij,n}(p)  \\
	& =  \sum_{k\geq 1} \(- \bTheta\)^{k}  = \frac{- \bTheta}{1+\bTheta}
\end{split}
\ee
Since we defined
\be
	\Sigma_{jj}  =  \prod_{k\geq 1}^{} e^{\Xi_{k(n+m)}(\bar q'_k)} - 1\,,
\ee
it follows that the soliton data on walls $\bar q'_k$ must be\footnote{To see this use the identity
\be
	\Sigma_{jj} = e^{\bar x_1+\bar x_2+\bar x_3+\dots} -1 = \frac{-\bTheta}{1+\bTheta}
	= \exp\(- \log(1+\bTheta)\) - 1 = \exp\(-\bTheta+\frac{\bTheta^2}{2} - \frac{\bTheta^3}{3} + \dots\) -1
\ee
}
\be\label{eq:Sigma-jj-exp-detour}
	\Xi_{jj,k(n+m)}(\bar q'_k) = \frac{1}{k} \(-\bTheta\)^k \,.
\ee

Once again we find that the requirement of flatness completely fixes the soliton data on the outgoing walls (the four infinite families $p'_k, r'_k, q'_k, \bar q'_k$) in terms of the soliton data of the incoming ones.
Next we wish to make some considerations about symmetry.
Reversing the orientation of $\wp,\wp'$ would give similar equations, related to the ones above by the replacements
\be\label{eq:symmetry-ij-ji}
	i\leftrightarrow j,
	\quad
	\Theta\leftrightarrow \bTheta,
	\quad
	p\leftrightarrow r,
	\quad
	q'\leftrightarrow \bar q',
	\quad
	p'_k\leftrightarrow r'_k,
	\quad
	m \leftrightarrow n
	\,,
\ee
and the flip of certain signs, to which we will return shortly.
Of course, consistency of nonabelianization requires that the soliton data of the outgoing $\CE$-walls is a fixed function of $\Xi_{ij,n}(p), \Xi_{ji,m}(r)$, regardless of how we choose $\wp,\wp'$.
It is quite amusing that such a symmetry is indeed realized on the soliton data of the single walls, whereas it is completely hidden when looking at components of the parallel transport (for example $\Sigma_{ii}$ and $\Sigma_{jj}$ turned out to be quite different, but the symmetry is nevertheless manifest in (\ref{eq:Sigma-ii-exp-detour}) and (\ref{eq:Sigma-jj-exp-detour})).
The realization of this symmetry on the soliton data of single walls is rather subtle, and relies crucially on the exponential nature of the detour rule (\ref{eq:detour-def}).
The fact this symmetry is realized is a highly nontrivial check on our construction of the nonabelianization map for exponential networks.
\footnote{
To be more precise, there is a perfect symmetry between (\ref{eq:Sigma-ii-exp-detour}) and (\ref{eq:Sigma-jj-exp-detour}) only after we take into account the presence of certain relative signs.
The signs seems to pose a problem because reversing the orientation of paths $\wp,\wp'$ one gets the following equations $\Xi_{ii,k(n+m)}(q'_k) = \frac{1}{k} \(-\Theta\)^k $ and $\Xi_{jj,k(n+m)}(\bar q'_k) = -\frac{1}{k} \(-\bTheta\)^k $.
Superficially, this appears to clash with (\ref{eq:Sigma-ii-exp-detour}) and (\ref{eq:Sigma-jj-exp-detour}), but this is not so.
The sign comes from the small arcs that arise in the concatenation of $ii$ (resp. $jj$) solitons among themselves, due to the fact that we are working with a \emph{twisted} flat connection subject to (\ref{eq:sign-twisting}).
As a result, carefully unpacking either equation for $\Xi_{ii,k(m+n)}$ and $\Xi_{jj,k(m+n)}$ gives identical expressions for $\mu(c),\mu(\bar c)$'s on $q'_k, \bar q'_k$ in terms of $\mu(a),\mu(b)$ on $p,r$.
}

A striking feature of the soliton data carried by walls $q'_k$ and $q'_k$ is the presence of \emph{fractional} degeneracies with denominator $k$. In view of the physical interpretation of these degeneracies in terms of soliton counting, the presence of rational numbers may appear puzzling.
In fact a similar situation is encountered in the context of $tt^*$ geometries of systems with \emph{collinear vacua}, where the $1/k$ factors are associated with the presence of multi-particle states \cite{Cecotti:2010qn}. More details on the physical origin of fractional degeneracies can be found in Appendix \ref{app:collinear-vacua}.

\subsubsection{$ii$-type walls as special phases}\label{subsec:periodic-ii}
\begin{figure}[h!]
\begin{center}
\includegraphics[width=0.3\textwidth]{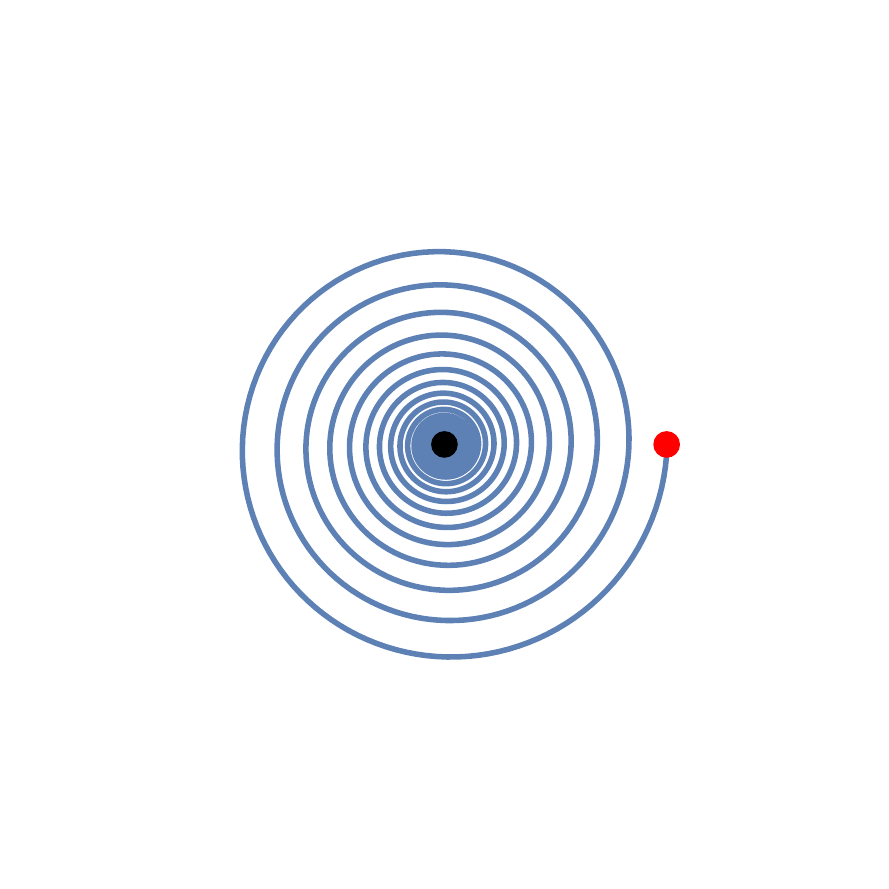}
\includegraphics[width=0.3\textwidth]{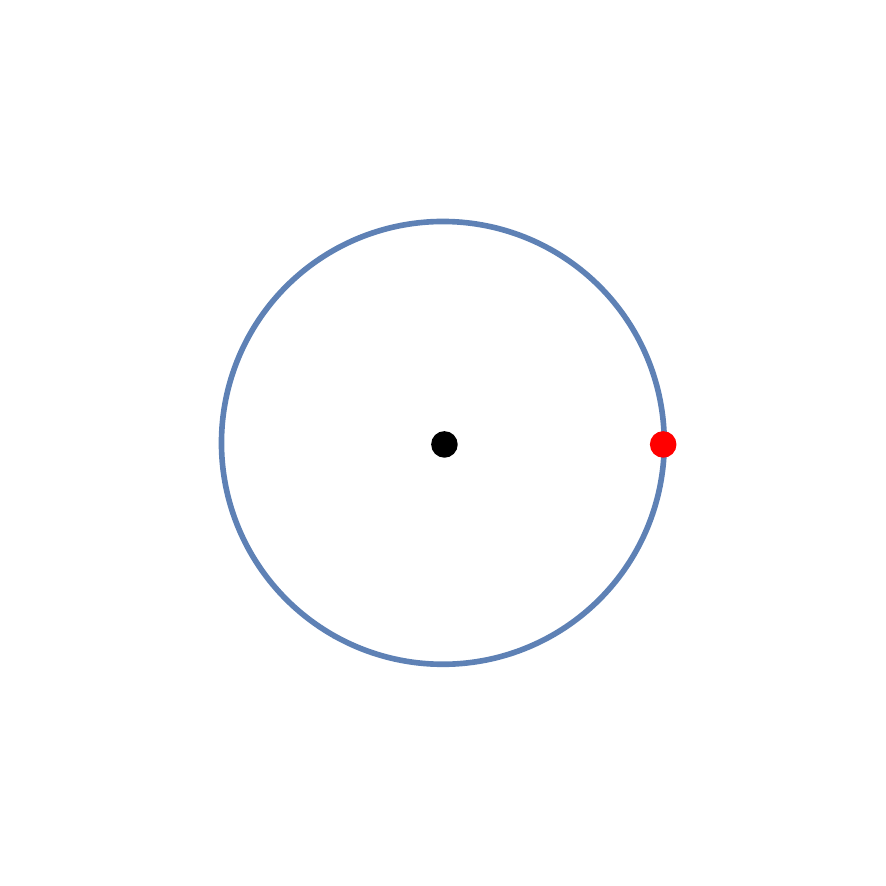}
\includegraphics[width=0.3\textwidth]{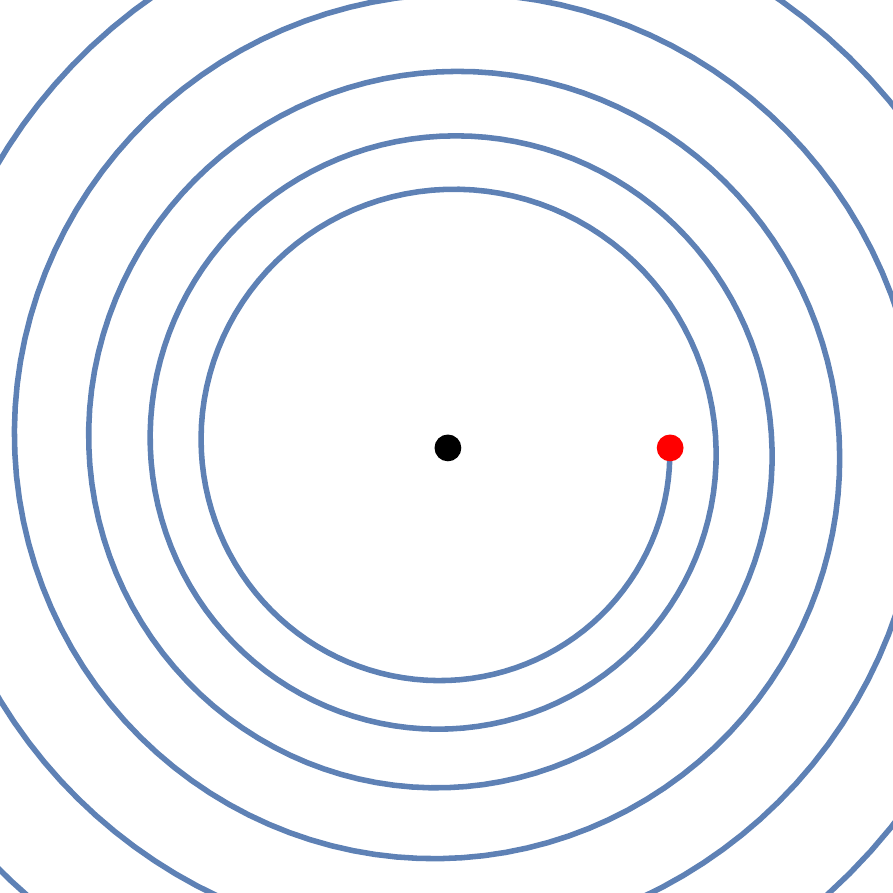}
\caption{Behavior of an $ii$-wall for $\vartheta<0$ (left), $\vartheta=0$ (center), and $\vartheta>0$ (right). The red dot indicates the starting point of integration $x_0$, which for simplicity is kept constant in $\vartheta$, the black dot indicates $x=0$. This behavior is universal, it does not depend on the geometry of the mirror curve.}
\label{fig:ii-solitons}
\end{center}
\end{figure}

We have seen above that walls of type $ii$ may be generated at junctions of other $\CE$-walls.
These are a special feature of exponential networks, which is not encountered in standard spectral networks, as first noted in \cite{Eager:2016yxd}.
The differential equation describing their geometry takes this form
\be
	2\pi i n \, \frac{d\log x}{d t} = e^{i\vartheta}\,.
\ee
from which it is clear that $ii$-type walls can only exist for logarithmic differentials.
Note that the geometry of $ii$-walls is completely insensitive to the geometry of the mirror curve $\Sigma$ because any dependence on $y_i(x)$ drops out.
Another characteristic property of $ii$-walls worth noting, is the  fact that when $\vartheta=0,\pi$ their shape is exactly circular on $\IC^*_x$. In fact let $x_0$ be the point where the wall originates (the position of the junction between $ij/ji$ walls), then the outgoing $ii$ wall has a trajectory that takes the form
\be
	x = x_0 \,\exp\(  \frac{t\,e^{i\vartheta}}{2\pi i n} \)\,,\qquad t\in \IR^+ \,.
\ee
Near the phase $\vartheta=0$ the behavior of an $ii$-wall is shown in Figure \ref{fig:ii-solitons}.

This behavior of $ii$ paths has some implications for nonabelianization, which will be important later.
Fix a path $\wp$ that crosses an $ii$ wall at the phase $\vartheta=0$, its parallel transport will pick up contributions from the solitons of type $ii$ described in (\ref{eq:Sigma-ii-exp-detour}), as well as from their extension around $x=0$, due to the periodic geometry of the wall. Let $a$ be an $ii$ soliton sourced at the junction, and let $a^{(+k)}$ be its continuation along the wall going $k$ times counterclockwise around the puncture. The parallel transport $F(\wp)$ will pick up detour contributions of the form $e^{\Xi}$ with
\be
	\Xi = \sum_{k\in \IN} \sum_{N\in \IZ} \sum_{\ell\geq 1} \sum_{a \in \Gamma_{ii,N,N+\ell(m+n)}} \mu(a) X_{a^{(k)}}\,.
\ee
Also note that there may be branch cuts both of logarithmic and square-root type ending on the puncture, which must be crossed by $a$ each time it is extended to get $a^{(k)}$.
This implies that $a^{(k)}$ will be charged in a different lattice than $a$, for example in $\Gamma_{jj,N',N'+\ell'(m+n)}$ or in $\Gamma_{ii,N',N'+\ell'(m+n)}$ for some $N', \ell'$ that depend on $k$. We will see an explicit instance of this below in Section \ref{sec:C3}.

Another important feature of periodic walls is that they support a novel kind of BPS solitons.
Fix some point $x$ along the periodic wall, there can be a path starting from (the lift of $x$ to) sheet $(i,N)$ and ending on sheet $(i,N+k)$ which simply goes around the puncture a number of times, without ever going through a branch point.
Such a path is a solution to the geometric BPS equations (see the discussion in Section \ref{sec:spectral-networks-review}) with boundary conditions fixed by the two vacua $(i,N)$ and $(i,N+k)$.
Since these are part of the soliton spectrum on the wall for the special phases $\vartheta=0,\pi$ their contribution to nonabelianization can be mostly neglected.
However, they are expected to appear in concatenations of open paths into closed paths (in the context of $\CK$-wall jumps, see below), as we will see in Section \ref{5dBPS}.
This the question arises of how to count them properly, if nonabelianization is (mostly) blind to them (at least when $\vartheta\neq 0,\pi$).
We will show in Section \ref{sec:ii-2-way-walls} that in fact contributions of $ii$-streets (and therefore of their solitons) always drop out from $\CK$-wall computations that capture 5d BPS states.

\subsection{Shift symmetry}\label{sec:shift-sym}
In subsection \ref{sec:flatness} we determined all soliton data on an exponential network $\CW(\vartheta)$ for a generic phase $\vartheta$.
A novel feature compared to standard spectral networks is that $\CE$-walls typically carry infinitely many solitons, significantly increasing the complexity of computations.
In this subsection we argue that the soliton data carried by each $\CE$-wall enjoys a certain symmetry, which effectively organizes infinitely many solitons into (often) finitely many orbits of the symmetry group. The presence of this symmetry will greatly simplify computations in concrete examples.

The soliton data of $\CW$ is determined combinatorially, starting with the solitons on primary walls given in (\ref{eq:simpleton}),  and proceeding with the soliton data on descendant walls as determined by the joint equations of subsections \ref{sec:ij-jk-one-way-joint} and \ref{sec:ij-ji-joints}.
It is already clear from (\ref{eq:simpleton}) that the soliton data of primary walls enjoys a shift symmetry.
We are going to show that this symmetry is preserved by the joint equations of both types $ij-jk$ and $ij-ji$.
Let us first give a precise definition of the symmetry, and then come back to its proof.

Recall that we defined two slighly different types of shift maps, in (\ref{eq:shift-map}) and (\ref{eq:shift-map-ii})
The shift map of the first type relates solitons solitons which are in different charge lattices of the \emph{same} $\CE$-wall. We claim that it is a symmetry of its soliton data in the following sense
\be\label{eq:shift-sym}
	\mu(a) = \mu(a^{(+k)}) \qquad \forall k\in\IZ \,.
\ee
The second type of shift map applies to $\CE$-walls of type $ii$, and relates them to solitons valued in charge lattices of \emph{different} $\CE$-walls.
We claim that it is a symmetry of tis soliton data in the following sense
\be\label{eq:shift-sym-ii}
	\mu(a) = \mu(a^{(i\to j, +k)}) \qquad \forall k\in\IZ \,.
\ee
As we learned from the $ij-ji$ joint analysis, the walls of types $(ii,k)$ and $(jj,k)$ are always created together, so if a wall of type $ii$ goes through a point $x\in C$, we also expect the presence of the wall of type $jj$. For this reason, the statement of this symmetry makes sense.

Let us now check these claims.
Consider $\CE$-walls $p,r$ shown in Figure \ref{fig:ij-jk-joint-homotopy}, and assume that they enjoy the shift symmetry (\ref{eq:shift-sym}).
From (\ref{eq:ij-jk-joint-a}) and (\ref{eq:ij-jk-joint-a}) it's clear that likewise walls $p',r'$ must enjoy the same symmetry.
For $q'$ we use (\ref{eq:ij-jk-joint-c})
\be
\begin{split}
	\mu(c,q')
	& = \sum_{\left(ab\sim c \left| \mathop{}^{a\in\Gamma_{ij,N,N+n}}_{b\in\Gamma_{jk,N+n,N+n+m}}\right.\right)}  (-1)^{w(c,ab)} \mu(a,p) \mu(b,r)
	\\
	& = \sum_{\left(ab\sim c \left| \mathop{}^{a\in\Gamma_{ij,N,N+n}}_{b\in\Gamma_{jk,N+n,N+n+m}}\right.\right)}  (-1)^{w(c^{(+\ell)},a^{(+\ell)}b^{(+\ell)})} \mu(a^{(+\ell)},p) \mu(b^{(+\ell)},r)
	\\
	& = \sum_{\left(ab\sim c \left| \mathop{}^{a\in\Gamma_{ij,N+\ell,N+n+\ell}}_{b\in\Gamma_{jk,N+n+\ell,N+n+m+\ell}}\right.\right)}  (-1)^{w(c,ab)} \mu(a,p) \mu(b,r)
	\\
	& \equiv \mu(c^{(+\ell)},q')
\end{split}
\ee
where in the second line we used the shift symmetry for $p,r$ and in the last line we simply relabeled the summation variables by absorbing the $(+\ell)$-shift in the soliton lattices.
This analysis reveals a simple property of the shift symmetry of the first type (\ref{eq:shift-sym}): the product of shift-symmetrc soliton generating functions also enjoys shift symmetry.

Next consider $\CE$-walls $p,r$ shown in Figure \ref{fig:ij-ji-joint-homotopy}, and assume that they enjoy the shift symmetry (\ref{eq:shift-sym}).
From definitions (\ref{eq:Theta-bTheta}) it follows that both $\Theta$ and $\bTheta$ are shift-symmetric in the sense of (\ref{eq:shift-sym}), because the symmetry is preserved by the product, as we have just seen.
This also applies to powers $\Theta^k$ and $\bTheta^k$, therefore from (\ref{eq:Sigma-ii-exp-detour}) and (\ref{eq:Sigma-jj-exp-detour}) we see that soliton data on $q'_k,\bar q'_k$ must enjoy the shift symmetry as well.
Likewise (\ref{eq:ij-ji-eqs-ij}) and (\ref{eq:ij-ji-eqs-ji}) are shift-symmetric because they are products of symmetric generating functions, ths proves that soliton data of walls $p'_k, r'_k$ enjoyes the symmetry of the first type (\ref{eq:shift-sym}) .

The last statement we need to prove is that soliton data on walls $q'_k$ and $\bar q'_k$ are related by the symmetry of the second type (\ref{eq:shift-sym-ii}).
According to their expressions (\ref{eq:Sigma-ii-exp-detour}) and (\ref{eq:Sigma-jj-exp-detour}), all we need to show is that $\Theta$ and $\bTheta$ are related by $(\ref{eq:shift-sym-ii})$.
This follows by direct inspection of their definitions:
\be
\begin{split}
	\Theta & = \Xi_{ij,n}(p)\Xi_{ji,m}(r) \\
	& = \sum_{N} \sum_{{\scriptsize \begin{array}{c} a\in\Gamma_{ij,N,N+n}\\ b\in\Gamma_{ji,N+n,N+n+m} \end{array}}} \mu(a,p)\mu(b,r) X_{ab} \\
	\bTheta & = \Xi_{ji,m}(r) \Xi_{ij,n}(p) \\
	& = \sum_{N} \sum_{{\scriptsize \begin{array}{c} \bar b\in\Gamma_{ji,N,N+m} \\   \bar a\in\Gamma_{ij,N+m,N+m+n}   \end{array} }} \mu(\bar a,p)\mu(\bar b,r) X_{\bar b\bar a}
\end{split}
\ee
The map $(\,\cdot\, )^{(i\to j,+0)}$ can be described explicitly as follows
\be
	(ab)^{(i\to j,+0)} = b^{(-n)}a^{(+m)}
\ee
This satisfies the requirement that $Z_{(ab)^{(i\to j,+0)}} = Z_{ab}$, since
\be
	Z_{(ab)^{(i\to j,+0)}} = Z_{b^{(-n)}a^{(+m)}} = Z_{b^{(-n)}} + Z_{a^{(+m)}} = Z_b+Z_a = Z_{ab}\,.
\ee
To see that it maps $\Theta$ into $\bTheta$ we should collect all terms within the same relative homology class
\be	
	\Theta
	= \sum_{N} \sum_{c\in \Gamma_{ii,N,N+m+n}}  \theta(c) X_c \,,
	\qquad
	\bTheta
	= \sum_{N} \sum_{\bar c\in \Gamma_{jj, N,N+m+n}}  \bar \theta(\bar c) X_{\bar c }\,,
\ee
where
\be\label{eq:theta-c}
	\theta(c) = \sum_{{\scriptsize\left(   \begin{array}{c} a\in\Gamma_{ij,N,N+n}\\ b\in\Gamma_{ji,N+n,N+m+n} \end{array}  \Big|  a^{}b^{}\sim c \right)}} \mu(a,p)\mu(b,r) \,,
\ee
\be\label{eq:-bar-theta-bar-c}
	\bar \theta(\bar c) = \sum_{{\scriptsize\left(   \begin{array}{c} \bar a\in\Gamma_{ij,N+m,N+m+n}\\ \bar b\in\Gamma_{ji,N,N+m} \end{array}  \Big|  \bar b \bar a \sim \bar c \right)}} \mu(\bar a,p)\mu(\bar b,r)  \,.
\ee
Now if we fix a $\bar c$ such that
\be\label{eq:c-cbar}
	\bar c = c^{(i\to j,+0)}
\ee
then there is a 1:1 map between
\be
	\left(   \begin{array}{c} a\in\Gamma_{ij,N,N+n}\\ b\in\Gamma_{ji,N+n,N+m+n} \end{array}  \Big|  a^{}b^{}\sim c \right)
	\leftrightarrow
	\left(   \begin{array}{c} \bar a\in\Gamma_{ij,N+m,N+m+n}\\ \bar b\in\Gamma_{ji,N,N+m} \end{array}  \Big|  \bar b \bar a \sim \bar c \right) \,.
\ee
This map is precisely
\be\label{eq:a-bar-b-bar}
	\bar a = a^{(+m)} \,, \qquad \bar b = b^{(-n)}\,.
\ee
This is a bijection because the shift-map is a bijection (it is invertible as discussed in Section \ref{sec:shift-map}).
Furthermore this map also guarantees that $ Z_{c}=Z_{a}+Z_{b}=Z_{a^{(+m)}}+Z_{b^{(-n)}} = Z_{\bar a}+Z_{\bar b} = Z_{\bar c}$.
The path $b^{(-n)} a^{(+m)}$ is  a soliton with charge in $\Gamma_{jj,N,N+m+n}$, i.e. it is valued in the same charge lattice as $\bar c$, and since its central charge is the same as that of $\bar c$, the two effectively coincide.
Now since $\mu(a^{(-m)},p)=\mu(a,p)$ and $\mu(b^{(+n)},r) = \mu(b,r)$ it follows from substitution of (\ref{eq:a-bar-b-bar}) into (\ref{eq:-bar-theta-bar-c}) that $\theta(c) = \theta(\bar c)$ when $c,\bar c$ are related as in (\ref{eq:c-cbar}).
This proves the statement that the soliton data of $\CE$-walls $q'_k,\bar q'_k$ (with same label $k$) are related by the second type of shift-symmetry (\ref{eq:shift-sym-ii}).

\subsection{Jumps of the nonabelianization map}\label{sec:Kwall}
The nonabelianization map of subsection \ref{sec:nonabel-map} is defined using the data of an exponential network $\CW$, which is defined by the equations developed in \ref{sec:flatness}.
In the derivation of these rules we always assumed that $\CW(\vartheta)$ is \emph{generic}, in the sense that there are no degenerate walls, except for $ii$ and $jj$ walls created at $ij-ji$ joints.
The soliton degeneracies carried by each wall are then determined combinatorially, and they depend essentially on the topology of $\CW$, rather than its geometry.

At certain values of the phase $\vartheta$ the network can however become degenerate, as observed initially in \cite{Gaiotto:2012rg} for spectral networks, and later in \cite{Eager:2016yxd} for exponential networks.
Let the critical phase be denoted by $\vartheta_c$, the critical network $\CW(\vartheta_c)$ contains pairs of $\CE$-walls of \emph{opposite types} that overlap partially or entirely, as well as (possibly) regular one-way walls.
We refer to these double-walls as \emph{two-way streets}.
$\CW(\vartheta_\mathrm{c})$ admits two natural resolutions into generic networks, obtained by perturbing the phase to $\vartheta_c^\pm = \vartheta_c \pm \epsilon$. Examples of these resolutions are depicted in Figures \ref{fig:2-way-ij-ji-jump} and \ref{fig:2-way-ii-jump}.

Since the topologies of $\CW({\vartheta_c^\pm})$ are different, the nonabelianization map that defines $F(\wp)$ must jump at $\vartheta_c$. The are known as \emph{$\CK$-wall jumps}.
Physically they are a manifestation of the mixing of BPS solitons supported on the codimension-2 defect, and the BPS spectrum of the bulk theory \cite{Gaiotto:2012rg} (also see Section \ref{sec:phys}).
We shall study this phenomenon by computing the jump in $F(\wp)$ for a path $\wp$ crossing some two-way street $p$.
Again we will find many similarities with spectral networks, the main novelty comes from the fact that each wall carries now infinitely many solitons.
A more analogous situation is the one of ADE spectral networks \cite{Longhi:2016rjt}, where the soliton data on each wall was found to come in representations of a Weyl-type symmetry.
In the case of exponential networks this is replaced by the shift symmetry studied in subsection \ref{sec:shift-sym}.
The $\CK$-wall jumps of exponential networks can indeed be described in a similar way to those of ADE networks \cite{Longhi:2016rjt}.
We will briefly describe the salient points, the interested reader should have no difficulties recovering the fine details by translating the treatment on ADE networks into the present setting.

\subsubsection{$ij$ 2-way walls}\label{sec:ij-ji-2-way-jumps}
\begin{figure}[h!]
\begin{center}
\includegraphics[width=0.75\textwidth]{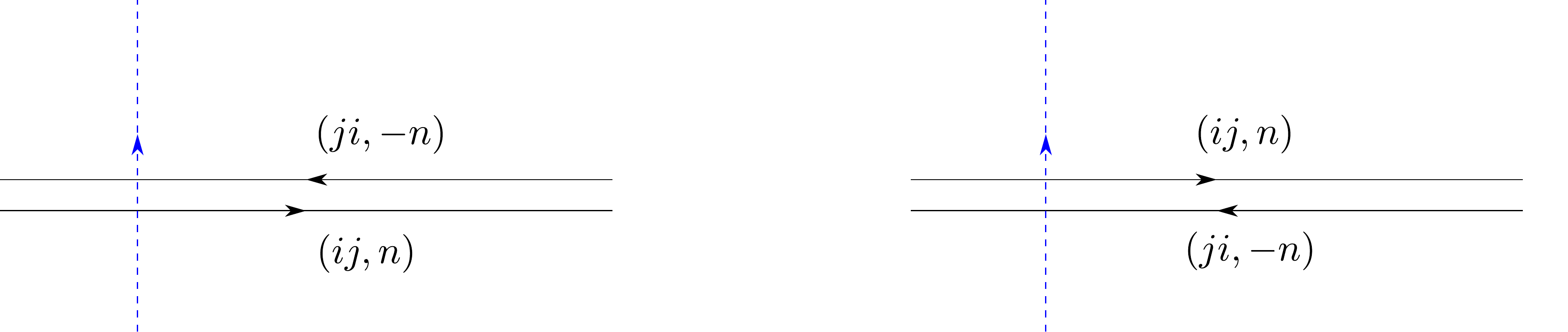}
\caption{Left: the path $\wp$ (dashed) crossing a two-way street of type $(ij,n)/(ji,-n)$ before the critical phase (American resolution). Right: after the critical phase (British resolution).}
\label{fig:2-way-ij-ji-jump}
\end{center}
\end{figure}

Consider a path $\wp$ crossing a two-way street $p$ of type $(ij,n)/(ji,-n)$, see Figure \ref{fig:2-way-ij-ji-jump}.

Let us split $\wp$ into $\wp_+\wp_-$, consisting respectively of the piece before the intersection with the $\CE$-wall, and the piece after the intersection.
Before the critical phase the parallel transport is
\be
\begin{split}
	F(\wp,\vartheta_c^{-})_{ii,N,N}
	& = X_{\wp^{(i,N)}_+}
	\left( 1 + \sum_{a\in\Gamma_{ij,N,N+n}} \sum_{b\in\Gamma_{ji,N+n,N}} \mu^{-}(a) \mu^{-}(b) X_{a} X_{b} \right)
	X_{\wp^{(i,N)}_-}
	\\
	F(\wp,\vartheta_c^{-})_{ij,N,N+n}
	& = X_{\wp^{(i,N)}_+}
	\left( 1 + \sum_{a\in\Gamma_{ij,N,N+n}}  \mu^{-}(a)  X_{a}  \right)
	X_{\wp^{(j,N+n)}_-}
	\\
	F(\wp,\vartheta_c^{-})_{ji,N,N-n}
	& = X_{\wp^{(j,N)}_+}
	\left( 1 + \sum_{b\in\Gamma_{ji,N,N-n}}  \mu^{-}(b)  X_{b}  \right)
	X_{\wp^{(i,N-n)}_-}
	\\
	F(\wp,\vartheta_c^{-})_{jj,N,N}
	& = X_{\wp^{(j,N)}}
\end{split}
\ee
where $\mu^{-}$ denote the soliton degeneracies determined by the topology of the network computed at the phase $\vartheta_c^{-}$.
After the critical phase the parallel transport becomes
\be
\begin{split}
	F(\wp,\vartheta_c^{+})_{ii,N,N}
	& = X_{\wp^{(i,N)}}
	\\
	F(\wp,\vartheta_c^{+})_{ij,N,N+n}
	& = X_{\wp^{(i,N)}_+}
	\left( 1 + \sum_{a\in\Gamma_{ij,N,N+n}}  \mu^{+}(a)  X_{a}  \right)
	X_{\wp^{(j,N+n)}_-}
	\\
	F(\wp,\vartheta_c^{+})_{ji,N,N-n}
	& = X_{\wp^{(j,N)}_+}
	\left( 1 + \sum_{b\in\Gamma_{ji,N,N-n}}  \mu^{+}(b)  X_{b}  \right)
	X_{\wp^{(i,N-n)}_-}
	\\
	F(\wp,\vartheta_c^{+})_{jj,N,N}
	& = X_{\wp^{(j,N)}_+}
	\left( 1 + \sum_{b\in\Gamma_{ji,N,N-n}} \sum_{a\in\Gamma_{jj,N-n,N}}  \mu^{+}(b)\mu^{+}(a)  X_{b} X_{a}   \right)
	X_{\wp^{(j,N)}_-}
\end{split}
\ee
To describe this jump, let us combine the soliton data into a new kind of generating function.
\be\label{eq:QN-def}
\begin{split}
	Q_N(p) =  1 + \sum_{\substack{a\in\Gamma_{ij,N,N+n}(p)\\ b\in\Gamma_{ji,N+n,N}(p)}}\mu(a)\mu(b)\,X_{{\rm cl}(ab)}
\end{split}
\ee
Note that we have omitted the superscript from $\mu^{\pm}$, since it turns out that $\mu^+(a) = \mu^{-}(a)$ (and similarly for $b$) \cite{Gaiotto:2012rg}.
Moreover shift symmetry implies that
\be
	Q_N(p) = Q_{N+1}(p)\,.
\ee
Note that unlike the generating functions that we have considered so far, $Q_N(p)$ depends on variables $X_\gamma$ associated with \emph{closed} homology classes.
By a genericity assumption\footnote{That is, by choosing the complex moduli of $\tSigma$ generically, in such a way that the phases of BPS central charges are nondegenerate.} we can restrict $Q_N(p)$ to depend on a single variable $X_{\gamma_c}$.
More precisely, for different $N$ the generating functions $Q_N(p)$ may contain formal variables associated with different {primitive} homology classes $\gamma_N$ on $\tSigma$. However the periods of the one-form $\lambda \sim Y(x)  \, d\log x$ around these cycles are all equal, because of shift symmetry (see discussion around (\ref{eq:Z-gamma})). We then identify all these cycles by taking a quotient on the homology lattice by $\ker Z$, and denote by $\gamma_c$ the equivalence class in this quotient.
For the physical meaning of this quotient see for example \cite{Gaiotto:2009hg, Longhi:2016rjt} and references therein.

To the two-way street $p$ we associate a set of integers $\alpha_{{\gamma}}(p)$ by factorizing $Q_N(p)$ as
\be\label{eq:Q_p_gamma_c}
	Q_N(p) = \prod_{k=1}^{\infty}\big(1 + X_{{k{\gamma}_{c}}}\big)^{\alpha_{{k{\gamma_{c}}}}(p)}, \quad \alpha_{{\gamma}}(p)\in\IZ\,.
\ee
Then a standard argument (see \cite{Gaiotto:2012rg}) shows that the jump of the parallel transport can be cast into the following form
\be\label{eq:K-wall-ij-ji-2-way}
	F(\wp,\vartheta_c^+) = \CK \(F(\wp,\vartheta_c^-)\)
\ee
where $\CK$ is a change of variables, provisionally defined as
\be
\begin{split}
	\CK(X_{\wp^{(i,N)}})
	& = X_{\wp^{(i,N)}} \, \prod_{k=1}^{\infty}\big(1 + X_{{k{\gamma}_{c}}}\big)^{- \alpha_{{k{\gamma_{c}}}}(p)}
	\\
	\CK(X_{\wp^{(j,N)}})
	& = X_{\wp^{(j,N)}} \, \prod_{k=1}^{\infty}\big(1 + X_{{k{\gamma}_{c}}}\big)^{\alpha_{{k{\gamma_{c}}}}(p)}
\end{split}	
\ee
while leaving other variables unchanged.

For later convenience, note that this jump can be written in a slightly different way.
Let $p$ be a two-way street formed by walls of type $(ij,n) / (ji,-n)$, the {canonical lift} of $p$ is the formal sum
\be\label{eq:canonical-lift}
	\pi^{-1}(p) =  \sum_{N\in\IZ} \(+ p_{(j,N+n)} - p_{(i,N)} \) \,.
\ee
Here $p_{(i,N)}$ is the lift of $p$ to sheet $(i,N)$ of $\tSigma$ and the positive / negative sign denotes the orientation of the lift, relative to that of the underlying $(ij,n)$ $\CE$-wall.
Since $\langle \pi^{-1}(p) ,  \wp^{(i,N)} \rangle = -1 = - \langle \pi^{-1}(p) ,  \wp^{(j,N)} \rangle$ for any $N$, the change of variables $\CK$ can be cast into the following more compact form
\be \label{eq:universal-K-wall-substitution}
	\CK(X_{a}) = X_{a} \, \prod_{k=1}^{\infty}\big(1 + X_{{k{\gamma}_{c}}}\big)^{\alpha_{ k \gamma}(p) \langle  \pi^{-1}(p) , a \rangle} \,.
\ee

\subsubsection{$ii$ 2-way walls}\label{sec:ii-2-way-walls}

Next we consider the $\CK$-wall jump of $F(\wp)$ for a path crossing a 2-way street of type $ii$.
Such a 2-way street is made of distinct one-way $\CE$-walls of opposite types and orientations, such as $(ii,\ell) / (ii,-\ell)$.
Let us focus for now on a single $\CE$-wall of type $(ii,\ell)$: this must be generated at a joint of type $ij-ji$, therefore it must always come together with an infinite family of walls of type $(ii,k(m+n))$, which overlap entirely.
In addition to this, there is another infinite family of walls of type $(jj,k(m+n))$ that also overlaps with the former. All walls in these two families have the same shape and orientation.
Going back to the two-way street, it consists of two copies of this setup: there is an infinite family of $\CE$-walls $(ii,k(m+n))/(jj,k(m+n))$  all running in the same direction, as well as an infinite family of walls of types $(ii,-k(m+n))/(jj,-k(m+n))$ that runs in the opposite direction.

\begin{figure}[h!]
\begin{center}
\includegraphics[width=0.75\textwidth]{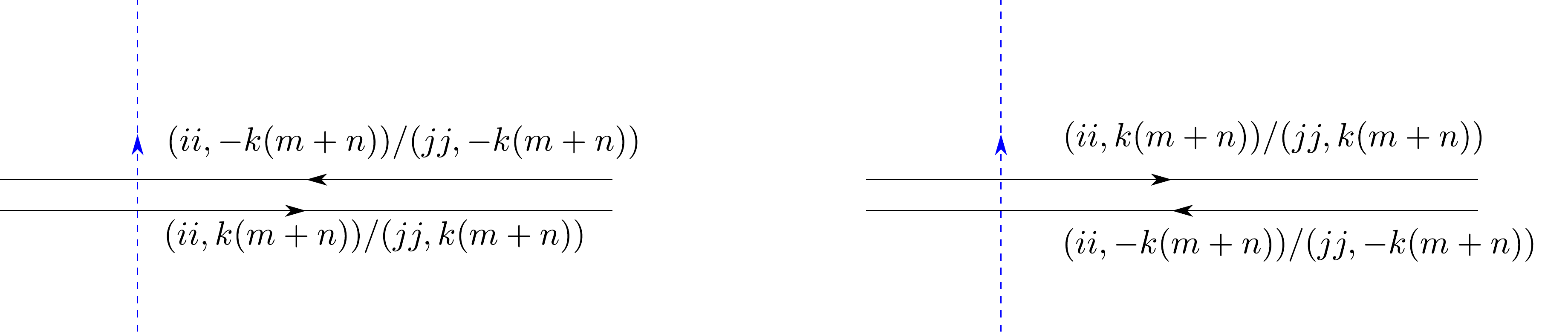}
\caption{Left: the path $\wp$ (dashed) crossing a two-way street of type $ii$ before the critical phase (American resolution). Right: after the critical phase (British resolution).}
\label{fig:2-way-ii-jump}
\end{center}
\end{figure}

Now let  $\wp$ be a path crossing a two-way street $p$ of type $(ii,n)/(ii,-n)$ as shown in Figure \ref{fig:2-way-ii-jump}.
Before the critical phase, the parallel transport is
\be
\begin{split}
	F(\wp,\vartheta_c^{-})_{ii,N,N+\ell(m+n)}
	& = X_{\wp^{(i,N)}_+}
	\(\prod_{\ell\geq 1} e^{\Xi^{-}_{ii,N,N+\ell(m+n)}} \)
	\(\prod_{\ell'\geq 1} e^{\Xi^{-}_{ii,N,N-\ell'(m+n)}} \)
	X_{\wp^{(i,N+\ell(m+n))}_-}
	\\
	F(\wp,\vartheta_c^{-})_{jj,N,N+\ell(m+n)}
	& = X_{\wp_+^{(j,N)}}
	\(\prod_{\ell\geq 1} e^{\Xi^{-}_{jj,N,N+\ell(m+n)}} \)
	\(\prod_{\ell'\geq 1} e^{\Xi^{-}_{jj,N,N-\ell'(m+n)}} \)
	X_{\wp^{(j,N+\ell(m+n))}_-}
	\\
	F(\wp,\vartheta_c^{-})_{ij,N,N'}
	& = 0
	\\
	F(\wp,\vartheta_c^{-})_{ji,N,N'}
	& = 0
\end{split}
\ee
After the critical phase the parallel transport becomes
\be
\begin{split}
	F(\wp,\vartheta_c^{+})_{ii,N,N+\ell(m+n)}
	& = X_{\wp^{(i,N)}_+}
	\(\prod_{\ell'\geq 1} e^{\Xi^{+}_{ii,N,N-\ell'(m+n)}} \)
	\(\prod_{\ell\geq 1} e^{\Xi^{+}_{ii,N,N+\ell(m+n)}} \)
	X_{\wp^{(i,N+\ell(m+n))}_-}
	\\
	F(\wp,\vartheta_c^{+})_{jj,N,N+\ell(m+n)}
	& = X_{\wp_+^{(j,N)}}
	\(\prod_{\ell'\geq 1} e^{\Xi^{+}_{jj,N,N-\ell'(m+n)}} \)
	\(\prod_{\ell\geq 1} e^{\Xi^{+}_{jj,N,N+\ell(m+n)}} \)
	X_{\wp^{(j,N+\ell(m+n))}_-}\,.
	\\
	F(\wp,\vartheta_c^{+})_{ij,N,N'}
	& = 0
	\\
	F(\wp,\vartheta_c^{+})_{ji,N,N'}
	& = 0
\end{split}
\ee
Note that the only differences between these expressions are the replacement of $\Xi^{-}_{(\cdots)} \to \Xi^{+}_{(\cdots)}$ and the permutation of the infinite products in parentheses.

Now we would like to ask if the two expressions can be related in some way like (\ref{eq:K-wall-ij-ji-2-way}).
First of all, we should ask whether the parallel transport changes at all going from $\vartheta_c^-$ to $\vartheta_c^+$.
Recall from subsection \ref{sec:ij-ji-2-way-jumps} that soliton data on one-way streets of type $ij/ji$ remains the same, i.e. $\mu^+(a) = \mu^-(a)$. Then recall from \ref{sec:ij-ji-joints} that all data on streets of type $ii/jj$ is determined entirely in terms of the $ij/ji$ soliton data of the walls that generate the joint.
Since the $ij / ji$ one-way data is unchanged, we conclude that the $ii/jj$ outgoing one-way data is also unchanged, implying that
$\Xi^+_{(\cdots)} = \Xi^-_{(\cdots)}$ in all the expressions.
Finally, a simple computation based on shift symmetry reveals that the two products in parentheses always commute with each other (we also provide a proof in \ref{app:q-commute}).
Since the only change in the parallel transport is the permutation of the two infinite products, it follows that the parallel transport of a path $\wp$ crossing only an $ii/jj$ two-way street does not change at all when the network jumps.

Going back to the main question, we can still describe the (non-)jumping of the formal parallel transport by a formula like (\ref{eq:K-wall-ij-ji-2-way}), by simply defining
\be
\begin{split}
	\CK(X_{\wp^{(i,N)}})
	& = X_{\wp^{(i,N)}} \, \prod_{k=1}^{\infty}\big(1 + X_{{k{\gamma}_{c}}}\big)^{0}
	= X_{\wp^{(i,N)}} \\
	\CK(X_{\wp^{(j,N)}})
	& = X_{\wp^{(j,N)}} \, \prod_{k=1}^{\infty}\big(1 + X_{{k{\gamma}_{c}}}\big)^{0}
	= X_{\wp^{(j,N)}} \\
\end{split}	
\ee
Furthermore, the vanishing exponent can once again be cast in terms of intersection pairing of
the lifts of $\wp$ and $p$ to $\tCW$.
The canonical lift of a two-way street of type $(ii,k(m+n))/(ii,-k(m+n))$ is
\be\label{eq:ii-street-lift}
	\pi^{-1}(p) =  \sum_{N\in\IZ} \( + p_{(i,N+n)} - p_{(i,N)} \)
\ee
Here $p_{(i,N)}$ is the lift of $p$ to sheet $(i,N)$ of $\tSigma$ and the positive / negative signs denote the orientation of the lifts relative to that of the underlying wall of type $(ii,k(m+n))$.
Unlike in the case of $ij$-walls, this formal sum now evaluates to zero!
This implies that we can once again describe the $\CK$-wall jump of $F(\wp)$ by the general formula (\ref{eq:universal-K-wall-substitution}).

An important difference from the case in $ij/ji$ two-way streets, is that now $\alpha_\gamma(p)$ is \emph{not} fixed by the K-wall jump of the formal parallel transport.
Since the intersection pairing always vanishes, $\alpha_\gamma(p)$ cannot be determined by comparing $F(\wp,\vartheta_c^\pm)$ this time:
the value of $\alpha_\gamma(p)$ for $ii/jj$ streets is undetermined by the K-wall jump.
Later on, we will fix these constants according to other criteria.
However it is worthwhile to stress that their value is immaterial for the validity of the K-wall formula: the general formula (\ref{eq:universal-K-wall-substitution}) is always valid regardless of how we choose to fix $\alpha_\gamma(p)$ for these particular  2-way streets.

\subsubsection{General $\CK$-wall formula and 5d BPS states}\label{sec:general-K-wall-formula}

In subsections \ref{sec:ij-ji-2-way-jumps} and \ref{sec:ii-2-way-walls} we proved that the jump of the formal parallel transport at $\vartheta_c$ is described by a universal change of variables
\be\label{eq:K-wall-formula}
	F(\wp,\CW_{\vartheta^{+}_{c}}) = \CK\big(F(\wp,\CW_{\vartheta^{-}_{c}})\big)\,,
\ee
where $\CK$ is a certain change of abelian parallel transport variables.
While we proved the formulae for paths $\wp$ intersecting only a single two-way street of type $ij/ji$ or $ii/jj$, the composition properties of the parallel transport $F(\wp,\vartheta_c^\pm) F(\wp',\vartheta_c^\pm) = F(\wp\wp',\vartheta_c^\pm)$ allow to extend formula (\ref{eq:universal-K-wall-substitution}) to an arbitrary path $\wp$.
The proof of this stament follows a standard argument that we shall not repeat here, since various versions of this idea can be found in the literature. See \cite{Gaiotto:2012rg} for the case of spectral networks, and \cite{Longhi:2016rjt} for the more similar case of ADE networks.

The upshot is that there exists a one-chain $\mathbf L$ on $\tSigma$ such that the parallel transport along an \emph{arbitrary} path $\wp$ jumps according to the following change of variables
\be\label{eq:K-wall-jump}
	\CK(X_{a}) = X_{a}\,\prod_{n\geq1}^{}\big(1+X_{{n{\gamma}_\mathrm{c}}}\big)^{\langle {\mathbf L}({n \gamma_\mathrm{c}}),\, a  \rangle} \,.
\ee
${\mathbf L}({n \gamma_\mathrm{c}})$ is defined as a weighted sum of canonical lifts of the 2-way streets, with weights determined by the soliton data:
\be\label{eq:L_gamma}
\begin{split}
	{\mathbf L}({\gamma})
	& := \sum_{p\in\CW_\mathrm{c}} \alpha_{{\gamma}}(p)\,\pi^{-1}(p) = \sum_N L_N(\gamma) \,.
\end{split}
\ee
It is then clear that, when $\wp$ is a short path crossing only a single two-way street, (\ref{eq:K-wall-jump})  reduces to (\ref{eq:universal-K-wall-substitution}).

This definition requires some clarification: as it is written it is ambiguous, because the weights $\alpha_\gamma(p)$ for $ii/jj$ 2-way streets are undetermined.
On the other hand we also showed that the precise value of $\alpha_\gamma(p)$ for these streets does not affect the validity of the $\CK$-wall formula, therefore we are free to fix these coefficients arbitrarily.
We choose to fix them in such a way that $L_N(\gamma)$ is a closed cycle: this implies the assumption that it is always possible to do so, and that this criterion fixes the coefficients.
Although we do not have a rigorous proof of this fact, we find it a fairly reasonable assumption based on concrete examples.

As homology classes $[L_N(\gamma)]$ should all fall in the same equivalence class after quotient described above (\ref{eq:Q_p_gamma_c})
\be
	[L_N(\gamma)] = [L_{N'}(\gamma)] \,.
\ee
A standard argument (see \cite{Gaiotto:2012rg, Longhi:2016rjt}) leads to the following formula for the degeneracies of bulk BPS states
\be\label{eq:BPS-index-formula}
	\Omega({\gamma}) = [L_N({\gamma})] \, /\,{\gamma}\,.
\ee
We will test this formula in a concrete example in the next Section.

\section{An Example}\label{sec:C3}

This section is entirely devoted to a detailed study of exponential networks for the mirror curve of the toric lagrangian brane of $\IC^3$.
We will work in a fixed choice of framing, chosen so that $\Sigma$ has two sheets.\footnote{This is the simplest choice, and it is made for illustration purposes. Our construction of nonabelianization maps applies also  to other choices of framing in which $\Sigma$ has more than one sheet.}
After a careful analysis of the covering maps $\pi,\tpi$, we study the exponential network at various phases.
We find a single $\CK$-wall jump which, despite the simplicity of this mirror curve, turns out to be quite complicated to analyze.
We conclude this section with comments on the implications of our results for the BPS spectrum of this theory, and on the 2d-4d limit $R\to 0$.

\subsection{Geometry}\label{sec:C3-geometry}

\subsubsection{Covering maps and trivialization}

The moduli space of a toric brane of  $\IC^3$ is captured by the mirror curve $\Sigma\subset \IC^\star \times \IC^\star$ described by the following equation
\be\label{eq:framed-curve}
	x y^f + y +1 = 0\,,
\ee
where $f$ accounts for  the dependence on framing \cite{Aganagic:2001nx}.
In order to best make contact with \cite{Eager:2016yxd} we will fix $f=-1$, and rewrite the curve as
\be
	y^2 + y +x = 0\,.
\ee
Presented as a covering of the $x$-plane, $\Sigma$ has two sheets
\be\label{eq:ypm}
	y_\pm = \frac{-1\pm\sqrt{1-4x}}{2}\,,
\ee
which meet at a square-root branch point located at $x=1/4$.
Since there are only two sheets, in this section we shall replace labels $ij$ for sheets of $\Sigma$ with $\pm$.
Going around the branch point once (either \emph{cw} or \emph{ccw}) exchanges $y_+$ and $y_-$, therefore an $\CE$-wall of type $(-+,n)$ determined by
\be
	(\log y_+ - \log y_- + 2\pi i n) \, \frac{d \log x}{dt} \in e^{i\vartheta} \IR^+
\ee
will pick up a monodromy that turns it into a wall of type $(+-,n)$.

The differential $\lambda = \log y \, d\log x$ further has logarithmic branching.
Near $x = 0$
\be\label{eq:log0C3}
	\lambda_+ \sim \log x \, d\log x  \qquad \lambda_- \sim d \log x
\ee
so the two branches have different behavior here.\footnote{This is in marked contrast with the situation in spectral networks, where the behavior of different branches of a Seiberg-Witten differential is qualitatively the same on all sheets: if there is a simple pole for one sheet at $x_0$, all other sheets also have a simple pole there.}
While  $\lambda_- \sim dx/x$ has a simple pole, $\lambda_+ \sim \frac{\log x}{x} dx$ has logarithmic branching and is therefore multi-valued around $x=0$.
This means that there is a logarithmic cut on sheet $+$ of $\Sigma$ starting above $x=0$ and running to infinity.
An $\CE$-wall of type $(-+,n)$
winding \emph{ccw} around $x=0$ will pick up a monodromy that shifts $n\to n+1$, whereas a wall of type $(+-,n)$ would pick up a shift $n\to n-1$.

Around $x=\infty$, in the local coordinate $w=1/x$ the two sheets of $\lambda$ behave as
\be
\begin{split}
	\lambda_\pm  & = - \log y_\pm \, d\log w \\
	& \sim \left[\frac{1}{2} \log w \pm \frac{1}{2} \left( \pi i + i \sqrt w +\dots \right)\right] \frac{dw}{w} \,. 
\end{split}
\ee
Both branches have the same logarithmic singularity now, so a wall of type  $(\pm\mp,n)$ or $(\pm\pm,n)$ will {not pick up} any monodromy from $\log w$.
There is however square-root branching of sheets at infinity, because taking $w\to w e^{2\pi i}$ takes
\be
	\lambda_+ - \lambda_- \to \lambda_- - \lambda_+ - 2\pi i \, \frac{dx}{x} \,.
\ee
Therefore a wall of type $(-+,n)$ turns into one of type $(+-,n-1)$ going (locally) \emph{ccw} around $x=\infty$.
The momodromy of other types of walls can be obtained from the behavior of single sheets, which we summarize below
\be
\begin{array}{lr}
	\lambda_+  \to \lambda_- - 2\pi i\ \frac{dx}{x} & (ccw) \\
	\lambda_+  \to \lambda_-  & (cw) \\
	\lambda_-  \to \lambda_+  & (ccw) \\
	\lambda_-  \to \lambda_+  + 2\pi i\ \frac{dx}{x} & (cw) \,.
\end{array}
\ee

\begin{figure}[h!]
\begin{center}
\includegraphics[width=0.5\textwidth]{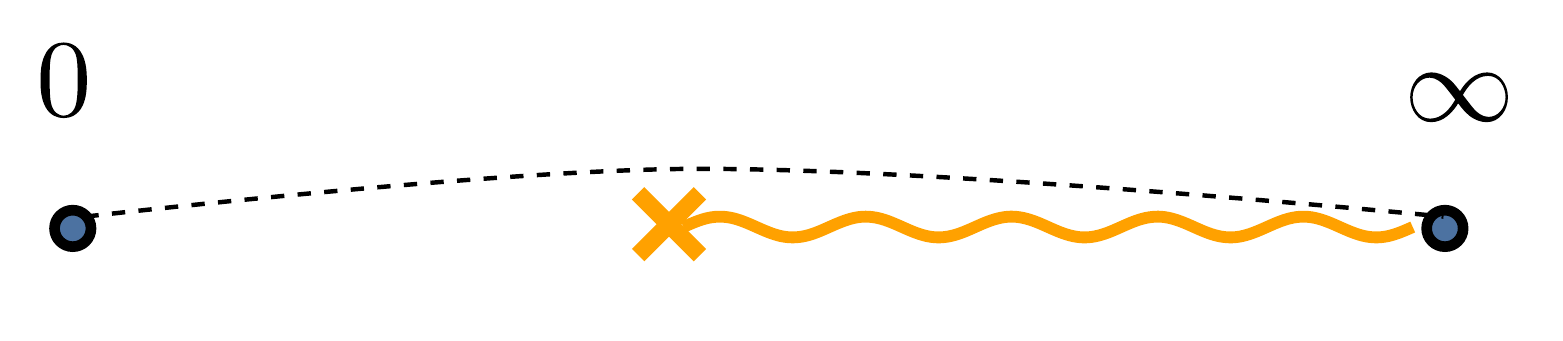}
\caption{Cuts for the trivialization of the mirror curve of $\IC^3$ in framing $p=-1$, shown on $\IC^*_x$. The wavy line is the square-root cut of $\pi$ while the dashed line is the logarithmic cut of $\tpi$ projected down to $\IC^*_x$.
The relative position of the cuts matters, it is consistent with the monodromies in Table (\ref{eq:C3-monodromies}).
}
\label{fig:C3-triv}
\end{center}
\end{figure}

When studying individual soliton paths, it is important to keep track of each sheet separately when the wall supporting the soliton crosses some cut (recall that the sheets of $\Sigma$ are identified with the vacua connected by a soliton).
Accordingly we introduce a slightly refined notation, replacing $(ij,n)$ with $(ij)_{N_L,N_R}$ with the understanding that $N_R = N_L + n$.
We can now summarize the monodromies of soliton charges  at the various branch points and punctures
\be\label{eq:C3-monodromies}
\begin{array}{c|c|c|c}
	\text{branch point / puncture} & \text{before} & \text{ccw} & \text{cw} \\
	\hline
	\multirow{4}{*}{$x=0$} & (++)_{N_L,N_R} & (++)_{N_L+1,N_R+1} & (++)_{N_L-1,N_R-1} \\
	& (--)_{N_L,N_R} & (--)_{N_L,N_R} & (--)_{N_L,N_R}  \\
	& (+-)_{N_L,N_R} & (+-)_{N_L+1,N_R} & (+-)_{N_L-1,N_R}  \\
	& (-+)_{N_L,N_R} & (-+)_{N_L,N_R+1} & (-+)_{N_L,N_R-1}  \\
	\hline
	\multirow{4}{*}{$x=\infty$} & (++)_{N_L,N_R} & (--)_{N_L-1,N_R-1} & (--)_{N_L ,N_R} \\
	& (--)_{N_L,N_R} & (++)_{N_L,N_R} & (++)_{N_L+1 ,N_R+1} \\
	& (+-)_{N_L,N_R} & (-+)_{N_L-1,N_R } & (-+)_{N_L ,N_R+1} \\
	& (-+)_{N_L,N_R} & (+-)_{N_L,N_R-1} & (+-)_{N_L+1 ,N_R} \\
	\hline
	\multirow{2}{*}{$x=\frac{1}{4}$} & (\pm\pm)_{N_L,N_R} & \multicolumn{2}{c}{(\mp\mp)_{N_L,N_R}} \\
	& (\pm\mp)_{N_L,N_R} &  \multicolumn{2}{c}{(\mp\pm)_{N_L,N_R}} \\
	\hline
\end{array}
\ee
We choose to trivialize $\pi\circ \tpi:\tSigma\to C$ by introducing a square-root cut between $x=1/4$ and $x=\infty$, and a logarithmic cut between $x=0$ and $x=\infty$, see Figure \ref{fig:C3-triv}.

\subsubsection{The exponential network at various phases}

\begin{figure}[h!]
\begin{center}
\fbox{\includegraphics[width=0.22\textwidth]{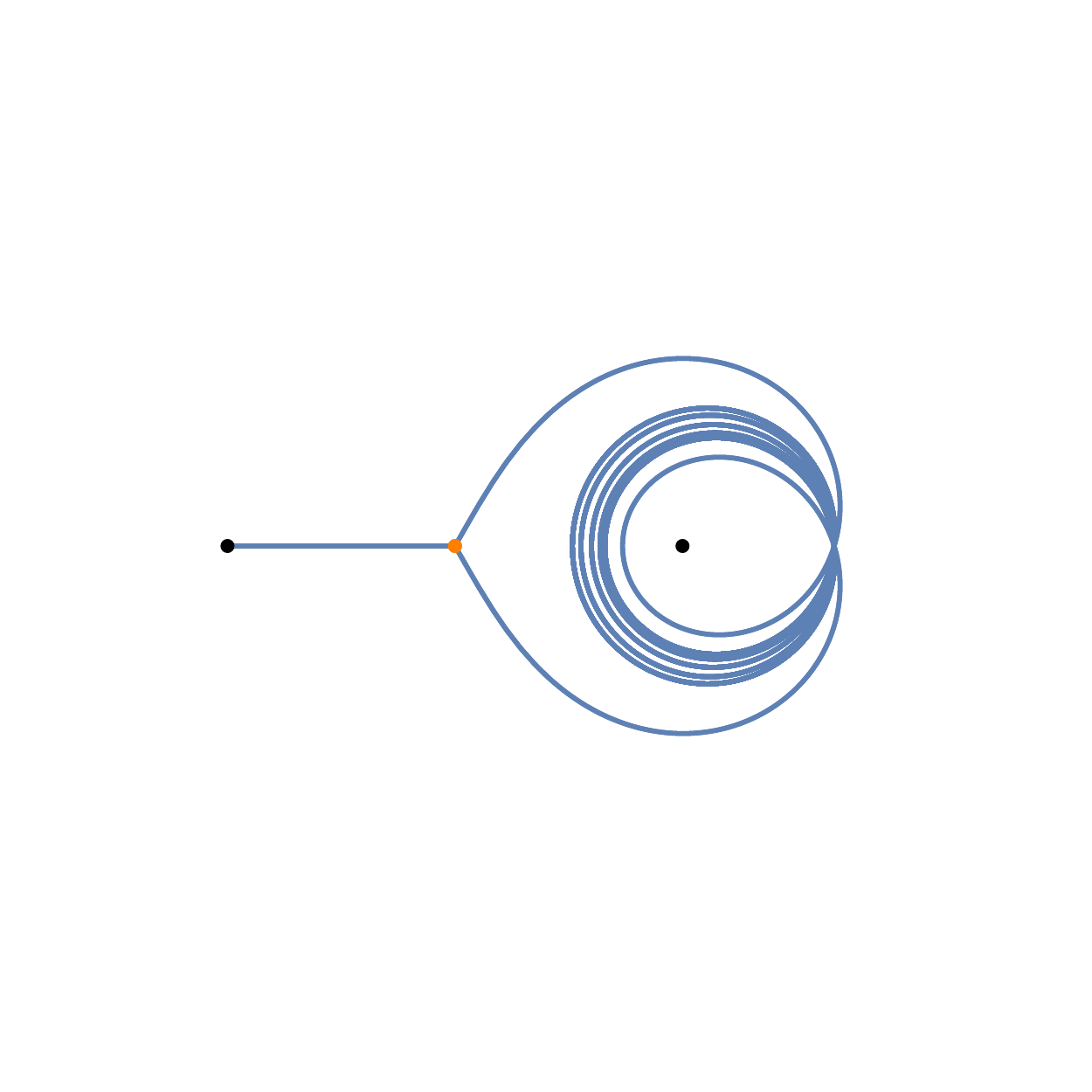}}
\fbox{\includegraphics[width=0.22\textwidth]{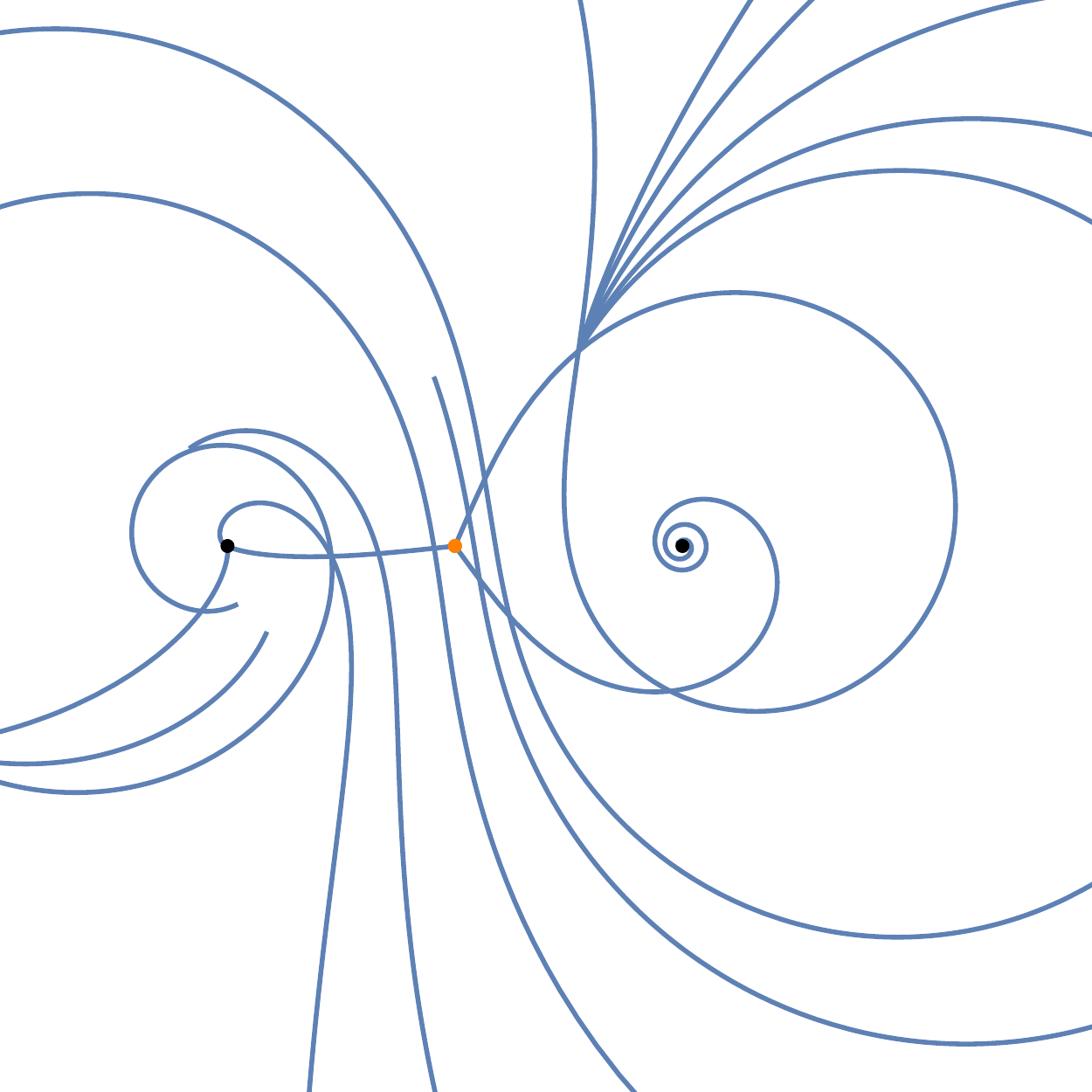}}
\fbox{\includegraphics[width=0.22\textwidth]{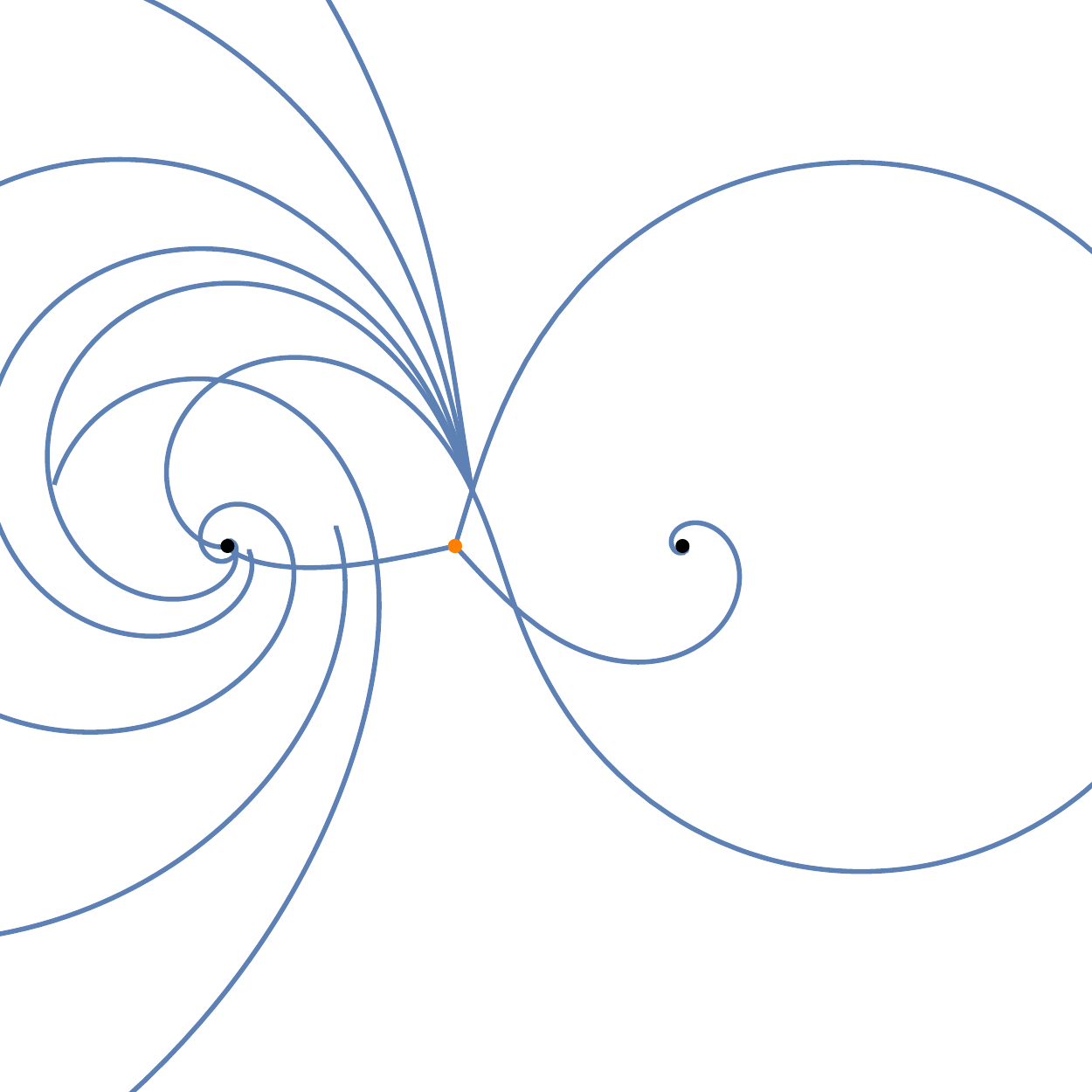}}
\fbox{\includegraphics[width=0.22\textwidth]{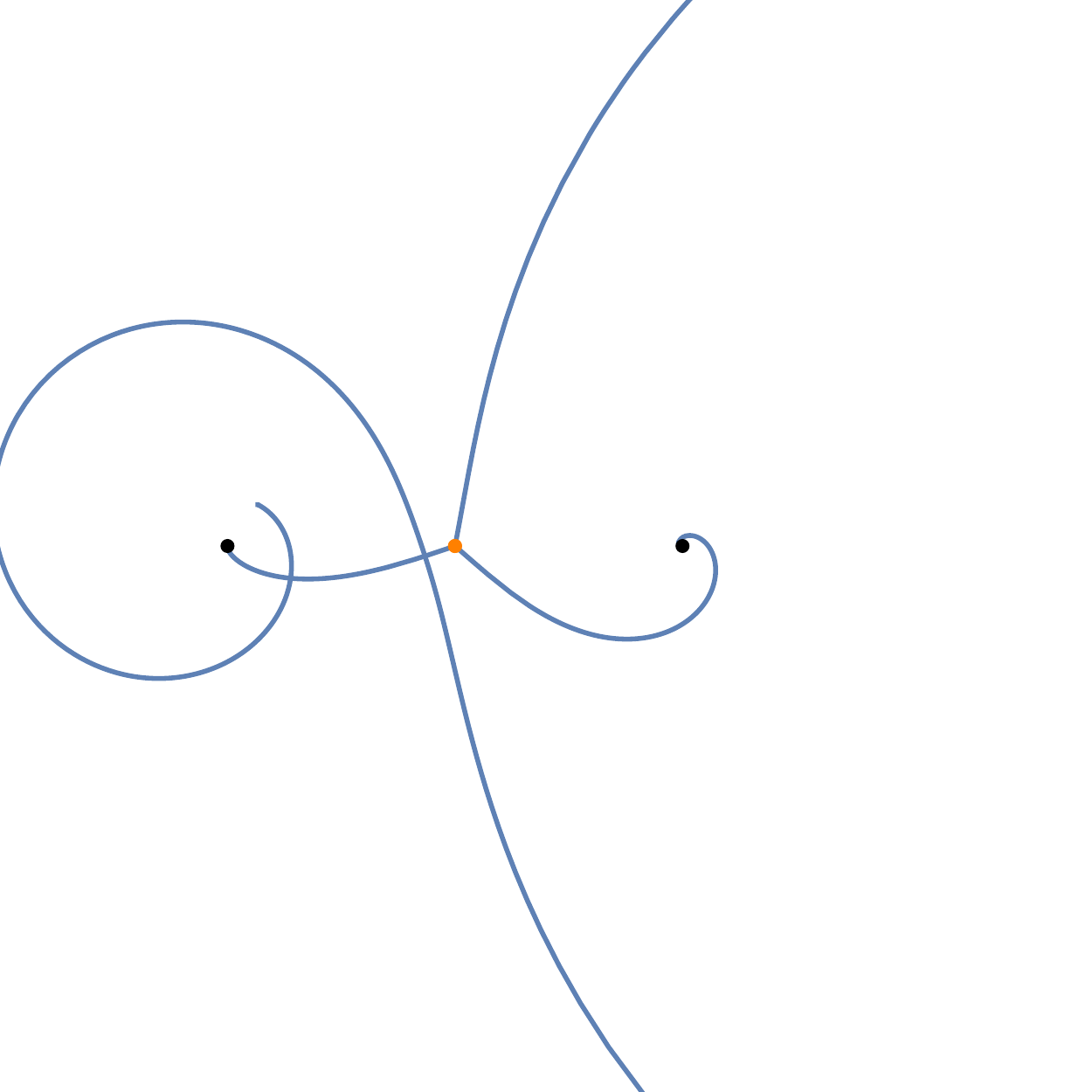}}
\\
\fbox{\includegraphics[width=0.22\textwidth]{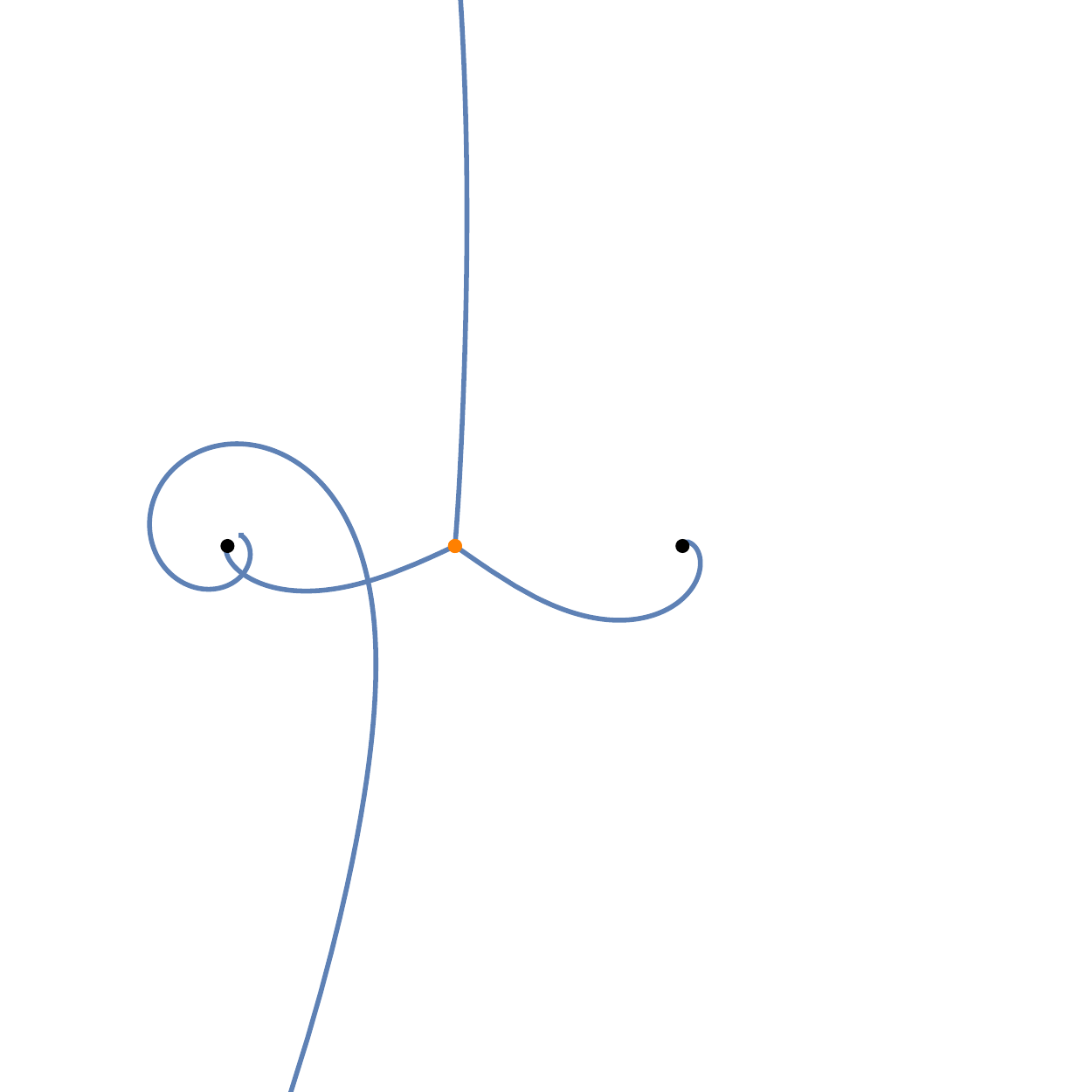}}
\fbox{\includegraphics[width=0.22\textwidth]{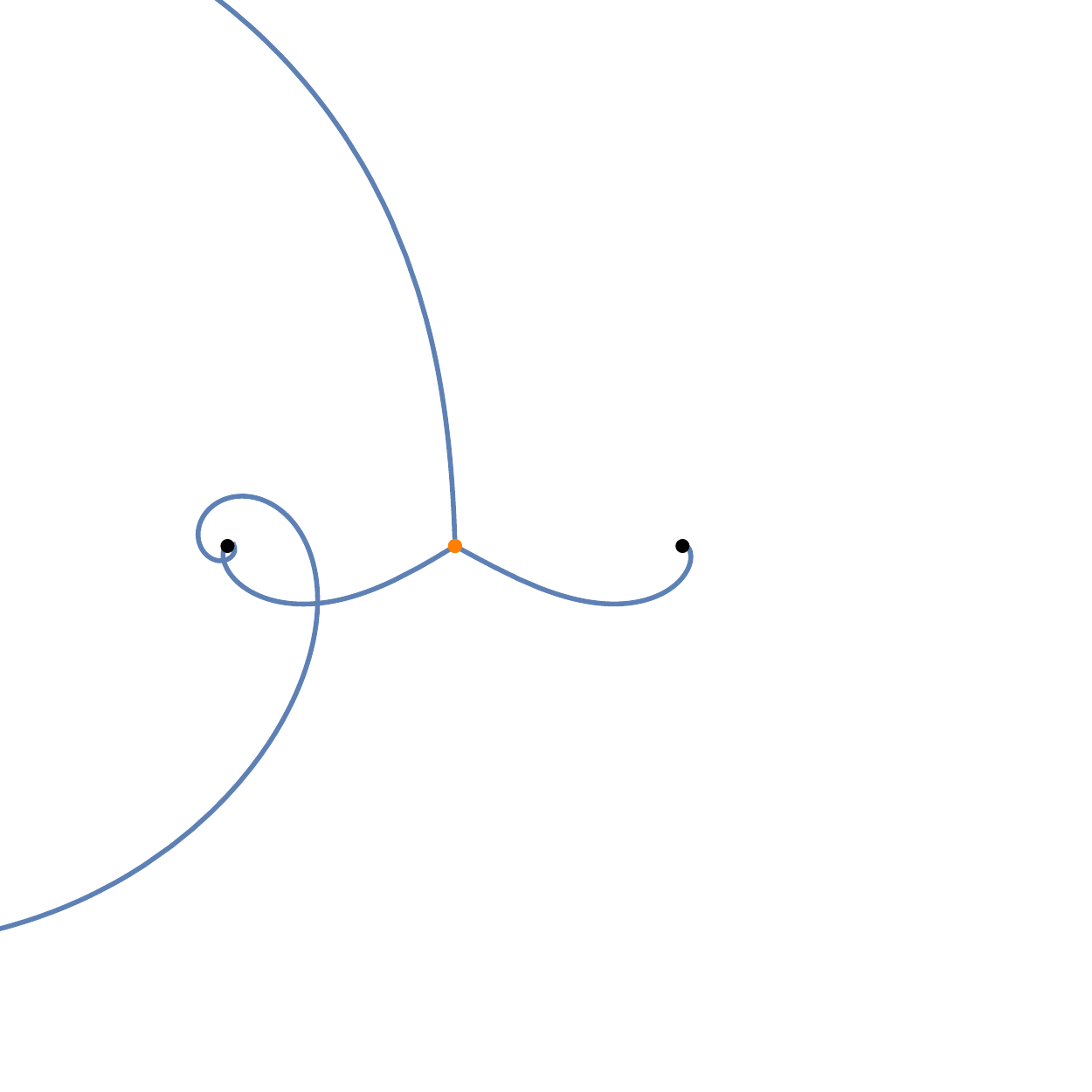}}
\fbox{\includegraphics[width=0.22\textwidth]{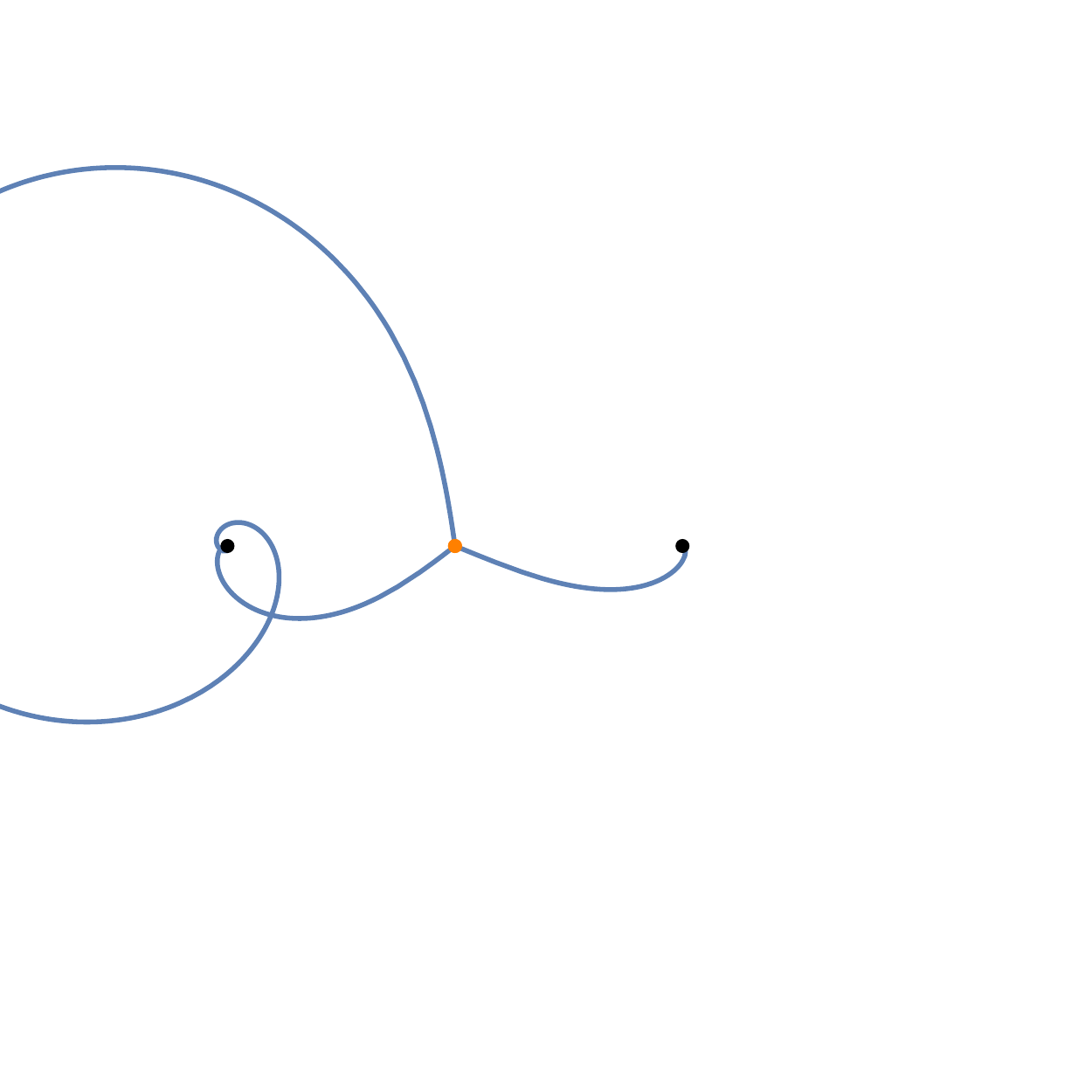}}
\fbox{\includegraphics[width=0.22\textwidth]{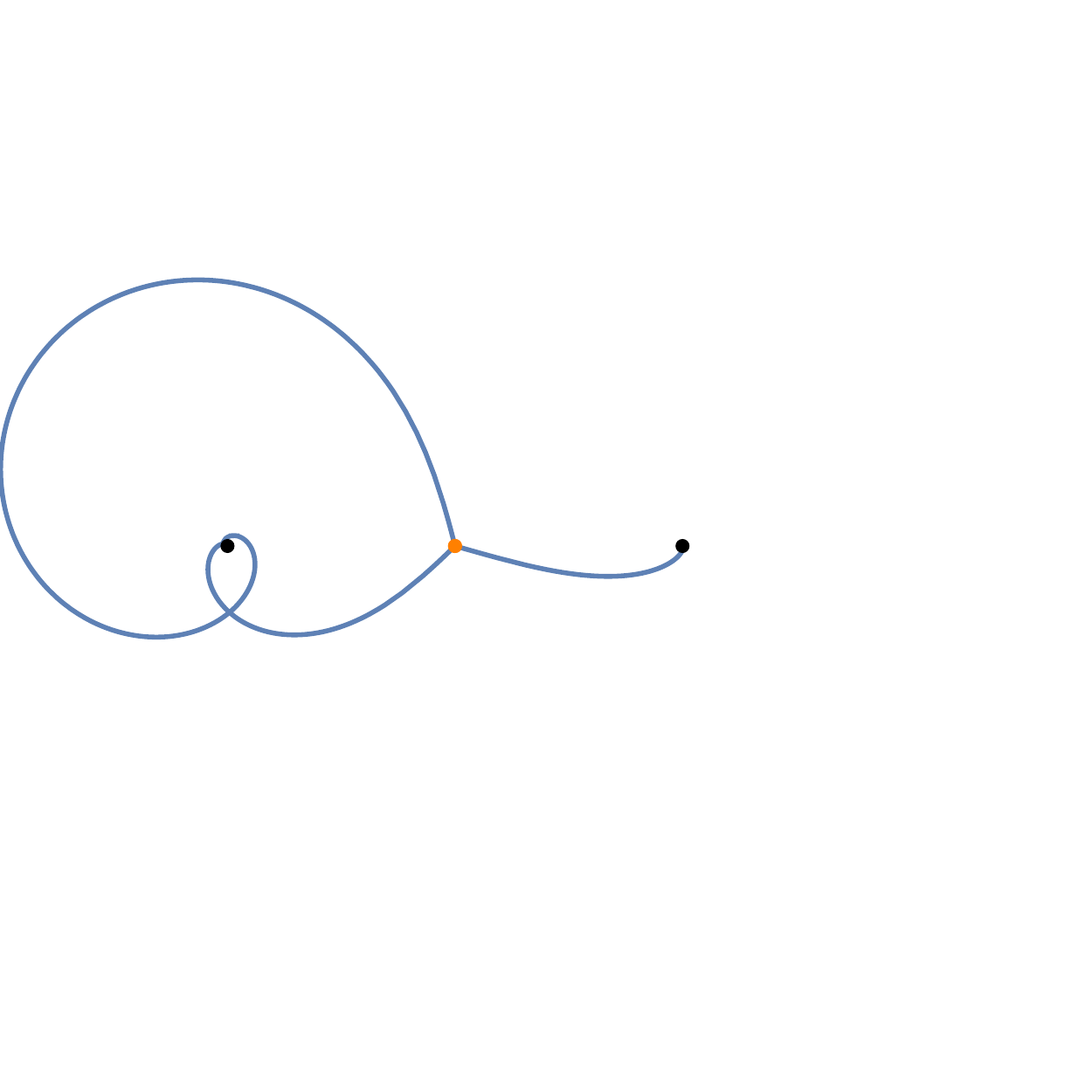}}
\\
\fbox{\includegraphics[width=0.22\textwidth]{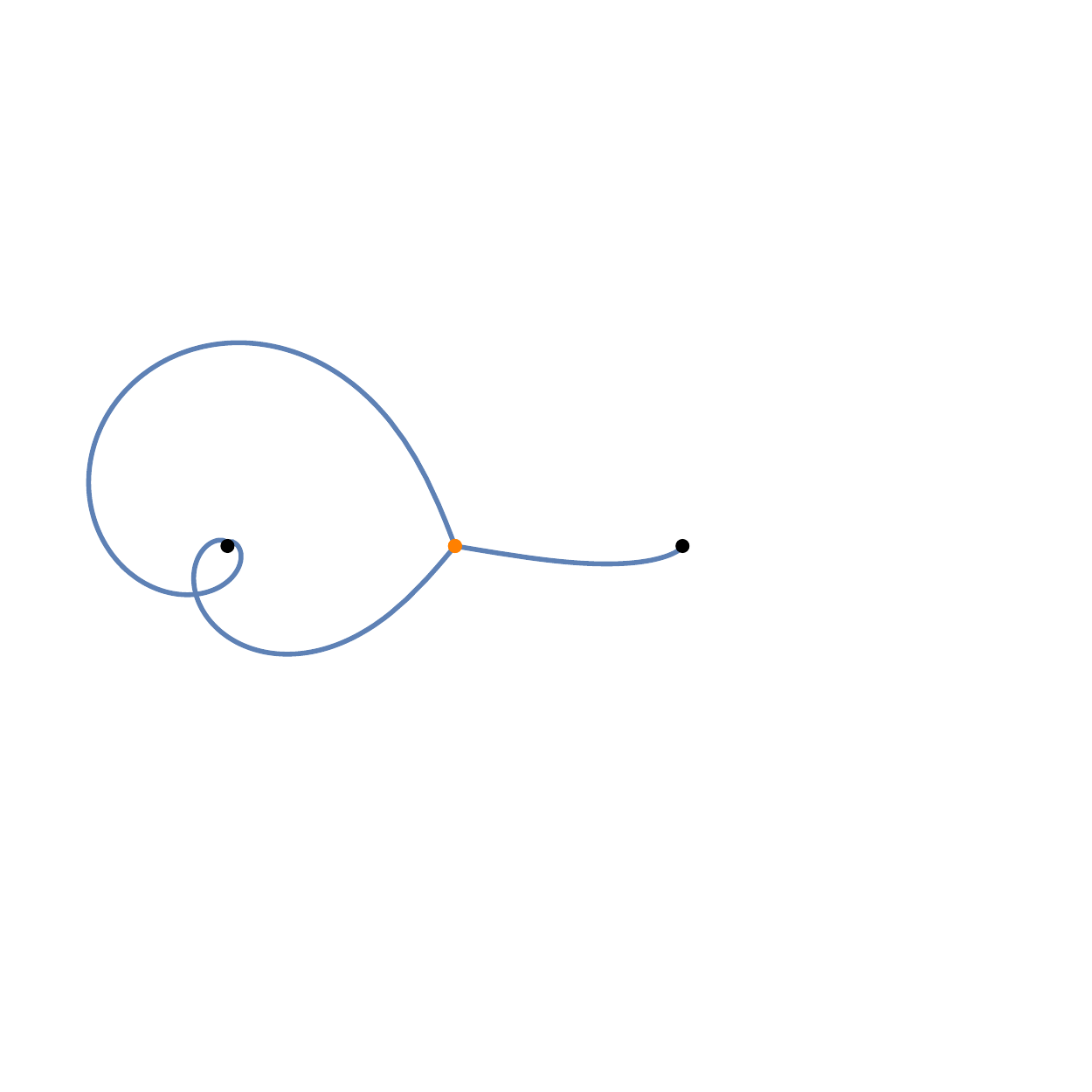}}
\fbox{\includegraphics[width=0.22\textwidth]{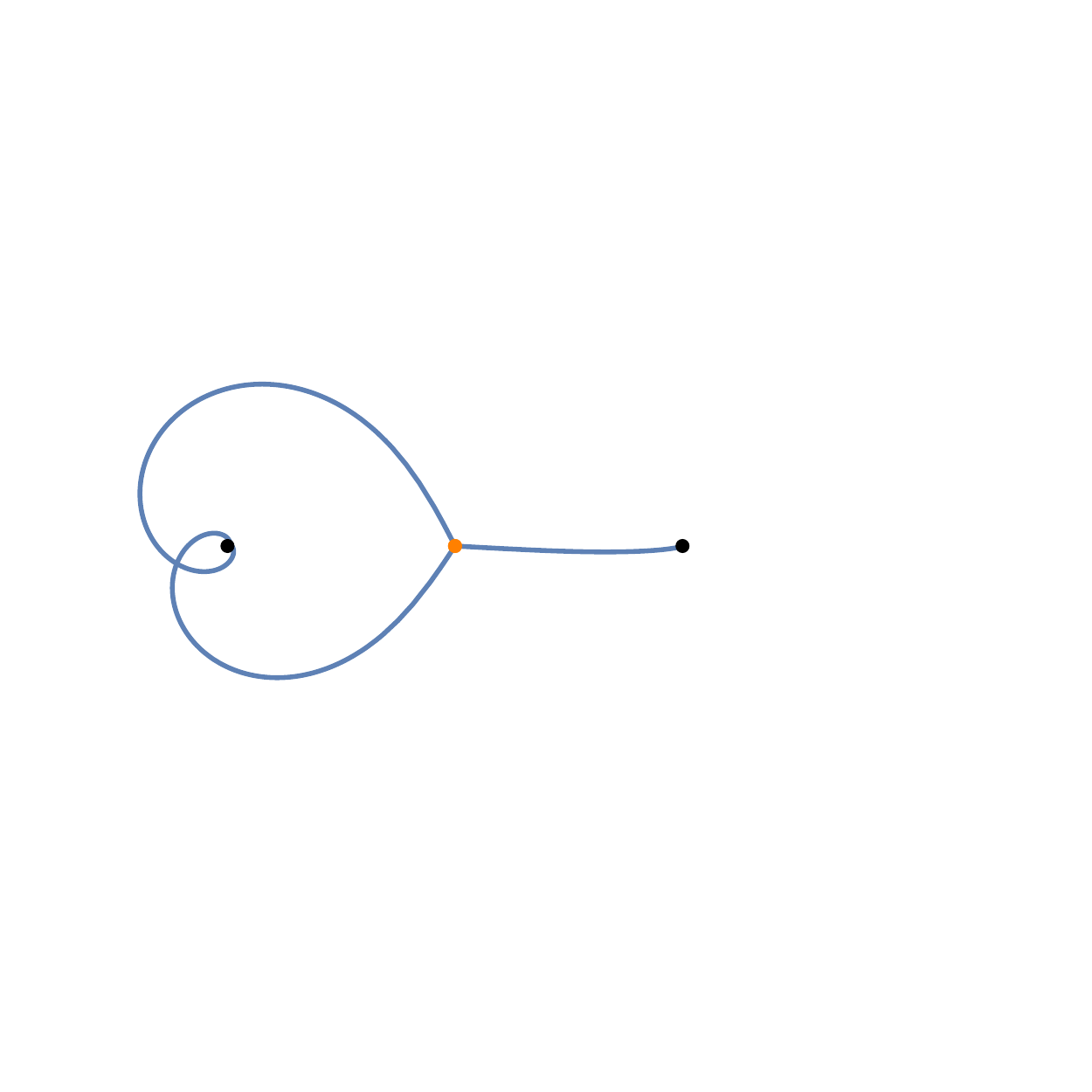}}
\fbox{\includegraphics[width=0.22\textwidth]{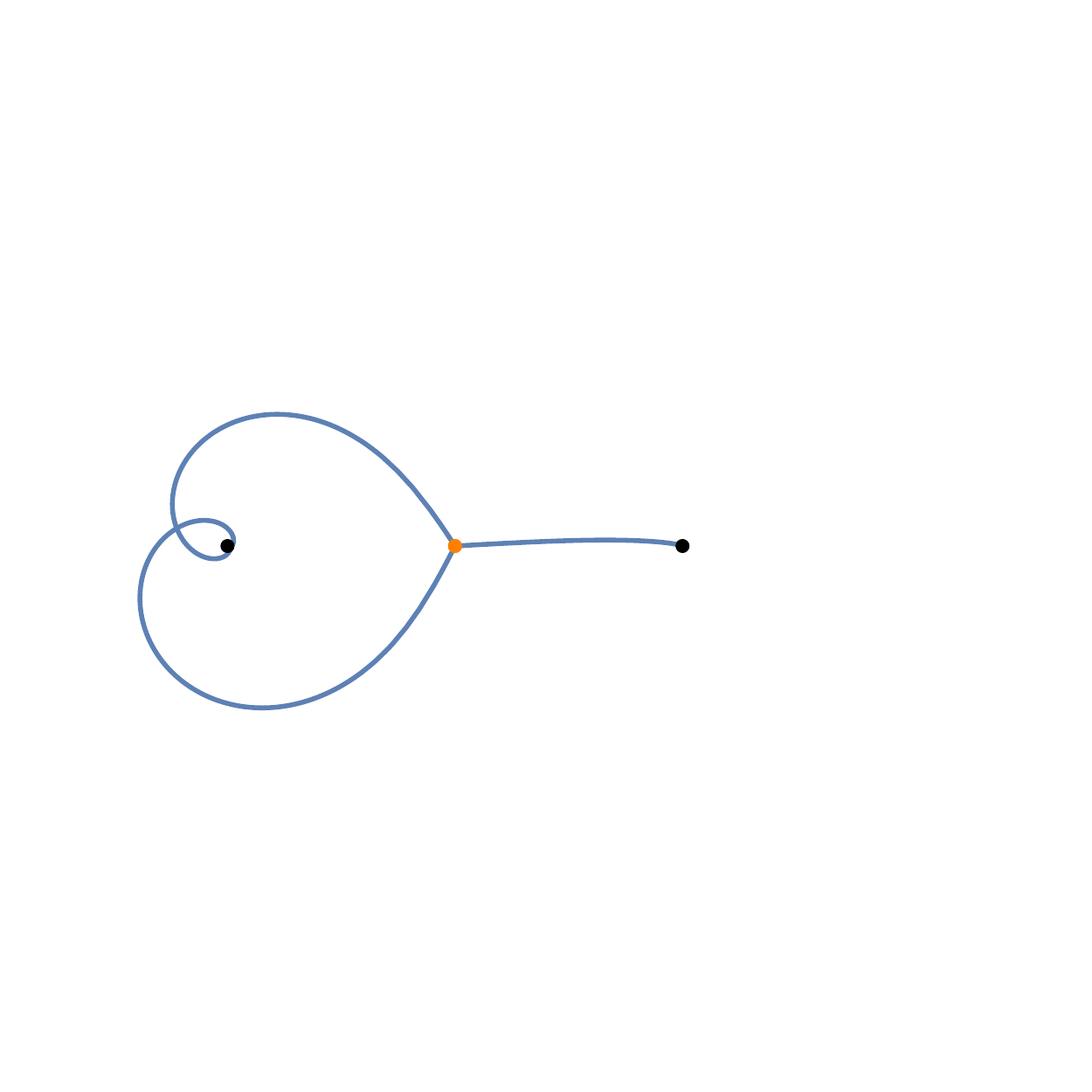}}
\fbox{\includegraphics[width=0.22\textwidth]{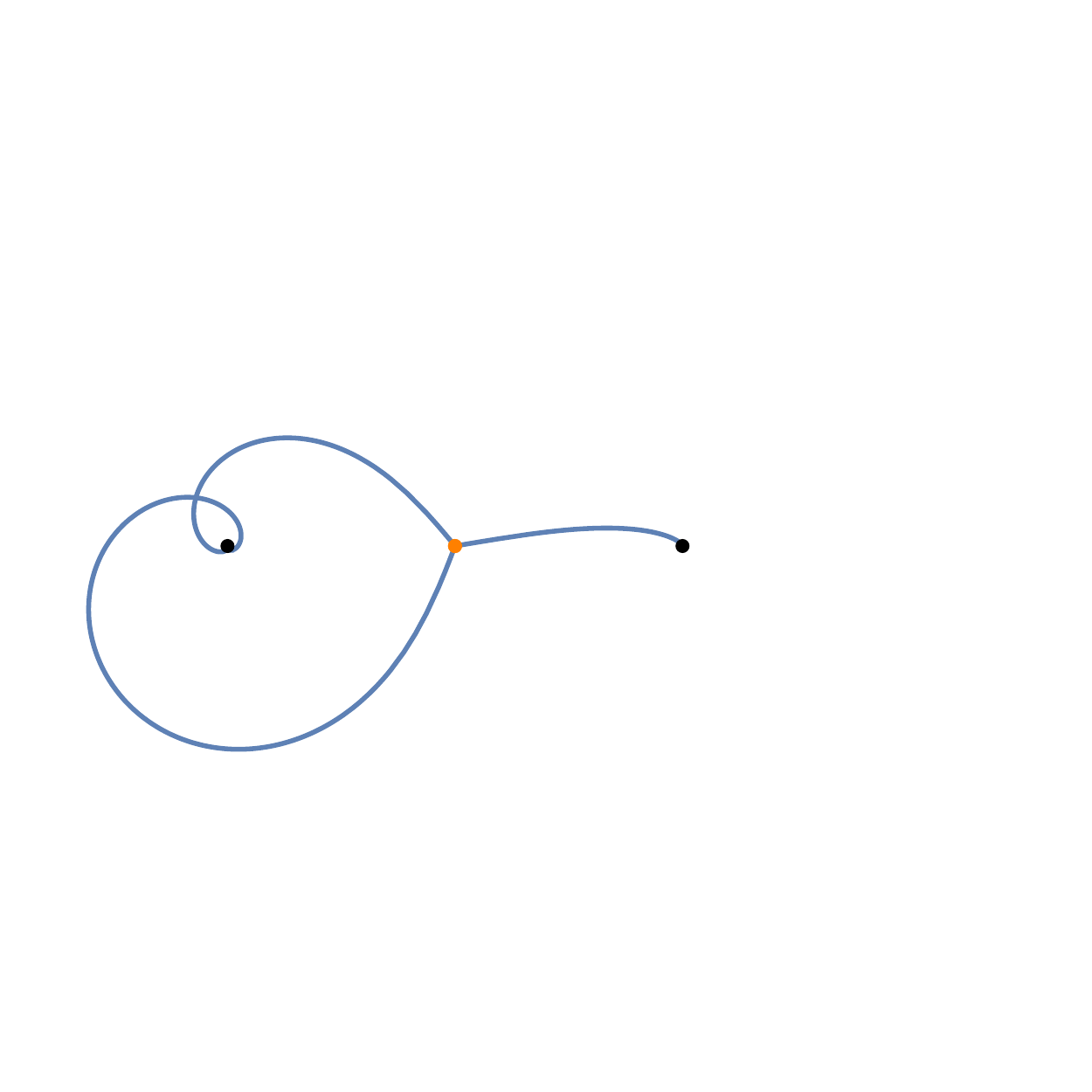}}
\\
\fbox{\includegraphics[width=0.22\textwidth]{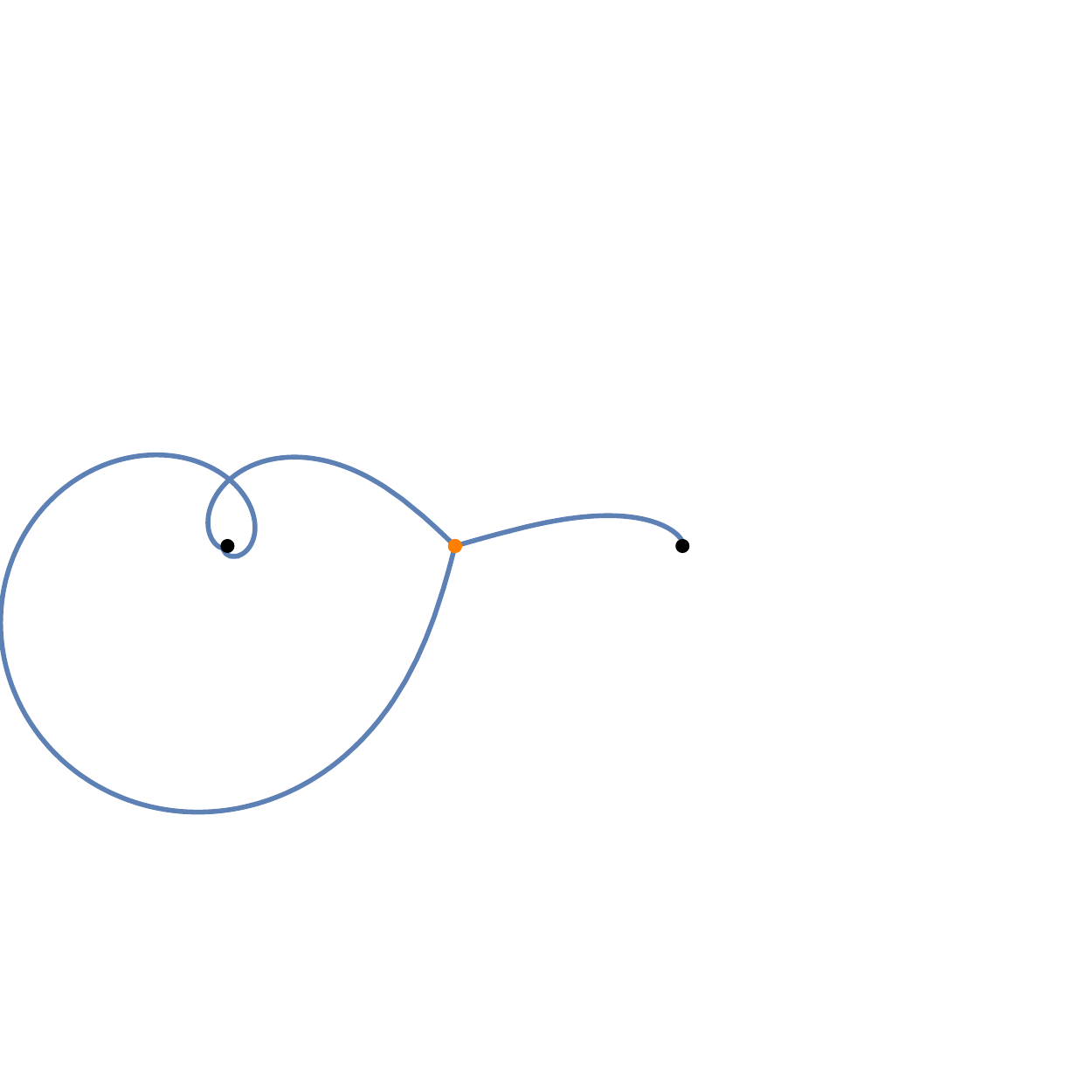}}
\fbox{\includegraphics[width=0.22\textwidth]{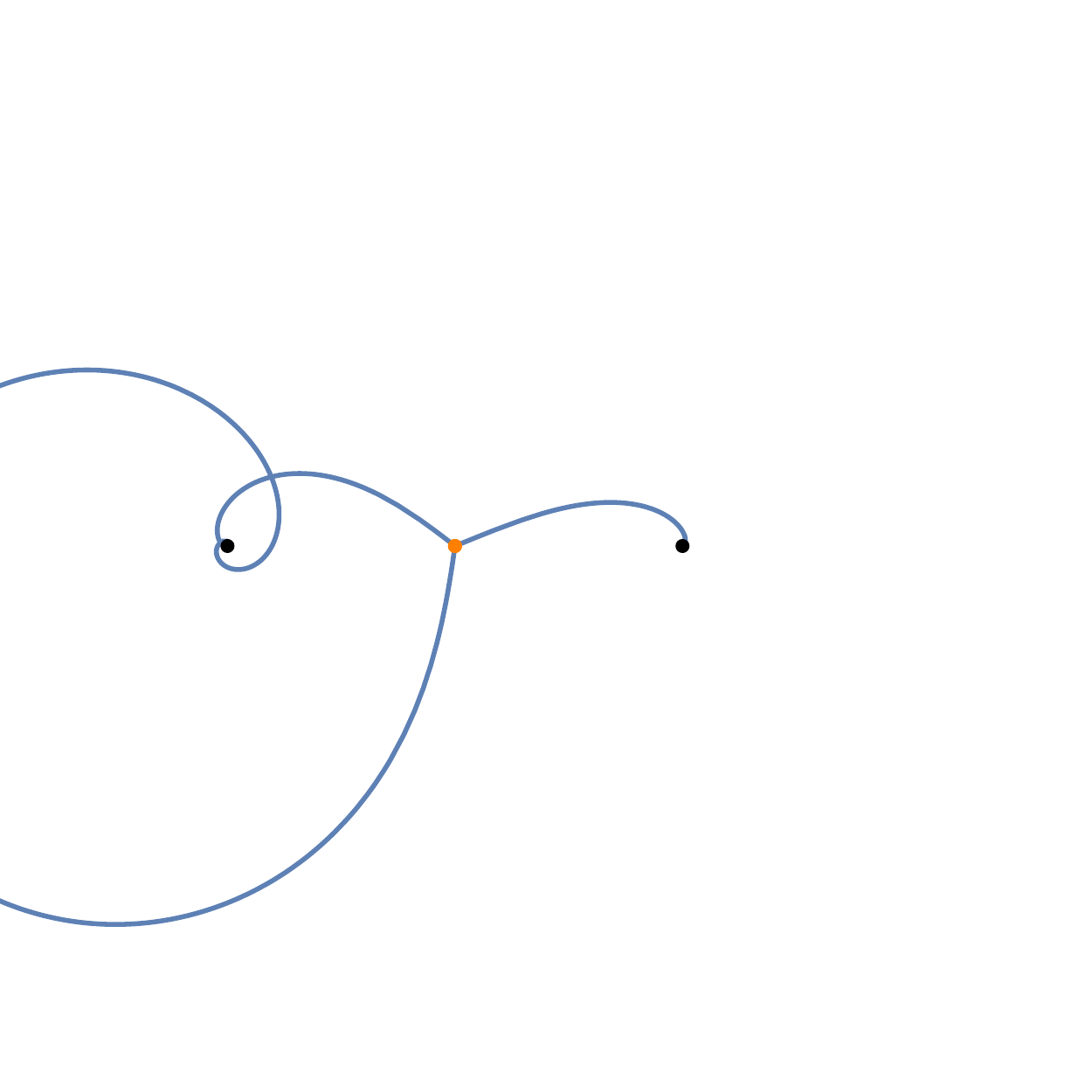}}
\fbox{\includegraphics[width=0.22\textwidth]{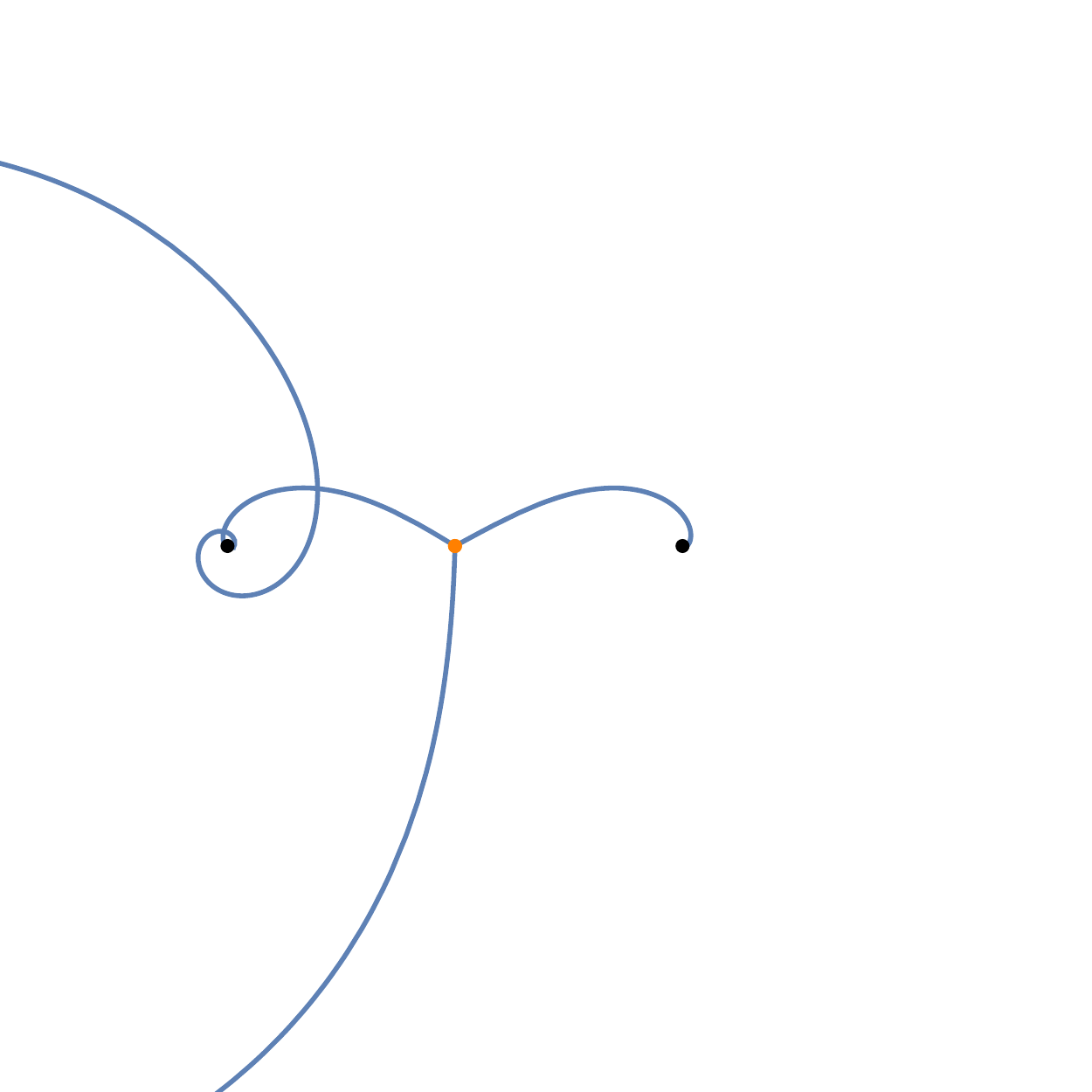}}
\fbox{\includegraphics[width=0.22\textwidth]{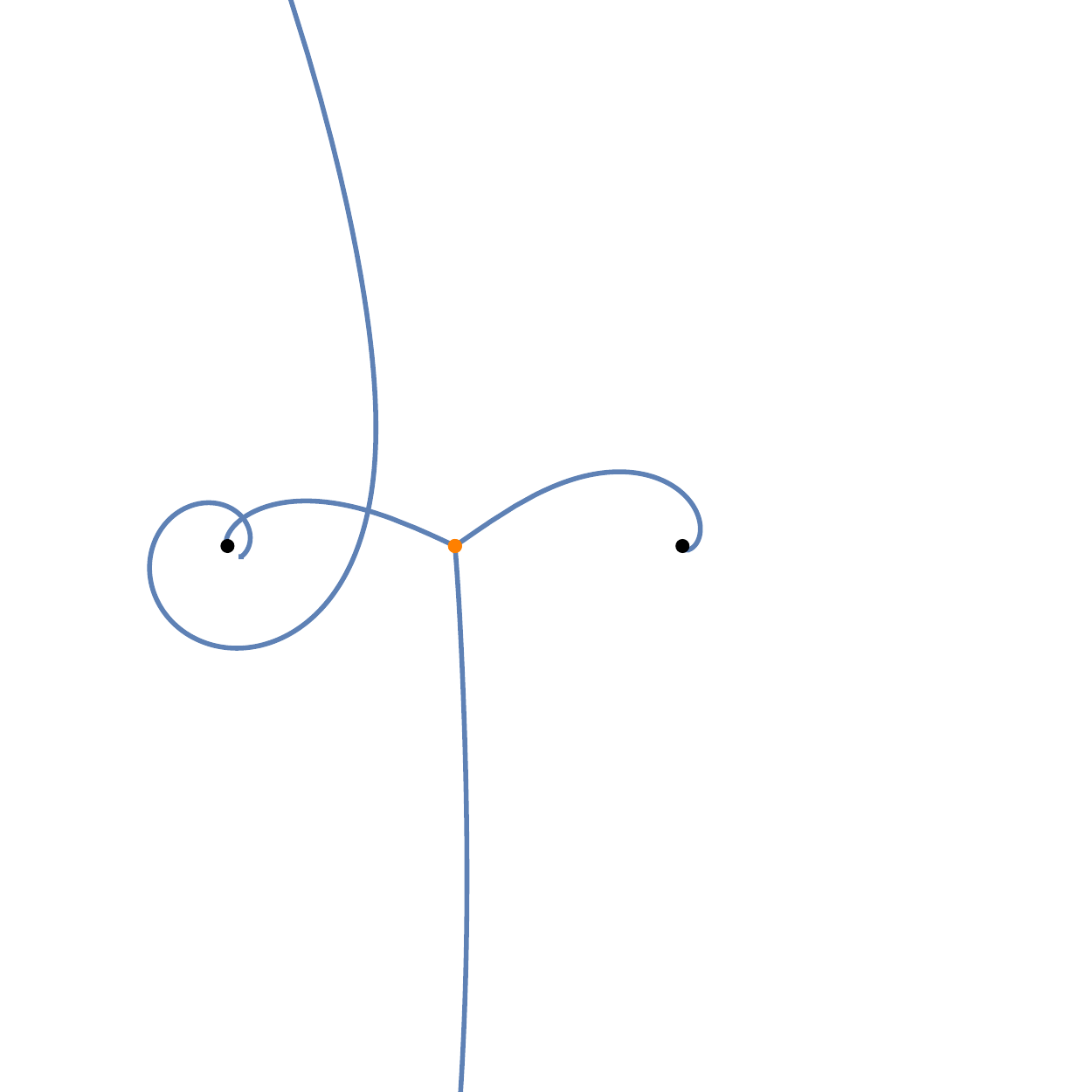}}
\\
\fbox{\includegraphics[width=0.22\textwidth]{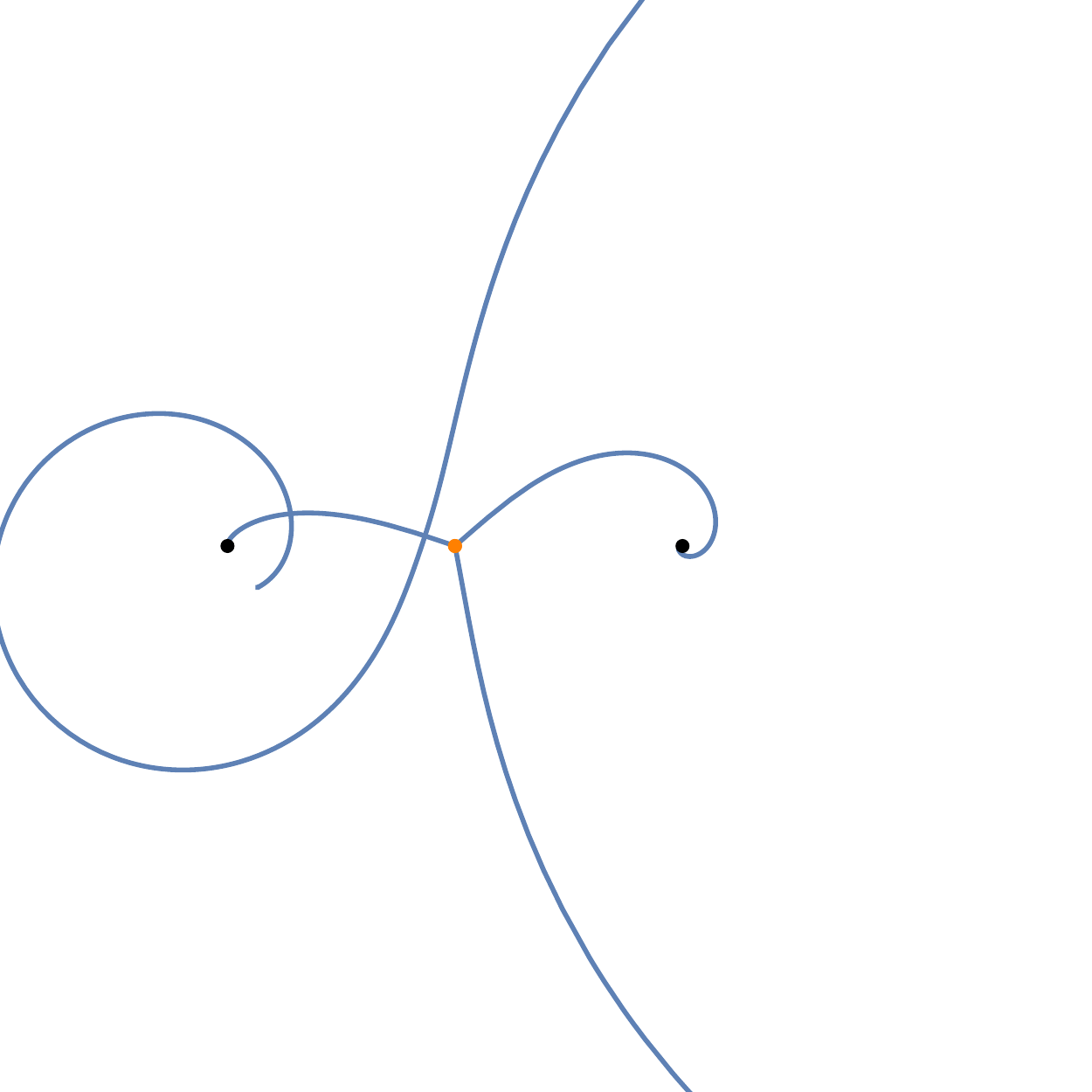}}
\fbox{\includegraphics[width=0.22\textwidth]{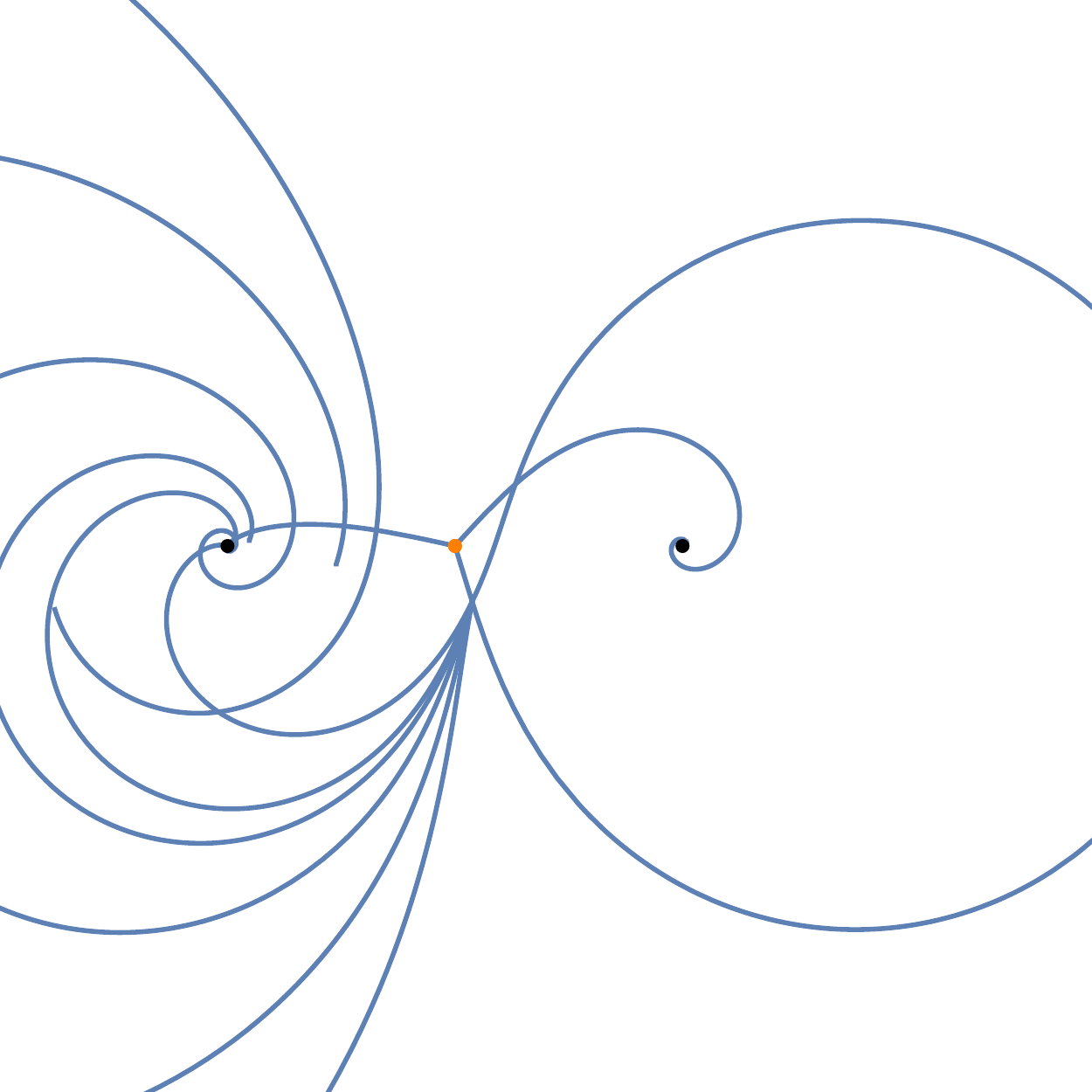}}
\fbox{\includegraphics[width=0.22\textwidth]{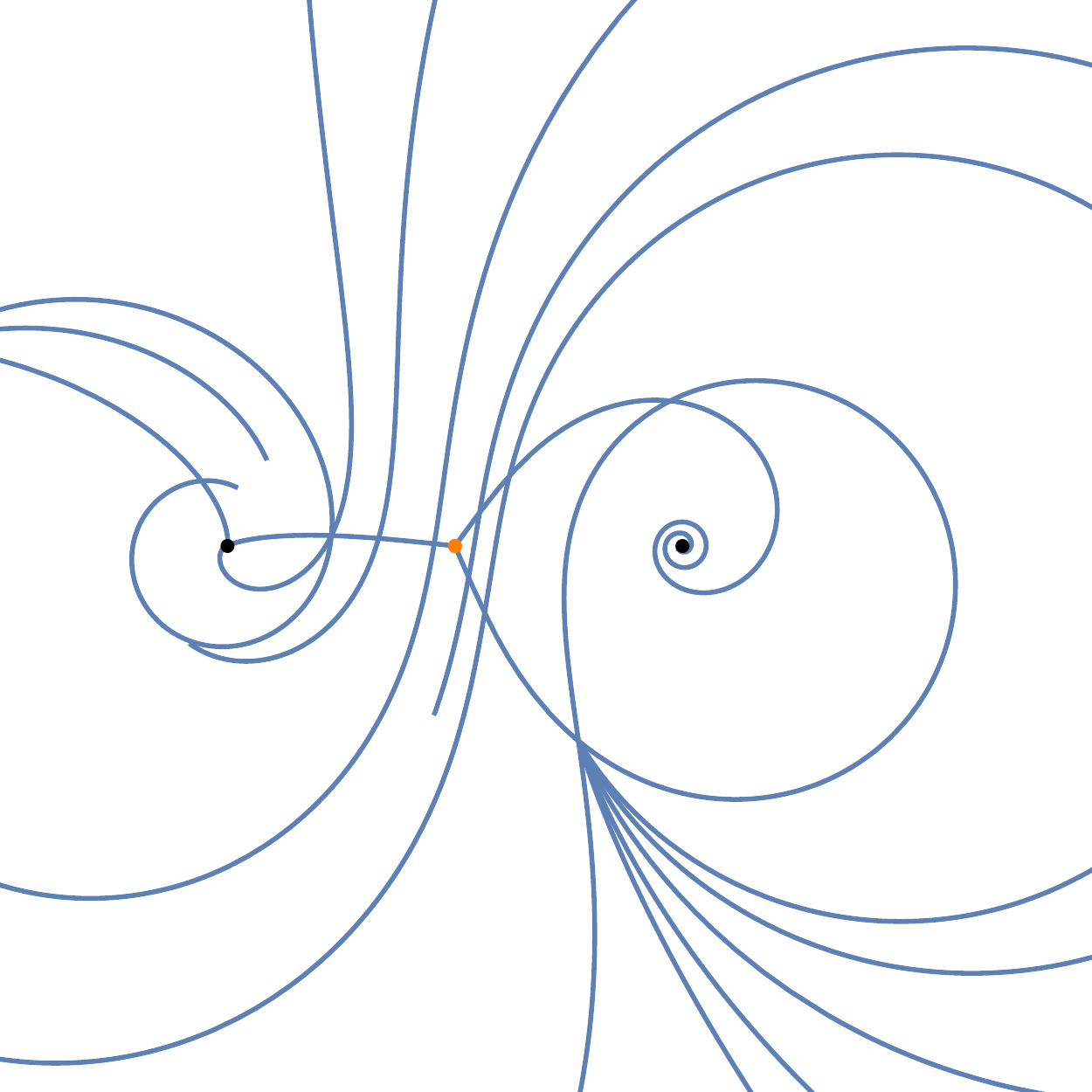}}
\fbox{\includegraphics[width=0.22\textwidth]{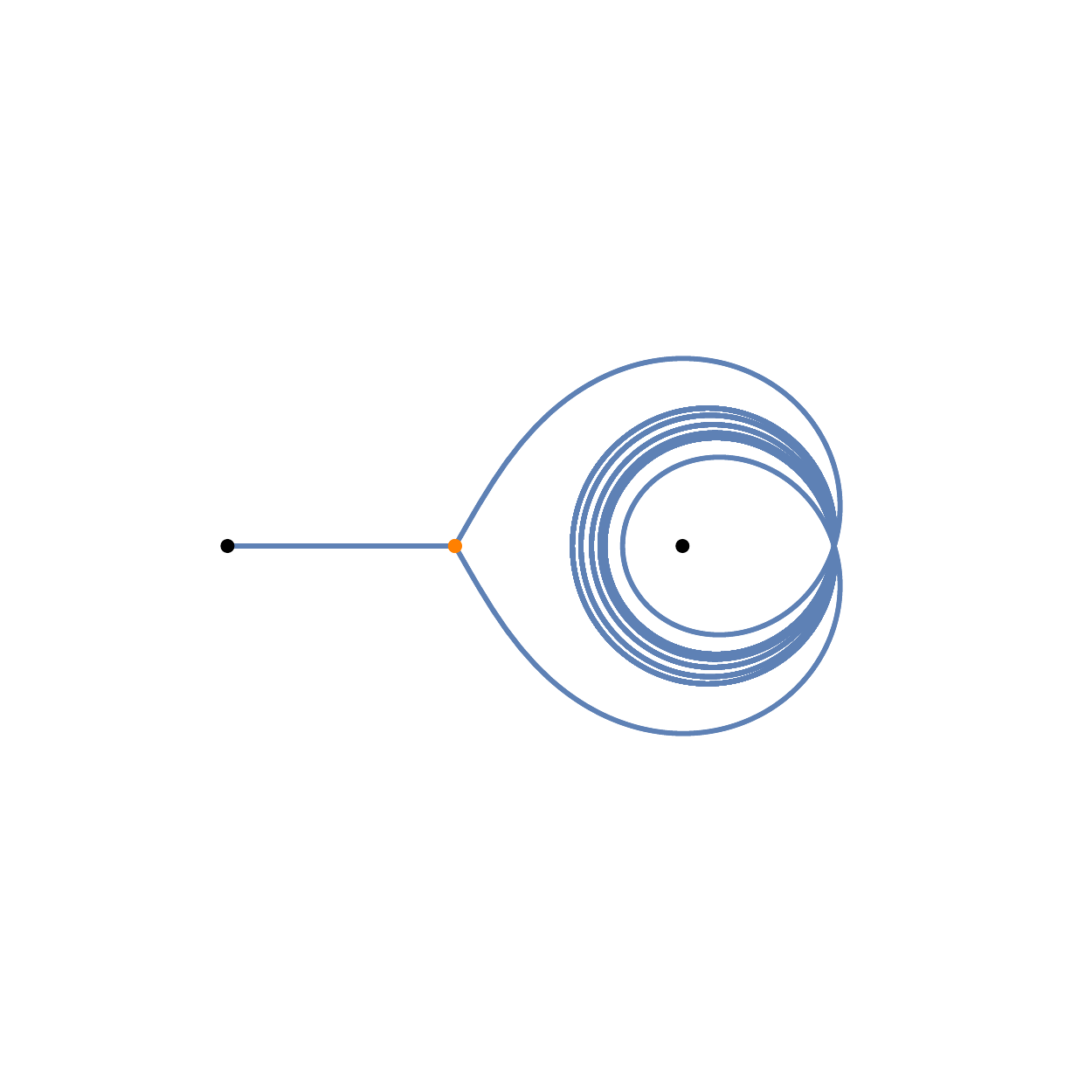}}
\caption{The exponential network for $\IC^3$ at framing $-1$. From $\vartheta=0$ (top left) to $\vartheta=\pi$ (bottom right).}
\label{fig:C3-network-plots}
\end{center}
\end{figure}

Here we describe the general features of the exponential network plotted for various values of the phase $\vartheta$.
Some details on plotting network can be found in Appendix \ref{app:plotting}.
There is only one square-root branch point, therefore the network is generated by exactly three primary walls.
As these walls propagate along the $x$-plane, they may mutually intersect or even self-intersect, giving rise to secondary walls.
Since the model only has two sheets, the only nontrivial joints are of $ij-ji$ types, featuring an infinite number of descendant walls.\footnote{This should be contrasted with the case of spectral networks. For example, the $AD_1$ Argyres-Douglas theory described in \cite{Gaiotto:2009hg} also features a single branch point, but the spectral network of that theory is very simple, consisting of just three primary walls and no descendants.}
The network is shown for various values of $\vartheta$ in Figure \ref{fig:C3-network-plots}.

\begin{figure}[h!]
\begin{center}
\includegraphics[width=0.35\textwidth]{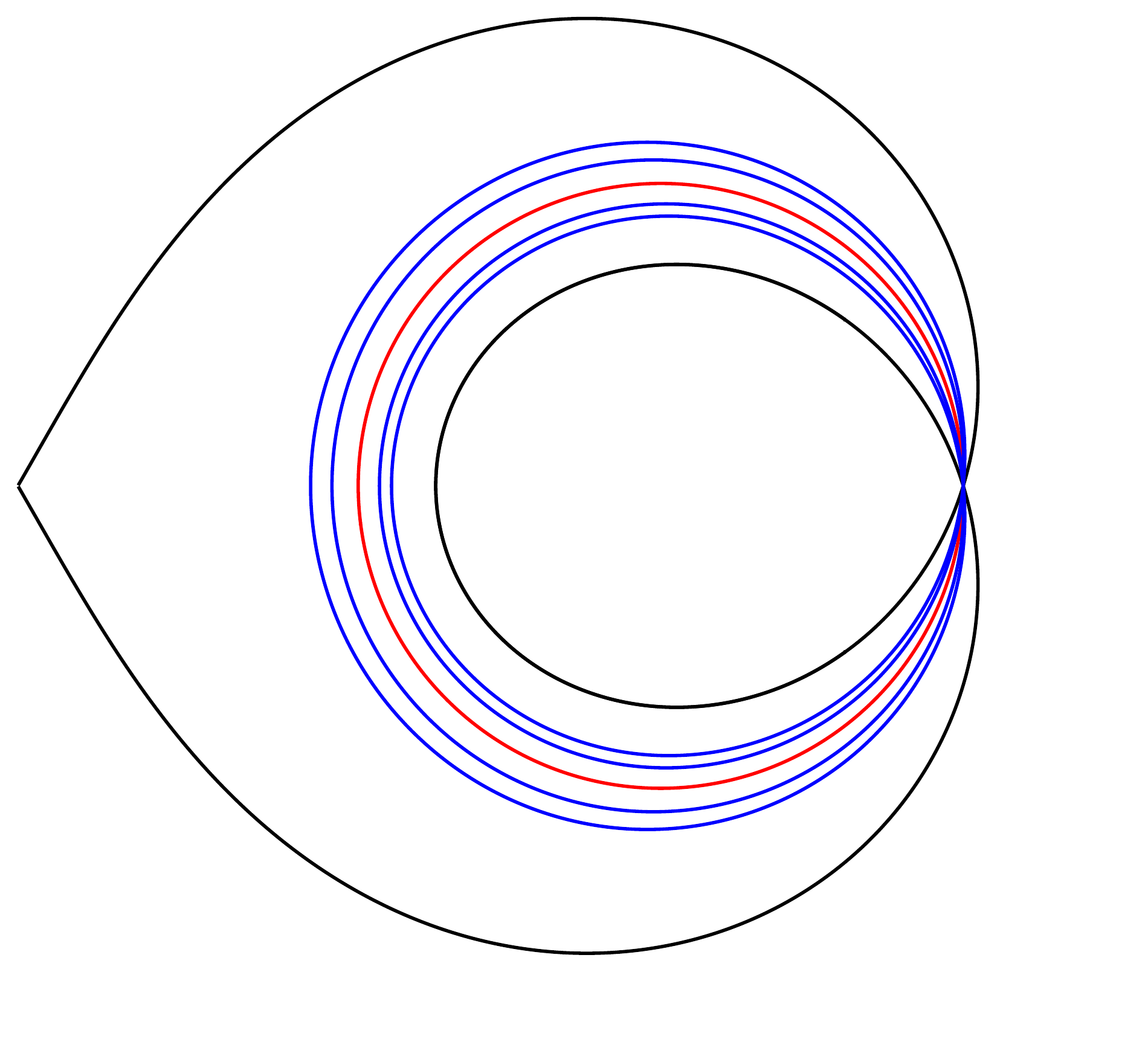}
\hspace{.06\textwidth}
\includegraphics[width=0.48\textwidth]{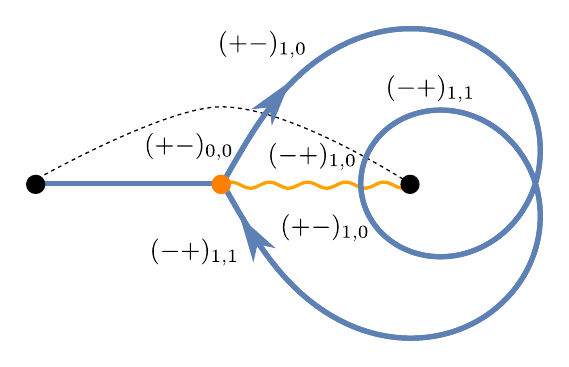}
\caption{Left: showing only the two-way streets of the critical network $\CW(\vartheta=\pi)$, with primary walls in black, secondary walls of types $ij/ji$ in blue, and secondary walls of type $ii/jj$ in red. Right: showing only primary walls and details of soliton charges supported on them.}
\label{fig:C3-Kwall}
\end{center}
\end{figure}

The network is non-degenerate as long as $0 < \vartheta < \pi$. There is a single $\CK$-wall jumps at $\vartheta=\pi$ (accompanied by the CPT-conjugate jump at $\vartheta=0$), shown in greater detail in Figure \ref{fig:C3-Kwall}.
At the critical phase we find \emph{infinitely many} two-way streets, this is the first hint that we may expect an infinite BPS spectrum of states with charge $n\gamma$ for a certain primitive $\gamma$.
We will confirm this expectation below, with a careful analysis of the $\CK$-wall jump of soliton data.

For later convenience let us be more precise about the generating charge $\gamma$ of the whole BPS spectrum of this model.
This arises naturally in the critical network shown in Figure \ref{fig:C3-Kwall}.
Two-way streets of primary walls are shown in black in the picture on the left, taking their lift gives
\be
	\pi^{-1} \(  -r_1 \cup -p_1 \cup -r_1' \cup -p'_1 \) =  \sum_N \gamma_N\,.
\ee
The sum over $N$ arises from the infinite logarithmic covering, in the picture on the right we provide the explicit form of $\gamma_{N=0}$. For example a label $(+-)_{1,0}$ next to an oriented wall means that we take its lift to sheet $(-,0)$ with the same orientation together with the lift to sheet $(+,1)$ with the opposite orientation. It is straightforward to verify that this gives a closed cycle, by $\gamma_N$ we denote its homology class on $\tSigma$.
The period of the differential $Y(x) \frac{dx}{x}$ on $\gamma_N$ is independent of $N$, therefore all these homology cycles are identified by the quotient by $\ker Z$ that defines the physical charge lattice. This equivalence class is the definition of $\gamma$
\be\label{eq:Z-quotient}
	[\gamma_N] = [\gamma_{N'}] \equiv \gamma\,.
\ee
The actual period is easy to compute, it evaluates to
\be\label{eq:Z-gamma}
\begin{split}
	Z_\gamma
	& = \oint_{\gamma_N} \lambda
	= -\frac{2\pi }{R} \,.
\end{split}
\ee
The normalization of $\lambda$ and the definition of $R$ were discussed around equation (\ref{eq:central-charge-def}).
The fact that $Z_\gamma$ is real and negative reflects the fact that $\gamma$ is the charge of a BPS state that appears at $\vartheta=\pi$.
It also turns out that $Z_\gamma$ is exactly one unit of KK momentum when the 3d-5d system is placed on a circle of radius $R$, as we will see later this fits naturally with our results on the BPS spectrum of this theory.

\subsection{Analysis of the $\CK$-wall jump}\label{sec:C3-K-wall}

We now focus on the $\CK$-wall jump that occurs at $\vartheta=\pi$. We will study the soliton content on the two-way streets that appear, and use it to compute the spectrum of bulk BPS states.

The critical network at $\vartheta=\pi$  features an infinite tower of 2-way streets, see Figure \ref{fig:C3-Kwall}.
There are two distinguished points in the sub-network of 2-way streets: the branch point (on the left, in the first picture), and the joint where all walls intersect.
We would like to consider two resolutions of this degenerate network, defined at $\vartheta = \pi \pm \epsilon$ (called respectively British and American), and study the soliton data for each of these.

In principle the results of \ref{sec:nonabel-map} completely fix the soliton data in either resolution. However in practice it is convenient to handle these computations without having to keep track of the single underlying $\CE$-walls.
Each two-way street of the network in Figure \ref{fig:C3-Kwall} is made of infinitely many $\CE$-walls running in either direction, but we can just as well cluster all the $\CE$-walls with the same orientation together, and model each 2-way street as a union of just two $\CE$-walls. This will help lighten notation.
%

\subsubsection{$ij-ji$ 2-way joints}\label{sec:ij-ji-2-way}
\begin{figure}[h!]
\begin{center}
\includegraphics[width=0.55\textwidth]{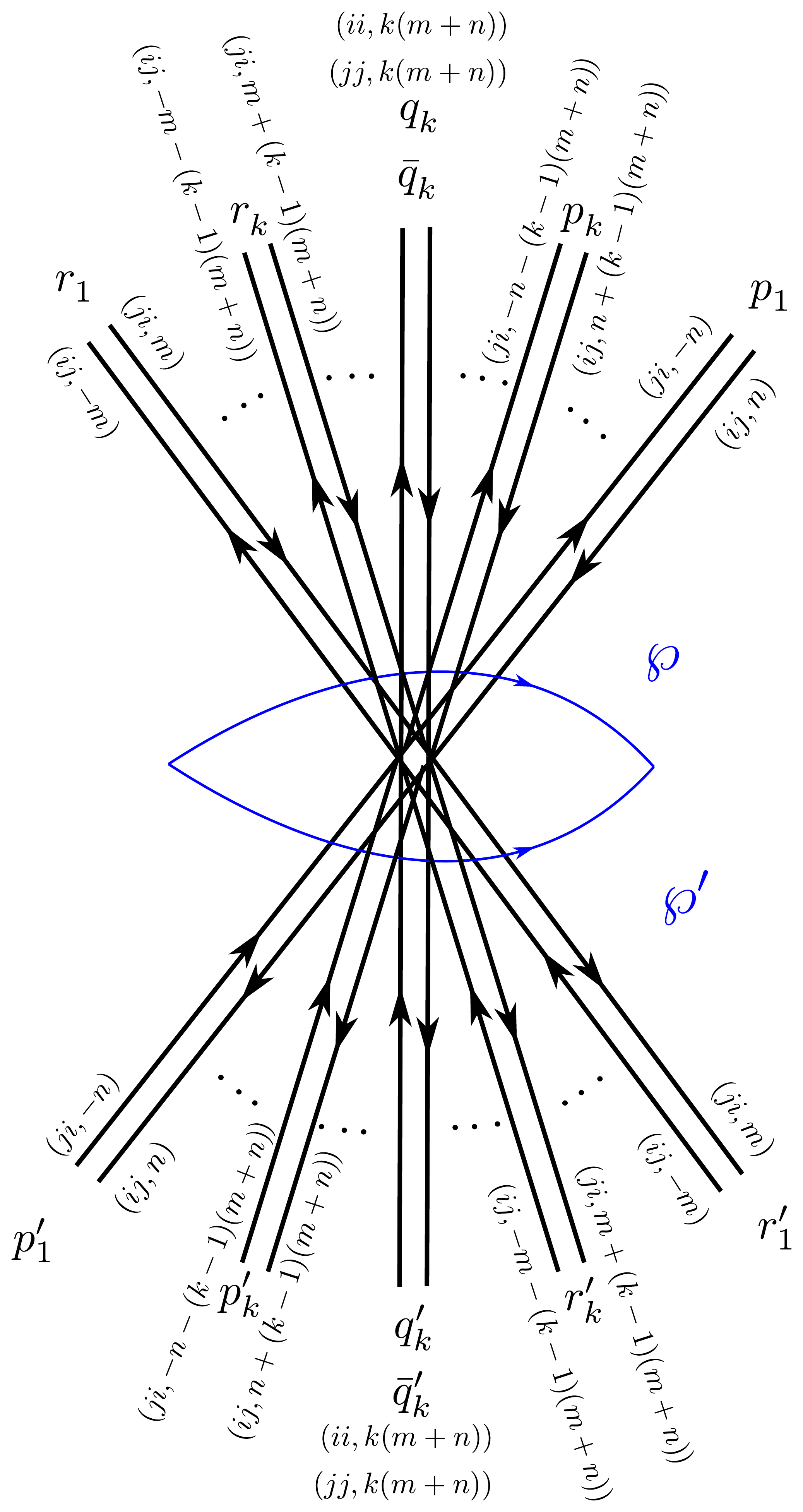}
\caption{$ij-ji$ joint of two-way streets in British resolution. This appears in the $\CK$-wall jump for exponential networks of $\IC^3$.}
\label{fig:ij-ji-2-way}
\end{center}
\end{figure}

Let us start from the most complex part of the problem: the propagation of soliton data across the two-way $ij-ji$ joint that appears within the critical network.
We label the generating functions of solitons (previously called $\Xi_{ij,n}$) as in Figure \ref{fig:ij-ji-2-way}, generating functions $\nu$ carry soliton paths that go \emph{into} the junction, while $\tau$ carry solitons that come \emph{out}.
The goal is to solve for the generating functions of out-going solitons in terms of the in-going ones.

The types of solitons carried by each wall are as follows
\be
\begin{array}{c|c|c}
	s & \tau_s & \nu_s \\
	\hline
	r_{k}	& (ij,-m-(k-1)(m+n)) & (ji,m+(k-1)(m+n)) \\
	q_k		& (ii, -k(m+n)) & (ii, k(m+n)) \\
	\bar q_k	& (jj, -k(m+n)) & (jj, k(m+n)) \\
	p_{k}	& (ji,-n-(k-1)(m+n)) & (ij,n+(k-1)(m+n)) \\
	\hline
	r'_{k}	& (ji,m+(k-1)(m+n)) &  (ij,-m-(k-1)(m+n)) \\
	q'_k		& (ii, k(m+n)) & (ii, -k(m+n)) \\
	\bar q'_k	& (jj, k(m+n)) & (jj, -k(m+n)) \\
	p'_{k}	& (ij,n+(k-1)(m+n)) & (ji,-n-(k-1)(m+n))  \\
	\hline
\end{array}
\ee
where for the sake of generality we introduced new sheet labels, which reduce to
\be\label{eq:label-change}
	i = + \qquad j=- \qquad
	n=-1\qquad m=0
\ee
with the choice of trivialization of subection \ref{sec:C3-geometry}.

The parallel transport equations for paths $\wp,\wp'$ shown in Figure \ref{fig:ij-ji-2-way} are
\be\label{eq:ij-ji-2-way-full-eqs-initial-form}
\begin{split}
	F(\wp) & =
	\(
		\prod_{k\geq 1}^{\nearrow} (1+\tau_{r_k})(1+\nu_{r_k})
	\)
	\cdot
	\(
		\prod_{k\geq 1} e^{\tau_{q_k}} e^{\tau_{\bar q_k}}e^{\nu_{q_k}} e^{\nu_{\bar q_k}}
	\)
	\cdot
	\(
		\prod_{k\geq 1}^{\searrow} (1+\tau_{p_k})(1+\nu_{p_k})
	\)
	\\
	F(\wp') & =
	\(
		\prod_{k\geq 1}^{\nearrow} (1+\nu_{p'_k})(1+\tau_{p'_k})
	\)
	\cdot
	\(
		\prod_{k\geq 1} e^{\tau_{q'_k}} e^{\tau_{\bar q'_k}}e^{\nu_{q'_k}} e^{\nu_{\bar q'_k}}
	\)
	\cdot
	\(
		\prod_{k\geq 1}^{\searrow} (1+\nu_{r'_k})(1+\tau_{r'_k})
	\) \,.
\end{split}
\ee
We will next turn to a detailed study of these equations.

\subsubsection{Factoring out $q,q',\bar q,\bar q'$ contributions}\label{sec:factoring-Q}
Writing down the parallel transport equations component-by-component, it turns out that most of the complexity in (\ref{eq:ij-ji-2-way-full-eqs-initial-form}) stems from the terms in the middle parentheses.
As it turns out, they can be actually factored out of the equations, leading to a great simplification.
Let us introduce
\be\label{eq:fQ-def}
	\fQ := \prod_{k\geq 1} e^{\tau_{q_k}} e^{\tau_{\bar q_k}}e^{\nu_{q_k}} e^{\nu_{\bar q_k}}
	\qquad
	\fQ' := \prod_{k\geq 1} e^{\tau_{q'_k}} e^{\tau_{\bar q'_k}}e^{\nu_{q'_k}} e^{\nu_{\bar q'_k}}
\ee
Let us recall that, since the $\CK$-wall appears at $\vartheta=0,\pi$ we also have the periodicity effects explained in Section \ref{subsec:periodic-ii} related to $ii/jj$ walls. This implies that each soliton $a$ comes with a semi-infinite tower of its extensions $a^{(k)}$ around the puncture. In writing (\ref{eq:fQ-def}) the various generating functions $\nu_{q},\tau_q$ are understood to \emph{include} extended copies of solitons generated in this way.
All factors in these products commute, so there are no ordering ambiguities. A proof can be found in Appendix \ref{app:q-commute}.
Therefore we can split $\fQ$ and $\fQ'$ into commuting pieces of different types
\be
\begin{split}
	\fQ & = \fQ_{ii} \fQ_{jj} = \fQ_{jj} \fQ_{ii}  \qquad
	\fQ' = \fQ'_{ii} \fQ'_{jj} = \fQ'_{jj} \fQ'_{ii}
\end{split}
\ee
with
\be\label{eq:def-fQ}
\begin{split}	
	\fQ_{ii} & = \prod_{k} e^{\tau_{q_k}} e^{\nu_{q_k}} = \sum_{\ell\in\IZ } \fQ_{ii,\ell} \qquad
	\fQ_{jj} = \prod_{k} e^{\tau_{\bar q_k}} e^{\nu_{\bar q_k}} = \sum_{\ell\in\IZ } \fQ_{jj,\ell} \\
	\fQ'_{ii} & = \prod_{k} e^{\tau_{q'_k}} e^{\nu_{q'_k}} = \sum_{\ell\in\IZ } \fQ'_{ii,\ell} \qquad
	\fQ'_{jj}  = \prod_{k} e^{\tau_{\bar q'_k}} e^{\nu_{\bar q'_k}} = \sum_{\ell\in\IZ } \fQ'_{jj,\ell}
\end{split}
\ee
where $\fQ_{ii,\ell}, \fQ_{jj,\ell}$ are generating functions of solitons with shifts of the logarithmic branch by $\ell(m+n)$, and can be expressed in terms of $\nu,\tau$'s (similarly for $\fQ'$).
Other components are just zero $\fQ_{ij}  = \fQ_{ji} = \fQ'_{ij}  = \fQ'_{ji}  = 0$.

Each terms $\fQ_{ii,\ell}$ can be expressed as a function of $\tau,\nu$ as follows.
Let $k$ be a non-negative integer, and consider a partition
\be
	(\underbrace{\lambda_1,\dots,\lambda_1}_{\delta_1},\underbrace{\lambda_2,\dots,\lambda_2}_{\delta_2},\dots,\underbrace{\lambda_n,\dots,\lambda_n}_{\delta_n})
	\qquad
	\text{s.t.}
	\qquad
	\sum_{i=1}^{n} \delta_i \lambda_i = k
\ee
Define
\be
\begin{split}
	\( \prod_{m\geq 1} e^{\nu_{q_m}} \)_{k}
	&=
	\sum_{\sum_{i} \delta_i \lambda_i = k} \, \frac{1}{\delta_1 !\,\cdots \delta_n!}\, (\nu_{q_{\lambda_1}})^{\delta_1} \cdots(\nu_{q_{\lambda_n}})^{\delta_n}
	\\
	\( \prod_{m\geq 1} e^{\tau_{q_m}} \)_{k}
	&= \sum_{\sum_{i} \delta_i \lambda_i = k} \, \frac{1}{\delta_1 !\,\cdots \delta_n!}\, (\tau_{q_{\lambda_1}})^{\delta_1} \cdots(\tau_{q_{\lambda_n}})^{\delta_n}
\end{split}
\ee
Then
\be\label{eq:fQ_ell-explicit-form}
	\fQ_{ii,\ell} =
	\left\{\begin{array}{lr}
	\sum_{s\geq 0} \( \prod_{m\geq 1} e^{\nu_{q_m}} \)_{s+\ell}   \( \prod_{m\geq 1} e^{\tau_{q_m}} \)_{s}
	& \ell > 0
	\\
	\sum_{s\geq 0} \( \prod_{m\geq 1} e^{\nu_{q_m}} \)_{s}   \( \prod_{m\geq 1} e^{\tau_{q_m}} \)_{s+\ell}
	& \ell < 0
	\\
	\sum_{s\geq 0} \( \prod_{m\geq 1} e^{\nu_{q_m}} \)_{s}   \( \prod_{m\geq 1} e^{\tau_{q_m}} \)_{s}
	& \ell = 0
	\end{array}
	\right.
\ee
For later convenience let us further split these generating series into
\be\label{eq:fQ-ii-N-ell}
	\fQ_{ii,\ell} = \sum_{N\in\IZ}  \fQ_{ii,N,N+\ell}
\ee
where the $N$-th term on the \emph{rhs} counts soliton paths connecting vacua $(i,N)$ and $(i,N+\ell(m+n))$.
Note these functions  satisfy a slightly non-standard relations, when multiplied with the standard soliton generating functions
\be
	\Xi_{ij}\fQ_{ii} = \Xi_{ij}
	\qquad
	\Xi_{ji}\fQ_{jj} = \Xi_{ji} \,.
\ee
These do not vanish due to the fact that $\fQ_{ii,0}$ includes the constant  $1$.

Let us now get back to the main claim we wish to prove, which is that $\fQ,\fQ'$ actually drop out of the equations.
In order to show this, we shall need two technical results.
The first result, proven in Appendix \ref{eq:fQ-rels}, states that given any shift-symmetric generating functions $\Xi_{ii}$ and $\Xi_{jj}$
\be\label{eq:ii-ii-comm}
	\fQ_{ii,\ell} \Xi_{ii,m} = \Xi_{ii,m} \fQ_{ii,\ell} \,
\ee
\be\label{eq:ii-jj-comm}
	\fQ_{ii,\ell} \Xi_{ij,m} = \Xi_{ij,m} \fQ_{jj,\ell}  \,.
\ee
The second result, whose proof is given in Appendix \ref{eq:fQ-equality}, states that
\be\label{eq:Qii-Qiip}
	\fQ_{ii}= \fQ'_{ii}\,, \qquad \fQ_{jj}= \fQ'_{jj}\,.
\ee

Since each component of the flatness equations (\ref{eq:ij-ji-2-way-full-eqs-initial-form}) is linear in the  factors $\fQ_{ii},\fQ_{jj}$ and $\fQ'_{ii},\fQ'_{jj}$, we can use properties (\ref{eq:ii-ii-comm})-(\ref{eq:ii-jj-comm}) to collect them as overall factors both in $F(\wp)$ and $F(\wp')$.
Moreover since $\fQ_{ii}, \fQ_{jj},\fQ'_{ii}, \fQ'_{jj}$ are invertible by definition, they can be cancelled on either side of the equations by virtue of (\ref{eq:Qii-Qiip}).
As a result we can rewrite (\ref{eq:ij-ji-2-way-full-eqs-initial-form}) by simply omitting the factors $\fQ,\fQ'$.
\be\label{eq:ij-ji-2-way-full-eqs}
\begin{split}
	F(\wp) & =
	\(
		\prod_{k\geq 1}^{\nearrow} (1+\tau_{r_k})(1+\nu_{r_k})
	\)
	\cdot
	\(
		\prod_{k\geq 1}^{\searrow} (1+\tau_{p_k})(1+\nu_{p_k})
	\)
	\\
	F(\wp') & =
	\(
		\prod_{k\geq 1}^{\nearrow} (1+\nu_{p'_k})(1+\tau_{p'_k})
	\)
	\cdot
	\(
		\prod_{k\geq 1}^{\searrow} (1+\nu_{r'_k})(1+\tau_{r'_k})
	\)
\end{split}
\ee
Note that due to the path-concatenation algebra of the generating functions involved in the flatness identity, this is a nontrivial simplification.
Henceforth we shall study this simplified form of the flatness equations.

\subsubsection{Breaking down the equations by path tilting}\label{sec:eqn-breakdown}
\begin{figure}[h!]
\begin{center}
\includegraphics[width=0.35\textwidth]{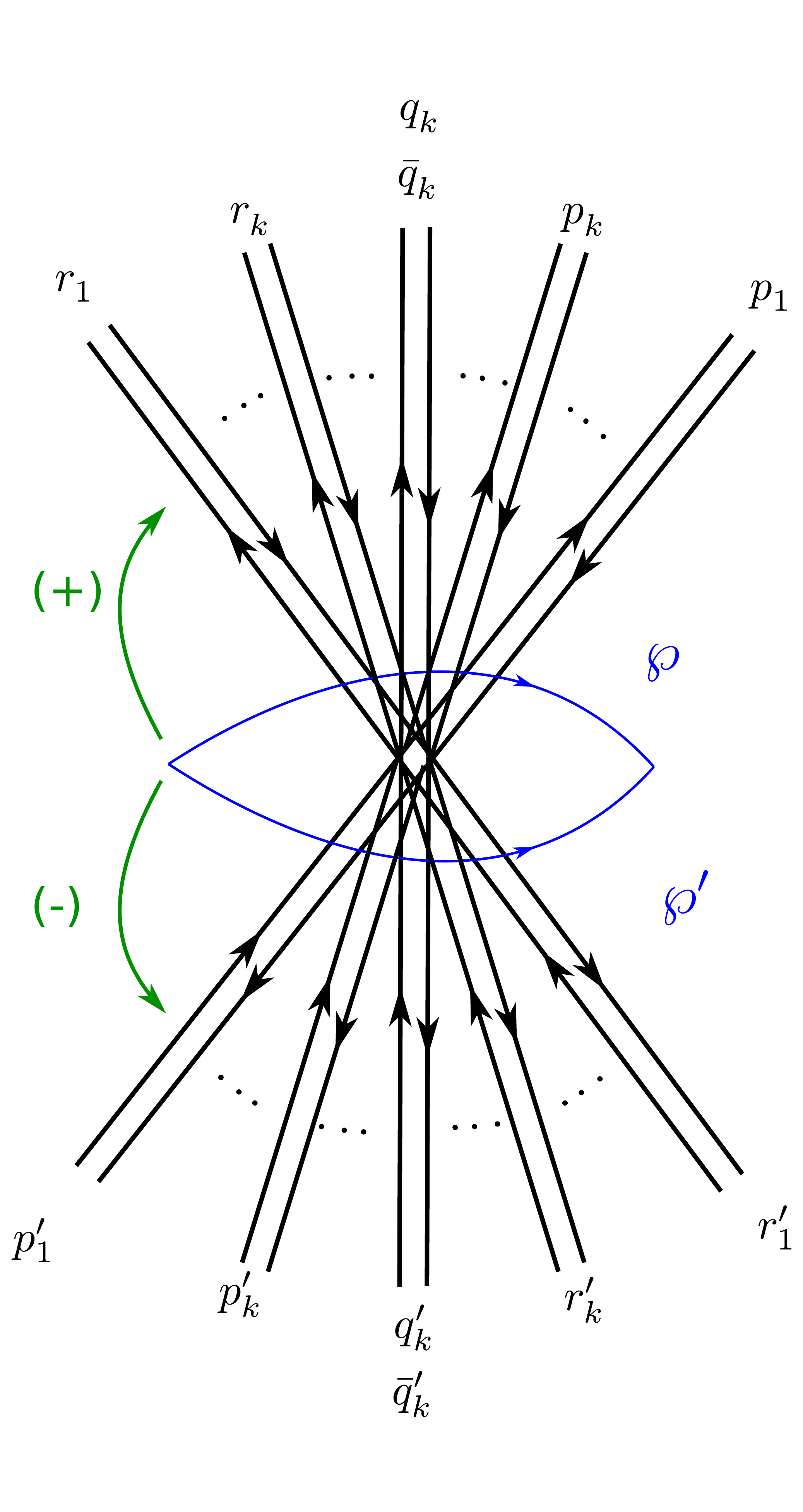}
\caption{Tilting the paths from Figure \ref{fig:ij-ji-2-way} gives equivalent flatness equations.}
\label{fig:ij-ji-2-way-tilted}
\end{center}
\end{figure}

Equations (\ref{eq:ij-ji-2-way-full-eqs}) still look rather daunting.
The most obvious approach to solving them would be to consider each component individually, however this turns out not to be particularly effective.
A better (but ultimately equivalent) strategy is to conjugate the equations by pieces of parallel transport, so that we produce new (equivalent) equations for \emph{tilted} paths.
More precisely, we replace $\wp,\wp'$ by $\wp_{(k)},\wp'_{(k)}$ defined by tilting the former clockwise ($k>0$) or counterclockwise ($k<0$) across $|k|$ 2-way streets. See Figure \ref{fig:ij-ji-2-way-tilted}.

Let us start with $k>0$, focussing on the following components of the parallel transport equations
(to lighten notation we suppress trivial terms like $X_{\wp^{(i)}}$)
\be
\begin{split}
	F(\wp_{(k)})_{ii,0} & = \prod_{s\geq k+1} Q(r_s)\, \qquad
	F(\wp'_{(k)})_{ii,0} = \prod_{s\geq k+1} Q(r'_s) \,.
\end{split}
\ee
where $Q(r_s)$ appearing here is related to $Q_N(r_s)$ defined in (\ref{eq:QN-def}) by $Q(r_s) = \sum_{N\in\IZ} Q_N(r_s)$.\footnote{Each $Q_N$ includes a `1' in the definition, they should be considered as identity operators restricted to the $(ii,N,N)$ sector. Then the sum $\sum_N$ is effectively a direct sum of operators.}
The two parallel transports given above must be equal for each $k$, therefore
\be
	{Q(r_s) = Q(r'_s)} \qquad s\geq 1\,.
\ee
Likewise, tilting with $k<0$ we obtain the analogous
\be
\begin{split}
	F(\wp_{(k)})_{jj,0} & = \prod_{s\geq k+1} Q(p_s) \,\qquad
	F(\wp'_{(k)})_{jj,0} = \prod_{s\geq k+1} Q(p'_s) \,.
\end{split}
\ee
Since this holds for each $k$, we find
\be
	{Q(p_s) = Q(p'_s)} \qquad s\geq 1\,.
\ee
The global structure of the critical network (see Figure \ref{fig:C3-Kwall}) further implies that $Q(r_s)=Q(p'_s)$.
Therefore we introduce
\be\label{eq:C3-Qs}
	Q_k := {Q(r_s)=Q(p_s)=Q(r'_s)=Q(p'_s)} \,.
\ee

Let us now turn to another component of the flatness equations, namely the sector $(ij,-m-k(m+n))$.
With $k>0$ we find
\be
\begin{split}
	F(\wp_{(k)})_{ij,-m-k(m+n)} & = \tau_{r_{k+1}} \(\prod_{s\geq 1} Q(p_s)\) \(\prod_{s= 1}^{k} Q(r'_s)\) \,,\\
	F(\wp'_{(k)})_{ij,-m-k(m+n)} & = \(\prod_{s\geq k+2} Q(r'_s) \) \nu_{r'_{k+1}}\,,
\end{split}
\ee
using (\ref{eq:C3-Qs}) this simplifies to
\be\label{eq:ij-ji-2-way-tau-r-k}
	{
	\tau_{r_{k}}  = Q_{k}^{-1} \(\prod_{s= 1}^{k-1} Q_s\) ^{-2} \nu_{r'_{k}}
	}\,.
\ee
Keeping the same tilted paths, but looking at sector $(ij,m+k(m+n))$ gives
\be
\begin{split}
	F(\wp_{(k)})_{ij,m+k(m+n)} & = \nu_{r_{k+1}} \(\prod_{s\geq k+2} Q(r_s)\) \,,\\
	F(\wp'_{(k)})_{ij,m+k(m+n)} & = \(\prod_{s\geq 1} Q(p'_s)\)  \tau_{r'_{k+1}}   \(\prod_{s=1}^{k} Q(r'_s) \)\,,
\end{split}
\ee
again this simplifies once we take into account (\ref{eq:C3-Qs}), leading to
\be\label{eq:ij-ji-2-way-tau-rp-k}
	\tau_{r'_{k}}  =  Q_k^{-1} \(\prod_{s=1}^{k-1} Q_s \)^{-2}  \nu_{r_{k}} \,.
\ee

Similarly, choosing paths with negative tilting $k<0$ and studying the $(ji,\pm(n+k(m+n)))$ components of the parallel transport, we obtain
\be\label{eq:ij-ji-2-way-tau-p-k}
	{
	\tau_{p_{k}}  =  Q_k^{} \(\prod_{s=1}^{k-1} Q_s \)^{2}  \nu_{p'_{k}} \,,
	}
	\qquad
	{
	\tau_{p'_{k}}  =  Q_k^{} \(\prod_{s=1}^{k-1} Q_s \)^{2}  \nu_{p_{k}} \,.
	}
\ee

In fact the two way streets junction appearing in this case for $\mathbb{C}^3$ appears to be sufficiently generic to deserve study in further details.
In appendix \ref{app:two-way-junction}, we give the full form of the parallel transport equations before implementing various properties, for example, periodicity
of the $\mathbb{C}^3$ network.

\subsubsection{Soliton data of primary walls}
We are now in a position to obtain explicit solutions for the soliton data on the $\CE$-walls $p_1,r_1, p'_1, r'_1$.
The global topology of the $\IC^3$ network implies
\be\label{eq:periodicity-C3}
\begin{split}
	\tau_{r_k} & \dot{=} \nu_{p_k'} \qquad \nu_{r_k}  \dot{=} \tau_{p_k'}  \qquad  k \geq 1 \\
	\tau_{p_k} & \dot{=} \nu_{r'_k} \qquad \nu_{p_k}  \dot{=} \tau_{r'_k} \qquad k\geq 2
\end{split}
\ee
where the $\dot{=}$ is a reminder that identifications are understood upon transporting a soliton along a loop around the puncture at infinity.

\begin{figure}[ht]
\begin{center}
\includegraphics[width=0.35\textwidth]{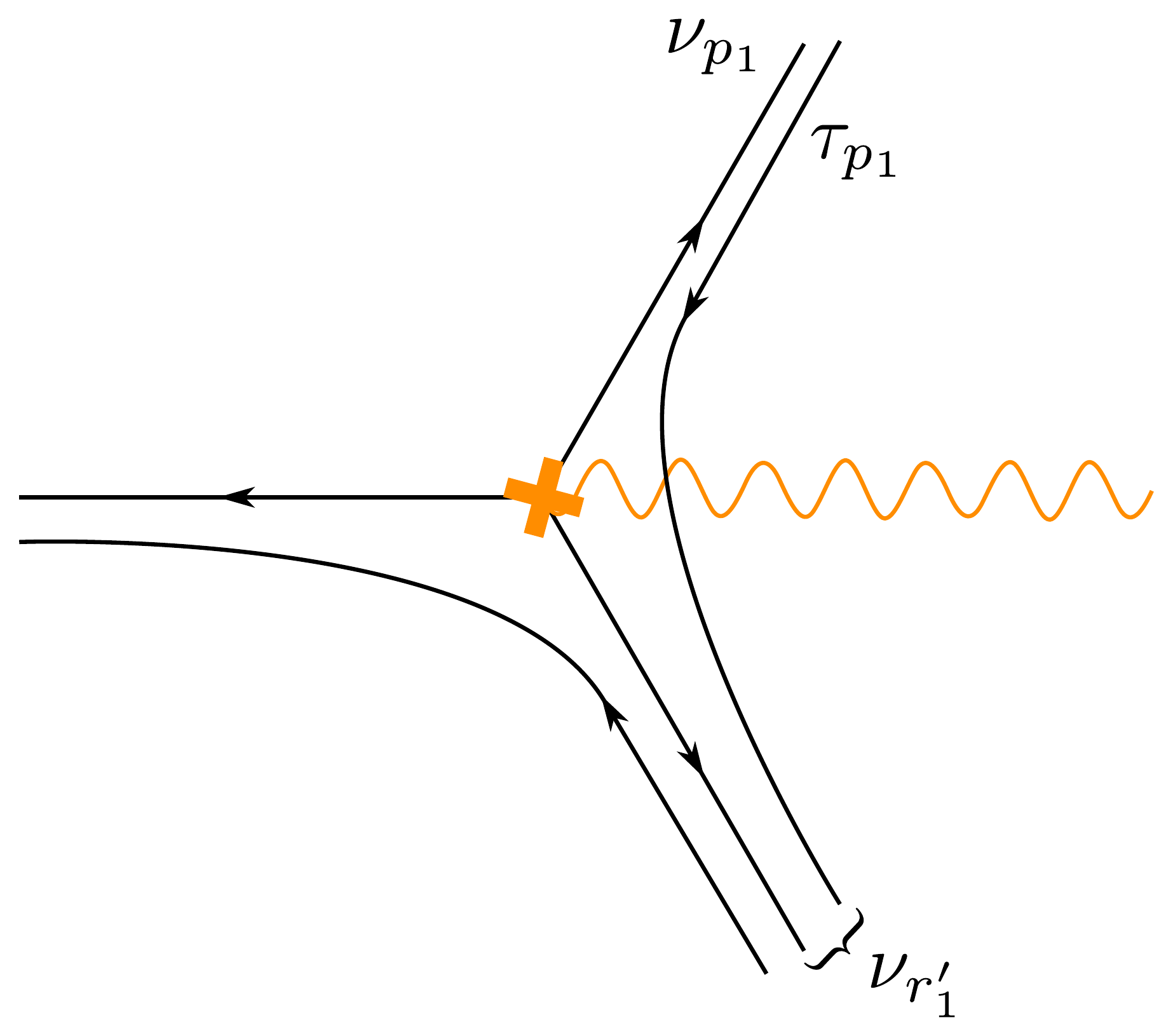}
\caption{Soliton propagation near the branch point of the network of $\IC^3$, just after the critical phase (British resolution).}
\label{fig:C3-branch-pt-detail}
\end{center}
\end{figure}

It follows that
\be\label{eq:Q-periodicity-C3}
\begin{split}
	Q(r_k) & = Q(p_k') \quad  k \geq 1 \\
	Q(p_k) & = Q(r'_k) \quad k\geq 2\,.
\end{split}
\ee
From homotopy invariance around the branch point (in the British resolution, see Figure \ref{fig:C3-branch-pt-detail}) it is straightforward to derive the following relations
\be\label{eq:branch-point-C3}
\begin{split}
	&
	\nu_{r_1'} = \tau_{p_1} + \sum_N X_{b_N}
	\qquad
	\nu_{p_1} = \sum_N X_{a_N}\,.
\end{split}
\ee
Solitons $a_N$ are sourced from the branch point along $p_1$ while $b_N$ are sourced form the branch point and run along $r_1'$.

Using the identities derived in subsection \ref{sec:eqn-breakdown}, we can compute $Q(p_1)$
\be\label{eq:Qp1-partial}
\begin{split}
	Q(p_1)
	& = 1 + \tau_{p_1} \nu_{p_1} \\
	& = 1 + \nu_{p_1'}  Q(p_1) \nu_{p_1} \\
	& = 1 + \tau_{r_1}  Q(p_1) \nu_{p_1} \\
	& = 1 + \nu_{r_1'}  Q(r'_1)^{-1} Q(p_1) \nu_{p_1} \\
	& = 1 + \(\(\sum_N X_{\gamma_N}\) \tau_{p_1}   + \sum_N X_{b_N} \) Q(r'_1)^{-1} Q(p_1) \nu_{p_1} \\
	& = 1 + \(\(\sum_N X_{\gamma_N}\) \tau_{p_1}\nu_{p_1}   + \sum_N X_{b_N a_N} \) Q(r'_1)^{-1} Q(p_1)\\
	& = 1 +\( \(\sum_N X_{\gamma_N}\) \(Q(p_1) - 1\)   + \sum_N X_{\gamma_N}  \)Q(r'_1)^{-1} Q(p_1)\\
	& = 1 + \(\sum_N X_{\gamma_N}\)  Q(p_1)^2 Q(r'_1)^{-1}  \,.
\end{split}
\ee
Here $\gamma_N$ is the closed cycle obtained by canonical lift (cf. subsection \ref{sec:ij-ji-2-way-jumps}) of two-way streets $p_1,r_1,p'_1,r'_1$ to the $N$-th sheet of the cover $\tSigma$.
The insertion of $\sum_N X_{\gamma_N}$ in front of $\tau_{p_1}$ is justified by the fact that the latter has been transported all the way along the 2-way wall by application of the various identities, and therefore each 2d-4d soliton (open path) inside $\tau_{p_1}$ should be transported. Moreover, given a path $a \in \Gamma_{ji,N,N-n}$, its transport will be of the type $X_{\gamma_N} X_a$ or $X_a X_{\gamma_{{N-n}}}$ where $\gamma_N \in\Gamma_{jj,N,N}$. Since concatenation of paths can only occur if the endpoints match, the whole sum over $N$ simply distributes one such concatenation for each soliton.

Now thanks to (\ref{eq:C3-Qs}) we immediately obtain
\be
	Q_1 = \(1- \(\sum_N X_{\gamma_N}\) \)^{-1}\,.
\ee
In terms of the generating functions $Q_N(p)$ defined in (\ref{eq:QN-def}), we have
\be
	Q_N(p_1) =
	Q_N(r_1) =
	Q_N(p'_1) =
	Q_N(r'_1) = \(1-  X_{\gamma_N} \)^{-1}\,.
\ee
Due to (\ref{eq:Z-quotient})  all $\gamma_N$ are identified, therefore we finally obtain
\be
	\alpha_\gamma(p_1) =
	\alpha_\gamma(r_1) =
	\alpha_\gamma(p'_1) =
	\alpha_\gamma(r'_1) =
	-1 \,.
\ee

We can also solve for the 2d-4d generating functions: the same manipulations as in (\ref{eq:Qp1-partial}) lead to
\be
	\tau_{p_1} = \(\sum_N X_{\gamma_N}\) \tau_{p_1} + \sum_N X_{b_N} \,,
\ee
therefore
\be
	\tau_{p_1} = \sum_{N} X_{b_N} \(1 - \sum_{N'} X_{\gamma_{N'}}\)^{-1} = \sum_{N} X_{b_N} \(1 - X_{N}\)^{-1}\,,
\ee
where use used the fact that $\gamma_N = a_N b_N\in \Gamma_{ii,N,N}$ can only concatenate with $b_M$ if $N=M$.
The remaining soliton data of primary walls can be obtained in a similar way, and reads
\be
	\nu_{p_1'}  = \tau_{p_1} Q_1^{-1} = \sum_{N} X_{b_N} \,,
\ee
\be
	\tau_{r_1} = \nu_{p_1'} = \sum_{N} X_{b_N} \,,
\ee
\be
	\nu_{r_1'} = \tau_{r_1} Q_1 = \sum_{N} X_{b_N} \(1 - X_{\gamma_N}\)^{-1} \,,
\ee
\be
	\tau_{p_1'} = \nu(p_1) Q_1 = \sum_N X_{a_N} (1-X_{\gamma_N})^{-1} \,,
\ee
\be
	\nu_{r_1} = \tau_{p_1'} = \sum_N X_{a_N} (1-X_{\gamma_N})^{-1} \,,
\ee
\be
	\tau_{r_1'} = \nu_{r_1} Q_1^{-1} = \sum_N X_{a_N} \,.
\ee

\subsubsection{Soliton data of descendant walls}\label{sec:descandant-solitons}

To compute $Q_k$ for $k>1$ we follow a different strategy, going back to the parallel transport equations.
Let us consider the horizontal paths $\wp,\wp'$ of Figure \ref{fig:ij-ji-2-way}, and study components $(ii,-m-n)$
\be
\begin{split}
	F(\wp)_{ii,-m-n} & = \tau_{r_1} \(\prod_{k\geq 2} Q(p_k)\)  \, \tau_{p_1}
				\\
	F(\wp')_{ii,-m-n} & = \sum_{s\geq 1}  \tau_{p'_{s}}\nu_{p'_{s+1}} \, \(\prod_{k\geq 1} Q(r'_k) \)
				+ : \sum_{s\geq 1}\(\prod_{k\neq s,s+1} Q(r'_k)\) \nu_{r'_{s+1}}\tau_{r'_s} :
\end{split}
\ee
Dividing each side by $\prod_{k\geq 1}Q_k$ brings the equality of these expressions into the form
\be\label{eq:ii-m-n}
	\frac{ \tau_{r_1}  \, \tau_{p_1}}{Q_1}
	=
	\sum_{k\geq 1} \( \tau_{p'_{k}}\nu_{p'_{k+1}}
	+ \frac{\nu_{r'_{k+1}}\tau_{r'_k}}{Q_{k}Q_{k+1}} \)\,.
\ee
On the other hand, combining (\ref{eq:ij-ji-2-way-tau-r-k})-(\ref{eq:ij-ji-2-way-tau-rp-k}) together with the periodicity properties (\ref{eq:periodicity-C3})
\be\label{eq:transport-nu-rp-k}
	\nu_{r'_{k+1}} = \tau_{r_{k+1}} Q_{k+1} Q_k^2\cdots Q_1^2
	\,\dot{=}\, \nu_{p'_{k+1}}Q_{k+1} Q_k^2\cdots Q_1^2  \,,
\ee
\be\label{eq:transport-tau-rp-k}
	\tau_{r'_{k}} = \nu_{r_{k}} Q^{-1}_{k} Q^{-2}_{k-1}\cdots Q_1^{-2}
	\,\dot{=}\, \tau_{p'_{k}} Q^{-1}_{k} Q^{-2}_{k-1}\cdots Q_1^{-2} \,,
\ee
therefore
\be\label{eq:suspect-equality}
	\frac{\nu_{r'_{k+1}}\tau_{r'_k}}{Q_{k}Q_{k+1}}
	\,\dot{=}\,
	\tau_{p'_{k}}\nu_{p'_{k+1}}  \,,
\ee
where the symbol $\,\dot{=}\,$ is a reminder that a transport of soliton endpoints has been employed, and needs to be properly taken into account. We show how to do this in Appendix \ref{app:soliton-transport}, the correct expression turns out to be
\be\label{eq:corrected-suspect-equality}
	\frac{\nu_{r'_{k+1}}\tau_{r'_k}}{Q_{k}Q_{k+1}}
	\,{=}\,
	-  \(\sum_N X_{\gamma_N}\)  \tau_{p'_{k}}\nu_{p'_{k+1}}  \,.
\ee
Therefore  (\ref{eq:ii-m-n}) becomes
\be\label{eq:tau1-tau1}
	\frac{ \tau_{r_1}  \, \tau_{p_1}}{Q_1}
	=
	\(1-\sum_N X_{\gamma_N}\)  \sum_{k\geq 1}  \tau_{p'_{k}}\nu_{p'_{k+1}}
	\,.
\ee
A similar analysis for components $(ii,m+n)$ of the parallel transport leads to a similar equation
\be\label{eq:nu1-nu1}
	\frac{ \nu_{r_1}  \, \nu_{p_1}}{Q_1}
	=
	\(1- \sum_N X_{\gamma_N}\)  \sum_{k\geq 1} \tau_{r'_{k+1}}\nu_{r'_{k}}
	\,.
\ee

Without loss of generality let us define $\theta_k,\bar\theta_k, C_k,\bar C_k$ as
\be\label{eq:recursion-eqs}
\begin{split}
	\tau_{p'_k} & = C_k \theta_k \tau_{p'_{k-1}}
	\qquad
	\nu_{p'_k} =   \bar\theta_k \nu_{p'_{k-1}}
	\\
	\nu_{r'_k} & = \bar C_k  \nu_{r'_{k-1}} \bar \theta_k
	\qquad
	\tau_{r'_k} = \tau_{r'_{k-1}} \theta_k
	\,.
\end{split}
\ee
$\theta_k$ is a formal series involving paths of type $(ii,m+n)$, while $\bar\theta_k$  carries paths of type $(jj,-m-n)$. $C_k, \bar C_k$ are functions of closed homology variables like $X_\gamma$, they commute with all other factors.
Note that periodicity relations among $r'_k, p'_k$ and $r_k, p_k$ in (\ref{eq:periodicity-C3}) fix the relation between $\tau_{p_k},\tau_{p_{k-1}}$ etc. in terms of $\theta_k, \bar\theta_k, C_k,\bar C_k$.
For convenience also introduce $\tilde\theta_k,\tilde{\bar{\theta}}_k$ defined by
\be
	\theta_k \Xi_{ij} = \Xi_{ij} \tilde\theta_k \,,
	\qquad
	\tilde{\bar\theta}_k \Xi_{ij} = \Xi_{ij} \bar\theta_k \,,
\ee
for any $\Xi_{ij}$ carrying paths of type $ij$ and enjoying shift symmetry.\footnote{Essentially the addition of $\,\tilde{}\,$ is an involution that switches $i\leftrightarrow j$, much like in (\ref{eq:ii-jj-comm}) this relies on the property of shift symmetry for each term in the equation.}

Combining (\ref{eq:ij-ji-2-way-tau-rp-k}) with (\ref{eq:periodicity-C3}) we obtain
\be
	{\tau_{r'_k}} {\tau_{p'_{k-1}}} = \(\sum_N X_{-\gamma_N}\)  Q_k^{-1} Q_{k-1}^{-1} {\tau_{r'_{k-1}}}  {\tau_{p'_k}} \,,
	\qquad
	{\nu_{p'_k}}  {\nu_{r'_{k-1}}}  = \(\sum_N X_{-\gamma_N}\)   Q_k^{-1} Q_{k-1}^{-1} {\nu_{p'_{k-1}}} {\nu_{r'_k}} \,,
\ee
where $X_{-\gamma_N} $ was introduced to keep track of the transport of soliton paths implicit in (\ref{eq:periodicity-C3}), and denoted by $\dot{=}$ in that expression.\footnote{A way to see why this corrections is necessary is by consistency with (\ref{eq:corrected-suspect-equality}).}
These relations fix $C_k, \bar C_k$ to be
\be
	C_k = \bar C_k =\(\sum_N X_{\gamma_N}\)  \, Q_kQ_{k-1} \,.
\ee
Moreover from
\be
	Q_k = 1+\tau_{p'_k}\nu_{p'_k}
	= 1 +\(\sum_N X_{\gamma_N}\) Q_k Q_{k-1} \theta_k \tilde{\bar\theta}_k (Q_{k-1}-1) \,,
\ee
it follows
\be\label{eq:theta-bar-theta}
	\theta_k\tilde{\bar\theta}_k
	=
	\frac{ \sum_N X_{-\gamma_N} }{Q_k Q_{k-1}} \frac{Q_{k}-1}{Q_{k-1}-1} \,.
\ee
Without loss of generality, we can express $\theta_k,\bar\theta_k$ in terms of a universal piece containing open paths times a $k$-dependent piece that only depends on $X_\gamma$
\be\label{eq:theta-f}
\begin{split}
	&\theta_k = \nu_{p_1}\nu_{r_1} \, f_k \,,
	\qquad
	\bar \theta_k = \tau_{p_1}\tau_{r_1} \, \bar f_k \,,
	\\
	&\tilde\theta_k = \nu_{r_1}\nu_{p_1} \, f_k \,,
	\qquad
	\tilde{\bar \theta}_k = \tau_{r_1}\tau_{p_1} \, \bar f_k\,.
\end{split}
\ee
Then (\ref{eq:theta-bar-theta}) turns into a relation for $f_k,\bar f_k$:
\be\label{eq:ffbar}
	f_k\bar f_k =  \frac{\sum_N X_{-\gamma_N}}{Q_k Q_{k-1}} \frac{Q_{k}-1}{Q_{k-1}-1} \frac{1}{\(Q_1 - 1 \)^{2}} \,.
\ee
Moreover, repeated application of (\ref{eq:recursion-eqs}) gives
\be
	\tau_{p'_k}\nu_{p'_{k+1}} = \tilde{\bar\theta}_{k+1} (Q_k - 1) \,,
	\qquad
	\tau_{r'_{k+1}}\nu_{r'_{k}} = \tilde \theta_{k+1} (Q_k - 1)  \,.
\ee
Substitution into the flatness equations (\ref{eq:tau1-tau1}) and (\ref{eq:nu1-nu1}) gives
\be\label{eq:f-fbar-eqs}
	\frac{1}{1-\sum_N X_{\gamma_N}} = Q_1 \sum_{k\geq 1} \bar f_{k+1}(Q_k-1) \,,
	\qquad
	\frac{1}{1-\sum_N X_{\gamma_N}} = Q_1 \sum_{k\geq 1} f_{k+1}(Q_k-1) \,.
\ee
from which we deduce $f_k = \bar f_k$.
Taking into account (\ref{eq:ffbar}), we propose the following solution
\be\label{eq:f-solution}
	f_k = \frac{1}{Q_{k-1}\(Q_1 - 1\)}	\,,
\ee
which implies
\be\label{eq:Qk-solution}
	Q_k = \(1 - \sum_N X_{\gamma_N}^{k}\)^{-1} \,.
\ee
It can be checked that this indeed solves both (\ref{eq:ffbar}) and (\ref{eq:f-fbar-eqs}).
In terms of the generating functions $Q_N(p)$ defined in (\ref{eq:QN-def}), we have
\be
	Q_N(p_k) =
	Q_N(r_k) =
	Q_N(p'_k) =
	Q_N(r'_k) = \(1-  X_{k\,\gamma_N} \)^{-1}\,.
\ee
These equations alone are not sufficient to fix all the $f_k$, one needs to consider other terms in the parallel transport equations as well. We checked that this proposal provides a nontrivial solution to  components $(ii,\pm(2m+2n))$ and $(jj,\pm(2m+2n))$ of the parallel transport equations .

Expression (\ref{eq:f-solution}) is our proposal to the solution to the flatness equations. This determines $\theta_k$ through (\ref{eq:theta-f}), and therefore determines all generating functions of solitons on walls $r_k, p_k, r'_k, p'_k$ through the relations (\ref{eq:recursion-eqs}) and the known expressions for $\tau_{r_1}, \nu_{r_1},\dots$ obtained in subsection \ref{sec:eqn-breakdown}.

\subsubsection{Full BPS spectrum}{\label{5dBPS}}

The BPS spectrum can be computed from the formula (\ref{eq:BPS-index-formula}).
By factoring $Q_N(p)$ for each 2-way street $p$ in the critical network, we found
\be
\begin{split}
	& \alpha_{k\gamma}(r_\ell) =\alpha_{k\gamma}(p_\ell) =\alpha_{k\gamma}(r'_\ell) =\alpha_{k\gamma}(p'_\ell) = -\delta_{k,\ell} \,.
\end{split}
\ee
As we explained in subsection \ref{sec:general-K-wall-formula} the analogue quantities for $ii/jj$ 2-way streets are ambiguous, but can be fixed by requiring that a certain 1-chain $L_N(k \gamma)$ defined in (\ref{eq:L_gamma}) is closed.
This can be achieved by taking the following values\footnote{Monodromy around the puncture at infinity includes a square-root type action, which switches $i \leftrightarrow j$.
Therefore the fact that we choose $\alpha$ differently for $q_i$ and $\bar q'_i$ may seem puzzling for two reasons: first these two-way streets have the same soliton content since they are the continuation of each other; second, the lift may appear to be ``broken'' at the branch cut. The first motivation is addressed by recalling that $\alpha_\gamma(q)$ are not defined in terms of soliton content. The second motivation is instead addressed by a careful analysis of the lifted paths: the reader can check that the different strands of lifted streets in (\ref{eq:L-gamma-higher}) concatenate correctly into closed cycles on $\tSigma$. An important role is played by the novel BPS solitons carried by periodic $ii$-walls, see Section \ref{subsec:periodic-ii}.}
\be
\begin{split}
	& \alpha_{k\gamma}(q_\ell) = \alpha_{k\gamma}(q'_\ell) = \delta_{k,\ell+1} \,, \\
	& \alpha_{k\gamma}(\bar q_\ell) = \alpha_{k\gamma}(\bar q'_\ell) = 0 \,.
\end{split}
\ee
With this choice we find,
\be
	\Omega(\gamma)
	= \frac{[L_N(\gamma)]}{\gamma}
	= \frac{[ \pi_N^{-1} \(  -r_1 \cup -p_1 \cup -r_1' \cup -p'_1  \) ] }{\gamma} = -1
\ee
for $k=1$, as well as
\be\label{eq:L-gamma-higher}
	\Omega(k\gamma)
	= \frac{[L_N(k\gamma)]}{k\gamma}
	= \frac{[ \pi_N^{-1} \(  -r_k \cup -p_k \cup -r_k' \cup -p'_k\cup q_{k-1}\cup q_{k-1}'   \) ]}{k\gamma} = -1\,.
\ee
for $k>1$. Here the notation $\pi^{-1}_N$ stands for the restriction of (\ref{eq:canonical-lift}) to a single closed cycle on $\tSigma$.

We thus find a tower of BPS states with central charges $k\cdot Z_\gamma$ for $k\geq 1$, all have the same degeneracy $\Omega=-1$.
The result we found has a straightforward interpretation: these are the KK modes of a single  BPS state of the five-dimensional bulk theory on a circle.
In fact, this is the expected BPS spectrum for the theory \cite{Gopakumar:1998ii}.
In fact in (\ref{eq:Z-gamma}) we computed the unit central charge to be one unit of KK momentum, therefore BPS states with charge $k\gamma$ are naturally identified with a half-tower of KK modes, while the corresponding anti-particles with charges $-k\gamma$ complete the other half of the KK tower.
In the limit $R\to 0$ all these BPS states should become infinitely massive and are expected to disappear from the spectrum. We discuss this limit next.

From the viewpoint of M theory on a circle, this spectrum captures KK modes of the massless fields in eleven-dimensional supergravity. Alternatively, from the viewpoint of Type IIA string theory it is the spectrum of boundstates of D0 branes.
The masses of D0 particles are in fact reproduced by the periods $Z_{n\gamma}$.

\subsection{2d-4d limit ($R\to 0$)} {\label{C34dlim}}
In order to define a limit $R\to 0$, recall that the exponential network probes the BPS spectrum of a 3d-5d system, which in our case is described by the 3d $\CN=2$ theory associated with a toric brane  $\IC^3$. This is a $U(1)$ gauge theory with massive a charged chiral multiplet.
Following \cite{Aganagic:2001uw, Aharony:2017adm} we view the theory on a circle as a 2d model with an infinite number of fields, most of them massive since they correspond to Fourier modes of the 3d fields.
The twisted effective superpotential is
\be\label{eq:twsted-eff-CW}
	\tCW = \frac{1}{2\pi R} \( \Li_2(e^{-2\pi R \, \sigma})    \)
	+ \pi R \(\kappa^\eff \sigma^2 + \frac{i}{R}\sigma \) + 2\pi R \, \zeta \sigma
\ee
where
\be
	\sigma = \phi+i A_2,\quad m = m^R+i A_2^{(F)} \,.
\ee
Here $\phi$ is the real adjoint scalar of the 3d vector-multiplet, $A$ is the gauge field, $m^R$ denotes the 3d real mass and $A_2^{(F)}$ is the flavor background holonomy.
As observed in \cite{Aharony:2017adm} there are several ways to take limits $R\to 0,\infty$.
We take $\zeta\sim 1/R$ since this keeps the masses of solitons finite in the 2d limit.

For finite $R$ the vacuum manifold is defined by $	\exp\({\partial\tCW}/{\partial\sigma}\) = 1$.
Introducing
\be
	x = e^{-2\pi R\,\zeta}, \quad
	y = e^{2\pi R\,\sigma}
\ee
the vacuum manifold is the algebraic curve
\be
	 y^{\kappa^\eff +1 } - y^{\kappa^\eff } + x y  = 0 \,.
\ee
Indeed this coincides with (\ref{eq:framed-curve}) upon identifying $p= -\kappa_{\eff}+1$.

Let's therefore take the model with $p=-1$ hence $\kappa_\eff = 2$.
Taking $R\to 0$ we keep the following combination finite
\be
	t = 2\pi R \, \zeta +  \log \(2\pi R \mu\)
\ee
this is identified with the complexified FI coupling of the 2d theory in the limit $R=0$.
The twisted superpotential in fact becomes
\be
\begin{split}
	\tCW  \to &
	\sigma \( \log \sigma/\mu - 1\) + \sigma \log(2\pi R\,\mu)
	+ \pi R \kappa^\eff \, \sigma^2 + i \pi \sigma
	+ 2\pi R\zeta \sigma\\
	& = \sigma(t+i\pi)
	+ \sigma  \( \log \sigma / \mu  -1 \)
\end{split}
\ee
this is the twisted effective superpotential of a 2d (2,2) GLSM with gauge group $U(1)$ and a massless chiral multiplet with charge $+1$. The 2d FI coupling is $t$, which is now probing the region of the mirror curve at
\be
	x = e^{- 2\pi R \zeta} =  e^{-t+\log(2\pi R\mu)} =2\pi R\mu\, e^{-t}  \mathop{\to}^{R\to 0} 0\,.
\ee
The limiting curve has in fact a single sheet, corresponding to the single vacuum of the model, located at
\be
	\sigma = \mu e^{t}\,.
\ee
It is interesting to study how this curve arise from the mirror curve: as mentioned above it simply corresponds to zooming near $\tilde x=0$. The mirror curve has two sheets\footnote{
For convenience we pass to coordinates $	\tilde x = -x $ and $	 \tilde y = -y $.
} $\tilde y_\pm(\tilde x)$, which behave as follows in the limit $R\to 0$
\be
\begin{split}
	\sigma_+
	= \frac{1}{2\pi R} \log \( -\tilde y_+\)
	 \mathop{\to}^{R\to 0} \infty
	\qquad
	\sigma_-
	= \frac{1}{2\pi R} \log \(\tilde y_-\)
	\mathop{\to}^{R\to 0}
	\mu e^{-t}
\end{split}
\ee
One vacuum therefore runs off to infinity, while the other remains at finite distance and defines our IR 2d limit. This kind of behavior was observed for some models in \cite{Aharony:2017adm}.

From the viewpoint of solitons supported on the exponential network, zooming near $x=0$  pushes the branch point (which is the source for primary walls) infinitely far away, and all solitons paths become extremely massive.
This  agrees with the observation that only one tower of the vacua $(-,N)$ remains at finite distance while the other $(+,N)$ flows to infinity, hence giving infinite mass to any solitons of type $(\pm\mp,n)$ interpolating between these towers of vacua.
The only solitons that may remain massless are instead of types $(\pm\pm,n)$.
However, the network predicts that there are no such BPS states near $x=0$ therefore we expect a trivial BPS spectrum in the $R\to 0$ limit.
Since the branch point is removed from the picture, there is no wall of the network remaining at finite distance in the $R\to 0$ limit, in agreement with the expectation of a trivial soliton spectrum on the codimension-two defect.

\subsection{$tt^*$ analysis} \label{sec:C3-tt-star}

Above we have recovered the expected BPS spectrum for the five-dimensional theory from the $\CK$-wall jump of the spectrum.
Now we would like to perform some checks on the claim that soliton data on the exponential network actually computes 3d-5d BPS states in agreement with the CFIV index.
For this purpose we analyze the 3d theory on the defect in its own right, studying the 3d BPS spectrum  using the techniques of three-dimensional $tt^*$ geometry directly. We then compare the result with predictions from the nonabelianization map.

\subsubsection{Shift symmetry} \label{sec:shift-sym-C3}

As reviewed in Section \ref{sec:3d-tt-star}, a fundamental assumption behind the development of $tt^*$ geometry in three dimensions is the existence of a shift-symmetry for the spectrum of solitons.
In the context of exponential networks we have indeed encountered such a symmetry arising in full generality, this was discussed in Section \ref{sec:shift-sym}.
Here we study the existence of shift symmetry from the viewpoint of the 3d quantum field theory. As we will see, its realization is much less trivial than it appears in the geometric setting -- this is another advantage of the latter.

Let us consider an LG model $(W,\mathbb{C}^{s})$ (i.e. a LG model with $s$ chirals), whose vacua, given by $\mathrm{Crit}(W)$ $(W,\mathbb{C})$, are isolated points in $\mathbb{C}^{n}$. The energy of a soliton interpolating between the vacua $p_{0},p_{1}\in \mathrm{Crit}(W)$ is bounded by
\begin{equation}\label{bpsbounde}
E_{01}\geq 2\left|\int_{-\infty}^{+\infty} d\sigma\partial_{\sigma}Y^{i}\partial_{i}W\right|.
\end{equation}
The BPS solitons are then solutions of the equation
\begin{equation}\label{mybpseq}
\partial_{t}Y^{i}=\alpha G^{i\bar{j}}\overline{\partial}_{\bar{j}}\overline{W}
\end{equation}
where $\alpha$ can be written as
\begin{equation}
\alpha:=\frac{\int_{-\infty}^{+\infty} d\sigma\partial_{\sigma}Y^{i}\partial_{i}W}{\left|\int_{-\infty}^{+\infty} d\sigma\partial_{\sigma}Y^{i}\partial_{i}W \right|}.
\end{equation}
They saturate the bound (\ref{bpsbounde}) and so, their central charge is given by
\begin{equation}
Z_{01}=\int_{-\infty}^{+\infty} d\sigma\partial_{\sigma}Y^{i}\partial_{i}W.
\end{equation}
for the cases that $W(Y)$ is given by a holomorphic single valued function in $Y\in \mathbb{C}^{s}$, then $Z_{01}$ simply reduces to the difference of critical values $\Delta W_{01}=W(p_{1})-W(p_{0})$, and
 $\alpha=\frac{\Delta W_{01}}{|\Delta W_{01}|}$, as in \eqref{solitoneq2} in section \ref{ttstarin2d} and so the phase $\alpha$ is completely fixed by the boundary conditions of the BPS equation.
However, in our cases we have multivalued superpotential on the $Y$-plane, for example \eqref{C3sup} and \eqref{conisup}. In such cases, one has to be choose the integration cycle for $Z_{01}$ appropriately.
For a path $\gamma$ running from $p_0$ to $p_1$, we write
\begin{equation}
Z_{01}=\int_{\gamma} dY^{i}\partial_{i}W.
\end{equation}
To lighten the discussion, we focus on the case $s=1$. Hence
\begin{equation}\label{1dimcc}
Z_{01}=\int_{Y_{0}}^{Y_{1}} dY\partial_{Y}W
\end{equation}
where $\{Y_{k}\}$ denote critical points (which can be infinitely many).
Since we want to consider the case where $W(Y)$ is multivalued, hence it has branch cuts in the $Y$-plane. In such a case, the value of (\ref{1dimcc}) depends crucially in the homotopy of
$\gamma$, on the $Y$-plane. Every time that the path $\gamma$ crosses one of these branch cuts, $W$ undergoes monodromy. An example of this is showed in Fig. 1.
\begin{figure}[h]
\centering
\includegraphics[width=4in]{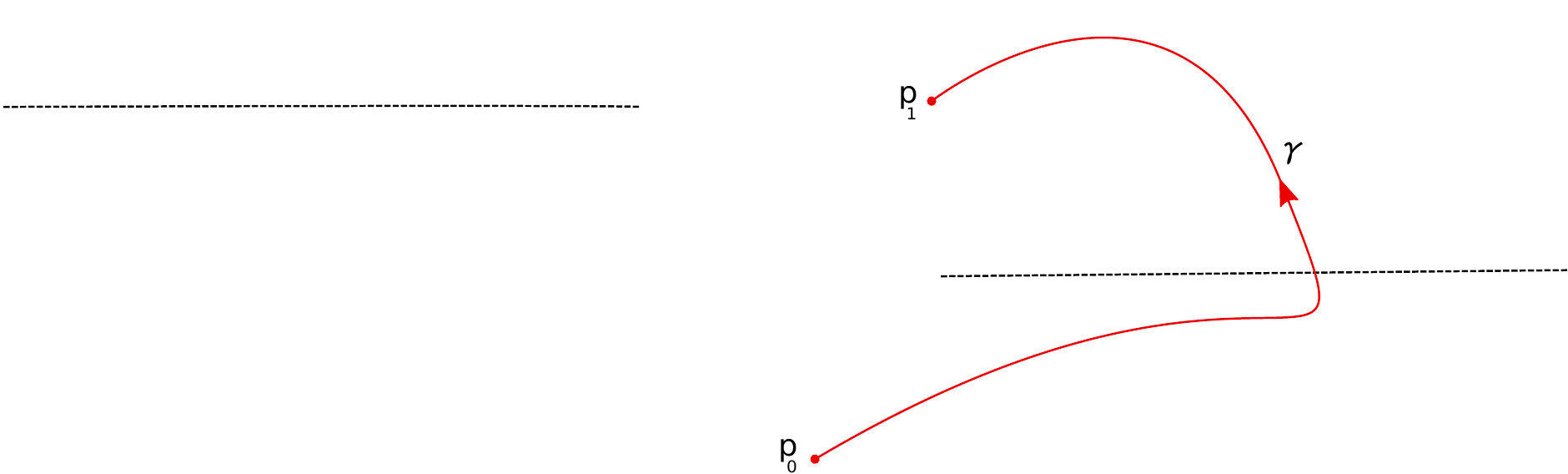}
\caption{An example of integration path in the $Y$-plane. Dashed lines indicate branch cuts of $W(Y)$.}
\label{fig:branchpath}
\end{figure}
If we denote $M_{b}$ the monodromy of $W$ when crossing the branch cut $b$ and $\gamma$ crosses the branch cuts in the order $b_{1},b_{2},\ldots, b_{k}$
(where $b_{j}$'s are not necessarily distinct), then the central charge of the soliton $s_{ij}(\gamma)$ i.e., the soliton interpolating between $Y_{i}$ and $Y_{j}$ via a
path $\gamma$ in the $Y$-plane, will be given by\footnote{A word of caution: if we are dealing with a vector valued function $F(z)$ that undergoes linear monodromy as
$z$ goes around the points $b_{\alpha}$, this means $F(\mathcal{L}_{b}z)=M_{b}\cdot F$ where $M_{b}$ is an invertible matrix, then the monodromy of a path tha goes
first around $b_{1}$ and then $b_{2}$ is given by $M_{1}M_{2}\cdot F$ i.e. the order of multiplication is reversed.}:
\begin{equation}
Z_{ij}(\gamma)=M_{b_{k}}\circ \ldots\circ M_{b_{1}}\circ W(Y_{j})-W(Y_{i})=M_{\gamma}\circ W(Y_{j})-W(Y_{i})
\end{equation}
where we think of $M_{b}$ as parallel transport operators. Hence, the BPS equation becomes much more complicated, since $\alpha$ no longer depends in the boundary
conditions but also on the integration path. We will concentrate in the case of collinear vacua, that is, we will have multiple 'towers of vacua' $\{Y_{\alpha,k}\}$ where
$\alpha=1,\ldots,q$, $k\in \mathbb{Z}$ and $Y_{\alpha,k}-Y_{\alpha,h}\sim (k-h)$.

 We will say that the theory enjoys a shift symmetry:
\begin{equation}
T:Y_{\alpha, k}\rightarrow Y_{\alpha, k+1}\qquad \text{ \ for all \ }\alpha,k
\end{equation}
if each soliton $((\alpha,k),(\beta,l);\gamma)$ is mapped to a soliton $((\alpha,k+1),(\beta,l+1);\gamma')$ with the same central charge i.e.:
\begin{equation}
Z_{(\alpha,k),(\beta,l)}(\gamma)=Z_{(\alpha,k+1),(\beta,l+1)}(\gamma')
\end{equation}
Note that this definition is equivalent to the shift map of Section \ref{sec:shift-map}, hence the paths $\gamma,\gamma'$ belongs to the relative homology groups:
\begin{equation}
\gamma\in \Gamma_{\alpha,\beta,k,l}\qquad \gamma'\in \Gamma_{\alpha,\beta,k+1,l+1}
\end{equation} in the following we will focus on the case of mirror $\mathbb{C}^{3}$.

\subsubsection*{Case of $\mathbb{C}^{3}$ at framing $f$}
The superpotential for $\mathbb{C}^3$ is given by \eqref{C3sup} which we repeat below for succinctness
\begin{equation}
W=(X-i\pi)Y+\frac{f}{2}Y^{2}+\mathrm{Li}_{2}(-e^{Y})\qquad f\in\mathbb{Z}
\end{equation}
and the vacua is given by
\begin{equation}
Y_{\alpha,k}=Y_{\alpha}+2\pi i k \qquad \alpha=1,\ldots,|f|-\frac{1}{2}(\mathrm{sgn}(f)-1)
\end{equation}
where $y_{\alpha}=\mathrm{exp}( Y_{\alpha})$ are the roots of
\begin{equation}\label{eq:C3-mirror-curve-f}
xy^{f}+y+1=0,
\end{equation}
which is the mirror curve for $\mathbb{C}^3$.
Before proceeding further, let is recapitulate some well-known facts about the dilogarithm function and its monodromy.

\subsubsection*{Dilogarithm and its monodromy}

The dilogarithm function has the integral representation
\be
\Li_2 (u) = -\int_0^u \log (1-u) d \log u.
\ee
There are two equivalent ways to express the monodromy \cite{Zagier:2007knq}. In the first picture, one assembles a vector with three entries $\Li_2,\log$ and $1$. Analytic continuation along a loop in $\mathbb{P}^1\backslash \{0,1,\infty\}$ leads to the
monodromy representation
\be
\begin{pmatrix}
\Li_2(u) \\ \log (u) \\ 1
\end{pmatrix}, \quad
M_0 = \begin{pmatrix}
1 & 0 & 0 \\
0&1&2\pi i \\
0&0&1
\end{pmatrix},\quad
M_1 = \begin{pmatrix}
1 & -2\pi i & 0 \\
0&1&0 \\
0&0&1
\end{pmatrix}.
\ee
which is a representation $M : \pi_1 (\mathbb{P}^1\backslash \{0,1,\infty\}) \mapsto {\textrm{GL}} (3,\mathbb{C})$.
Going around a loop in counterclockwise direction, enclosing both $0,1$, one gets a monodromy
\be
\begin{pmatrix}
\Li_2(u) \\ \log (u) \\ 1
\end{pmatrix}
\mapsto M \cdot \begin{pmatrix}
\Li_2(u) \\ \log (u) \\ 1
\end{pmatrix}, \quad
M =
\begin{pmatrix}
1 & -2\pi i & 0 \\
0&1&2\pi i \\
0&0&1
\end{pmatrix}
\ee
For the record, we define the group commutator as $\mathfrak{M} = M_0 M_1 M_0^{-1} M_1^{-1}$.
Going around $k$ times in the counterclockwise loop around both $0$ and $1$ gives an action by
\be
M^k =
\begin{pmatrix}
1 & -2\pi i k & 2\pi^2 (k^2-k) \\
0&1&2\pi i k\\
0&0&1
\end{pmatrix}\implies
\begin{matrix}
 \Li_2 (u) \mapsto \Li_2(u) - 2\pi i k \log (u) + 2\pi^2 (k^2-k) \\
\log u \mapsto \log (u) + 2\pi i k \\
1 \mapsto 1
\end{matrix}
\ee

In the alternative picture, consider the complexified Heisenberg group $H_{\mathbb{C}}$ of upper triangular $3 \times 3$ matrices with ones in diagonal.
The representative element is
\be
L(u) =
\begin{pmatrix}
1 & -\Li_1 (u) & -\Li_2 (u) \\
0&1& \log (u)\\
0&0&1
\end{pmatrix}
\ee
and left operations with $H_{\mathbb{Z}}$ generates the multi-valuedness. The left-multipliers are
\be
h_0 =
\begin{pmatrix}
1 & 0 & 0 \\
0&1&2\pi i \\
0&0&1
\end{pmatrix}, \quad
h_1 =
\begin{pmatrix}
1 & 2\pi i & 0 \\
0&1& 0 \\
0&0&1
\end{pmatrix}.
\ee
One can view this as the following. Starting from
$(a,b| \alpha) \rightarrow (r,s) = (e^a, e^b)$ a bundle over $\mathbb{C}^*_r \times \mathbb{C}^*_s$ with fiber $(2\pi i)^2\mathbb{Z}\backslash \mathbb{C}_\alpha$ (isomorphic to $\mathbb{C}^*$
by $\alpha \rightarrow S := e^{\alpha/2\pi i}$)
\be
\begin{matrix}
H_{\mathbb{Z}}\backslash H_{\mathbb{C}} \\
\downarrow \\
\mathbb{C}^*_r \times \mathbb{C}^*_s =  (2\pi i)^2\mathbb{Z}\backslash \mathbb{C}_{a,b}
\end{matrix}.
\ee
One now just needs to pullback to get the Heisenberg bundle $\mathfrak{H}$ over $\mathbb{P}^1\backslash \{0,1,\infty\}$
\[ \begin{tikzcd}
\mathfrak{H} \arrow{r}{} \arrow[swap]{d}{(2\pi i)^2 \mathbb{Z}\backslash\mathbb{C}} &  H_{\mathbb{Z}}\backslash H_{\mathbb{C}} \arrow{d}{} \\%
\mathbb{P}^1\backslash \{0,1,\infty\} \arrow{r}{1-z,z}&\mathbb{C}^* \times \mathbb{C}^*
\end{tikzcd}
\]
A section of $\mathfrak{H}$ has the form $H_{\mathbb{Z}} (-\Li_1 z , \log z| \alpha)$ and $L(z)$ is a flat section.

\subsubsection*{Shift symmetry for soliton masses}

Now getting back to $\mathbb{C}^3$, $Y_\alpha$'s are functions of $X$ and are in general multivalued. So the prescription is as follows. Start with a reference point
$X_0$ and a set of reference values $Y_\alpha (X_0)$. Then
\begin{equation}
Y_{\alpha}(X)=\int_{X_{0}}^{X}\partial_{u}Y_{\alpha}(u)
\end{equation}
where the integration is over some path. The choice of path is what we mean by the choice of prescription. We will use this approach whenever we have to evaluate multivalued functions. Hence this is relevant when manipulating critical values of $W$, as we will see shortly.
We will consider the value of $X$ as fixed, so we will assume a prescription as chosen to evaluate $Y_{\alpha}(X)$ but it will not be relevant for us. Also, if we move on paths $\Gamma(x)_{n,m}$ in the $X$-plane,
 it will act by just interchanging the critical points $Y_{\alpha,k}(X)$.

What will be important for us is to define the critical values $W(Y_{\alpha,k})$. As it stands, this equation is not a priori well defined for the reasons we discussed before. Nevertheless, we can still define the problem of counting the solitons joining vacua, for a given value of central charge, in a completely unambiguous way. Start by defining
\begin{equation}
W_{\alpha}(a,b):=W(Y_{\alpha,0})+ 2\pi i a Y_{\alpha,0}+2\pi^{2}b\qquad a,b\in \mathbb{Z}
\end{equation}
where some choice of prescription has been made to evaluate $W(Y_{\alpha,0})$, but we will see this is indeed irrelevant when stating the soliton counting problem.
The above formula is a consequence of the fact that when one defines $W_\alpha$, one has to choose a specific path which starts form $Y_\alpha(X_0)$ as was mentioned
above. Because of the monodromy of the dilogarithm now, it depends on how one crosses the respective branch cuts and what ``word'' of monodromy group generator shows up.
However, the statement is that, for any such path, there exists a pair of such integers $a,b$.
 This also defines the lattices of points:
\begin{equation}
\mathbb{L}_{\alpha}:=\{W_{\alpha}(a,b)\}_{a,b\in \mathbb{Z}}.
\end{equation}
The reason for this is the following: suppose we choose a prescription to evaluate $W$ at $Y_{\alpha,k}=Y_{\alpha,0}+2\pi i k$. Then, we get
\begin{equation}
\label{eqsupg}
W(Y_{\alpha,k})=W(Y_{\alpha,0})+f(2\pi^{2}(k^{2}-k)-2\pi i kY_{\alpha,0})+2\pi i s Y_{\alpha,0}-2\pi^{2}(s^{2}-s)\qquad s\in \mathbb{Z}
\end{equation}
where the second term comes from the evaluation of $\mathrm{Li}_{2}(-\exp(Y_{\alpha,k}))$ due to the action of monodromy as stated above. As we can see this as exactly
the form of a point in $\mathbb{L}_{\alpha}$. If we change the prescription to evaluate $\mathrm{Li}_{2}$ (which is the only choice to be made), it will just be another point on
$\mathbb{L}_{\alpha}$. Now we are ready to state the soliton counting problem. First, if we want to count the solitons interpolating between $Y_{\alpha,k}$ and $Y_{\beta,l}$
we must solve the equation (\ref{mybpseq}) with phase
\begin{equation}
\alpha_{(\alpha,\beta)}[(a,b),(c,d)]:=\frac{W_{\beta}(a,b)-W_{\alpha}(c,d)}{|W_{\beta}(a,b)-W_{\alpha}(c,d)|}
\end{equation}
and with boundary conditions $Y(-\infty)=Y_{\alpha,k}$, $Y(\infty)=Y_{\beta,l}$ for all $W_{\beta}(a,b)\in \mathbb{L}_{\beta}$ and $W_{\alpha}(c,d)\in \mathbb{L}_{\alpha}$.
This will give us all the solitons with central charges $W_{\beta}(a,b)-W_{\alpha}(c,d)$. Hence, we will argue for shift symmetry, if the solitons interpolating between $Y_{\alpha,k}$ and $Y_{\beta,l}$ ,
 with fixed central charge, satisfy the same counting problem as solitons interpolating between $Y_{\alpha,k+1}$ and $Y_{\beta,l+1}$.\\
\begin{figure}[h]
\centering
\includegraphics[width=4in]{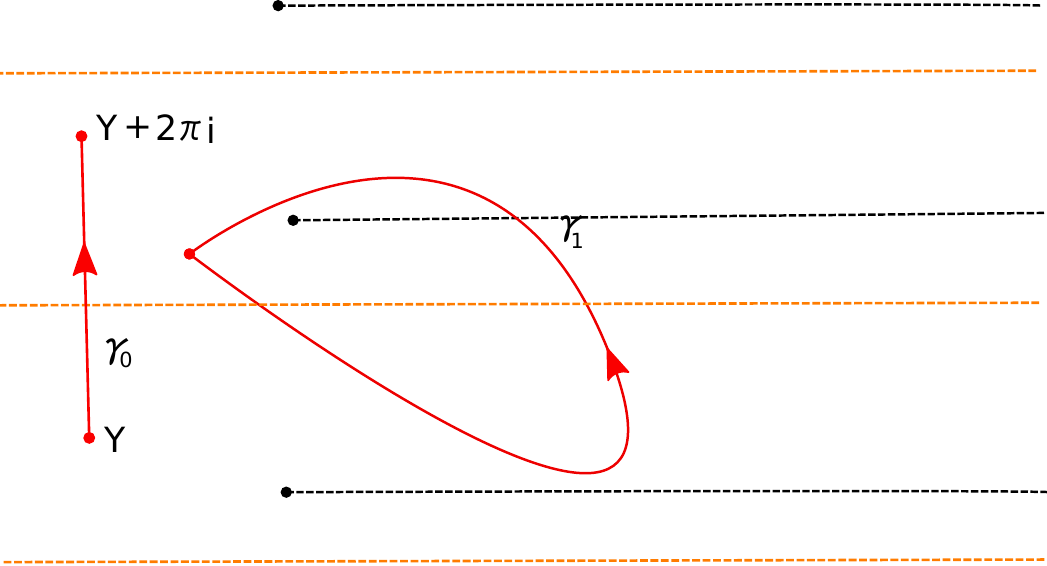}
\caption{The branch cuts of $\mathrm{Li}_{2}$ on the $Y$-plane: dashed black line are the branch cuts at $-i\pi+2\pi i \mathbb{Z}$ . The orange lines are ${\mathrm{Im}}(Y)=2\pi \mathbb{Z}$ (they are branch cuts on $y$-plane). $\gamma_{0}$ and $\gamma_{1}$ is the path around $y=0$ and $y=-1$, respectively.}
\label{fig:brancutdilog}
\end{figure}

For the simplest case of $f=0$, there is only a single root, so we can drop the index $\alpha$ and write
\begin{equation}
Y_{k}=i\pi+\log(1+x)+2\pi i k \qquad k\in \mathbb{Z}
\end{equation}
here, the central charge depends only on the differences $a-c$ and $b-d$ and so, we have an explicit realization of the shift symmetry.
The case with $f \neq 0,1$ are more difficult though because of multiple roots.

To understand the latter cases, let us start by considering the generic form of the superpotential, provided our prescription :
\begin{equation}
W(Y_{\alpha,0})-W(Y_{\beta,0})+2\pi i aY_{\alpha,0}-2\pi i c Y_{\beta,0}-2\pi^{2}(b-d).
\end{equation}
This clearly means that there is a symmetry which we dub as the ``shift symmetry'' given $(b,d) \mapsto (b+1,d+1)$.
Let us clarify the last point a bit further. Given two values $Y_{\alpha,0}$ and $Y_{\beta,0}$, following specific paths, computing
monodromies coming from dilogarithm appropriately, one can build up the superpotential at $Y_{\alpha,k}$ and $Y_{\beta,l}$ following
\eqref{eqsupg}. Of course, these representations are ambigous and tacitly depend on the choice of paths which we reiterate again. Our
goal is to find the solutions to the BPS equation \eqref{mybpseq}. For most of the choices of $\alpha$ and $\beta$, there does not exist
any solution. But the crucial point is that, if there is one solution for some $\alpha,\beta$, one gets an infinite set of solution due to freedom
coming from $s$ in \eqref{eqsupg}. In terms of Lefschetz thimbles, this has a clear interpretation. In fact, these are the objects which
are captured by our exponential network. Given vacua labeled by $\alpha$ and $\beta$, if the thimbles intersect for some $k$ and $l$, one can
find a family of such Lefschetz thimbles whose intersections contribute to the the degeneracy of the same 3d CFIV index. This precisely fits our computations
using exponential networks (and its application to $\mathbb{C}^3$).

A parallel way to describe shift symmetry in the soliton masses is directly using the exponential network set up. Start with the wall emanating from the branch point $(ij,0)$, or vis-a-vis
$(ij,N,N)$ type soliton for $N\in \mathbb{Z}$. Now one can compute
\be
W(x,Y_j^{(N)}) - W(x,Y_i^{(N)}) = \int_a \lambda = \int_{\pi(a)} (\log y_j - \log y_i ) d \log x,
\ee
where $\pi$ is the projection map defined before.

As before, the difference does not depend on $N$ and moreover, now we give an explicit realization of it in terms of the chain integral. Individually, the monodromy
of the dilogarithm does indeed affect how one goes from $W(x,Y_i^{(N)})$ to $W(x,Y_i^{(N+1)})$. But as long as we define $ W(x,Y_j^{(N)}) = W(x,Y_i^{(N)})
+ \int_{a_N} \lambda$, {\it for all} $N$, the soliton mass is independent of this $N$. The exponential network provides the definition of such paths $a_N$.

\subsubsection{$tt^*$ metric for toric lagrangian theory of $\IC^3$ }

We now study the $tt^*$ equations and compute their predictions for the soliton spectrum of the 3d theory on the  defect engineered by an M5 brane wrapped on the toric Lagrangian brane of $\IC^3$.
For the purpose of comparing with predictions from the nonabelianization map, we will work with framing $f=-1$.

The fundamental field of the model is $Y$, with superpotential
\be
	W = (X-i\pi)  Y +\frac{f}{2}Y^2 + \Li_2(-e^{Y})
\ee
$X$ is the deformation parameter and $Y$ is the operator by which we deform the theory.
Critical points are labeled by $(i,N)$ with $i=\pm$ and  $N\in\IZ$
\be
	Y_{\pm,N} =\log y_\pm(x) + 2\pi i \, \,,
	\qquad y_\pm(x) = \frac{-1\pm\sqrt{1-4x}}{2}\,.
\ee

The chiral ring is the ring of functions of $Y$, generically denoted $\phi(Y)$, modulo $dW=0$.
This means that
\be
	y^2 + y = -x
\ee
is proportional to the identity operator.
\be
	\mathcal{R} = \{\phi(Y)\} / (e^{2Y} +e^Y + e^X =  0)\,.
\ee
Operators in the \emph{chiral} ring are {holomorphic} functions $\phi(Y)$. They are therefore entirely characterized by their values at the critical points
\be
	\phi(Y) \quad\leftrightarrow \quad \{(\phi)_{\pm,N}\}_{N\in\IZ} := \{\phi(Y_{\pm,N})\}_{N\in\IZ} \,.
\ee
Ring multiplication is the multiplication of such functions.

Choosing a basis $\phi_\alpha$, we could write the topological metric as (see \cite[eq. (2.10)]{Cecotti:1991me})
\be
	\eta_{\alpha\beta} = \sum_{i=\pm, N\in\IZ} \frac{ \phi_\alpha(Y_{i,N}) \phi_\beta(Y_{i,N}) }{{\fH_W}(Y_{i,N})}
\ee
It is convenient to work in the \emph{point basis}  which is defined by
\be
	\phi_\alpha(Y_{i,N}) \quad \to\quad a_{j,M}(Y_{i,N}) = \delta_{ij}\delta_{MN}
\ee
or in more shorthand notation $(a_{j,M})_{i,N} = \delta_{ij}\delta_{MN}$. That is, $a_{i,N}(Y)$ is a certain function of $Y$ such that it is $1$ at vacuum $Y_{i,N}$ and zero at all other vacua.
The metric in the point basis is simply
\be
	\eta_{(i,N)(j,M)}
	= {\eta_i}   \delta_{ij}\delta_{MN}  =
	\(\begin{array}{cc}
	\eta_+ & \\
	 & \eta_-
	\end{array}\)_{ij} \otimes \mathbb{I}_{MN}\,.
\ee
where
\be
	\eta_\pm = \frac{1}{\fH_W({Y_{\pm,N}})} = -\frac{1}{2} \(1\pm \frac{1}{\sqrt{1-4x}}\)
\ee
is in fact independent of $N$. We also introduced $x=e^X$, and absorbed $2\pi R$ into $X,Y$ compared to the discussion of the previous subsection, in order to lighten notation.

We can express the deformation operator $\phi_X$ in the point basis easily
\be
\begin{split}
	\phi_X & := Y  = \sum_{i=\pm, N\in\IZ}  Y_{i,N} a_{i,N}
\end{split}
\ee
in components this reads
\be
\begin{split}
	(\phi_X)_{(i,N),(j,M)} &
	= Y_{i,N} \delta_{ij}\delta_{MN} \\
	&=
	\(\begin{array}{cc}
	\log y_+(x) & \\
	 & \log y_-(x)
	\end{array}\)_{ij} \otimes \mathbb{I}_{MN}
	\ + \
	2\pi i N \, \mathbb{I}_{ij} \otimes \mathbb{I}_{MN}
\end{split}
\ee
This is a good opportunity to work out the chiral ring structure constants for $\phi_\lambda$ in the point basis:
\be
	\phi_X a_{(i,N)} = C_{X ,({i,N)}}^{\phantom{\lambda ,(i,N)\,} {(j,M)}} a_{j,M} = Y_{i,N} \delta_i^j \delta_N^M a_{(j,M)}
\ee
Therefore we can express in point-basis
\be\label{eq_C-operator}
	(C_\lambda)^{(j,M)}_{(i,N)} =
	Y_{i,N} \delta_{i}^{j}\delta^{M}_{N}
	=
	\(\begin{array}{cc}
	\log y_+(x) + 2\pi i N & \\
	 & \log y_-(x) + 2\pi i N
	\end{array}\)_{i}^{j}   \otimes \mathbb{I}_{N}^M	\,.
\ee

As argued above, $W$ is invariant under shift symmetry
\be
	T: Y \to Y+2 \pi i
\ee
This symmetry is reflected by the $tt^*$ metric $g_{i\bar j}$.
In the point basis it acts on vacua as
\be
	T \cdot |i,N\rangle  =  |i,N+1\rangle,,
\ee
and implies
\be
\begin{split}
	g_{(i,N),(\bar j,\bar M) } & =  g_{(i,N+1),(\bar j,\overline{M+1})} \,.
\end{split}
\ee
Then the metric only depends on the difference $ g_{i,\bar j}(N-M)$.
To take advantage of this symmetry it is convenient to introduce the Bloch basis
\be\label{eq:bloch-basis-C3}
	|i,\theta\rangle = \sum_{k\in\IZ} e^{i k \theta} |i,k\rangle
\ee
where $|k\rangle$ is the $k$-th vacuum in the point-basis.
Then\footnote{The definition by $g(\theta) = \langle\theta | g| \theta'\rangle $ would produce an extra infinite sum over $\delta$ functions, but restricting the range of $\theta$ within $[0,2\pi[$ gets rid of it.}
\be
	g_{i \bar j}(\theta) = \sum_{n\in\IZ} e^{i n\theta} \, g_{ i \bar j}(n)  \,.
\ee
as a side remark, note that $n$ here is the same $n$ that labels $\CE$-walls in the exponential network as $(ij,n)$.

The $tt^*$ metric is hermitian, this imposes constraints on its components.
To understand what this means in the Bloch basis, note that conjugation acts as follows $(g_{i\bar j} )^*  = \langle \bar j | i\rangle^* =  \langle j | \bar i \rangle = g^*_{\phantom{*}\bar i j}$ therefore
 \be
 	g_{\phantom{*}\bar i j}^*(\theta) = \sum_n e^{i n \theta} g^*_{\phantom{*}\bar i j} (n)
	=\sum_n e^{(-i) n (-\theta)} g^*_{\phantom{*}\bar i j} (n)
	= [g_{i\bar j}(-\theta)]^* \,.
 \ee
Likewise transposition of the matrix in point basis becomes
 \be
 	g_{\phantom{T}\bar j i}^{T}(\theta) = \sum_n e^{i n \theta} g^T_{\phantom{*}\bar j i} (-n)
	=\sum_n e^{i (- n) (-\theta)} g^T_{\phantom{*}\bar j i} (-n)
	= [g_{i\bar j}(-\theta)]^T \,.
 \ee
Therefore overall the hermitean conjugate of $g$ then takes the usual matrix form,
\be
	g^\dagger(\theta) = [g(\theta)]^\dagger.
\ee
Thus hermiticity implies that we can parametrize the metric as
\be
	g(\theta) = \(\begin{array}{cc}
	A(\theta) & B(\theta) \\
	B^*(\theta) & D(\theta)
	\end{array}\)\qquad
\ee
where $A,D$ are real functions of $X,\theta$, while $B$ is complex.
Finally, given the definition of $g$ in terms of inner products of vacuum states and their adjoints, it follows that $A$ and $D$ must be positive functions.
Additional constraints come from CPT invariants in the form of the reality condition (\ref{eq:tt-star-reality}), however in the present case these are not particularly useful and will be omitted.

The $tt^*$ equations (\ref{eq:tt-star-eqs-g}) in the case of a single variable read
\be\label{eq:C3-tt-star-eqs-g}
	\partial_{\bar X}\(g \partial_X g^{-1}\) = [C_X, g C^\dagger_{\bar X} g^{-1}]
\ee
To study them we need to express $C_X$ given in (\ref{eq_C-operator}) in Bloch basis (\ref{eq:bloch-basis-C3})
\be
\begin{split}
	C_X(\theta) & =  \(\begin{array}{cc}  \log y_+(x) + 2\pi \, \frac{\partial}{\partial\theta} & \\ & \log y_-(x)  + 2\pi \,\frac{\partial}{\partial\theta}  \end{array}\)\,, \\
	C^\dagger_{\bar X}(\theta) & =  \(\begin{array}{cc}  \log (y_+(x))^* - 2\pi \, \frac{\partial}{\partial\theta} & \\ & \log (y_-(x))^*  - 2\pi\, \frac{\partial}{\partial\theta}  \end{array}\)\,.
\end{split}
\ee

Let us now study these equations in a specific regime, near $X = +\infty$ where massive vacua are well-separated.
Recall that this corresponds to the puncture on the right in Figure \ref{fig:C3-network-plots}.
From the exponential network perspective we see immediately why this is a convenient regime to analyze: there should be only two (towers of) solitons ending there. On the contrary, had we chosen to study the region near $x=0$ we may expect infinitely many solitons.
In the point basis, the asymptotic behavior of $g$ is expected to take the following general form (see \cite{Cecotti:1991me})
\be\label{eq:C3-metric-soliton-sum}
\begin{split}
	g_{i\bar i} &\sim \frac{1}{|\fH_W(Y_i)|} \\
	g_{i\bar j} &= (g_{i\bar i}  g_{j\bar j})^{1/2} \sum_{a} \frac{(\alpha_{ij}(a))^{1/2}}{(4\pi z_{i\bar j}(a))^{1/2}}  {e^{-2 z_{i\bar j}(a)}} \qquad (i\neq j)
\end{split}
\ee
where
\be
	z_{i\bar j}(a) = 2 \left| W(Y_{j}) - W(Y_i) + Z_f(a) \right|
\ee
is the mass of a soliton with charge $a$, and the sum runs over all solitons in the topological sector $i\to \bar j$.
In our parametrization this means that
\be
\begin{split}
	& A(\theta) \sim |\eta_+| \to \frac{1}{2} \qquad
	D(\theta) \sim |\eta_-| \to \frac{1}{2} \\
	& B(\theta) \sim  \frac{(|\eta_+\eta_-|)^{1/2}}{(4\pi)^{1/2}} \alpha(\theta) \frac{e^{-2 z_{+-}}}{(z_{{+-}})^{1/2}}
\end{split}
\ee
where in the second line the sum over $a$ disappeared and has been replaced by $\alpha(\theta)$, by virtue of the fact that there is a single tower of solitons of type $(+-)$ based on network observations, and that they all have the same central charge.\footnote{There is also a tower of $(-+)$ solitons, encoded by $B^*$}
Since all solitons interpolate between vacua shifting the logarithmic branch by the same amount, which we shall denote by  $n$, we expect that $\alpha(\theta) = \alpha_0 e^{i n \theta}$ for  some fixed constant $\alpha_0$.

As explained at length above, the central charge of these solitons should be computed with care, by  finding the appropriate analytic continuation of the dilogarithm so that soliton masses are shift-symmetric.
In point basis, the critical values of the superpotential are
\be
	W_{\pm}^{(k)} = -(i\pi -X) \(\log y_\pm(x) + 2\pi i k\)
	-\frac{1}{2} \(\log y_\pm(x) + 2\pi i k\)^2  + \Li_2\( - y_\pm(x) e^{2\pi i k} \) \,.
\ee
A careful choice of the branch of $Li_2$, engineered to cancel the quadratic dependence on $k$, leads to the following leading behavior for the central charge at large $X>0$:
\be
	W_{+}^{(N)} -W_{-}^{(M)}  \stackrel{X\to+\infty}{\sim} 2\pi i (N-M)\, X \,.
\ee
Here we take $N-M=n$ for all solitons in the tower, where this is precisely the same $n$ appearing in the exponent of $\alpha$. Therefore
\be
	z_{{+-}} = 4 \pi   n \,|X|  \,.
\ee
Therefore
\be
\begin{split}
	B(\theta) & \sim \frac{(|\eta_+\eta_-|)^{1/2}}{(4\pi)^{1/2}} \, \alpha_0 \, e^{i n \theta}\, \frac{e^{-8 \pi  n |X|}}{(4\pi  n |X|)^{1/2}}
	 = \frac{1}{2} \,  \tilde \alpha_0 \frac{e^{i n \theta -8 \pi \, n |X|}}{|X|^{1/2}}
\end{split}
\ee
where we absorbed $4\pi i n^{1/2}$ into $\tilde \alpha_0$.

We have used data from exponential networks to predict the asymptotic form of the $tt^*$ metric in the IR regime where $X\to+\infty$.
Next we would like to check that this is in fact a solution to the differential equation (\ref{eq:C3-tt-star-eqs-g}) defined by the geometry of the chiral ring.
To do that we also need the asymptotic behavior of $C_X$, this is immediately obtained by noting that
\be
	\log y_\pm(x)  \sim \frac{X}{2} \,.
\ee
Then $C_X$ becomes asymptotically proportional to the identity and equal to
\be
\begin{split}
	C_X(\theta) & =  \frac{X}{2} + 2\pi\, \frac{\partial}{\partial\theta} \,,
	\qquad%
	C^\dagger_{\bar X}(\theta) =  \frac{\bar X}{2} - 2\pi\, \frac{\partial}{\partial\theta}\,.
\end{split}
\ee
Here $\bar X$ denotes the complex conjugate of $X$.

The \emph{lhs} of the $tt^*$ differential equation is
\be
	\partial_{\bar X} (g \partial_X g^{-1}) =  -4\pi^2 n^2  \frac{e^{-8\pi n |X|}}{|X|^{1/2}}
	 \(\begin{array}{cc}
	0  &   \tilde \alpha_0 \, {e^{i n \theta}} \\
	 \, \tilde \alpha_0^{*} \, {e^{-i n \theta}}  &  0
	\end{array}\) + \dots
\ee
up to sub-leading corrections in $1/|X|$.
The \emph{rhs} is instead
\be
	[C_X , g C^\dagger_{\bar X} g^{-1}]
	=
	-4\pi^2\, n^2  \frac{e^{-8\pi n |X|}}{|X|^{1/2}}   \(\begin{array}{cc}
	0  &   \tilde \alpha_0 \, e^{i n \theta} \\
	\tilde \alpha_0^* \, e^{-i n \theta}  &  0
	\end{array}\)
	+ \dots
\ee
again up to sub-leading corrections in $1/|X|$.
Therefore to leading order we find a perfect match, confirming the expectation that the geometry supports a single tower of solitons of type $(+-)$, and a single tower of solitons of type $(-+)$ near $X\to +\infty$.

\section{Discussion and future directions}\label{sec:discussion}

In this work we developed new tools to study BPS states in M theory compactifications on Calabi-Yau threefolds.
In addition to being of interest for the physical problem of BPS counting, we hope that our construction may provide a new angle on related questions of enumerative geometry.
We conclude with remarks on the new questions raised by this work, and with suggestions for future directions.

\subsubsection*{More Calabi-Yaus}

The most natural next step would be to apply nonabelianization to more interesting toric threefolds.
Our main example, $\mathbb{C}^3$, does not have any non-trivial two- or four-cycles, which leads to a rather uninteresting spectrum featuring only D0 branes.
In \cite{Eager:2016yxd} it was argued that exponential networks should be sensitive to boundstates including also D2 and D4 branes.
\footnote{Contributions from D4 branes can arise on a similar footing as D2 branes, via a symplectic rotation \cite{Dijkgraaf:2006um}.}
This can also be argued on field-theoretic grounds.
While D2 would appear as instanton-dyons of the 5d gauge theory, the D4 would correspond to magnetic monopole strings.
Going down to four dimensions, the two types of states gain similarities: the former descend to BPS particles, while the strings wrapping the circle become magnetic monopoles. Therefore both (more precisely, their KK modes) are expected to be captured by the 2d-4d wall-crossing described by the nonabelianization map.

It would be especially interesting to carry out the computations in the conifold, to compare  with  \cite{Jafferis:2008uf, Morrison:2011rz}.
Other interesting cases include local $\mathbb{P}^2$ and elliptic fibrations over Hirzebr\"uch surfaces or del Pezzo surfaces. The 5d gauge theories engineered by the latter admit Lagrangian descriptions and the chamber structure of their Coulomb branches have been studied long ago \cite{Seiberg:1996bd, Intriligator:1997pq, Douglas:1996xp}.
Compactifying these theories on a circle leads to interesting questions on wall-crossing, regarding the relation between 5d and 4d BPS spectra \cite{Kachru:2018nck}. We expect that these questions may be appropriately approached through our framework.
Another application of computing such BPS spectra would be the study of  D-instanton corrections to metrics on hypermultiplet moduli spaces of type II string theory \cite{Haghighat:2011xx, Haghighat:2012bm, Alexandrov:2017mgi}.\footnote{This was one of the initial motivations for this work.}

\subsubsection*{Soliton data in field theory: kinky vortices}

Another pressing question that was postponed throughout this work regards the physical interpretation of the soliton data in the nonabelianization map.
In section \ref{sec:phys}  we introduced the picture of 3d-5d systems as the main motivation for the claim that the ($\CK$-wall jumps of the) nonabelianization map really count 5d BPS states.
However this argument strictly relies on considering 3d-5d systems on a circle, which can be viewed as 2d-4d systems of the related Kaluza-Klein modes. Through this perspective we can view wall-crossing phenomena involving 3d-5d BPS states as emerging from the better-understood 2d-4d wall-crossing of \cite{Gaiotto:2011tf}. A satisfactory field-theoretic description of 2d-4d states and their interactions with 4d BPS states is indeed available,
While we believe that this picture is correct for the purpose of counting 5d BPS states (more properly, their KK modes), it would be more satisfactory to have a intrinsic three-dimensional interpretation of soliton data.

The M5 brane wrapping the Lagrangian $L$ engineers a 3d $\CN=2$ theory $T[L]$, which can be described as a $U(1)$ gauge theory with charged chiral matter.
In general, such theories feature a Higgs branch with BPS vortex states.
When $T[L]$ is placed on a spatial circle $\IR_t\times \IR_x \times S^1$, there are vortex-like BPS field configurations that are static along the time direction $\IR_t$, and resemble vortices on the cylinder $\IR_x \times S^1$.
At the two ends of the cylinder, the fields are required to approach vacuum configurations.
When the circle shrinks, the vacua $|i\rangle$ coincide with those of the 2d $\CN=(2,2)$ theory obtained in this limit.
The BPS spectrum of the 2d theory contains $\mu_{ij}$ kinks interpolating between vacua $|i\rangle$ and $|j\rangle$.
At finite radius however the vacuum field configurations are labelled by an extra integer $|i,N\rangle$, related to the holonomy of the gauge field at the end of the cylinder.
The 2d kinks are lifted to vortices on a cylinder, labelled by the pair of vacua $|i,N_L\rangle, |j, N_R\rangle$ at the endpoints of the cylinder. Physical properties of these BPS states, such as their vortex flux and their mass can be shown to depend only on $N_R-N_L$.
The similarities between these field configurations and the soliton data of the nonabelianization map are rather compelling, and we expect that these \emph{``kinky vortices''} (vortices with kink-like features) are the physical field configurations  counted by the latter.

\subsubsection*{Soliton data in string theory: M2 branes and domain walls for toric Lagrangians}

Another interpretation of the soliton data is from the viewpoint of string theory.
As noted by \cite{Eager:2016yxd} the string theoretic setting in which Exponential Networks arise presents some important differences from the class $\CS$ setting of Spectral Networks.
In the context of spectral networks, a soliton interpolating between two vacua (two points on the Seiberg-Witten curve) may be identified with a domain wall between semi-infinite M2 branes placed at those points.
This picture doesn't immediately carry over to Exponential Networks.

The Seiberg-Witten curve is replaced by the mirror curve, which encodes the Coulomb branch geometry of the 5d gauge theory engineered by the toric Calabi-Yau $X$.
The mirror curve is also a moduli space for the Lagrangian brane on which the M5 defect brane is wrapped: points on $\Sigma$ can be interpreted as configurations for $L$ and its gauge bundle.
In this picture, solitons of the nonabelianization map, which are defined as open paths connecting two points on $\Sigma$, get naturally interpreted as BPS domain walls connecting two configurations of $L$.

Let us consider this geometry in the context of Type IIA string theory, with a D4 brane placed on $L$.
The effective worldvolume dynamics on the D4 is described by a 2d $\CN=(2,2)$ model, which appears as a surface defect theory within the ambient 4d $\CN=2$ theory engineered by $X$.
The 2d theory has a discrete set of massive vacua, each of whom corresponds to a configuration for $L$ and its gauge bundle.
A BPS kink is engineered by a D2 brane wrapping a holomorphic curve in $X$ and ending on the D4.
In spacetime, the D2 extends along a time-like worldline in $\IR^2$ corresponding to a kink between vacua $|i\rangle$ and $|j\rangle$.
To understand how the D4 is deformed by the presence of the kink it is best to view its worldvolume as fibered of the the spacetime $\IR^2$.
With the soliton kink localized near $x=0$, the D4 brane configuration evolves from a configuration $L_i$ for  $x\to -\infty$ to $L_j$ for $x\to+\infty$.\footnote{
A related interpretation of open paths on the mirror curve was proposed in \cite{Gukov:2016njj}, where they are interpreted as large gauge transformations in the Chern-Simons gauge theory on $L$.
}
Lifting these configurations on a circle to five dimension, introduces contributions from massive KK modes which deform $\tCW_\eff$ and the corresponding vacuum geometry.
D2 branes lift to M2 branes located at a point on the space-like cylinder $S^1_R\times\IR_{x}$, these are the ``kinky vortices'' described above.

It is important to distinguish these BPS states from the usual counting of worldsheet instantons computed by the open topological string wavefunction.
The latter is defined at a fixed choice Lagrangian modulus, whereas D2 branes counted by nonabelianization interpolate between different configurations of the D4 brane.
Nevertheless a connection between the two should exists: a certain sector of D2 branes engineering domain walls on a D4 brane appears to be counted by the (square of the) open topological string partition function \cite{Aganagic:2005dh}. In our language, these should correspond to solitons of type $(ii,n)$


\subsubsection*{Calabi-Yau BPS graphs: quivers and wall-crossing invariants}

In \cite{Gabella:2017hpz} a relation between spectral networks and BPS quivers was proposed, in the context of class $\CS$ theories.
The quiver is dual to a distinguished spectral network, defined at points in the Coulomb moduli space where all central charges are maximally aligned. Tuning the phase $\vartheta$ to the common phase of all central charges gives a degenerate spectral network known as ``BPS graph''.
In \cite{Longhi:2016wtv} it was further shown that the topology of the BPS graph encodes the Kontsevich-Soibelman wall-crossing invariant.
The existence of a BPS graph depends on whether the moduli space contains (at least) one point where all central charges have the same phase.
In the case of class $\CS$ theories the existence of such points is partially established, thanks to mathematical theorems for $A_1$ theories, and by direct inspection for certain higher-rank theories.

The notion of BPS graph carries over naturally to exponential networks, where the Coulomb branch should be replaced, roughly, by the complexified Kahler moduli space of $X$.
It would be very interesting to find explicit examples of such BPS graphs for some toric Calabi-Yau threefolds.
One immediate application of CY BPS graphs would be to derive the dual BPS quivers.
An indication that this should be possible comes from the work of \cite{Eager:2016yxd}, where certain BPS states arising from exponential networks were identified with specific quiver representations.
Another application of CY BPS graphs would be to study the wall-crossing invariants of 5d BPS spectra by extending work of \cite{Longhi:2016wtv} to five dimensions and Calabi-Yau geometries.

\subsubsection*{Applications to knot theory}

An interesting generalization of our work arises from relaxing the restriction to the toric Lagrangian branes of  \cite{Aganagic:2000gs}.
Indeed, the same field-theoretic arguments that led us to formulate the counting of 3d-5d BPS states with exponential networks would apply to other choices of Lagrangian branes.
An especially interesting class is provided by knot conormals in the resolved conifold \cite{Ooguri:1999bv}.
The 3d $\CN=2$ field theory arising from wrapping an M5 brane on a knot conormal $L_K$ is generally more complicated than a simple $U(1)$ GLSM.
On the other hand, several key aspects of these theories have been recently elucidated and UV descriptions of their 3d-5d systems are available for large classes of knots \cite{Dimofte:2010tz, Fuji:2012nx, Ekholm:2018eee}.
The critical set of the twisted superpotential of the 3d theory $T[L_K]$ coincides with the augmentation variety of the knot $K$ \cite{Aganagic:2013jpa}, and its geometry encodes information on holomorphic disks ending on $L_K$ in a configuration parameterized by a point on the curve.

The nonabelianization map associates new topological data to the geometry of the mirror curve, namely its soliton data.
This differs markedly from standard open Gromov-Witten invariants, which are counted by the open topological string wavefunction. \footnote{An obvious reason is that a soliton of the exponential network is associated to \emph{two} points on the mirror curve, and thus appears much like a domain wall for the Lagrangian $L_K$. This is in contrast to the worldsheet instantons, which are counted for $L_K$ in a single configuration (a single point on $\Sigma$).}
It would be interesting to explore what kind of information the soliton data encodes about the knot $K$.
A natural interpretation is that it should count holomorphic maps between two copies of the knot, corresponding to the Lagrangians $L_K$ and $ L_{K'}$ parametrized by the two points on $\Sigma$ connected by the soliton.

\appendix

\section{Collinear vacua and fractional indices}\label{app:collinear-vacua}

The aim of this section is to check some of the predictions about the appearance of fractional CFIV indices in systems with collinear vacua, which we derived in Section \ref{sec:ij-ji-joints}.
For this purpose it is actually unnecessary to study $tt^*$ geometry of a 3d system, because features such as collinear vacua are also encountered in a larger class of models, which go by the name of \emph{periodic} $tt^*$ geometries \cite{Cecotti:2013mba}.

One of the well-studied models for periodic $tt^*$ geometry is that of the free chiral 2d multiplet with the twisted mass $m_{\mathrm {twisted}} = 4\pi i \mu$.
The twisted effective superpotential is given by
\be
\widetilde{\mathcal{W}}_{\mathrm{eff}} = \mu (\log \mu - 1).
\ee
This theory admits a Landau-Ginzburg description involving a periodic field $Y$ with the superpotential
\be
W(Y) = \mu Y - e^Y,
\ee
with the periodic identification
\be
Y \sim Y + 2\pi i.
\ee
This LG theory has a single vacuum given by
\be
Y = \log \mu.
\ee
Plugging this back to $W$ returns indeed $\widetilde{\mathcal{W}}_{\mathrm{eff}}$. The target space $\mathcal{M}$ of this LG model is non-simply connected
with the topology of $\mathbb{R}\times S^1$. Periodicity of $Y$ means that the vacuum is unique,
in particular the different branches corresponding to $\log \mu + 2\pi i N$  with $N \in \mathbb{Z}$ are physically identified.

The distinction between standard $tt^*$ and periodic $tt^*$ geometry stems from the fact that the latter consists of studying  deformations of the theory associated with flavor symmetries, instead of
chiral operators.
In particular, one should view the term $\mu Y$ in the superpotential as playing a role similar to the standard FI term $t\sigma$ of the $U(1)$ GLSM.
Then as we would study $tt*$ geometry over the space of $t$ parameters,  we will  vary $\mu$ in this case.

The quantity
\be
dW = \partial_\mu W\, d\mu = Y d\mu
\ee
is a one form on $\mathcal{M}$ whose periods compute the twisted masses of the theory, similar to the periods of the Seiberg-Witten differential in the context of 2d-4d systems, where they appear as the twisted masses of the theory.
Since $\partial_\mu W$ is multi-valued, $dW$ is closed but not exact. Its periods correspond to the shifts of $\widetilde{\mathcal{W}}_{\mathrm{eff}}$ by the twisted mass :
\be
\Delta W = \int dW = \int Y d\mu = \int \log \mu d\mu = \mu(\log\mu - 1) + 2\pi i \mu n, \quad n \in \mathbb{Z},
\ee
where $n$ is the number of times we integrate around $\mu = 0$ in the $\mu$-plane.

One can view the ambiguity in $W$ in two ways actually. On the one hand, it simply originates from the multi-valuedness of the twisted superpotential $\widetilde{\mathcal{W}}_{\mathrm{eff}} = W (Y = \log \mu)$,
as a function of $\mu$. On the other hand from the LG viewpoint, it is modeled by the periodicity of the variable $Y$, inducing the multivalued behavior of $W$.
This brings forth the fact that the mirror theory must be modeled by a periodic variable, reflecting the fact that the original theory (free massive chiral) had a multivalued twisted superpotential $\widetilde{\mathcal{W}}_{\mathrm{eff}}$.

Moving to the LG description, the target space is $\mathcal{M}$ and $W(Y)$ is multivalued on it. This means that the unique vacuum vacuum is replicated infinitely many times on the $W$-plane, along a tower with spacing $2\pi i  \mu$.
An alternative way to view this is pass to the universal cover $\widetilde{\mathcal{M}} \mapsto \mathcal{M}$, where $W$ is single valued. In this way the single vacuum $|i\rangle$ gets lifted to a whole tower  on the (cover of the) $Y$-plane
\be
	|i,N\rangle \qquad
	\leftrightarrow
	\qquad
	Y = \log\mu + 2\pi i N\,.
\ee

In this basis of vacua we can compute the $C_\mu$ operator corresponding to the deformation by the operator $Y$: it is a diagonal matrix with size the number of vacua (infinite in this basis), whose diagonal elements are the  critical values of the superpotential
\be
	\langle i,N| \,C_\mu\, |i,N\rangle = \(\partial_\mu W\)_{Y = \log\mu+2\pi i N} =\log\mu+2\pi i N \,.
\ee

It is natural to pass to a Fourier-transformed description (Bloch-waves)
\be
\label{thetvac}
	|i,x\rangle = \sum_{N\in\IZ} e^{2\pi i \, N \, x} \, |i,N\rangle\,.
\ee
with $x\sim x+1$ valued in $S^1 \sim \IR/\IZ$.
In this basis the $C$ matrix becomes a differential operator acting on $|i,x\rangle$ vacua
\be
	C_\mu = {\partial_x} + \log\mu
\ee
In this way the $tt^*$ geometry is formulated in terms of a \emph{three-dimensional} parameter space $\IC\times S^1$ with coordinates $\mu,\bar\mu,x$.
This situation is slightly different from standard $tt^*$ in that the parameter space has odd real dimension (instead of even), thanks to the extra coordinate $x$ arising in the Bloch basis.
To write down the  $tt^*$ equations let us momentarily revert to the $|i,N\rangle$ basis.
Then `$i$' takes only one value (corresponding to the variation along $\mu$) and
\be
	C_\mu = 2\pi i \cdot {\bf 1} \otimes N_\mu + B_\mu \otimes {\bf 1}
\ee
where $N_\mu |\mu,N\rangle = N |\mu,N\rangle$ is an infinite-dimensional diagonal matrix with eigenvalues $N$, and $B_\mu |\mu,N\rangle = \log\mu \, |\mu,N\rangle$ a $1\times 1$-dimensional matrix (since there is a single tower of vacua).

Within the tower of vacua there is no distinguished one, since they are all equally spaced and because we are always free to add a constant shift to the superpotential. Therefore the system has a {shift symmetry} group $\IZ$.
This implies that physical quantities can only depend on the difference $N-N'$ and not on $N$ itself. In particular one expects
\be
	\langle \bar j,\bar N  | i ,N\rangle  = g_{i\bar j,N,\bar N}(t,\bar t)  = g_{i\bar j,\bar N-N}(t,\bar t)
\ee

Now switching to the Bloch basis, we can compute\footnote{The computation produces an infinite sum of delta-functions $ \sum_{k}\delta({x-y-k})$ but recall that $x,y$ are valued in $\IR/S^1$.}
\be\label{eq:bloch-metric}
\begin{split}
	\langle\bar j, y| i, x\rangle
	& = \delta({x-y}) \sum_{n} e^{- 2\pi i n y}g_{i\bar j,n}( \mu,\bar  \mu)\\
	& = g_{i\bar j}( \mu,\bar  \mu,x) \delta({x-y} )
\end{split}
\ee
where we defined $n$ as $\bar N=N+n$ and $g_{i\bar j}(t,\bar t,x)$ as the Fourier transform of  $g_{i\bar j,n}(t,\bar t)$.

In standard 2d $tt^*$ geometry the differential equations for the metric in holomorphic gauge take the form
$\partial_{\bar i} (g \partial_j g^{-1}) = \left[ C_j, g (C_i)^\dagger g^{-1}\right]$. Taking the trace yields
then $	\partial_{\bar i}\partial_{j} \log\det g = 0$.
It is shown in \cite{Cecotti:2013mba} that a similar argument in the case of periodic $tt^*$ leads to
\be\label{eq:harmonic-g-periodic}
	\det g(\mu,\bar\mu,x) = g_{\mu\bar\mu}(\mu,\bar\mu,x) = \exp\( L(\mu,\bar\mu,x) - a(\mu) - \bar a(\bar \mu) \)
\ee
with $L$ a harmonic function which is periodic in $x$ and $a,\bar{a}$ are holomorphic and anti-holomorphic functions of $\mu$.
To write down the $tt^*$ equations note that
\be
	C_\mu = \partial_x + \log\mu
\ee
\be
	\bar C_{\bar \mu}
	= g \(C_\mu\)^\dagger g^{-1}
	= e^{L-a-\bar a} \(-\partial_x +\log\bar\mu\) e^{-L+a+\bar a}
	= - \partial_x -\partial_x L +\log\bar\mu
\ee
\be
	D_{\bar\mu} = \partial_{\bar\mu}
\ee
\be
	D_{\mu} = - e^{L-a-\bar a} \partial_\mu e^{-L+a+\bar a} = \partial_\mu + \partial_{\mu}L - \partial_\mu a
\ee
Since $g(\mu,\bar\mu,x) = g(|\mu|,x)$ is a real and positive function,  $L = L(|\mu|,x)$ must also be real, and that $\bar a(\bar\mu) = (a(\mu))^*$.

The reality condition reads
\be
	1 = M M^*
	= (\eta^{-1} g)(\eta^{-1} g)^*
	= e^{L(|\mu|,x)+L(| \mu|,-x)} e^{-a(\mu)-\bar a(\bar \mu) -c.c.}
\ee
where we used the fact that we are always working in a basis in which the topological metric $\eta=1$.
Taking into account the restriction to $|\mu|$-dependence mentioned above, we learn that
\be
	 a ( \mu) = 0 = \bar a (\bar \mu) \,,
\qquad
	L(|\mu|,x) + L(|\mu|,-x) = 0\,.
\ee
We can then expand $L$ as follows
\be
	L(|\mu|,x) = \sum_{m\geq 1} a_m \sin(2\pi m x) \ell_m(|\mu|)
\ee
with $a_m,\ell_m \in\IR$.
From the $tt^*$ equations it follows that $L$ must be harmonic
\be
	\frac{1}{|\mu|}\frac{\partial}{\partial |\mu|}\(|\mu| \frac{\partial}{\partial |\mu|}L\) + 4 \frac{\partial^2}{\partial x^2} L = 0
\ee
which implies
\be
	\frac{\partial^2}{\partial |\mu|^2} \ell_m + \frac{1}{|\mu|}\frac{\partial}{\partial |\mu|}\ell_m - (4\pi m)^2 \ell_m = 0
\ee
which is the Bessel differential equation, with {imaginary variable}. Its solutions are the {modified} Bessel functions
\be
	\ell_{m}(|\mu|) = K_0(4\pi m|\mu|)
\ee
To fix the $a_m$ one needs to consider UV and IR boundary conditions.

To fix the boundary conditions we use the CFIV index, which is related to the metric in a simple way in the one variable case.
According to \cite{Cecotti:1992rm} the CFIV index in canonical coordinates is
\be
\begin{split}
	Q & = -\frac{1}{2} g \, |\mu|  \partial_{|\mu|}g^{-1} \\
	& = -\frac{1}{2} \sum_{m\geq 1} a_m \, (-4\pi m) \,\sin(2\pi m x)\, |\mu|\, K_1(4\pi m|\mu|).
\end{split}
\ee

Now let's use the UV limit first.
To do that we need some basic observations, which are most convenient in the point-basis.
Since there is a single tower of vacua let us lighten notation $|i,N\rangle\to |N\rangle$ from now on. Define the chiral operators in point basis operators as usual by
\be
	l_N|M\rangle = \delta_{N,M}|M\rangle\,.
\ee
Clearly they have the following algebra
\be
	l_N l_M = \delta_{N,M} l_N\,.
\ee
The Bloch operators are ``dual'' to the Bloch states defined above, and they are
\be
	\mathcal{O}_x = \sum_{N} e^{2\pi i N x} l_N
\ee
It is straightforward to see that they satisfy
\be
\begin{split}
	\mathcal{O}_x |N\rangle & = e^{2\pi i x} |N\rangle \\
	\mathcal{O}_x  \mathcal{O}_{x'}  & = \mathcal{O}_{x+x'} \,.
\end{split}
\ee
Moreover, let $T$ be the shift-symmetry operator:
\be
	T |N\rangle = |N+1\rangle
\ee
Then we have
\be
	\mathcal{O}_x \,T\, |N\rangle = e^{2\pi i x}\, T\,\mathcal{O}_x \,|N\rangle
\ee
Therefore if we define $\omega_x, \tau$ by $\mathcal{O}_x = e^{ \omega_x}, T = e^{\tau}$, they have the following commutation relations (by Baker-Campbell-Hausdorff formula)
\be
	[\omega_x,\tau] = 2\pi i x
\ee
then we can compute the shift-symmetry charge of $\mathcal{O}_x$ as
\be
	[\tau,\mathcal{O}_x] = 2\pi i x \, \mathcal{O}_x\,.
\ee
The operator with minimal charge is $\mathcal{O}_{0^+}$ whereas the one with maximal charge is $\mathcal{O}_{1^-}$.
In fact, these are two distinguished operators:
\be
\begin{split}
	\mathcal{O}_0\, |N\rangle = \sum_{M} \ell_M \, |N\rangle = |N\rangle \qquad
	&\Rightarrow
	\qquad
	\mathcal{O}_0 = 1
	\\
	\mathcal{O}_1\, |N\rangle = \sum_{M} e^{2\pi i M }\ell_M \, |N\rangle = e^{2\pi i N } \, |N\rangle \qquad
	&\Rightarrow
	\qquad
	\mathcal{O}_1 = e^X = \partial^2W/\partial X^2 = {\mathfrak{H}}(X)
	\\
\end{split}
\ee
The second identification can be derived by noting that upon shifting our $W$ by a constant $-\mu(\log\mu-1)$, it evaluates to $W(X=X_N) = (\mu X - e^X - \mu(\log\mu-1))_{X=X_N}= 2\pi i N\mu$ at $X_N= \log\mu+2\pi iN$.
The shift-symmetry prevents different $\mathcal{O}_x$ from mixing with each other under RG flow.

Now in the UV limit the CFIV index should approach the (shifted) R-charge of chiral primaries.
Concretely, if we define the UV limit of $Q$ to be $q(x)$:\footnote{Note that $\lim_{z\to 0} z K_1(a z) = a^{-1}$.}
\be\label{eq:Q-UV-limit}
	q(x) = \lim_{|\mu|\to 0} Q(|\mu|,x) = \frac{1}{2}\sum_{m\geq 1} a_m \sin(2\pi m x)
\ee
then the R-charge should be
\be
	\hat q(x) =q(x) - q(0)
\ee
so that $\hat q(0)=0$ which is the charge of the identity operator $\mathcal{O}_{x=0}$.
Then, since the chiral ring has the structure $\mathcal{O}_x \mathcal{O}_{x'} = \mathcal{O}_{x+x'}$ we expect $\hat q$ to be linear, hence $q$ too. Moreover we also know that $Q$ is odd,
since it only contains $\sin(2\pi m x)$ (this came up because of the reality constraint). The only function that is linear and odd on the circle is
\be
	q(x) = \alpha(2\pi x - \pi)
\ee
for some $\alpha$. Note that this is indeed discontinuous at $x=1$, as expected.
Comparing with (\ref{eq:Q-UV-limit}) we can use this to fix $a_m$:
\be
	2\pi x-\pi = -2\sum_{m\geq 1}\frac{\sin 2\pi m x}{m}
	\qquad
	\Rightarrow
	\qquad
	a_m = -4\alpha/m
\ee
Now we should also compare with the IR limit of the CFIV index. This gives the constraint
\be
	\alpha = K/(2\pi)
\ee
therefore the UV limit of the CFIV index is
\be
	q(x) = K\frac{2\pi x-\pi}{2\pi}
\ee
and the discontinuity is $q(x=1)-q(x=0) = K$. Since the discontinuity should equal the central charge $\hat c$ at the UV fixed point, this is $K=1$ for the free 2d chiral.

From the IR interpretation we also learn something else: taking the inverse Fourier transform we can compute the CFIV index for solitons interpolating between vacua $|i,N\rangle$ and $|i,N+n\rangle$.
Notice that these are precisely the $ii$-solitons that are encountered in the $ij-ji$ one-way joint of exponential networks! See Section \ref{sec:ij-ji-joints}.
By shift-symmetry the answer only depends on $n$ and not on $N$
\be
	Q_{n}(|\mu|) = \int_0^{1} {d x} \, e^{2\pi i n x }\, Q(|\mu|,x) = -\frac{{\rm sign}(n)}{4 i} a_{|n|} |n \mu| K_1(|n\mu|)
\ee
since on general grounds this should be in the IR
\be
	Q_n \simeq -i \mu^{(CFIV)}_{n} e^{i\pi f_{n}} \frac{1}{2\pi} (M_n )K_1(M_n)
\ee
with $M_n$ the mass of the soliton we are considering (hence $M_n = n|\mu|$) and $\mu^{(CFIV)}_{n} $ the usual CFIV index.
This gives
\be
	a_n = {\rm sign}(n) \, \frac{2}{\pi} \mu^{(CFIV)}_{n}
\ee
Combining with what we found from the UV limit this entails
\be
	\mu^{(CFIV)}_{n} = {\rm sign}(n) \frac{\pi}{2} \(-\frac{2 K}{\pi n}\) =  - {\rm sign}(n)  \frac{K}{ n}
\ee
Note that we know that $K=1$ for $n=1$ hence $\mu_{n=1} = \pm 1$. But for higher $n$ we can get fractional $\mu^{(CFIV)}_{n}$ for $ii$-type solitons.

A more geometric picture in \cite{Cecotti:2010qn} was given. It was stressed that in presence of collinear vacua, the soliton multiplicities $\mu_{ij}$
as defined by the IR asymptotics of the index $Q_{ij}$ belong to the Lie algebra $\mathfrak{sl} (n,\mathbb{Q})$. For several collinear vacua, it was argued that
appearance of $1/n$ in the CFIV index is in fact quite natural and expected.
This provides a nontrivial check for the results of computations through the exponential networks
of the CFIV indices. In particular, the appearance of $1/n$ for the CFIV indices of the $ii$-type solitons that we derived in Section \ref{sec:ij-ji-joints}.

\section{Plotting}\label{app:plotting}
To plot the networks shown in Section \ref{sec:C3}, we adopt the following change of coordinates
\be
	w = \frac{x}{x+1/4} \qquad x = \frac{w/4}{1-w} \,,
\ee
which takes the puncture at $x=\infty$ to $w=1$ and the branch point at $x=1/4$ to $w=1/2$.
The curve is then
\be
	H(y,w) = y^2 + y + \frac{w/4}{1-w}=0 \,.
\ee
Note that  $d\log x = (1-w)^{-1} d\log w$, therefore the equation of a $(ji,n)$ $\CE$-wall should now be plotted as the solution to the following differential equation
\be\label{eq:trajec}
	(\log y_i - \log y_j +2\pi i n) \frac{1}{1-w}\frac{d\log w}{dt} = e^{i\vartheta} \,.
\ee
The \emph{lhs} requires some care when plotting, because of multi-valuedness.
We get around this difficulty by a trick of \cite[App.A]{Eager:2016yxd}, considering instead an equivalent set of differential equations
\be\label{eq:int-eqs}
\left\{
\begin{split}
\dot w & = e^{i\vartheta} \frac{w(1-w)}{v_i(w)-v_j(w)+2\pi i n}\\
\dot v_i & = -\frac{\partial_w H(y,w)}{\partial_v H(y,w)} \dot w
\end{split}
\right.
\ee
with $v = \log y$.
In order to draw trajectories on the $w$-plane, we need to fix the boundary conditions for this differential equation.
This may be achieved by taking $(w,y)$ on the curve: the ramification point is at
\be
	w_* = 1/2\qquad y_* = -1/2 \,,
\ee
therefore we pick three starting points at $w_m = w_* + \delta w_m$ with
\be\label{eq:start-w}
	\delta w_m = \left( \frac{3}{4} \sqrt\kappa y_* w_*(1-w_*) e^{i\vartheta}\,  (\Delta t) \right)^{2/3} \, e^{\frac{2\pi i}{3} m }\qquad m=0,1,2
\ee
together with
\be\label{eq:start-y}
	\log y_{m,\pm} = \log\left(		y_*  \pm	\frac{3}{4}\frac{y_* w_* (1-w_*) e^{i\vartheta} (\Delta t)}{w_m-w_*}	\right)
\ee
and where
\be
	\kappa = -\frac{1}{2} \(\frac{\partial^2 H}{\partial y^2}\)_{(w_*,y_*)}
	\(\frac{\partial H}{\partial w}\)^{-1}_{(w_*,y_*)}\,.
\ee

\section{Proof of commutativity of $ii/jj$-soliton generating functions}\label{app:q-commute}
Here we prove that soliton generating functions of type $ii$ (and $jj$ for $i\neq j$) always commute among themselves.
For concreteness, let us consider the generating functions featured in (\ref{eq:fQ-def}), our goal is to show that there are no ordering ambiguities in those infinite products.

It is clear that $[\tau_{q_k},\tau_{\bar q_\ell}] = [\nu_{q_k},\nu_{\bar q_\ell}]=[\tau_{q_k},\nu_{\bar q_\ell}]=[\nu_{q_k},\tau_{\bar q_\ell}]=0$ because $ii$ paths do not concatenate with $jj$ paths.
So we only need to show that $[\nu_{q_k},\nu_{q_\ell}]=[\tau_{q_k},\tau_{q_\ell}]=[\tau_{q_k},\nu_{q_\ell}]=0$ and their counterparts with $q\to\bar q$. The proof is essentially the same for each case, so let us focus on the first one.
Given the incoming data on $q_k, q_\ell$ in the general form
\be
\begin{split}
	\nu_{q_k} & = \sum_N \sum_{a\in\Gamma_{ii,N,N+k(m+n)}} \mu(a,q_k) X_a  \,,
	\qquad
	\nu_{q_\ell} = \sum_N \sum_{b\in\Gamma_{ii,N,N+\ell(m+n)}} \mu(b,q_\ell) X_b\,, \\
\end{split}
\ee
their products are simply
\be
\begin{split}
	\nu_{q_k}\nu_{q_\ell} & = \sum_N \sum_{{\scriptsize \begin{array}{c}
		a\in\Gamma_{ii,N,N+k(m+n)}\\
		b\in\Gamma_{ii,N+k(m+n),N+(k+\ell)(m+n)}
		\end{array}}}
		\mu(a)\mu(b) X_{ab}
	\\
	& =
	\sum_N \sum_{c\in\Gamma_{ii,N,N+(k+\ell)(m+n)}} \fm(c) X_c \,,
	\\
	\nu_{q_\ell}\nu_{q_k} & = \sum_N \sum_{{\scriptsize \begin{array}{c}
		a\in\Gamma_{ii,N+\ell(m+n),N+(\ell+k)(m+n)}\\
		b\in\Gamma_{ii,N,N+\ell(m+n)}
		\end{array}}}
		\mu(a)\mu(b) X_{ba}
	\\
	& =
	\sum_N \sum_{c'\in\Gamma_{ii,N,N+(k+\ell)(m+n)}} \fm'(c') X_{c' } \,.
\end{split}
\ee
On both sides there are generating series of solitons in the same lattices $c,c'\in\Gamma_{ii,N,N+(k+\ell)(m+n)}$. There is a 1:1 correspondence between the two sets of solitons and their degneracies, in fact, at fixed $N$
\be
\begin{split}
	\fm(c) & = \sum_{\({\scriptsize \begin{array}{c}
		a\in\Gamma_{ii,N,N+k(m+n)}\\
		b\in\Gamma_{ii,N+k(m+n),N+(k+\ell)(m+n)}
		\end{array}\Big| ab\sim c}\)}
		\mu(a)\mu(b) \,,
	\\
	\fm'(c') & = \sum_{\({\scriptsize \begin{array}{c}
		a\in\Gamma_{ii,N+\ell(m+n),N+(\ell+k)(m+n)}\\
		b\in\Gamma_{ii,N,N+\ell(m+n)}
		\end{array}\Big| ba \sim c'}\)}
		\mu(a)\mu(b) \,.
\end{split}
\ee
Given paths $a,b$ such that $ab=c$ there is a corresponding pair of paths $a^{(+\ell(m+n))}, b^{(-k(m+n))}$ such that
\be
	a^{(+\ell(m+n))} \in\Gamma_{ii,N+\ell(m+n),N+(\ell+k)(m+n)}\,,
	\qquad
	b^{(-k(m+n))} \in \Gamma_{ii,N,N+\ell(m+n)} \,,
\ee
and $c' = b^{(-k(m+n))} a^{(+\ell(m+n))}$ has central charge
\be
	Z_{c'}
	= Z_{b^{(-k(m+n))} } + Z_{a^{(+\ell(m+n))}}
	= Z_{b } + Z_{a}
	= Z_{ab}
	= Z_c\,,
\ee
because the shift map preserves central charge, and because $c$ is uniquely fixed by its soliton type $(ii,N,N+(k+l)(m+n))$ and by its central charge. This establishes a bijection between the sets
\be
	\({\scriptsize \begin{array}{c}
		a\in\Gamma_{ii,N,N+k(m+n)}\\
		b\in\Gamma_{ii,N+k(m+n),N+(k+\ell)(m+n)}
		\end{array}\Big| ab\sim c}\)
	\mathop{\longleftrightarrow}^{1:1}
	\({\scriptsize \begin{array}{c}
		a\in\Gamma_{ii,N+\ell(m+n),N+(\ell+k)(m+n)}\\
		b\in\Gamma_{ii,N,N+\ell(m+n)}
		\end{array}\Big| ba \sim c'}\) \,.
\ee
Moreover by shift symmetry
\be
	\mu(a^{(+\ell(m+n))}) = \mu(a) \,, \qquad
	\mu(b^{(-k(m+n))}) = \mu(b) \,,
\ee
therefore
\be
	\fm(ab) = \fm'(ba) \,.
\ee
This concludes the proof that all factors commute.

\section{Flatness equations for a two-way street joint}\label{app:two-way-junction}
Here we give the detailed version of the flatness equations studied in subsection  \ref{sec:eqn-breakdown}.
Referring to Figure  \ref{fig:ij-ji-2-way}, one can compute the following (we used the commutation rules for $\fQ$ and $\fQ'$
in (\ref{eq:ii-jj-comm})).

For the paths indicated in Figure \ref{fig:ij-ji-2-way} the $ij$ components of the parallel transports are
\be
\begin{split}
F(\wp)_{ij} &=\, \fQ_{ii} \(\sum_{k\ge 1} \(\prod_{\beta=1}^{k}Q(r_k)\)  \frac{\tau_{r_k}}{Q(r_k)}\)  \(\prod_{\alpha=\infty}^1 Q(p_\alpha)\)
\\ &
+ \fQ_{ii}  \(\prod_{\alpha=1}^\infty Q(r_\alpha)\)\( \sum_{k\ge 1}\frac{\nu_{p_{k}}}{Q(p_k)} \(\prod_{\beta=k}^1 Q(p_\beta)\)  \)
\\ &
+ F_1(\wp)_{ij} + F_2(\wp)_{ij}+ F_3(\wp)_{ij} + F_4(\wp)_{ij}
\end{split}
\ee
The last four terms ``higher order'' contributions and their form is as follows
\be
\begin{split}
 F_1(\wp)_{ij}  & = \fQ_{ii} \bigg[\sum_{\substack{k\ge 1\\ l > k}}   \(\prod_{\beta=1}^k Q(r_\beta)\)\frac{\tau_{r_k}}{Q(r_k)} \(\nu_{r_l} \tau_{r_{l+1}}\)\bigg] \(\prod_{\alpha=\infty}^1 Q(p_\alpha)\)
\\ &
+  \fQ_{ii} \bigg[ \sum_{\substack{k\ge 1\\ l > k}}  \(\prod_{\beta=1}^k Q(r_\beta)\) \frac{\tau_{r_k}}{Q(r_k)} (\nu_{r_l}Q(r_{l+1}) \tau_{r_{l+2}})
\\ & \quad
+ \sum_{\substack{k\ge 1\\ l_1 > k\\ l_2>l_1+1}}  \(\prod_{\beta=1}^k Q(r_\beta)\) \frac{\tau_{r_k}}{Q(r_k)}(\nu_{r_{l_1}}\tau_{r_{l_1+1}}\nu_{r_{l_2}}\tau_{r_{l_2+1}})\bigg]  \(\prod_{\alpha=\infty}^1 Q(p_\alpha)\) + \dots,
\end{split}
\ee
\be
\begin{split}
F_2(\wp)_{ij} &=\, \fQ_{ii} \(\prod_{\alpha=1}^\infty Q(r_\alpha)\) \bigg[  \sum_{\substack{k\ge 1\\ l>k}} (\nu_{p_{l+1}}\tau_{p_l}) \frac{\nu_{p_k}}{Q(p_k)} \(\prod_{\beta=k}^1 Q(p_\beta)\) \bigg]
\\ &
+ \fQ_{ii}\(\prod_{\alpha=1}^\infty Q(r_\alpha)\) \bigg[  \sum_{\substack{k\ge 1\\ l>k}} (\nu_{p_{l+2}}Q(p_{l+1}) \tau_{p_l}) \frac{\nu_{p_k}}{Q(p_k)} \(\prod_{\beta=k}^1 Q(p_\beta)\)
\\ & \quad
+ \sum_{\substack{k\ge 1\\ l_1 > k\\ l_2>l_1+1}}  (\nu_{p_{l_2+1}}\tau_{p_{l_2}}) (\nu_{p_{l_1+1}}\tau_{p_{l_1}})  \frac{\nu_{p_k}}{Q(p_k)} \(\prod_{\beta=k}^1 Q(p_\beta)\)
+ \dots \bigg].
\end{split}
\ee
\be
\begin{split}
F_3(\wp)_{ij} &=\,  \fQ_{ii} \sum_{\substack{k,k'\ge 1\\ l >k}} \(\prod_{\beta=1}^k Q(r_\alpha)\) \frac{\tau_{r_k}}{Q(r_k)}(\nu_{r_l}\tau_{r_{l+1}})\(\frac{\tau_{p_{k'+1}}\nu_{p_{k'}}}{Q(p_{k'+1})Q(p_{k'})}\)\(\prod_{\alpha=\infty}^1 Q(p_\alpha)\)
\\ &
+ \fQ_{ii} \sum_{\substack{k,k'\ge 1\\ l >k}} \(\prod_{\beta=1}^k Q(r_\alpha)\)  \frac{\tau_{r_k}}{Q(r_k)}
\bigg\{(\nu_{r_l}Q(r_{l+1})\tau_{r_{l+2}} + \sum_{l'>l+1} (\nu_{r_{l}}\tau_{r_{l+1}} \nu_{r_{l'}}\tau_{r_{l'+1}}) \bigg\}
\\ & \qquad\qquad \qquad\times
\(\frac{\tau_{p_{k'+1}}\nu_{p_{k'}}}{Q(p_{k'+1})Q(p_{k'})}\)\(\prod_{\alpha=\infty}^1 Q(p_\alpha)\)
\\&
+ \fQ_{ii} \sum_{\substack{k,k'\ge 1\\ l >k}}  \(\prod_{\beta=1}^k Q(r_\alpha)\)  \frac{\tau_{r_k}}{Q(r_k)} (\nu_{r_l}\tau_{r_{l+1}})
\bigg\{\frac{\tau_{p_{k'+2}}\nu_{p_{k'}}}{Q(p_{k'+2})Q(p_{k'+1})Q(p_{k'})}
\\ & \quad
+ \sum_{l'>k'+1} \frac{\tau_{p_{l'+1}}\nu_{p_{l'}}}{Q(p_{l'+1})Q(p_{l'})}\frac{\tau_{p_{k'+1}}\nu_{p_{k'}}}{Q(p_{k'+1})Q(p_{k'})}\bigg\}
\(\prod_{\alpha=\infty}^1 Q(p_\alpha)\) + \dots,
\end{split}
\ee
\be
\begin{split}
F_4(\wp)_{ij}
& = \fQ_{ii} \sum_{\substack{k,k'\ge 1\\ l' >k'}}  \(\prod_{\beta=1}^\infty Q(r_\alpha)\)\frac{\tau_{r_k}\nu_{r_{k+1}}}{Q(r_k)Q(r_{k+1})}
{\nu_{p_{l'+1}}\tau_{p_{l'}}}\frac{\nu_{p_{k'}}}{Q(p_{k'})}\(\prod_{\alpha=k'}^1 Q(p_\alpha)\)
\\ &
+ \fQ_{ii} \bigg[ \sum_{\substack{k,k'\ge 1\\ l' >k'}} \(\prod_{\beta=1}^\infty Q(r_\alpha)\) \bigg\{ \frac{\tau_{r_k}\nu_{r_{k+2}}}{Q(r_k)Q(r_{k+1})Q(r_{k+2})}
\\ &
+ \frac{\tau_{r_k}\nu_{r_{k+1}}}{Q(r_k)Q(r_{k+1})}  \sum_{l>k} \frac{\tau_{r_l}\nu_{r_{l+1}}}{Q(r_l)Q(r_{l+1})} \bigg\} \bigg]
{\nu_{p_{l'+1}}\tau_{p_{l'}}}\frac{\nu_{p_{k'}}}{Q(p_{k'})}\(\prod_{\alpha=k'}^1 Q(p_\alpha)\) \bigg]
\\ &
+ \fQ_{ii} \bigg[ \sum_{\substack{k,k'\ge 1\\ l' >k'}} \(\prod_{\beta=1}^\infty Q(r_\alpha)\)\frac{\tau_{r_k}\nu_{r_{k+1}}}{Q(r_k)Q(r_{k+1})}
\\ & \times
\bigg\{ (\nu_{p_{l'+2}} Q(p_{l'+1})\tau_{p_{l'}}) +(\nu_{p_{l'+1}}\tau_{p_{l'}}) \sum_{l''>l'+1} (\nu_{p_{l'+1}}\tau_{p_{l'}}) \bigg\} \frac{\nu_{p_{k'}}}{Q(p_{k'})}\(\prod_{\alpha=k'}^1 Q(p_\alpha)\) \bigg] + \dots
\end{split}
\ee
The $ij$ component of the parallel transport for the complementary path $\wp'$ is
\be
\begin{split}
F(\wp')_{ij} &= \,\fQ'_{ii} \sum_{k\ge 1}  \(\prod_{\alpha=\infty}^{k+1} Q(r'_\alpha)\) \nu_{r'_k}
+ \fQ'_{ii} \sum_{k\ge 1} \frac{\tau_{p'_k}}{Q(p'_k)}\(\prod_{\alpha=k}^\infty Q(r'_\alpha)\)
\\ &
+ F'_1(\wp')_{ij} + F'_2(\wp)_{ij} + F'_3(\wp')_{ij} + F'_4(\wp')_{ij}
\end{split}
\ee
where again the non-linear contributions are collected in the last four terms, which read
\be
\begin{split}
F'_1(\wp')_{ij}&=\, \fQ'_{ii} \sum_{\substack{k\ge1\\ l>k}}  \frac{\tau_{p'_k}}{Q(p'_k)} \frac{\nu_{p'_l}\tau_{p'_{l+1}}}{Q(p'_l)Q(p'_{l+1})} \(\prod_{\alpha=k}^\infty Q(p'_\alpha)\)
\\ &
+ \fQ'_{ii} \sum_{\substack{k\ge 1 \\ l_1>k \\ l_2>l_1}} \frac{\tau_{p'_k}}{Q(p'_k)} \frac{\nu_{p'_{l_2}} \tau_{p'_{l_2+1}}\nu_{p'_{l_1}} \tau_{p'_{l_1+1}}}{Q(p'_{l_1})Q(p'_{l_1+1})Q(p'_{l_2})Q(p'_{l_2+1})} \(\prod_{\alpha=k}^\infty Q(p'_\alpha)\)
\\ &
+\fQ'_{ii} \sum_{\substack{k\ge1\\ l>k}}  \frac{\tau_{p'_k}}{Q(p'_k)} \frac{\nu_{p'_l}\tau_{p'_{l+2}}}{Q(p'_l)Q(p'_{l+1})Q(p'_{l+2})} \(\prod_{\alpha=k}^\infty Q(p'_\alpha)\) + \dots,
\end{split}
\ee
\be
\begin{split}
F'_2(\wp')_{ij} &=\,\fQ'_{ii}\sum_{\substack{k\ge 1\\ l>k}}  \(\prod_{\alpha=\infty}^k Q(r'_k)\) \frac{\nu_{r'_{l+1}}\tau_{r'_l}}{Q(r'_{l+1})Q(r'_l)}\frac{\nu_{r'_k}}{Q(r'_k)}
\\ &
+ \fQ'_{ii} \sum_{\substack{k\ge 1\\ l_1>k \\ l_2>l_1+1}}\(\prod_{\alpha=\infty}^k Q(r'_\alpha)\)\frac{\nu_{r'_{l_2+1}}\tau_{r'_{l_2}}\nu_{r'_{l_1+1}}\tau_{r'_{l_1}}}{Q(r'_{l_1})Q(r'_{l_1+1})Q(r'_{l_2})Q(r'_{l_2+1})} \frac{\nu_{r'_k}}{Q(r'_k)}
\\ &
+ \fQ'_{ii} \sum_{\substack{k\ge 1\\ l>k}}  \(\prod_{\alpha=\infty}^k Q(r'_\alpha)\) \frac{\nu_{r'_{l+2}}\tau_{r'_l}}{Q(r'_{l+2})Q(r'_{l+1})Q(r'_l)}\frac{\nu_{r'_k}}{Q(r'_k)} + \dots
\end{split}
\ee
\be
\begin{split}
F'_3(\wp')_{ij} &= \,\fQ'_{ii} \sum_{\substack{k,k'\ge1 \\ l>k}}  \frac{\tau_{p'_k}}{Q(p'_k)}\frac{\nu_{p'_l}\tau_{p'_{l+1}}}{Q(p'_l)Q(p'_{l+1})}\(\prod_{\alpha=k}^\infty Q(p'_\alpha)\) (\tau_{r'_{k'+1}}\nu_{r'_{k'}})
\\ &
+ \fQ'_{ii} \bigg[ \sum_{\substack{k,k'\ge1 \\ l>k}}  \frac{\tau_{p'_k}}{Q(p'_k)}\frac{\nu_{p'_l}\tau_{p'_{l+2}}}{Q(p'_l)Q(p'_{l+1})Q(p'_{l+2})} \(\prod_{\alpha=k}^\infty Q(p'_\alpha)\) (\tau_{r'_{k'+1}}\nu_{r'_{k'}})
\\ &
+ \sum_{\substack{k,k'\ge1 \\ l_1>k\\ l_2>l_1+1}}  \frac{\tau_{p'_k}}{Q(p'_k)}\frac{\nu_{p'_{l_1}}\tau_{p'_{l_1+1}}}{Q(p'_{l_1})Q(p'_{l+1})}\frac{\nu_{p'_{l_2}}\tau_{p'_{l_2+1}}}{Q(p'_{l_2})Q(p'_{l_2+1})}\(\prod_{\alpha=k}^\infty Q(p'_\alpha)\) (\tau_{r'_{k'+1}}\nu_{r'_{k'}})
\\ &
+ \sum_{\substack{k,k'\ge1 \\ l>k}}  \frac{\tau_{p'_k}}{Q(p'_k)}\frac{\nu_{p'_l}\tau_{p'_{l+1}}}{Q(p'_l)Q(p'_{l+1})}\(\prod_{\alpha=k}^\infty Q(p'_\alpha)\) (\tau_{r'_{k'+2}}Q(r'_{k'+1})\nu_{r'_{k'}})
\\ &
+  \sum_{\substack{k,k'\ge1 \\ l>k \\ l'>k'}}  \frac{\tau_{p'_k}}{Q(p'_k)}\frac{\nu_{p'_l}\tau_{p'_{l+1}}}{Q(p'_l)Q(p'_{l+1})}\(\prod_{\alpha=k}^\infty Q(p'_\alpha)\)  (\tau_{r'_{l'+1}}\nu_{r'_{l'}})  (\tau_{r'_{k'+1}}\nu_{r'_{k'}})
\bigg] +  \dots
\end{split}
\ee
\be
\begin{split}
F'_4(\wp')_{ij} \, & = \, \fQ'_{ii} \sum_{\substack{k,k'\ge1 \\ l>k}}  (\tau_{p'_{k'}}\nu_{p'_{k'+1}}) \frac{\nu_{r'_{l+1}}\tau_{r'_{l}}}{Q(r'_{l+1}) Q(r'_{l})} \(\prod_{\alpha=\infty}^k Q(r'_\alpha)\)\frac{\nu_{r'_{k}}}{Q(r'_{k})}
\\ &
+ \fQ'_{ii} \bigg[  \sum_{\substack{k,k'\ge1 \\ l>k}}  (\tau_{p'_{k'}}\nu_{p'_{k'+1}}) \frac{\nu_{r'_{l+2}}\tau_{r'_{l}}}{Q(r'_{l+2})Q(r'_{l+1}) Q(r'_{l})} \(\prod_{\alpha=\infty}^k Q(r'_\alpha)\)\frac{\nu_{r'_{k}}}{Q(r'_{k})}
\\ &
+ \sum_{\substack{k,k'\ge1 \\ l_1>k\\ l_2>l_1+1}}(\tau_{p'_{k'}}\nu_{p'_{k'+1}})   \frac{\nu_{r'_{l_2+1}}\tau_{r'_{l_2}}}{Q(r'_{l_2+1}) Q(r'_{l_2})} \frac{\nu_{r'_{l_1+1}}\tau_{r'_{l_1}}}{Q(r'_{l_1+1}) Q(r'_{l_1})}\(\prod_{\alpha=\infty}^k Q(r'_\alpha)\)\frac{\nu_{r'_{k}}}{Q(r'_{k})}
\\ &
+ \sum_{\substack{k,k'\ge1 \\ l>k \\ l' > k'}}  (\tau_{p'_{k'}}\nu_{p'_{k'+1}}) (\tau_{p'_{l'}}\nu_{p'_{l'+1}})\frac{\nu_{r'_{l+1}}\tau_{r'_{l}}}{Q(r'_{l+1}) Q(r'_{l})} \(\prod_{\alpha=\infty}^k Q(r'_\alpha)\)\frac{\nu_{r'_{k}}}{Q(r'_{k})}
\\ &
+  \sum_{\substack{k,k'\ge1 \\ l>k}}  (\tau_{p'_{k'}}Q(p'_{k'+1})\nu_{p'_{k'+2}}) \frac{\nu_{r'_{l+1}}\tau_{r'_{l}}}{Q(r'_{l+1}) Q(r'_{l})} \(\prod_{\alpha=\infty}^k Q(r'_\alpha)\)\frac{\nu_{r'_{k}}}{Q(r'_{k})} \bigg] + \dots
\end{split}
\ee

Due to the cumbersome nature of these equations, in the following, we will only report the analysis for the $ii$-type components. The other two cases $ji$ and $jj$ can be computed in the same fashion.
Next we consider the  $ii$ components of the parallel transports for paths $\wp,\wp'$. We have
\be
F(\wp)_{ii} \, = \, F_1(\wp)_{ii} + F_2(\wp)_{ii} + F_3(\wp)_{ii} + F_4(\wp)_{ii},
\ee
where
\be
\begin{split}
F_1(\wp)_{ii} & = \,   \fQ_{ii}\(\prod_{\alpha=1}^\infty Q(r_\alpha)\) + \fQ_{ii} \tau_{r_1} \(\prod_{\alpha=\infty}^2 Q(p_\alpha)\) \tau_{p_1}
\\ &
+  \fQ_{ii} \tau_{r_1}  \sum_{k\ge 2} \(\prod_{\alpha=\infty}^{k} Q(p_\alpha)\)\frac{\tau_{p_k}}{Q(p_k)}
+ \fQ_{ii} \sum_{k\ge 2}  \(\prod_{\alpha=1}^{k} Q(r_\alpha )\)\frac{\tau_{r_k}}{Q(r_k)} \(\prod_{\beta=\infty}^2 Q(p_\beta)\)\tau_{p_1}
\\ &
+\fQ_{ii} \sum_{k,l\ge 2}  \(\prod_{\alpha=1}^{k} Q(r_\alpha)\)\frac{\tau_{r_k}}{Q(r_k)} \(\prod_{\beta=\infty}^{l} Q(p_\beta)\)\frac{\tau_{p_l}}{Q(p_l)}
\end{split}
\ee
\be
\begin{split}
F_2(\wp)_{ii} &= \, \(\prod_{\alpha=1}^\infty Q(r_\alpha)\)  \fQ_{ii} \bigg[  \sum_{k\ge 1} \frac{\tau_{r_k}\nu_{r_{k+1}}}{Q(r_k) Q(r_{k+1})}
\\ & \qquad
+  \bigg\{ \sum_{k\ge 1}  \frac{\tau_{r_k}\nu_{r_{k+2}}}{Q(r_k) Q(r_{k+1})Q(r_{k+2}))}
+ \sum_{\substack{k_1\ge 1 \\ k_2 > k_1+1}}\frac{\tau_{r_{k_1}}\nu_{r_{k_1+1}}\tau_{r_{k_2}}\nu_{r_{k_2+1}}}{Q(r_{k_1}) Q(r_{k_1+1}) Q(r_{k_2})Q(r_{k_2+1})} \bigg\}
\\ & \qquad
+  \bigg\{ \sum_{k\ge 1}  \frac{\tau_{r_k}\nu_{r_{k+3}}}{Q(r_k) Q(r_{k+1}Q(r_{k+2})Q(r_{k+3}))}
\\ &
+ \sum_{\substack{k_1\ge 1 \\ k_2 > k_1+1}}\frac{\tau_{r_{k_1}}\nu_{r_{k_1+1}}\tau_{r_{k_2}}\nu_{r_{k_2+2}}}{Q(r_{k_1}) Q(r_{k_1+1}) Q(r_{k_2})Q(r_{k_2+1})Q(r_{k_2+2}))}
\\ & \qquad
+  \sum_{\substack{k_1\ge 1 \\ k_2 > k_1+2}}\frac{\tau_{r_{k_1}}\nu_{r_{k_1+2}}\tau_{r_{k_2}}\nu_{r_{k_2+1}}}{Q(r_{k_1}) Q(r_{k_1+1}) Q(r_{k_1+2}) Q(r_{k_2})Q(r_{k_2+1}))}
\\ & \qquad
+  \sum_{\substack{k_1\ge 1 \\ k_2 > k_1+1 \\ k_3 > k_2+1}}\frac{\tau_{r_{k_1}}\nu_{r_{k_1+1}}\tau_{r_{k_2}}\nu_{r_{k_2+1}}\tau_{r_{k_3}}\nu_{r_{k_3+1}}}{Q(r_{k_1}) Q(r_{k_1+1})  Q(r_{k_2})Q(r_{k_2+1}))Q(r_{k_3})Q(r_{k_3+1})}
 \bigg\}
+ \dots \bigg] ,
\end{split}
\ee
\be
\begin{split}
F_3(\wp)_{ii} & =\, \fQ_{ii}\(\prod_{\alpha=1}^\infty Q(r_\alpha)\)\bigg\{ \sum_{k\ge 1} \nu_{p_{k+1}}\tau_{p_k}
+  \bigg[ \sum_{k\ge 1} \nu_{p_{k+2}}Q(p_{k+1})\tau_{p_k} + \sum_{\substack{k_1\ge 1\\ k_2 > k_1+1}} \nu_{p_{k_2+1}}\tau_{p_{k_2}}\nu_{p_{k_1+1}}\tau_{p_{k_1}} \bigg]
\\ &
+ \bigg[\sum_{k\ge 1} \nu_{p_{k+3}}Q(p_{k+2})Q(p_{k+1})\tau_{p_k} + \sum_{\substack{k_1\ge 1\\ k_2 > k_1+2}} \nu_{p_{k_2+1}}\tau_{p_{k_2}}\nu_{p_{k_1+2}}Q(p_{k_1+1})\tau_{p_{k_1}}
\\ &
+  \sum_{\substack{k_1\ge 1\\ k_2 > k_1+1}} \nu_{p_{k_2+2}}Q(p_{k_2+1})\tau_{p_{k_2}}\nu_{p_{k_1+1}}\tau_{p_{k_1}}
+  \sum_{\substack{k_1\ge 1\\ k_2 > k_1+1 \\ k_3>k_2+1}} \nu_{p_{k_3+1}}\tau_{p_{k_3}}  \nu_{p_{k_2+1}}\tau_{p_{k_2}}\nu_{p_{k_1+1}}\tau_{p_{k_1}}  \bigg] \bigg\}
+ \dots ,
\end{split}
\ee
\be
\begin{split}
F_4(\wp)_{ii} & = \, \(\prod_{\alpha=1}^\infty Q(r_\alpha)\) \fQ_{ii}\bigg[  \sum_{k,l\ge 1}\frac{\tau_{r_k}\nu_{r_{k+1}}}{Q(r_k) Q(r_{k+1})} \nu_{p_{l+1}}\tau_{p_l}
\\ &
+ \bigg\{\sum_{k,l\ge 1}\frac{\tau_{r_k}\nu_{r_{k+2}}}{Q(r_k) Q(r_{k+1})Q(r_{k+2})} \nu_{p_{l+1}}\tau_{p_l}
+ \sum_{\substack{k_1,l\ge 1 \\ k_2> k_1+1}} \frac{\tau_{r_{k_1}}\nu_{r_{k_1+1}}\tau_{r_{k_2}}\nu_{r_{k_2+1}}}{Q(r_{k_1}) Q(r_{k_1+1}) Q(r_{k_2})Q(r_{k_2+1}))}  \nu_{p_{l+1}}\tau_{p_l}
\\ &
+ \sum_{\substack{k,l_1\ge 1 \\ l_2> l_1+1}}\frac{\tau_{r_k}\nu_{r_{k+1}}}{Q(r_k) Q(r_{k+1})} \nu_{p_{l_2+1}}\tau_{p_{l_2}}\nu_{p_{l_1+1}}\tau_{p_{l_1}}
+ \sum_{k,l\ge 1} \frac{\tau_{r_k}\nu_{r_{k+1}}}{Q(r_k) Q(r_{k+1})} \nu_{p_{l+2}} Q(p_{l+1})\tau_{p_l} \bigg\} + \dots\bigg].
\end{split}
\ee
For the complementary path, we split the terms again.
\be
F(\wp')_{ii} \, = \, F'_1(\wp')_{ii}+ F'_2(\wp')_{ii} + F'_3(\wp')_{ii} + F'_4(\wp')_{ii},
\ee
where
\be
\begin{split}
 F'_1(\wp')_{ii} & =\, \fQ'_{ii}\bigg[\(\prod_{\alpha=\infty}^1  Q(r'_\alpha)\) +  \tau_{p'_1} \(\prod_{\alpha=2}^\infty Q(p'_\alpha)\)\tau_{r'_1}
\\ &
+ \sum_{k\ge 2} \frac{\tau_{p'_k}}{Q(p'_k)} \(\prod_{\alpha=k}^\infty Q(p'_\alpha)\)  \tau_{r'_1}
+   \tau_{p'_1} \(\prod_{\alpha=2}^\infty Q(p'_\alpha)\)   \sum_{k\ge 2} \frac{\tau_{r'_k}}{Q(r'_k)} \(\prod^{1}_{\beta=k} Q(r'_\beta)\)
\\&
+ \sum_{k,l\ge 2}  \frac{\tau_{p'_l}}{Q(p'_l)} \(\prod_{\alpha=l}^\infty Q(p'_\alpha)\)  \frac{\tau_{r'_k}}{Q(r'_k)}  \(\prod^{1}_{\beta=k} Q(r'_\beta)\) \bigg],
\end{split}
\ee
\be
\begin{split}
F'_2(\wp')_{ii} & =\, \fQ'_{ii}\bigg\{\sum_{k\ge 1} \tau_{p'_k}\nu_{p'_{k+1}}
+  \bigg[ \sum_{k\ge 1} \tau_{p'_k}Q(p'_{k+1})\nu_{p'_{k+2}}
 + \sum_{\substack{k_1\ge 1\\ k_2 > k_1+1}}\tau_{p'_{k_1}}  \nu_{p'_{k_1+1}}\tau_{p'_{k_2}}\nu_{p'_{k_2+1}}\bigg]
\\ &
+\bigg[ \sum_{k\ge 1} \tau_{p'_k}Q(p'_{k+1})Q(p'_{k+2}) \nu_{p'_{k+3}} + \sum_{\substack{k_1\ge 1 \\ k_2>k_1+1}} (\tau_{p'_{k_1}}  \nu_{p'_{k_1+1}})(\tau_{p'_{k_2}}Q(p'_{k_2+1})\nu_{p'_{k_2+2}})
\\ &
+  \sum_{\substack{k_1\ge 1 \\ k_2>k_1+2}} (\tau_{p'_{k_1}}Q(p'_{k_1+1})\nu_{p'_{k_1+2}}) (\tau_{p'_{k_2}}\nu_{p'_{k_2+1}})
\\ &
+ \sum_{\substack{k_1\ge 1 \\k_2>k_1+1\\ k_3>k_2+1}}
 (\tau_{p'_{k_1}}  \nu_{p'_{k_1+1}}) (\tau_{p'_{k_2}}  \nu_{p'_{k_2+1}}) (\tau_{p'_{k_3}}  \nu_{p'_{k_3+1}}) \bigg]
+\dots\bigg\} \(\prod_{\alpha=\infty}^1 Q(r'_\alpha)\),
\end{split}
\ee
\be
\begin{split}
F_3(\wp')_{ii} & = \, \(\prod_{\alpha=\infty}^1 Q(r'_\alpha)\) \fQ'_{ii} \bigg[  \sum_{k\ge 1} \frac{\nu_{r'_{k+1}}\tau_{r'_k}}{Q(r'_k) Q(r'_{k+1})}
\\ &
+  \bigg\{ \sum_{k\ge 1}  \frac{\nu_{r'_{k+2}}\tau_{r'_k}}{Q(r'_k) Q(r'_{k+1})Q(r'_{k+2})}
+ \sum_{\substack{k_1\ge 1 \\ k_2 > k_1+1}}\frac{\nu_{r'_{k_2+1}}\tau_{r'_{k_2}}\nu_{r'_{k_1+1}}\tau_{r'_{k_1}}}{Q(r'_{k_1}) Q(r'_{k_1+1}) Q(r'_{k_2})Q(r'_{k_2+1}))} \bigg\}
\\ &
+\bigg\{\sum_{k\ge 1} \frac{\nu_{r'_{k+3}}\tau_{r'_k}}{Q(r'_{k+3})Q(r'_{k+2})Q(r'_{k+1})Q(r'_{k})}
\\ &
+ \sum_{\substack{k_\ge 1 \\ k_2>k_1+1}} \frac{\nu_{r'_{k_2+2}}\tau_{r'_{k_2}}\nu_{r'_{k_1+1}}\tau_{r'_{k_1}}}{Q(r'_{k_2+2})Q(r'_{k_2+1})Q(r'_{k_2})Q(r'_{k_1+1})Q(r'_{k_1})}
\\ &
+  \sum_{\substack{k_\ge 1 \\ k_2>k_1+2}} \frac{\nu_{r'_{k_2+1}}\tau_{r'_{k_2}}\nu_{r'_{k_1+2}}\tau_{r'_{k_1}}}{Q(r'_{k_2+1})Q(r'_{k_2})Q(r'_{k_1+2})Q(r'_{k_1+1})Q(r'_{k_1})}
\\ &
+  \sum_{\substack{k_\ge 1 \\ k_2>k_1+2\\ k_3>k_2+1}} \frac{\nu_{r'_{k_3+1}}\tau_{r'_{k_3}}\nu_{r'_{k_2+1}}\tau_{r'_{k_2}}\nu_{r'_{k_1+1}}\tau_{r'_{k_1}}}{Q(r'_{k_3+1})Q(r'_{k_3})Q(r'_{k_2+1})Q(r'_{k_2})Q(r'_{k_1+1})Q(r'_{k_1})}
+ \dots \bigg],
\end{split}
\ee
\be
\begin{split}
F'_4(\wp')_{ii} & = \, \(\prod_{\alpha=\infty}^1 Q(r'_\alpha)\)\fQ'_{ii}\bigg[  \sum_{k,l\ge 1} \tau_{p'_k}\nu_{p'_{k+1}}\frac{\nu_{r'_{l+1}}\tau_{r'_l}}{Q(r'_l) Q(r'_{l+1})}
\\ &
+ \bigg\{ \sum_{k,l\ge 1} (\tau_{p'_k}Q(p'_{k+1})\nu_{p'_{k+2}})\frac{\nu_{r'_{l+1}}\tau_{r'_l}}{Q(r'_{l+1})Q(r'_l)}
+  (\tau_{p'_k}\nu_{p'_{k+1}})\frac{\nu_{r'_{l+2}}\tau_{r'_l}}{Q(r'_{l+2})Q(r'_{l+1})Q(r'_l)}
\\ &
+ \sum_{\substack{k_1,l\ge 1\\ k_2>k_1+1}}(\tau_{p'_{k_1}}\nu_{p'_{k_1+1}})(\tau_{p'_{k_2}}\nu_{p'_{k_2+1}}) \frac{\nu_{r'_{l+1}}\tau_{r'_l}}{Q(r'_l) Q(r'_{l+1})}
\\ &
+  \sum_{\substack{k,l_1\ge 1\\ l_2>l_1+1}}(\tau_{p'_{k_1}}\nu_{p'_{k_1+1}}) \frac{\nu_{r'_{l_2+1}}\tau_{r'_{l_2}}\nu_{r'_{l_1+1}}\tau_{r'_{l_1}}}{Q(r'_{l_2}) Q(r'_{l_2+1})Q(r'_{l_1}) Q(r'_{l_1+1})}\bigg\}
+ \dots \bigg]
\end{split}
\ee

From these (and also similar $ji$, $jj$ components), one can check the claims in the section \ref{sec:eqn-breakdown}, after factoring out $\fQ$ and $\fQ'$ and implementing various properties of $\mathbb{C}^3$
directly.

\section{Factorization of $ii/jj$ contributions in the $\CK$-wall jump of $\IC^3$}\label{app:C3-Kwall-details}

Here we provide proofs to some technical identities used in subsections \ref{sec:factoring-Q} and \ref{sec:descandant-solitons} in the study of the  $\CK$-wall jump of the exponential network of $\IC^3$.

\subsection{Multiplication rules for $\fQ, \fQ'$}\label{eq:fQ-rels}
This subsection provides a derivation of  (\ref{eq:ii-ii-comm}) and (\ref{eq:ii-jj-comm}).

Both types of  shift symmetry (\ref{eq:shift-sym}) and (\ref{eq:shift-sym-ii}) apply to the soliton data in $\CE$-walls $q_k,q'_k,\bar q_k, \bar q'_k$,  because they are $\CE$-walls of type $ii$.
Let $\fq$ be their coefficients, namely
\be
\begin{split}
	\fQ_{ii} & = \sum_\ell \fQ_{ii,\ell} = \sum_{\ell,N} \sum_{a\in \Gamma_{ii,N,N+\ell(m+n)}} \fq_{ii,\ell}(a) X_a\\
	\fQ_{jj} & = \sum_k \fQ_{jj,k} = \sum_{k,M} \sum_{b\in \Gamma_{ii,M,M+k(m+n)}} \fq_{jj,k}(b) X_b
\end{split}
\ee
the shift symmetry implies the following identities
\be\label{eq:ii-shift-sym-fQ}
\begin{split}
	\fq_{ii,\ell}\(a^{(+r)}\) & = \fq_{ii,\ell}(a), \qquad
	\fq_{jj,\ell}\(b^{(+r)}\) = \fq_{jj,\ell}(b), \qquad\\
	& \fq_{ii,\ell}(a) = \fq_{jj,\ell}(a^{(i\to j,+r)})\,.
\end{split}
\ee
Note that these symmetries never mix $\fQ_{ii,\ell}$ with $\fQ_{ii,k}$ if $\ell\neq k$.

Now let $c\in \Gamma_{ij,N,N+k}$ be a generic path. We can consider left-multiplication by $\fQ_{ii}$
\be\label{eq:fQ-ii-c}
\begin{split}
	\fQ_{ii,\ell} X_{c}
	& = \sum_{M} \sum_{a\in\Gamma_{ii,M,M+\ell(m+n)}} \fq_{ii,\ell}(a) X_a X_c
	= \sum_{a\in\Gamma_{ii,N-\ell(m+n), N}} \fq_{ii,\ell}(a) X_{ac} \,,
\end{split}
\ee
as well as right-multiplication by $\fQ_{jj}$
\be\label{eq:c-fQ-jj}
\begin{split}
	X_{c} \fQ_{jj,\ell}
	& = \sum_{M} \sum_{b\in\Gamma_{jj,M,M+\ell(m+n)}} \fq_{jj,\ell}(b) X_c X_b
	=  \sum_{b\in\Gamma_{jj,N+k,N+k+\ell(m+n)}} \fq_{jj,\ell}(b) X_{cb} \,.
\end{split}
\ee
The soliton paths appearing in the resulting generating functions are of respective types
\be
	ac \in \Gamma_{ij,N-\ell(m+n) , N+k}
	\qquad
	cb \in \Gamma_{ij,N,N+k+\ell(m+n)}\,,
\ee
therefore shift symmetry (\ref{eq:ii-shift-sym-fQ}) implies
\be\label{eq:fQ-ab-shift-map}
	\fq_{ii,\ell}(a) = \fq_{jj,\ell}(b) \qquad \text{for}\qquad  b = a^{(i \to j, +(k+\ell(m+n)))}\,.
\ee
This merely says that the terms of series (\ref{eq:fQ-ii-c}) and (\ref{eq:c-fQ-jj})  are in 1:1 correspondence. But these sums are \emph{not} equal, because the that are related to each other correspond to paths from different charge lattices.

Let us replace $X_c$ with a generating function $\Xi_{ij,k}$ which enjoys shift symmetry of type (\ref{eq:shift-sym})
\be
	\Xi_{ij,k} = \sum_N\sum_{c\in\Gamma_{ij,N,N+k}} \mu(c) X_c
\ee
then
\be\label{eq:ii-jj-comm-proto}
\begin{split}
	\fQ_{ii,\ell} \Xi_{ij,k}
	& = \sum_N
	\sum_{{\scriptsize
	\begin{array}{c}	
		a\in\Gamma_{ii,N, N+\ell(m+n)}\\
		c\in\Gamma_{ij,N+\ell(m+n), N+\ell(m+n)+k}
	\end{array}
	}}
	\fq_{ii,\ell}(a) \mu(c) X_{ac}
	\\
	\Xi_{ij,k} \fQ_{jj,\ell}
	& =  \sum_N
	\sum_{{\scriptsize
	\begin{array}{c}	
		b\in\Gamma_{jj,N+k, N+k+\ell(m+n)} \\
		c'\in\Gamma_{ij,N, N+k}
	\end{array}
	}}
	\fq_{jj,\ell}(b) \mu(c') X_{c'b}  \,.
\end{split}
\ee
Using the 1:1 map $c'=c^{(-\ell(m+n))}$ together with the map (\ref{eq:fQ-ab-shift-map})
we obtain
$X_{ac}=X_{c'b}$ and $\fq_{ii,\ell}(a) \mu(c) = \fq_{jj,\ell}(b) \mu(c') $.
Therefore the two expressions in (\ref{eq:ii-jj-comm-proto}) are immediately seen to be identical, proving (\ref{eq:ii-jj-comm}).

A similar argument prove (\ref{eq:ii-ii-comm}): consider a generic $\Xi_{ii,m}$
\be
\begin{split}
	\Xi_{ii,m} & = \sum_{M}\sum_{b\in\Gamma_{ii,M,M+m}} \mu(b) X_b \,,
\end{split}
\ee
enjoying the shift symmetry $\mu(b)  = \mu(b^{(+k)})$.
Then a simple computation yields
\be
\begin{split}
	\fQ_{ii,\ell} \Xi_{ii,m}
	& = \sum_{N}\sum_{a\in\Gamma_{ii,N,N+\ell}} \sum_{b\in\Gamma_{ii,N+\ell,N+\ell+m}} \fq_{ii,\ell}(a) \mu(b) X_{a b} \\
	& \\
	\Xi_{ii,m} \fQ_{ii,\ell}
	& = \sum_{N} \sum_{b'\in\Gamma_{ii,N-m,N}} \sum_{a'\in\Gamma_{ii,N,N+\ell}}  \fq_{ii,\ell}(a') \mu(b') X_{b' a'}\,.
\end{split}
\ee
Now shift symmetry says that there is a 1:1 map between soliton paths $b'\in \Gamma_{ii,N-m,N}$ and $b \in \Gamma_{ii,N+\ell,N+\ell+m}$ denoted by ${b'}{}^{(+(\ell+m))} = b$, and that this map is a symmetry of the soliton data $\mu({b'}{}^{(+(\ell+m))} ) = \mu(b)$. This proves (\ref{eq:ii-ii-comm}).

\subsection{Proof of $\fQ_{ii} = \fQ_{ii}'$}\label{eq:fQ-equality}

Here we provide a proof of (\ref{eq:Qii-Qiip}), the argument relies on two key facts: the periodic identification of $q,\bar q$ with $\bar q',q'$ and the shift symmetry property of their soliton generating functions.

Let us sketch the rough idea for the argument, a more detailed proof will follow.
Shift symmetry of type (\ref{eq:shift-sym-ii}) relates $\tau_{q_k} \leftrightarrow \tau_{\bar q_k}$ as well as $\tau_{q'_k} \leftrightarrow \tau_{\bar q'_k}$.
The global topology of the critical network, shown in Figure \ref{fig:C3-Kwall}, implies that $q_k,q'_k$ walls are periodically identified.
Therefore the relations between $\tau$-type soliton generating functions get translated into relations between $\nu_{q_k} \leftrightarrow \nu_{\bar q_k}$ and $\nu_{q'_k} \leftrightarrow \nu_{\bar q'_k}$.
Combining all of these together, implies that the coefficients of the following generating functions are mapped into each other
\be
	\fQ_{ii} \leftrightarrow \fQ_{jj},, \qquad
	\fQ'_{ii} \leftrightarrow \fQ'_{jj}\,.
\ee
Periodic identification of streets finally establishes the desired relation between $\fQ_{ii}\leftrightarrow \fQ'_{ii}$ and $\fQ_{jj}\leftrightarrow \fQ'_{jj}$.

In table (\ref{eq:C3-monodromies}) we summarized the monodromy of sheets around the puncture at infinity for the covering $\tSigma$. Written in terms of (\ref{eq:label-change}) it reads
\be
	CW: \qquad
	(i,N)\to (j, N-m)\,,
	\quad
	(j,N)\to (i,N-n)\,
\ee
\be
	CCW: \qquad
	(i,N)\to (j, N+n)\,,
	\quad
	(j,N)\to (i,N+m)\,.
\ee
Going back to Figure \ref{fig:ij-ji-2-way}, let us consider the generating function of solitons $\tau_{q_k}$: these are solitons on wall $q_k$ supported at some $x\in C$ close to the junction, and coming out of it.
The pairs of walls $q_k,\bar q_k$ winds around the puncture at infinity and feeds back into walls $q'_k, \bar q'_k$, which attach to the joint from below.
This implies that the solitons carried by $\tau_{q_k},\tau_{\bar q_k}$ (coming out from above the joint) must determine those carried by $\nu_{q'_k}, \nu_{\bar q'_k}$ (going into the joint from below).

Let's consider a soliton of $\tau_{q_k}$
\be
	a \in \Gamma_{ii,N,N-k(m+n)}
\ee
whose endpoints lie just above the junction, on sheets $(i,N)$ and $(i,N-k(m+n))$.
Transporting this soliton path's endpoints \emph{ccw} along $q_k$ around the puncture involves crossing both the logarithmic cut and the square-root cut.
The monodromy computed above implies that $a$ becomes a soliton of type
\be
	\Gamma_{jj, N+n , N-m-(k-1)(m+n)}
\ee
therefore $a$ actually becomes a contribution for $\nu_{\bar q'_k}$ (as opposed to $\nu_{q'_k}$).
In fact, when $a$ is transported around the puncture, it gets concatenated with two ``extension'' paths attached to its endpoints
\be
	a \quad \to\quad
	(\gamma_{\infty}^{ij,N\to N+n})^{-1} \cdot a \cdot \gamma_{\infty}^{ij,N-k(m+n) \to N-m-(k-1)(m+n)} \,.
\ee
Here $\gamma_{\infty}^{ij,N_L,N_R}$ is a path winding \emph{ccw} around infinity starting from above the junction on sheet $(i,N_L)$ and ending back above the junction on sheet $(j,N_R)$.
This concatenation is exactly the definition of the shift map (\ref{eq:shift-map-ii}). Therefore the above equation can be written more simply as
\be
	a \quad \to\quad a^{(i\to j, +n)} \,.
\ee
These are thus the contributions to $\nu_{\bar q'_k}$:
\be
\begin{split}
	\nu_{\bar q'_k} & = \sum_N \sum_{a\in \Gamma_{ii,N,N-k(m+n)}} \mu(a) X_{(\gamma_{\infty}^{ij,N\to N+n})^{-1} \cdot a \cdot \gamma_{\infty}^{ij,N-k(m+n) \to N-m-(k-1)(m+n)}} \\
	& = \sum_N \sum_{a^{(i\to j, +n)} \in \Gamma_{jj,N+n,N-m-(k-1)(m+n)}} \mu(a^{(i\to j, +n)} )  X_{a^{(i\to j, +n)} } \\
	& \mathop{=}^{(\ref{eq:shift-sym-ii})} \tau_{\bar q_k} \,.
\end{split}
\ee
where in the last step we used the relation between soliton data of out-going $ii$ and $jj$ walls derived in the context of the one-way $ij-ji$ joint in Section \ref{sec:ij-ji-joints}.
In a similar fashion we derive
\be
	\nu_{q'_k} = \tau_{q_k} \qquad
	\nu_{q_k} = \tau_{q'_k} \qquad
	\nu_{\bar q_k} = \tau_{\bar q_k}  \,.
\ee
Taken together with the definitions (\ref{eq:def-fQ}) these imply (\ref{eq:Qii-Qiip}).

\subsection{Transport of soliton generating functions around the puncture at infinity}\label{app:soliton-transport}

Let's analyze the transport involved in (\ref{eq:transport-nu-rp-k}) in detail.
$\nu_{r'_{k+1}}$ carries solitons of type $(ij,-m-k(m+n))$, the prototype being a path starting on sheet $(i, N)$ and ending on $(j, N-m-k(m+n))$.
In the first equality we simply move the endpoints across the junction, by an infinitesimal amount.
In the second equality instead we transport the endpoints along $r_{k+1}\sim p'_{k+1}$ in the \emph{ccw} direction.
This involves attaching two paths running around infinity:
the first one $\alpha_1$ starts on sheet $(j,N+n)$ (starting point for a soliton in $\nu_{p'_{k+1}}$), runs \emph{cw} around the puncture, and ends on sheet $(i,N)$ (starting point of a soliton in $\tau_{r_{k+1}}$);
the second one $\alpha_2$ starts on sheet $(j,N-m-k(m+n))$ (endpoint of a soliton in $\tau_{r_{k+1}}$), runs \emph{ccw} around the puncture, and ends on sheet $(i, N-k(m+n))$ (endpoint of a soliton in $\nu_{p'_{k+1}}$).

Let's also analyze the transport involved in (\ref{eq:transport-tau-rp-k}) in detail.
$\tau_{r'_{k}}$ carries solitons of type $(ji,m+(k-1)(m+n))$, the prototype being a path starting on sheet $(j, N)$ and ending on $(i, N+m+(k-1)(m+n))$.
In the first equality we simply move the endpoints across the junction, by an infinitesimal amount.
In the second equality instead we transport the endpoints along $r_{k+1}\sim p'_{k+1}$ in the \emph{ccw} direction.
This involves attaching two paths running around infinity:
the first one $\beta_1$ starts on sheet $(i,N+m)$ (starting point for a soliton in $\tau_{p'_{k}}$), runs \emph{cw} around the puncture, and ends on sheet $(j,N)$ (starting point of a soliton in $\nu_{r_{k}}$);
the second one $\beta_2$ starts on sheet $(i,N+m+(k-1)(m+n))$ (endpoint of a soliton in $\nu_{r_{k}}$), runs \emph{ccw} around the puncture, and ends on sheet $(j, N+k(m+n))$ (endpoint of a soliton in $\tau_{p'_{k}}$).

Now to compute the effect of these transports of soliton endpoints on (\ref{eq:suspect-equality}) we wish to compute their contribution to the central charge.
It helps to consider a shifted version of the prototype paths $\alpha_i,\beta_i$, to properly take into account canceling of different contributions.
Shifting a path changes its central charge as
\be
\begin{split}
	Z_{\alpha_1^{(+k)}} & = Z_{\alpha_1} + \frac{1}{2\pi R} \int_{\alpha_1} (2 \pi i \, k) dx/x
	= Z_{\alpha_1} - k Z_{\gamma} \,,
	\\
	Z_{\alpha_2^{(+k)}} & = Z_{\alpha_1} + \frac{1}{2\pi R} \int_{\alpha_2} (2 \pi i \, k) dx/x
	=  Z_{\alpha_2} + k Z_{\gamma} \,,
\end{split}
\ee
where the integral is really taken along the projection $\pi\circ\tpi(\alpha_i)$, and relative sign difference is due to opposite orientations. The same holds for $\beta_1,\beta_2$.
Then notice
\be
\begin{split}
	Z_{\alpha_1^{(-n)}} + Z_{\beta_2^{(-n-(k-1)(m+n))}} & =
	\oint_{cw} \lambda_{(j,N)\to(i,N+m-n)}
	+
	\oint_{ccw} \lambda_{(i,N+m-n)\to (j,N)}
	\\
	& =
	\(\oint_{cw} - \oint_{cw}\) \lambda_{(j,N)\to(i,N+m-n)}
	\\
	& =
	0
\end{split}
\ee
\be
\begin{split}
	Z_{\alpha_2^{(+k(m+n))}} + Z_{\beta_1} & =
	\oint_{ccw} \lambda_{(j,N)\to(i,N)}
	+
	\oint_{cw} \lambda_{(i,N)\to (j,N)}
	\\
	& =
	\(\oint_{ccw} - \oint_{ccw}\) \lambda_{(j,N)\to(i,N)}
	\\
	& =
	0 \,.
\end{split}
\ee
Therefore
\be
\begin{split}
	Z_{\alpha_1} + Z_{\alpha_2} + Z_{\beta_1} + Z_{\beta_2}
	& =
	\( Z_{\alpha_1}^{(-n)} +\(- n\) Z_{\gamma} \)
	+
	\( Z_{\alpha_2}^{(k(m+n))} - \(k(m+n)\) Z_{\gamma} \)
	\\
	&
	+
	Z_{\beta_1}
	+
	\( Z_{\beta_2}^{(-n-(k-1)(m+n))} - \(-n-(k-1)(m+n)\) Z_{\gamma} \)
	\\
	& = -(m+n) Z_{\gamma} \\
	& = Z_{\gamma} \,.
\end{split}
\ee
Recall that $Z_\gamma$ is the period of an infinite number of homology cycles on $\tSigma$, denoted by $\gamma_N$ in (\ref{eq:Z-quotient}). The generating functions in (\ref{eq:suspect-equality}) in fact contain an infinite tower of solitons labeled by $N\in\IZ$, and each of these gets transported separately by the addition of $\gamma_N$.
We deduce that the correct form of (\ref{eq:suspect-equality}) is actually
\be
	\frac{\nu_{r'_{k+1}}\tau_{r'_k}}{Q_{k}Q_{k+1}}
	\,{=}\,
	- \(\sum_N X_{\gamma_N}\) \tau_{p'_{k}}\nu_{p'_{k+1}} \,.
\ee
The sign has been added to account for the twisting of the flat connection (cf. discussion in subsection \ref{sec:nonabel-map}).

\bibliography{biblio}{}
\bibliographystyle{JHEP}

\end{document}